\documentclass[a4paper,11pt]{article}
\usepackage{jcappub} 
\usepackage{aas_macros}
\usepackage{amssymb}
\usepackage{amsmath}
\usepackage{graphicx}
\usepackage{dcolumn}
\usepackage[numbers]{natbib}
\usepackage{hyperref}
\usepackage{color}
\usepackage{ulem}
\usepackage{etoolbox}
\usepackage{orcidlink}
\definecolor{bazaar}{rgb}{0.6, 0.47, 0.48}
\definecolor{auburn}{rgb}{0.43, 0.21, 0.1}
\definecolor{antiquefuchia}{rgb}{0.57, 0.36, 0.51}
\definecolor{ao}{rgb}{0.0, 0.5, 0.0}
\definecolor{blush}{rgb}{0.87, 0.36, 0.51}
\definecolor{dp}{rgb}{0.5, 0, 0.5}

\newcommand{\new}[1]{{\textcolor{black} {#1}}}
\newcommand{\newer}[1]{{\textcolor{black} {#1}}}
\newcommand{\llrw}[1]{{\textcolor{black} {#1}}}

\title{Addressing the core-cusp and diversity problem of dwarf and disk galaxies using cold collisionless DARKexp theory}

\author[a\orcidlink{0000-0002-6039-8706}]{Liliya L. R. Williams,}
\author[b\orcidlink{0000-0002-4571-2306}]{Jens Hjorth,}
\author[a\orcidlink{0000-0003-0605-8732}]{Evan D. Skillman}

\affiliation[a]{School of Physics and Astronomy, University of Minnesota, 116 Church Street SE, Minneapolis, MN 55455, USA}
\affiliation[b]{DARK, Niels Bohr Institute, University of Copenhagen, Jagtvej 155A, 2200 Copenhagen, Denmark}

\emailAdd{llrw@umn.edu}
\emailAdd{jens@nbi.ku.dk}
\emailAdd{skill001@umn.edu}

\abstract{Dwarf galaxies are observed to have linearly rising rotation curves, which indicate constant density cores in their centers. These do not arise naturally from dark matter only simulations, giving rise to the ``core-cusp'' problem.  Hydrodynamic simulations incorporating baryonic feedback can create cored profiles, but this solution requires some fine tuning.  Additionally, there is a related ``diversity'' problem in which simulations cannot reproduce the large range in rotation curve shapes for galaxies of similar mass. Here we investigate, \newer{for the first time,} whether a theoretical model based on statistical mechanics (DARKexp) can produce flat density cores with its single shape parameter.  We find that these theoretical profiles are able to fit the observed rotation curves of galaxies with last measured velocities in the range $\sim 20$--200 km s$^{-1}$ by producing fits to 96 SPARC catalog  galaxies covering this range.  We additionally show that the projected stellar density distributions of ultrafaint dwarfs that show cores are also well fit, \newer{and that wave, or fuzzy dark matter is unable to fit the full range of SPARC galaxies.} Thus, DARKexp appears to be able to address the core-cusp problem and the diversity of rotation curves in these dark matter dominated galaxies with cold collisionless dark matter alone, i.e., without introducing baryonic feedback, \newer{or alternative nature of dark matter}.}


\begin{document}
\maketitle
\flushbottom

\section{Introduction}

It has been nearly a century since the existence of dark matter was established through astrophysical observations and analysis \citep{Zwicky1933,Rubin1970,Ostriker1974,Einasto1974}. However its particle nature still remains unknown. Lack of laboratory detections of the simplest form of dark matter particles, WIMPs, and astrophysical problems on small, $\lesssim\,$kpc, spatial scales, have led to a wider range of candidates being proposed: self-interacting, warm, and very light mass particles \cite[e.g.,][]{Spergel2000,Hu2000,Bullock2017,Hui2021}, or modified gravity \citep[e.g.,][]{Famaey2012,Milgrom2014}.

Because astrophysics is the only setting where dark matter has been definitively detected owing to its gravitational effects, testing dark matter properties also falls on astrophysics, at least for now. Microscopic properties of dark matter particles---mass and interaction cross-section---determine the macroscopic structure of dark-matter halos. Galaxies and clusters of galaxies are great dark matter testbeds, especially those that are dominated by dark matter, so that the effects of baryons, which are subject to forces other than gravity, can be neglected. 

Dwarf galaxies are believed to be dark matter dominated, and are observed to have linearly, or approximately linearly rising rotation curves, implying that their density profiles have flat or nearly flat central density distributions \citep{Schaller2015,Oman2015,Lelli2016}. Rotation curves are the preferred way to get at the radial dark matter distribution, because they are based on dynamics of the combined mass, baryonic as well as dark.  For ultrafaint dwarfs (UFDs) observations can only deliver surface stellar density profiles, which are flat in the central regions. UFDs are believed to be strongly dark matter dominated at all radii, down to the very center. It is quite likely that the stars trace the dark matter distribution.

High-resolution N-body simulations of dark-matter halos make a prediction as to the radial profile of relaxed self-gravitating systems if dark matter is cold and collisionless: the central density slope, $\rho\propto r^\gamma$, has $\gamma=-1$ \citep{Dubinski1991,Navarro1997}, referred to as cuspy NFW profiles. Even though some dispersion in the central density slope has been observed in simulated halos and sub-halos \citep{Navarro2010,Ludlow2013,Vera-Ciro2013,Hjorth2015}, flat density profiles, also known as cores, are not seen. This is in stark contrast to observations of dwarfs, where \llrw{cores are common} \citep[e.g.,][]{Oh2015,Carlsten2021,Montes2024}. 

\llrw{Inclusion of bursty star formation and stellar feedback allows some galaxies to be matched by simulations \citep[e.g.,][]{DiCintio2013,Oman2015,Read2016,Collins2021,Collins2022,Bouche2022}, and possibly solve the Too Big to Fail problem \citep{Onorbe2015}. However, only galaxies with a limited range of stellar-to-halo ratio, around $M_*/M_{\rm halo}\sim 5\times 10^{-3}$, achieve cores \citep{Azartash2024}. It also appears that multiple episodes of star formation may be required to create cores, more than observations indicate \citep{Mostow2024}. A further problem for simulations is to reproduce the diversity of rotation curve shapes, observed for galaxies with similar maximum rotational velocity \citep{Oman2015,Sales2022}.}

The apparent \llrw{inconsistency} between observations vs.\ N-body simulations, and the inability of baryonic feedback within hydrodynamic simulations to solve the problem \llrw{in a consistent way for a wide range of disk galaxies} is one of the prime reasons for considering dark matter properties other than cold and/or collisionless: self-interacting, warm, or wave \citep{Bullock2017,Boldrini2021,Straight2025}, or even modified gravity theories.

\begin{figure*}
    \centering
    \includegraphics[trim={0cm 5cm 0cm 2cm},clip,width=0.47\textwidth]{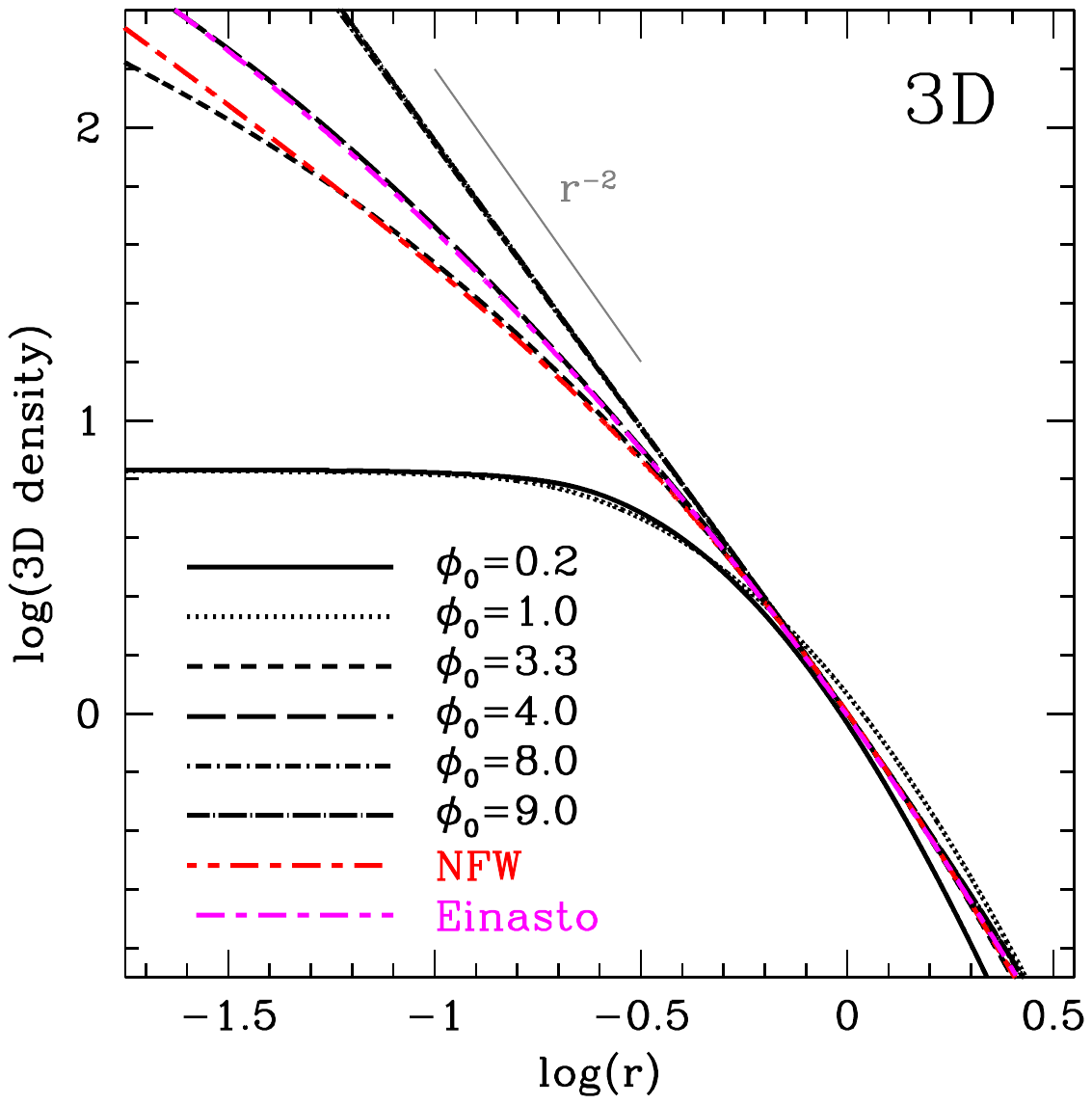}
    \includegraphics[trim={0cm 5cm 0cm 2cm},clip,width=0.47\textwidth]{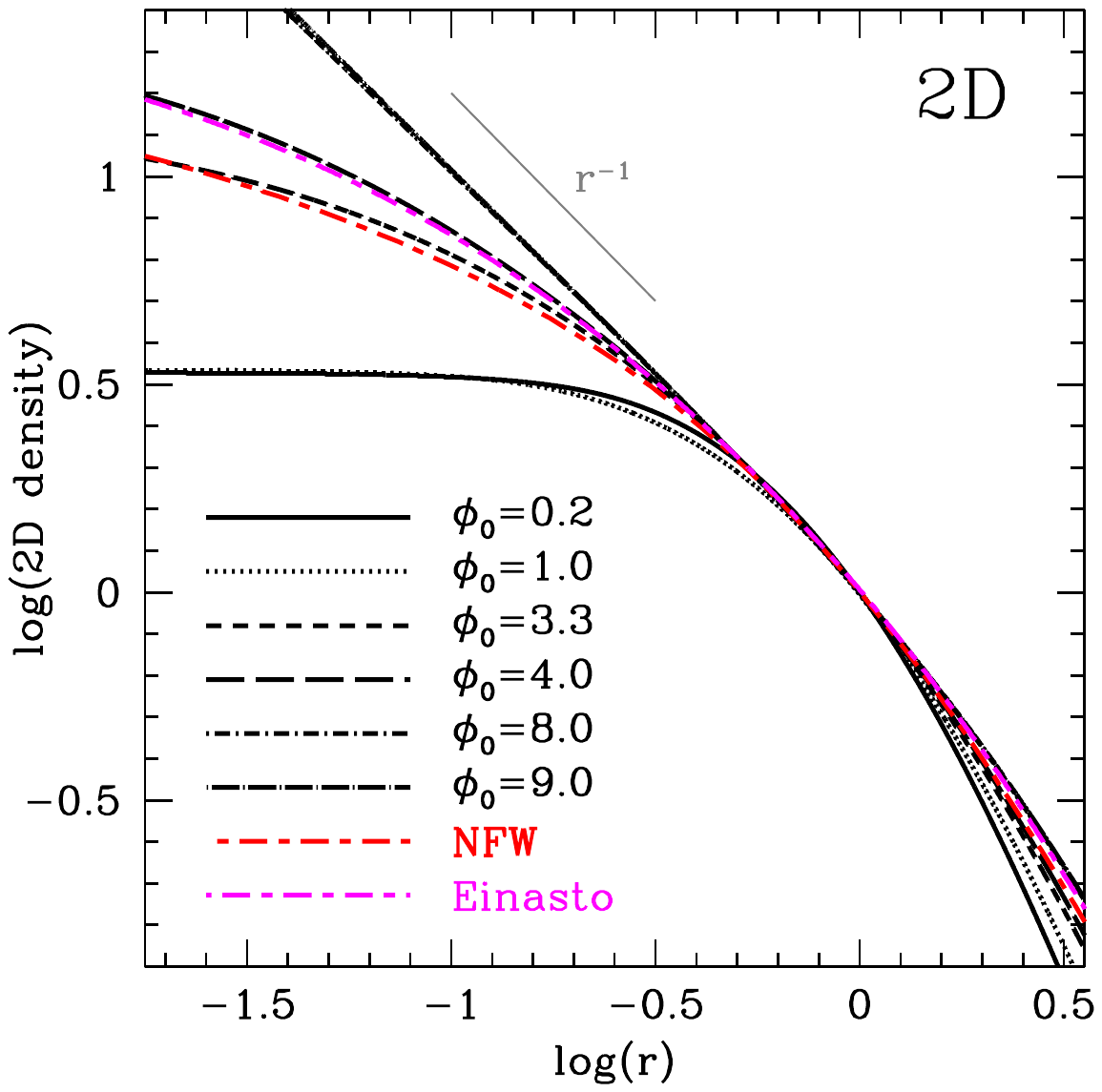}
    \caption{Shapes of 3D ({\it left}), and sky-projected 2D ({\it right}) density profiles of DARKexp halos of different $\phi_0$. DARKexp of 
    $\phi_0=0.2$ and $1.0$ were rescaled in radius and density so that their central profiles coincide, to illustrate their similarity. With this rescaling it is apparent that these models are very similar except for somewhat different outer slopes. DARKexp profiles of high $\phi_0\gtrsim 8$ behave similarly to each other, and have central 3D slope that scale as $r^{-2}$ (gray line segment), resulting in flat rotation curves. DARKexp $\phi_0\approx 3.2-3.3$ are well approximated by NFW (red) in this radial range. DARKexp $\phi_0\approx 4$ is well approximated by Einasto profile of index $0.17$ (magenta).}
    \label{fig:phi0comp}
\end{figure*}

In this paper we revisit the issue of the consistency of the galaxies' observed properties with cold collisionless dark matter by comparing rotation curves of a range of galaxies and projected stellar density profiles of UFD, described in Section~\ref{sec:data}, to the theoretical predictions. 
Our dark matter model, DARKexp, is described in Section~\ref{sec:theory}; we emphasize that our model is not based on simulations of any kind, but solely on analytical theory. \newer{We also demonstrate, in Appendix~\ref{sec:FDM}, that simply producing cored profiles is not necessarily sufficient to fit the rotation curves of low-mass galaxies by examining the predictions from fuzzy/wave dark matter theory.} \llrw{Section~\ref{sec:fits} presents the fits to 96 galaxy rotation curves, and 6 surface brightness profiles of UFD. Section~\ref{sec:sum} summaries our findings. Appendix~\ref{sec:MW} presents the fits to the Milky Way galaxy.}

\section{Data}\label{sec:data}

We use two types of data, rotation curves of nearly a 100 dwarf and regular disk galaxies, and sky-projected stellar number density data from six Ultra Faint Dwarfs.

The rotation curves are taken from the SPARC ({\it Spitzer} Photometry and Accurate Rotation Curves) database of 175 late-type galaxies (spirals and irregulars) with {\it Spitzer} photometry at 3.6 $\mu$m, which traces the stellar mass distribution, and high-quality {H}{I}+H$\alpha$ data \citep{Lelli2016,Lelli2016a}. {H}{I} gas is a very good tracer of the rotation curves as it is dynamically cold, has nearly circular orbits and extends beyond the stellar distribution.  SPARC galaxies span a broad range in luminosity, surface brightness, rotation velocity, and Hubble type, and according to the authors of the database, the SPARC catalog forms a representative sample of disk galaxies in the nearby universe.

From the 175 galaxies in the full catalog, we exclude those whose last recorded rotational velocity $v_{\rm last}\gtrsim 200$ km s$^{-1}$, as these massive galaxies tend to be baryon dominated near the center. Furthermore, we select only galaxies whose rotation curves span at least a decade in radius; shorter ranges do not provide sufficient modeling constraints. We make an exception for the Milky Way (which is not a part of the SPARC catalog); see Appendix~\ref{sec:MW}. We also fit the surface density profiles of six Local Group UFDs near M31, analyzed by \cite{Richstein2024} that appear to have flat density cores in their stellar distribution. 

\begin{figure*}[h!]
    \centering
    \vskip-1.75cm
    \includegraphics[width=0.237\linewidth]{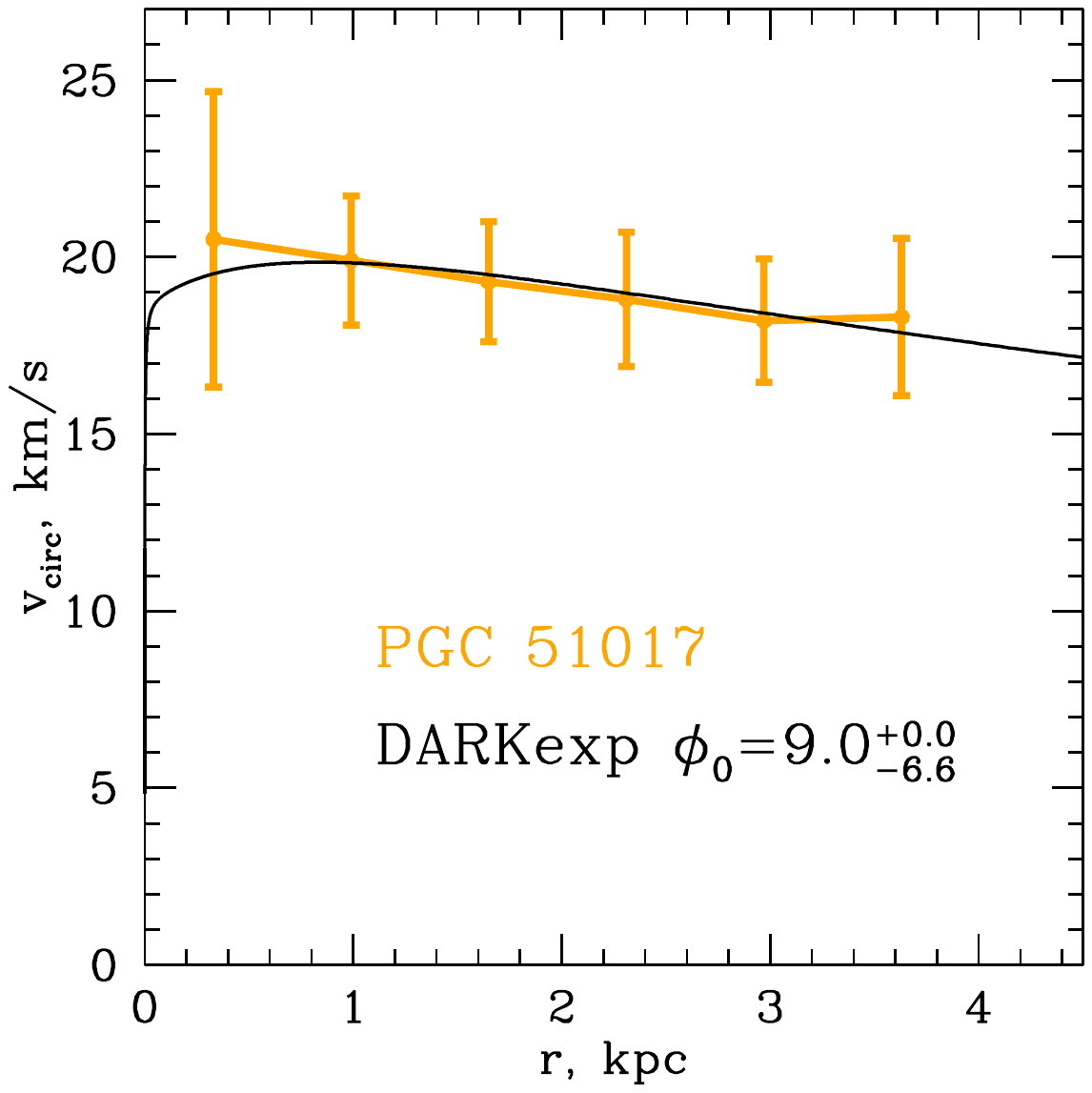}
    \includegraphics[width=0.237\linewidth]{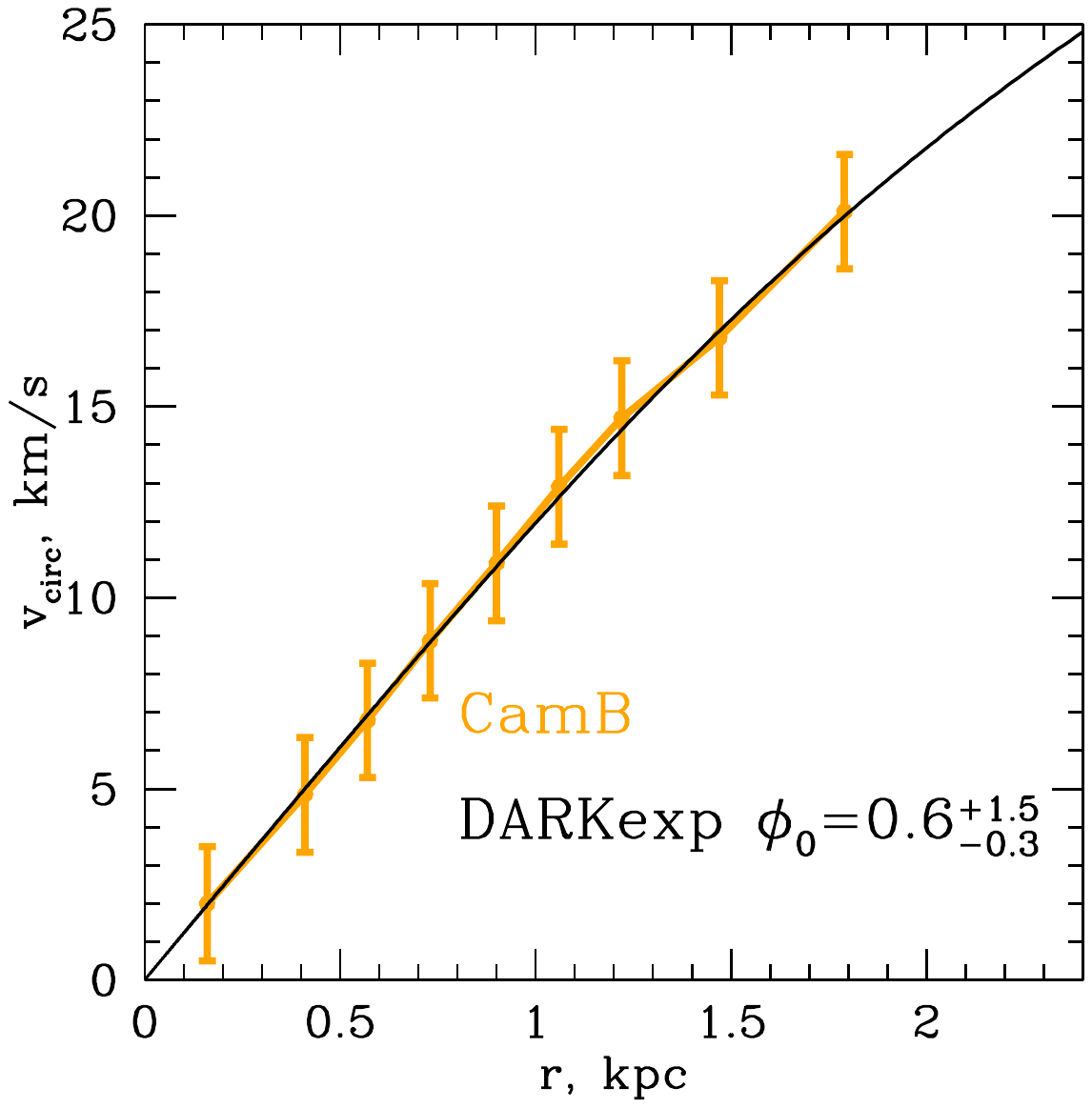}
    \includegraphics[width=0.237\linewidth]{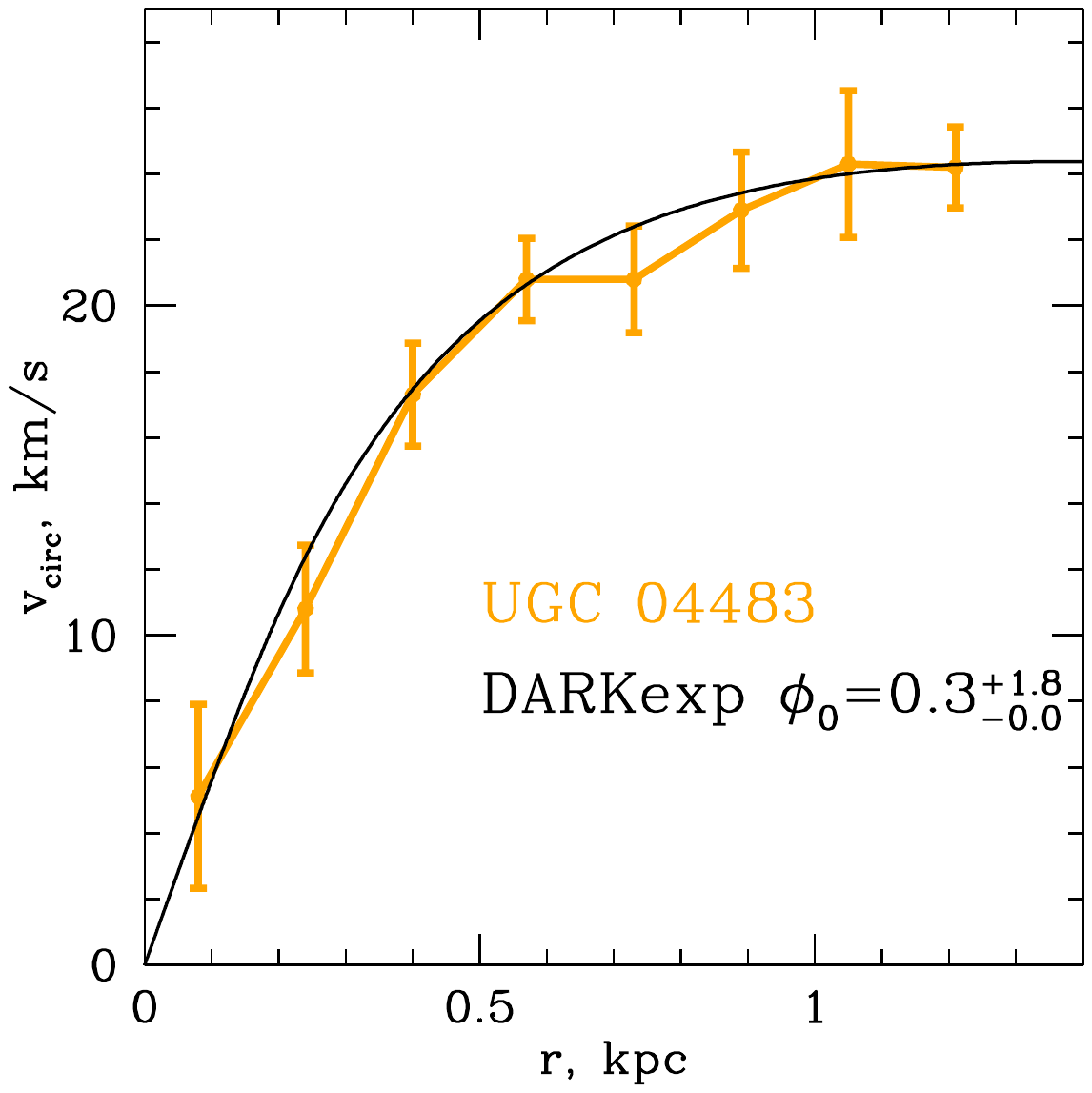}    \includegraphics[width=0.237\linewidth]{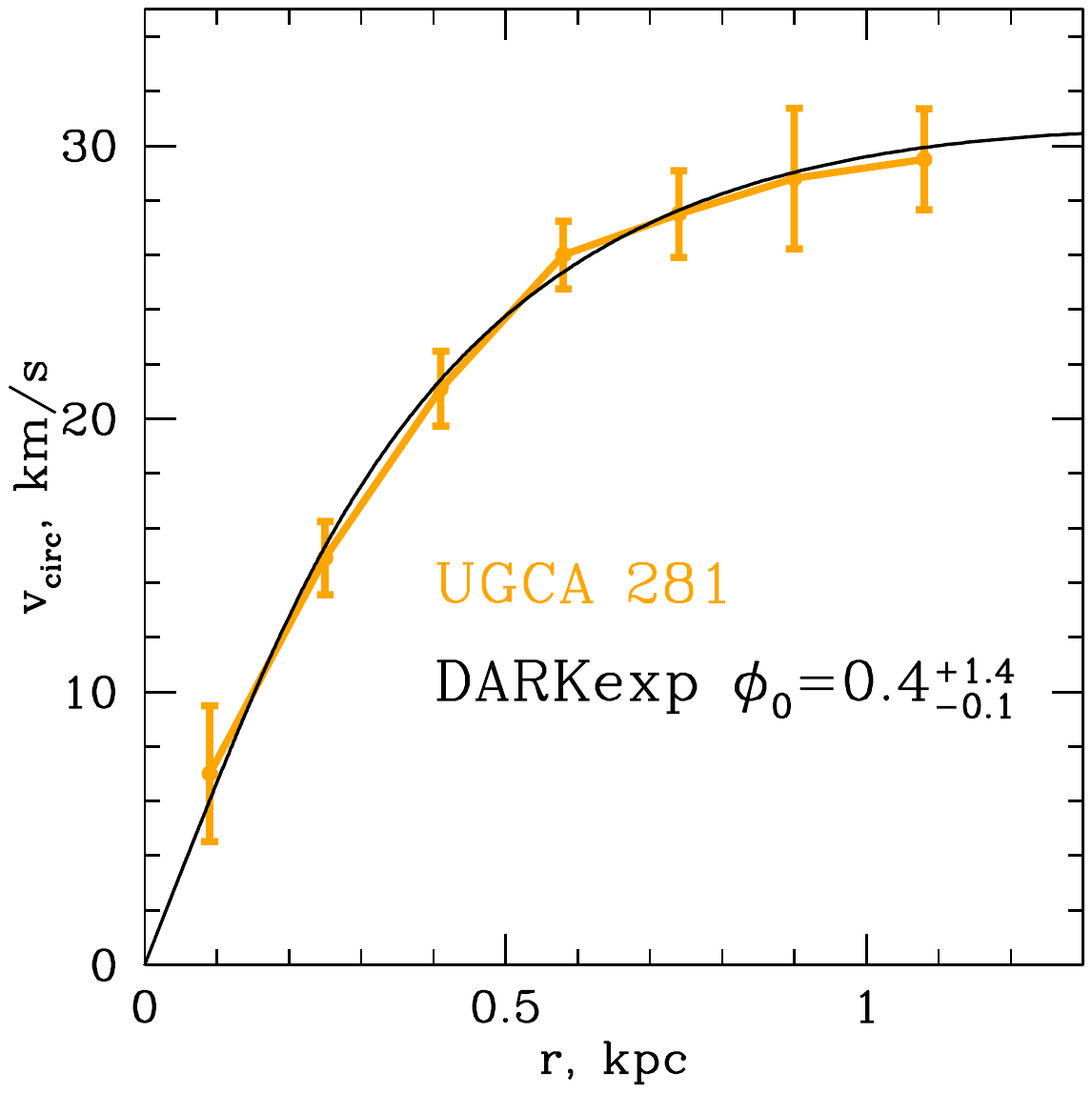}
            \vskip-1.65cm
    \includegraphics[width=0.237\linewidth]{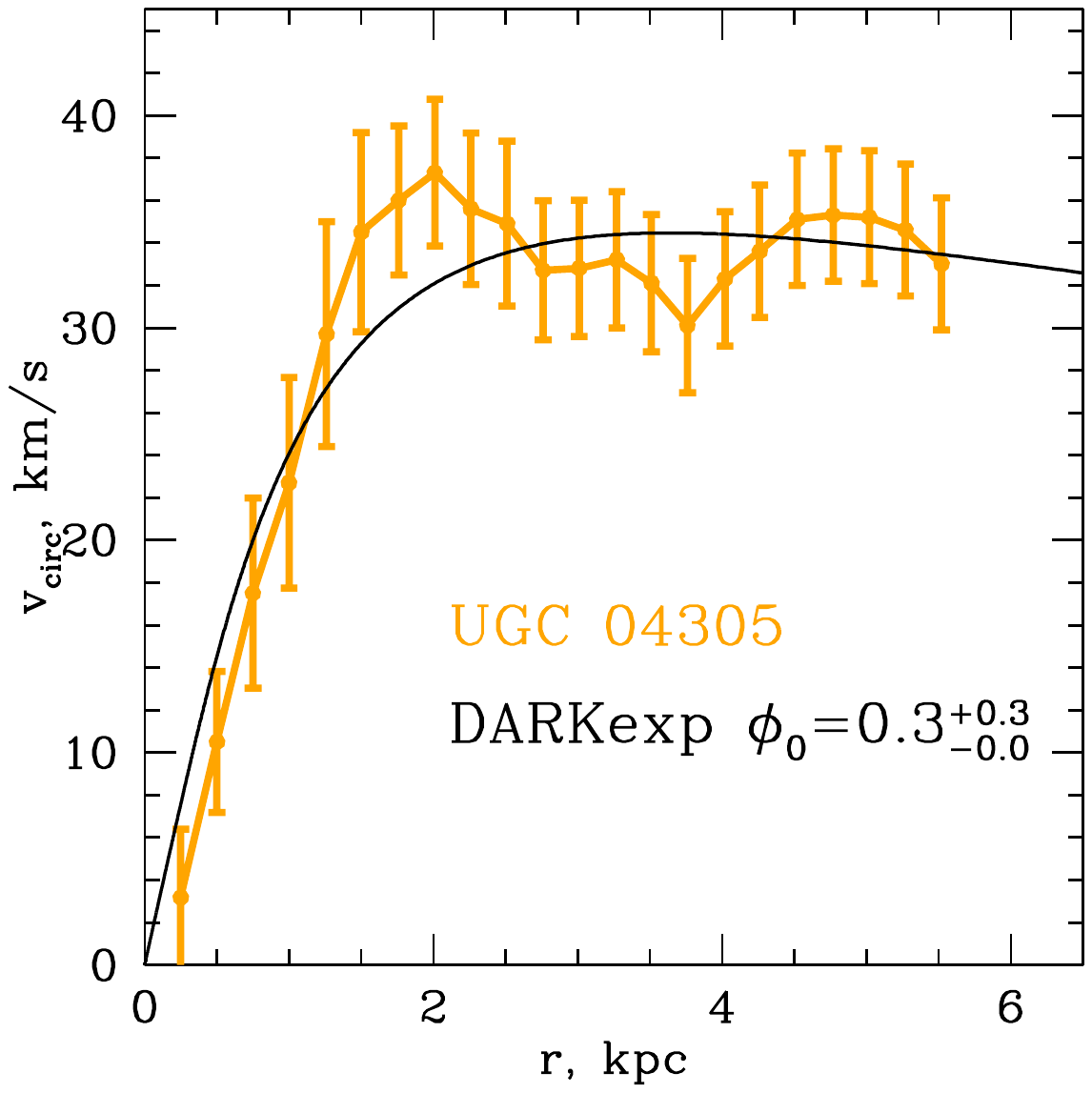}
    \includegraphics[width=0.237\linewidth]{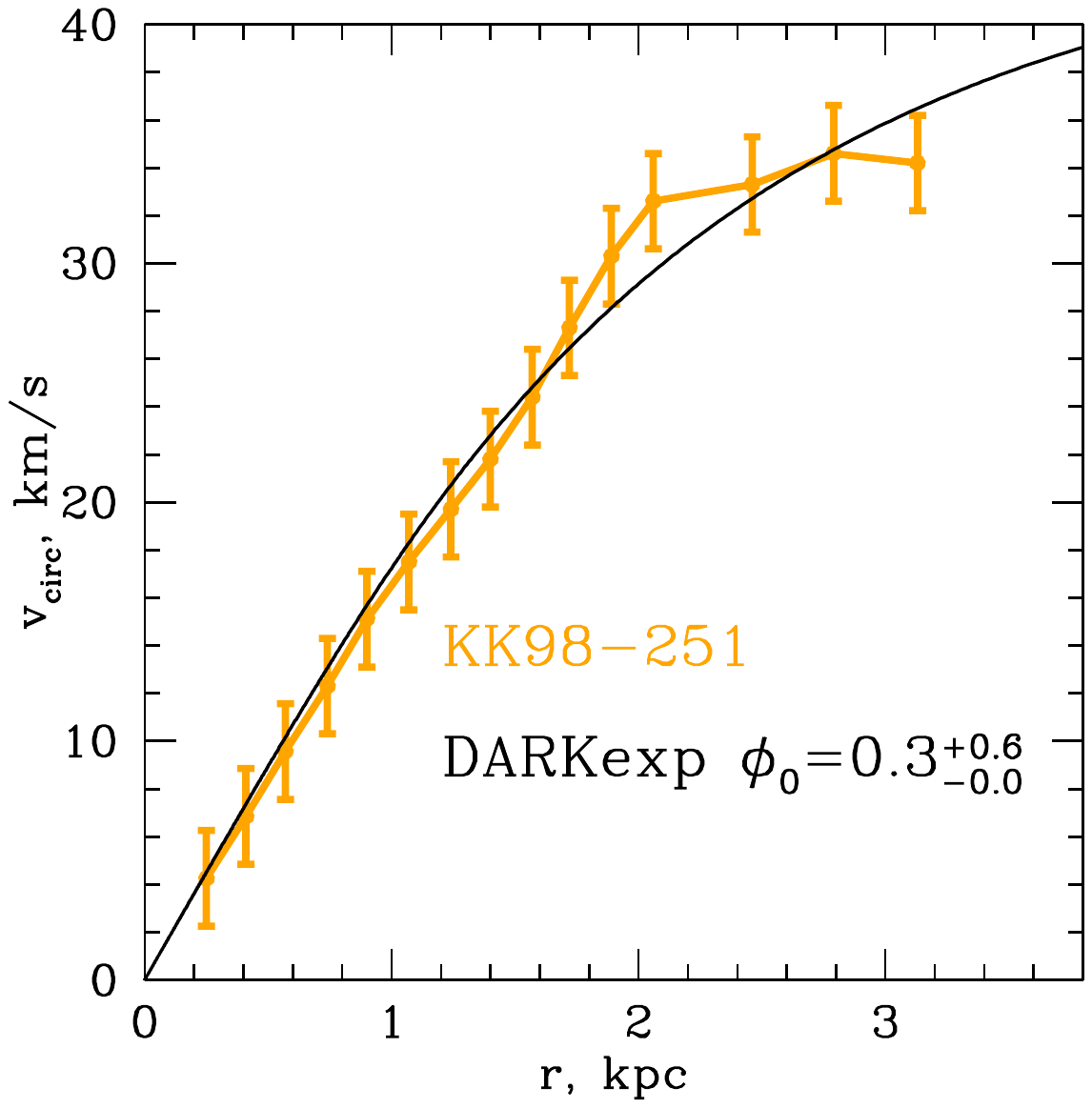}
    \includegraphics[width=0.237\linewidth]{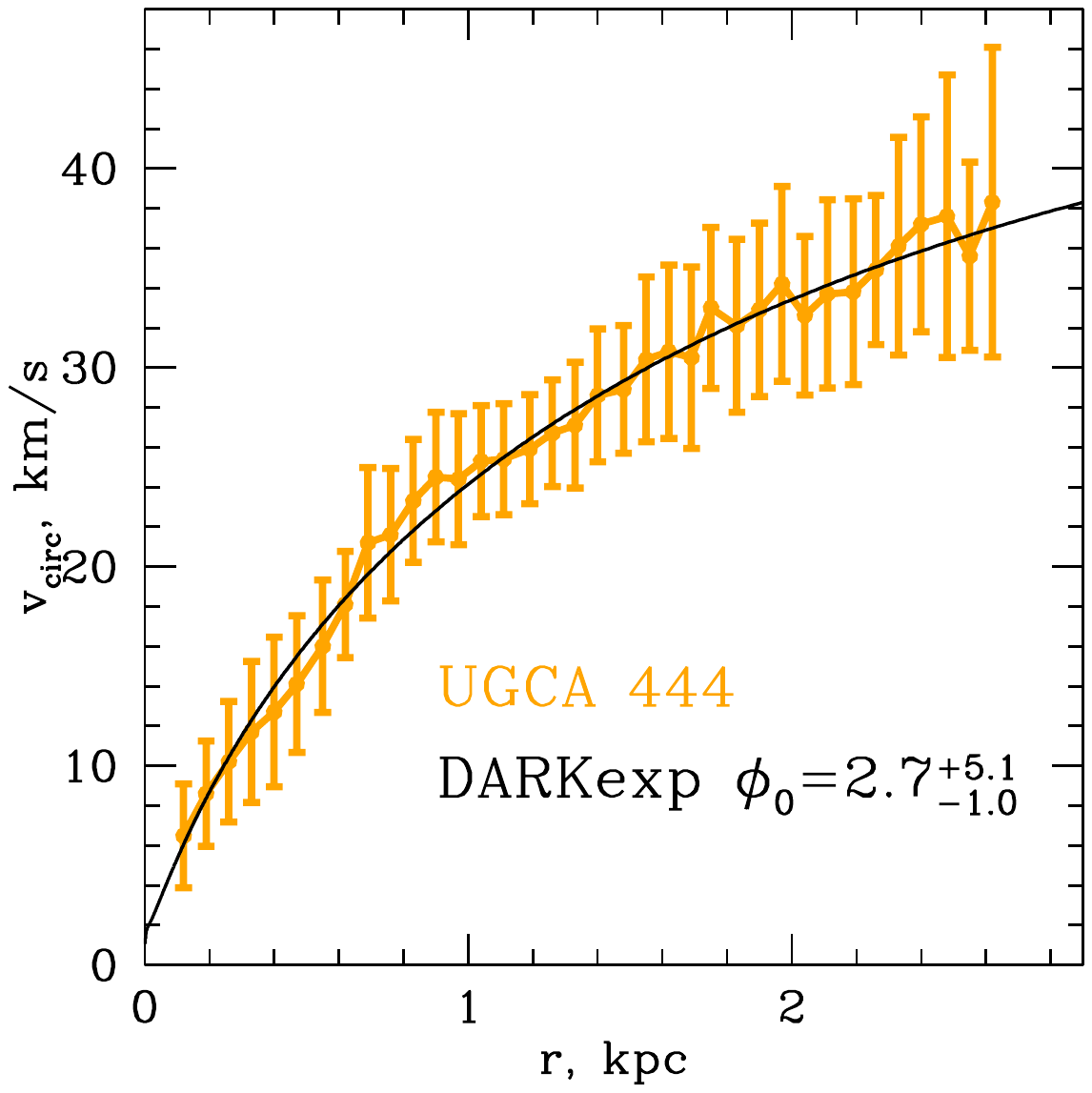}
    \includegraphics[width=0.237\linewidth]{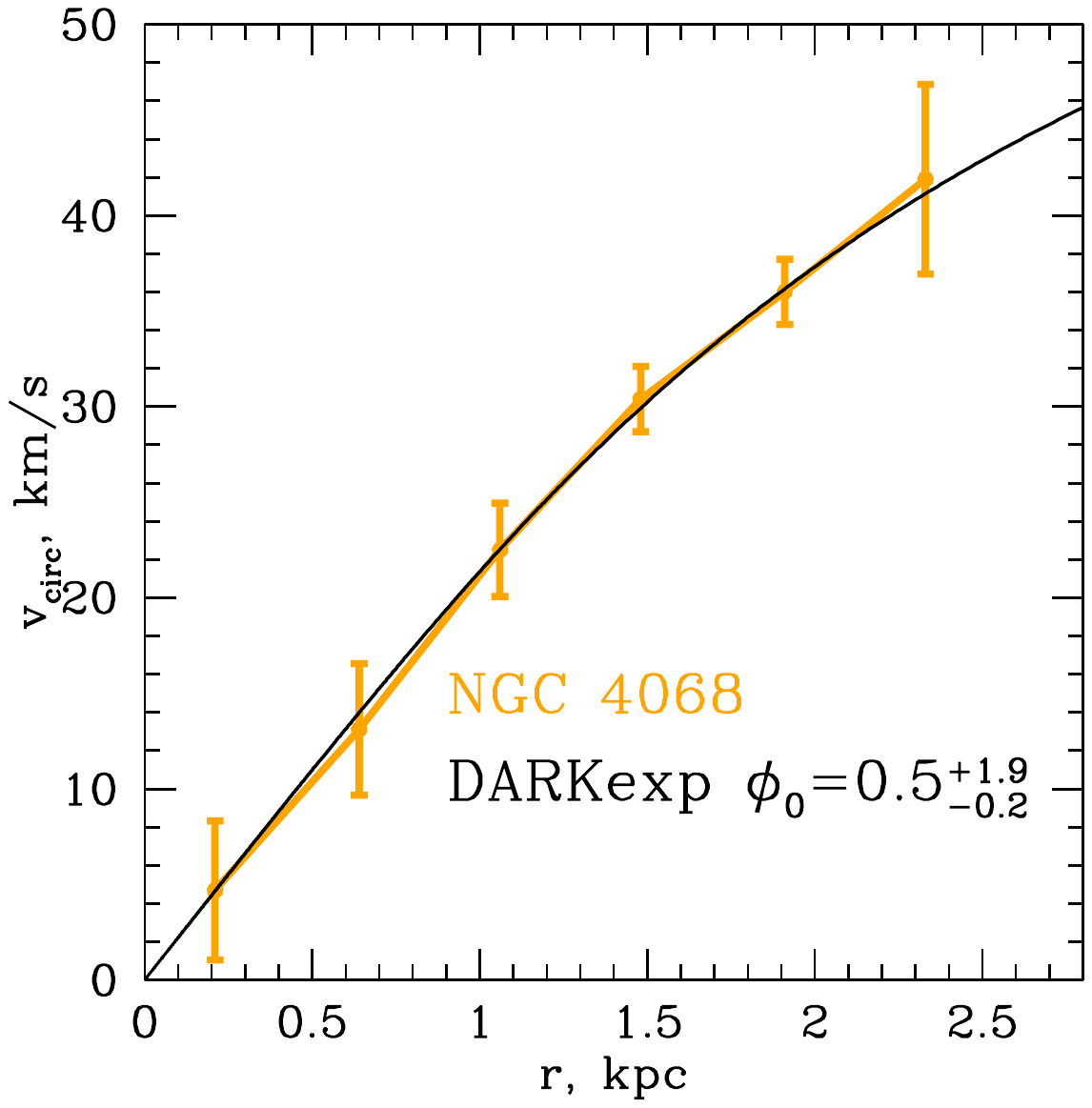}
         \vskip-1.65cm
    \includegraphics[width=0.237\linewidth]{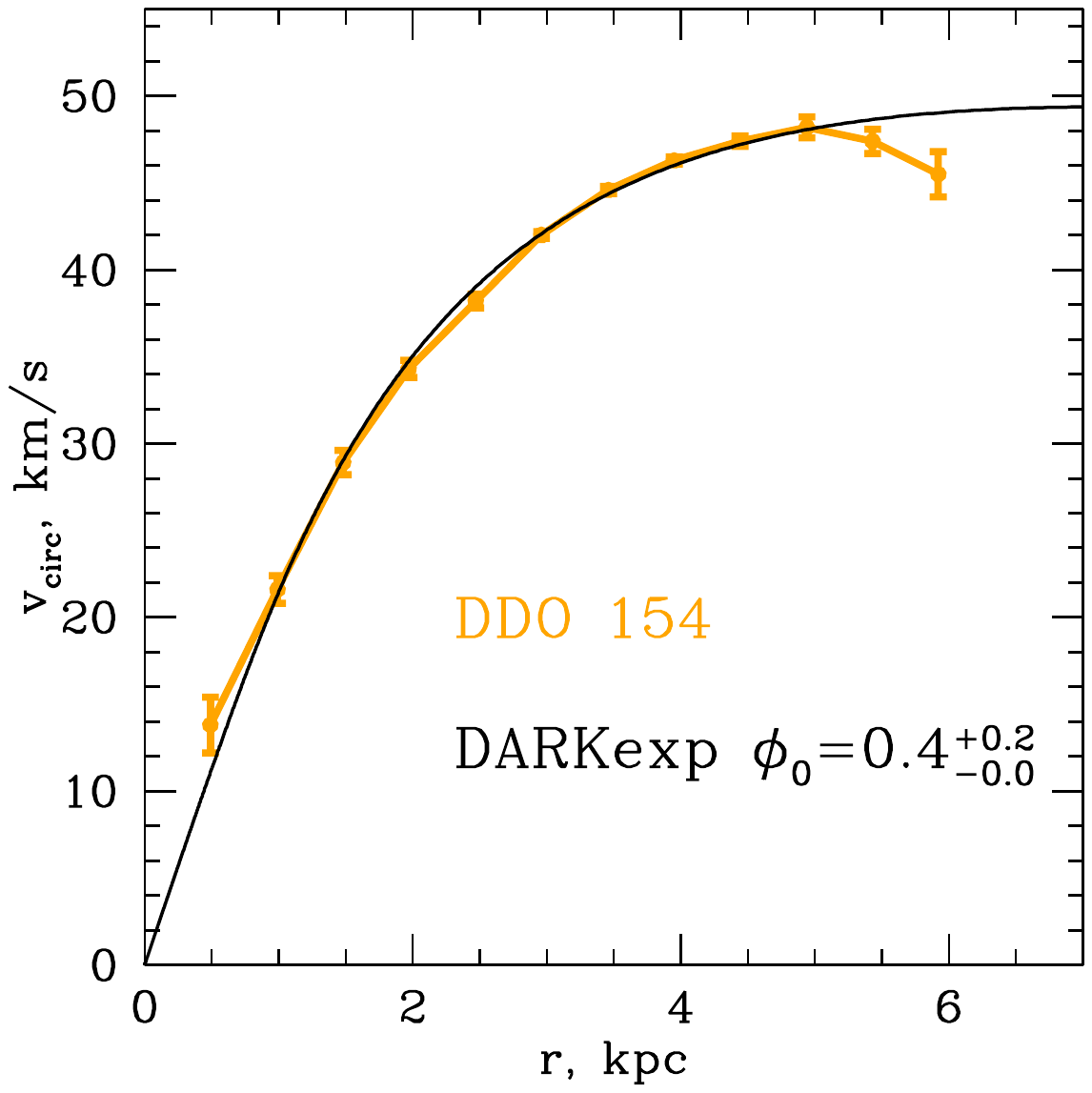}
    \includegraphics[width=0.237\linewidth]{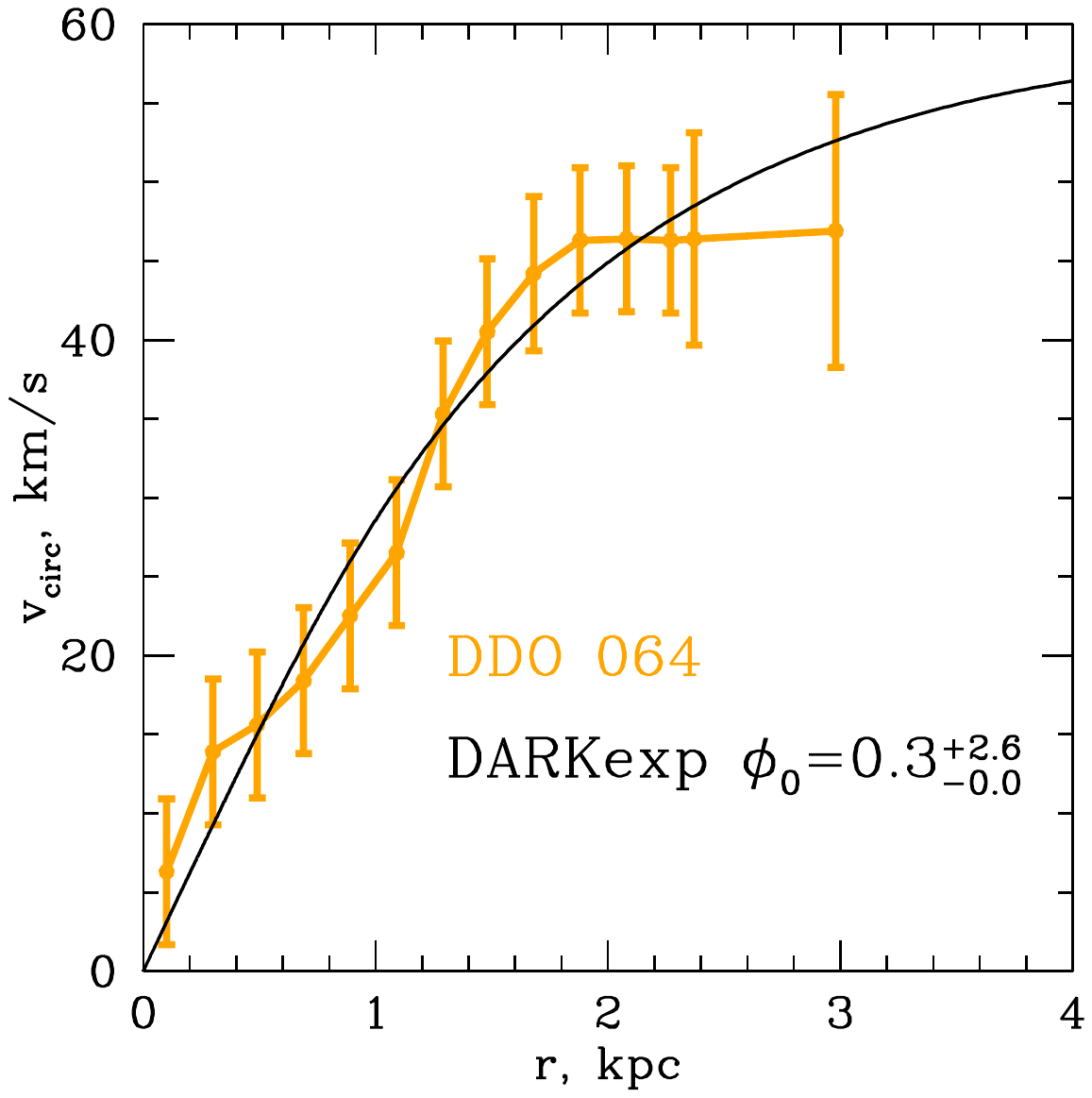}
    \includegraphics[width=0.237\linewidth]{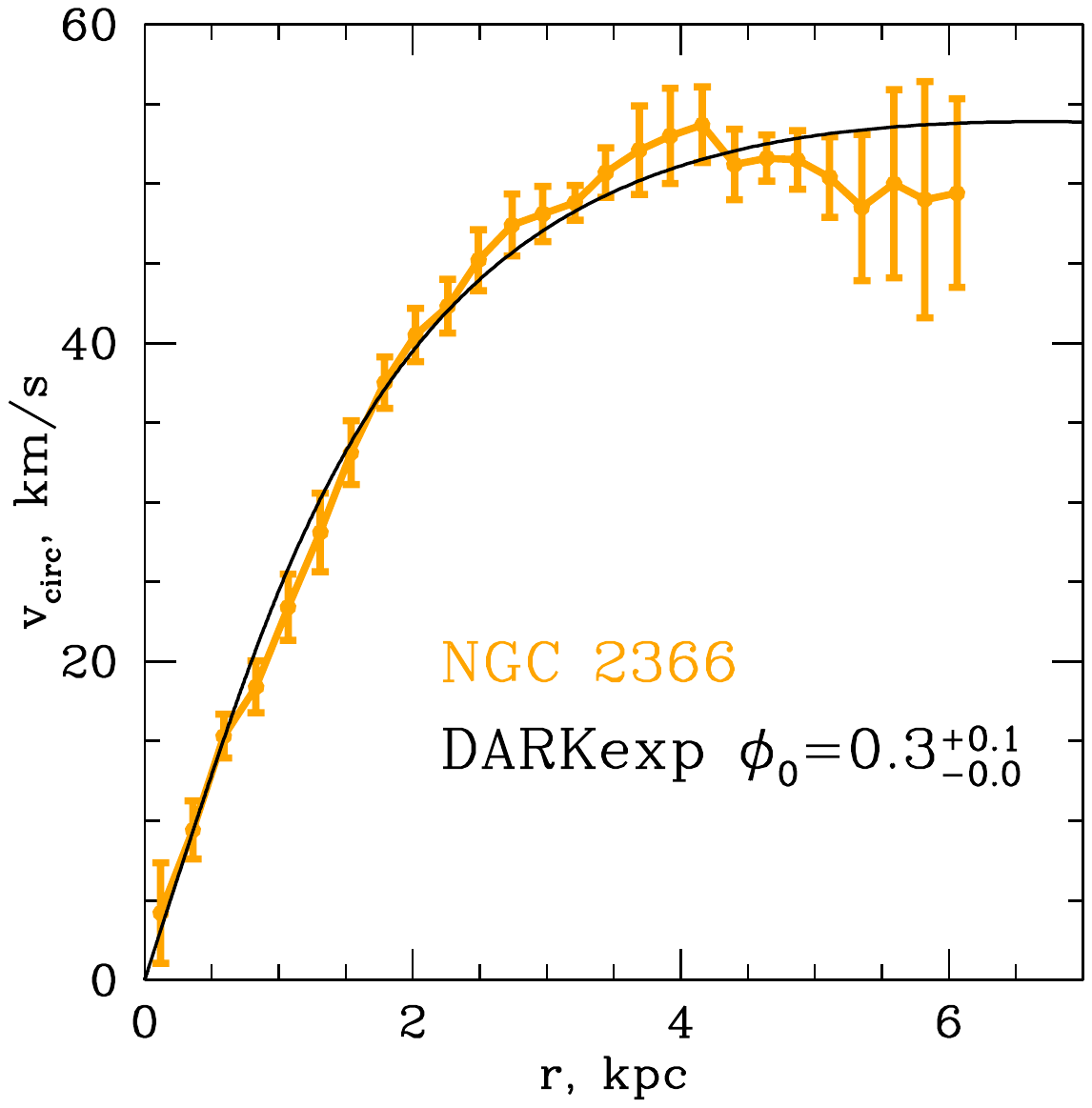}
    \includegraphics[width=0.237\linewidth]{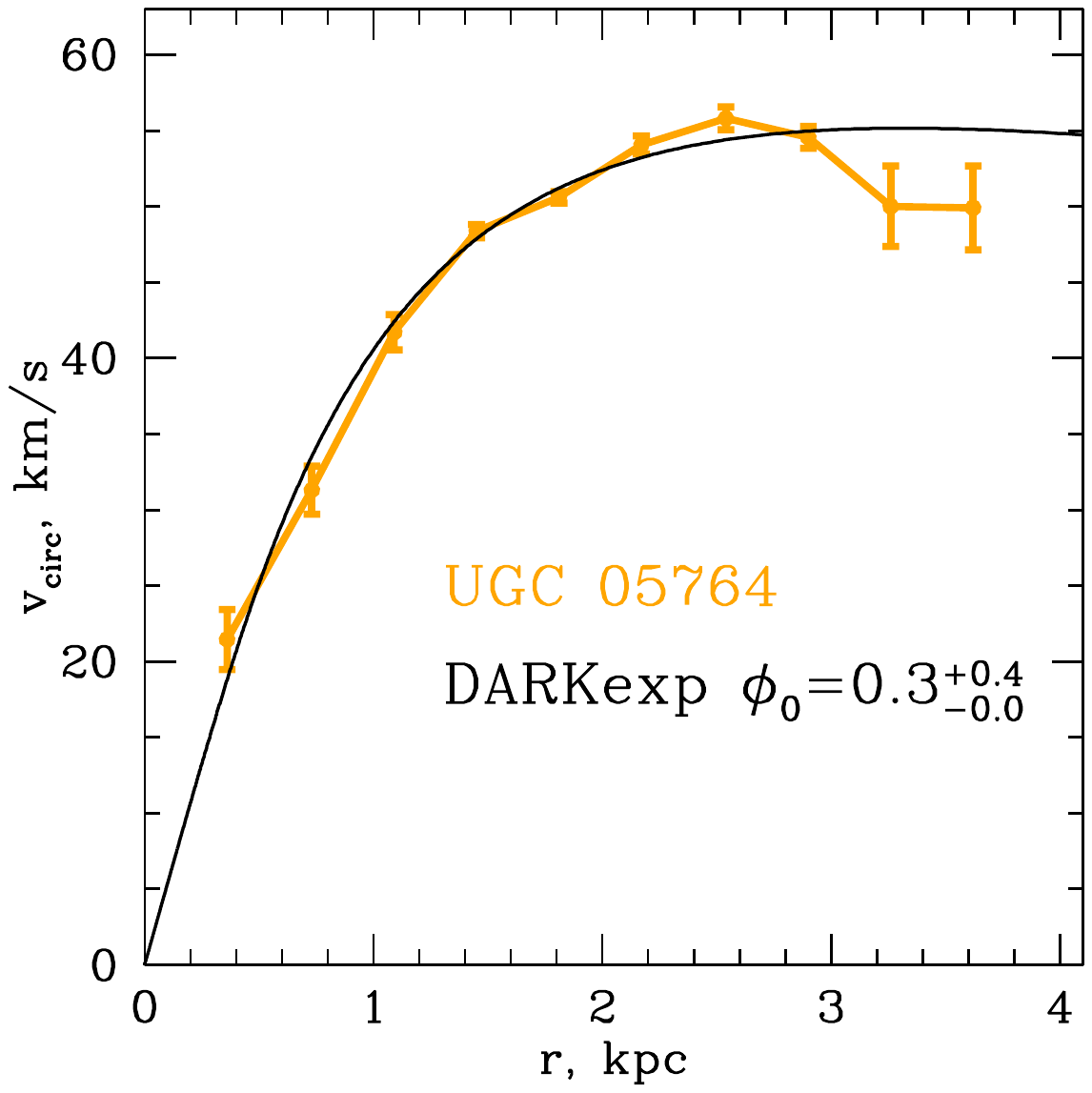}  
        \vskip-1.65cm
    \includegraphics[width=0.237\linewidth]{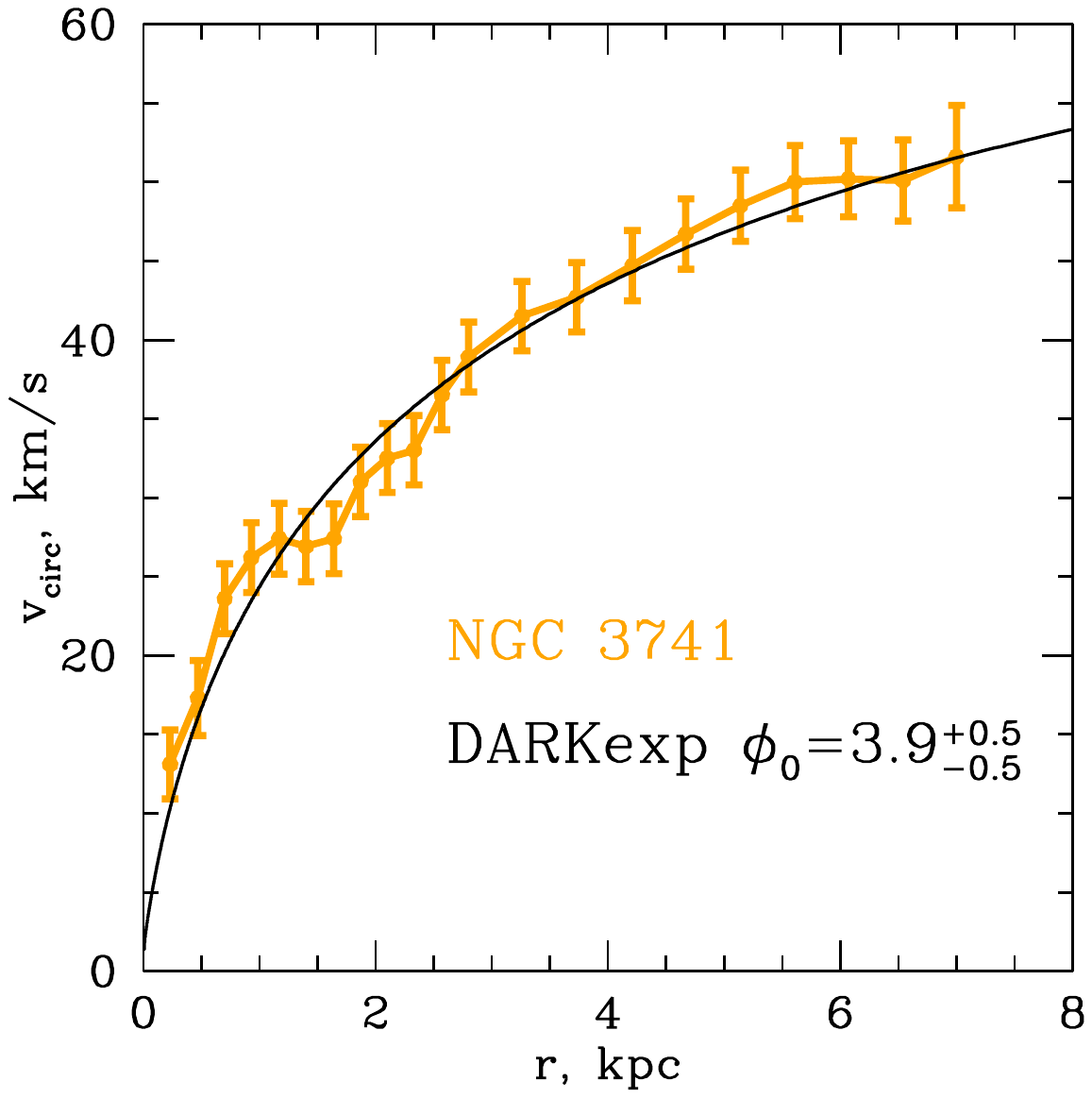}
    \includegraphics[width=0.237\linewidth]{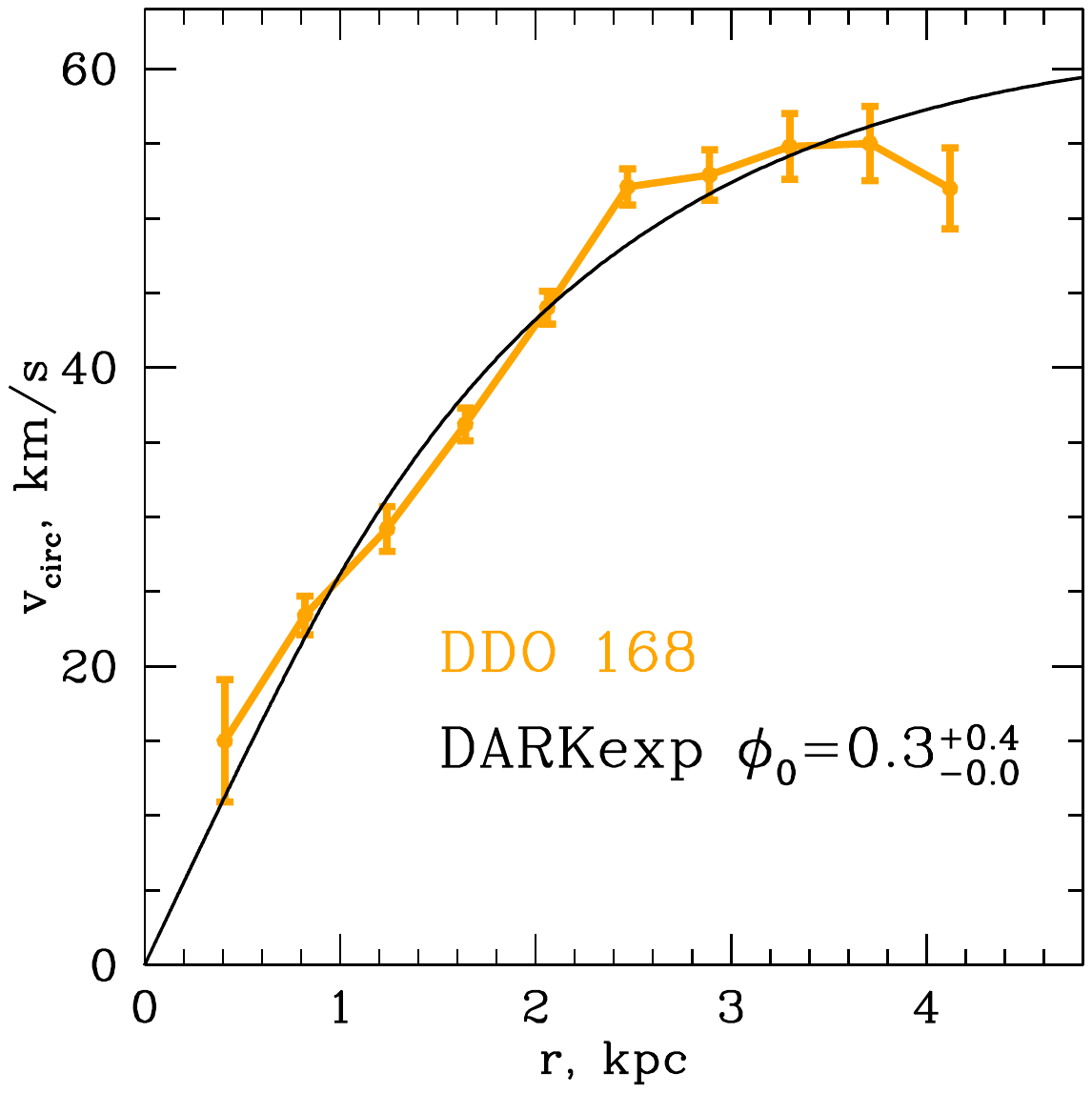}
    \includegraphics[width=0.237\linewidth]{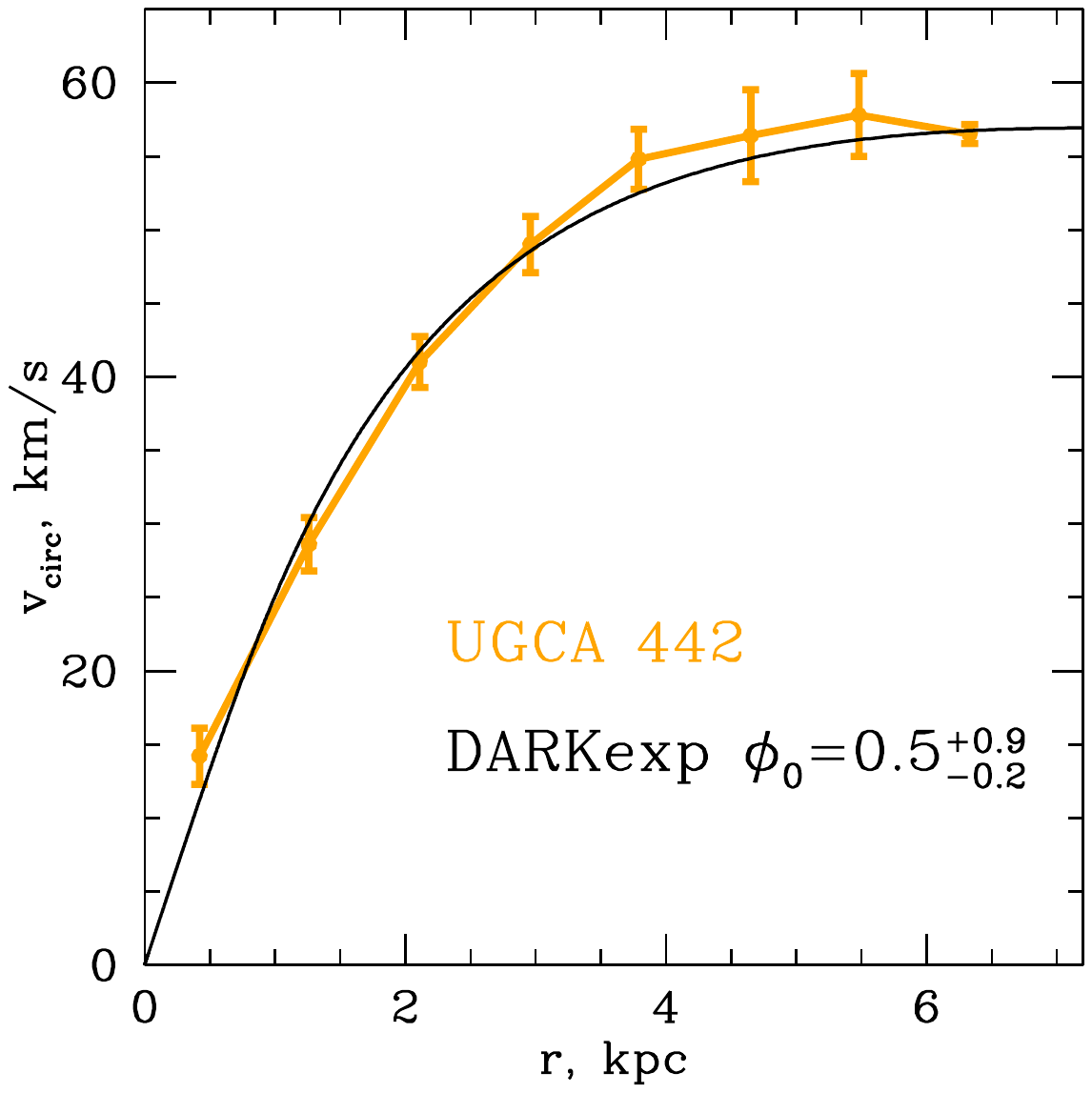}    \includegraphics[width=0.237\linewidth]{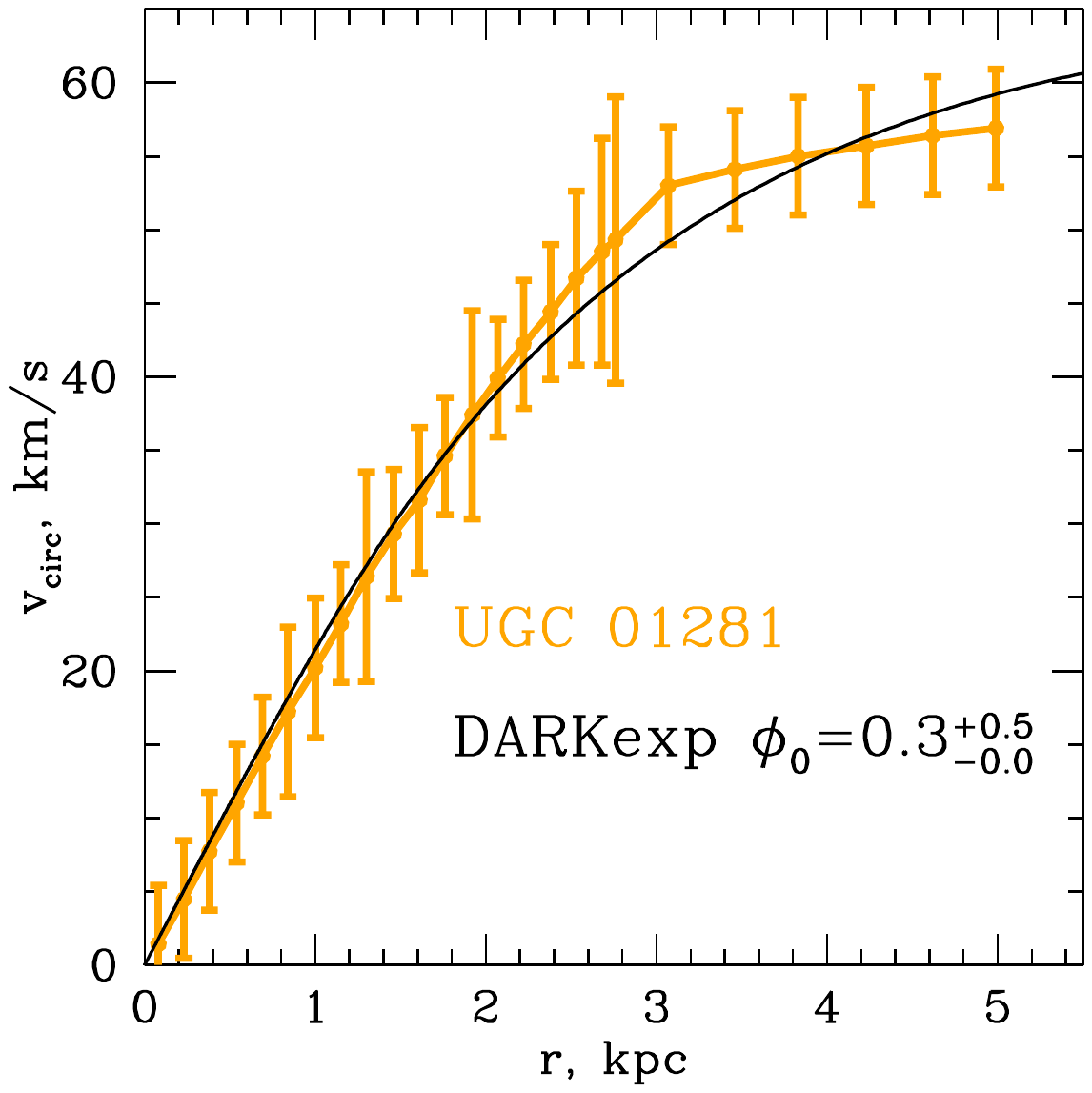}
            \vskip-1.65cm
    \includegraphics[width=0.237\linewidth]{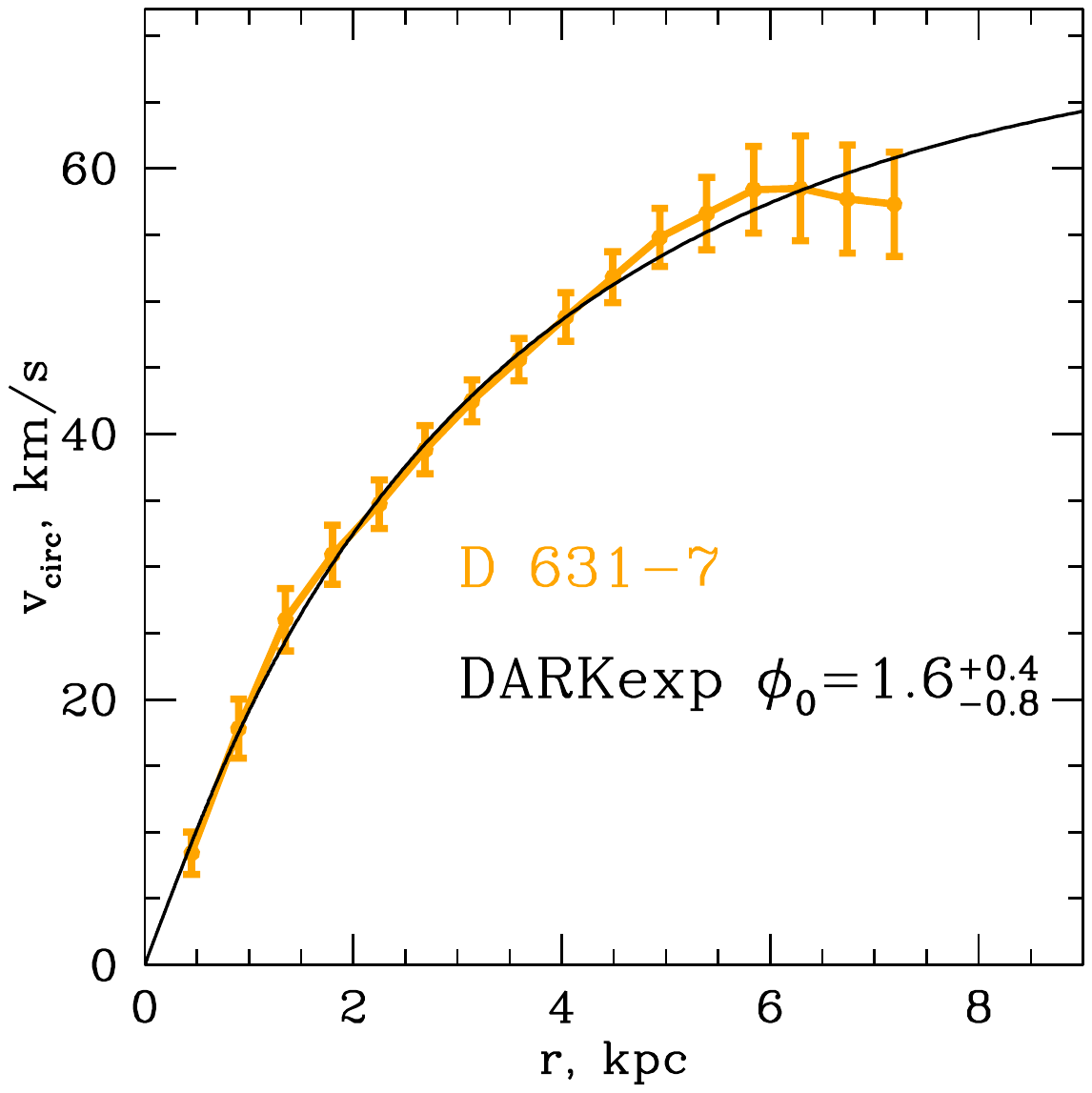}
    \includegraphics[width=0.237\linewidth]{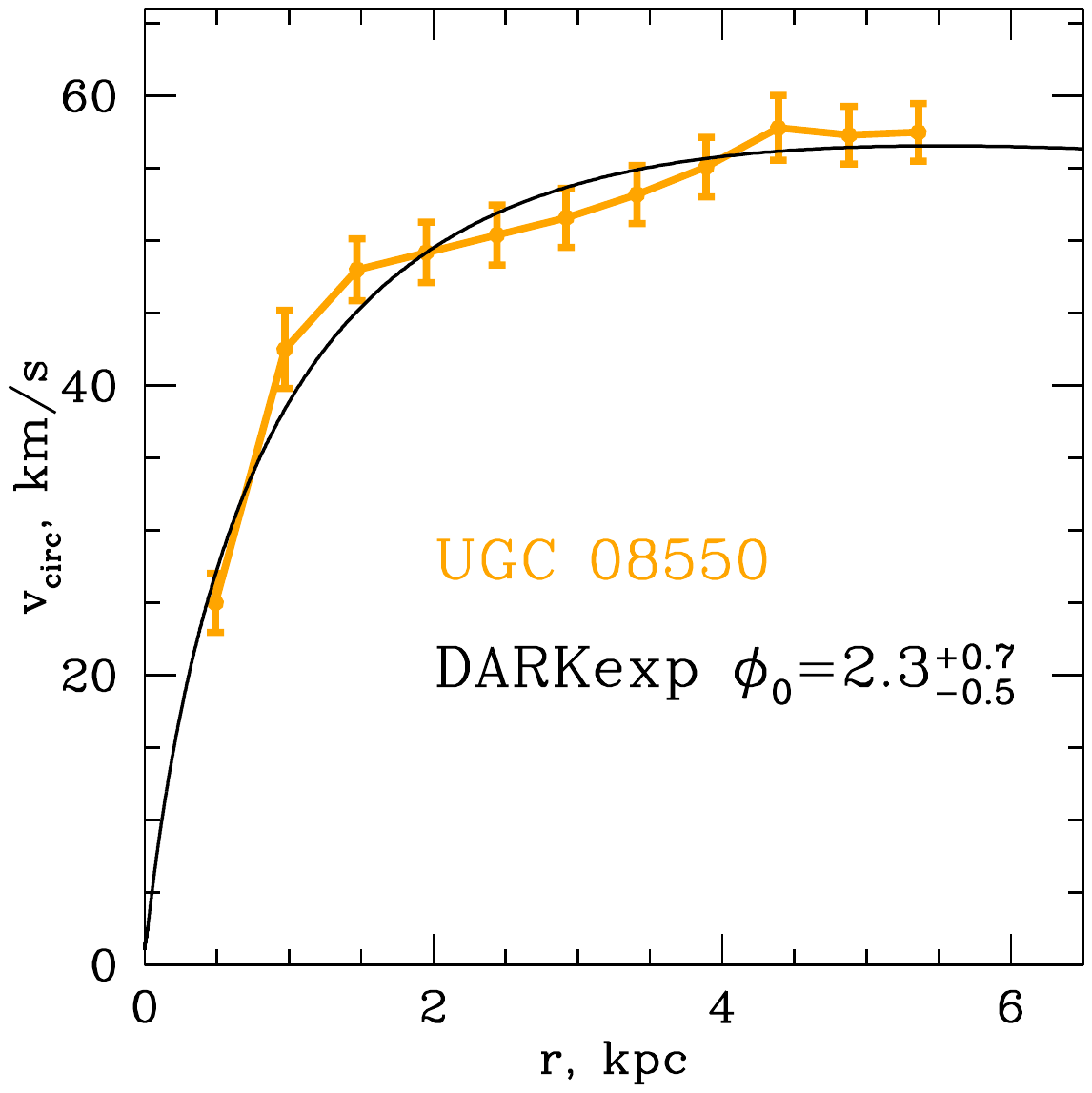}
    \includegraphics[width=0.237\linewidth]{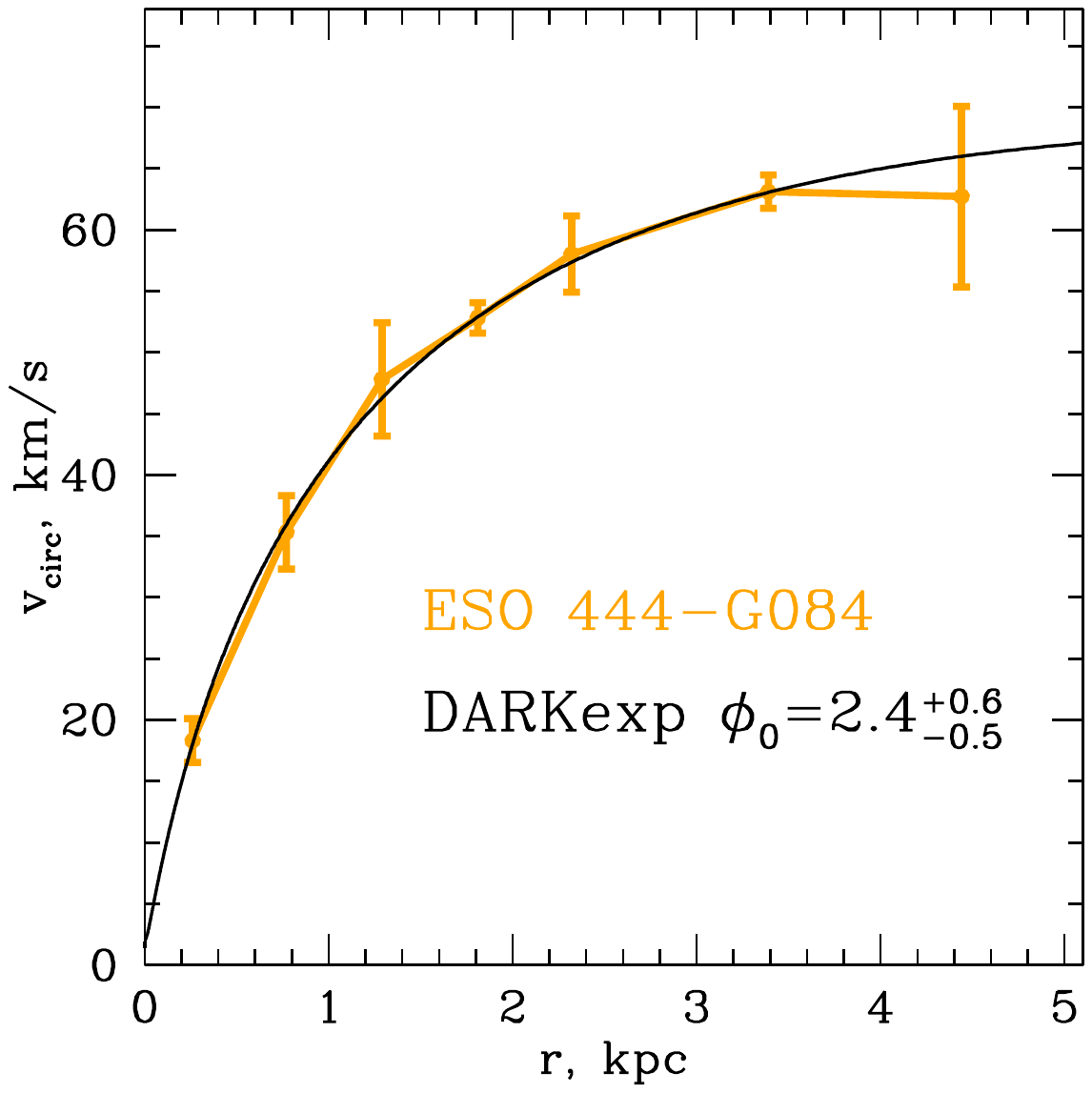}
    \includegraphics[width=0.237\linewidth]{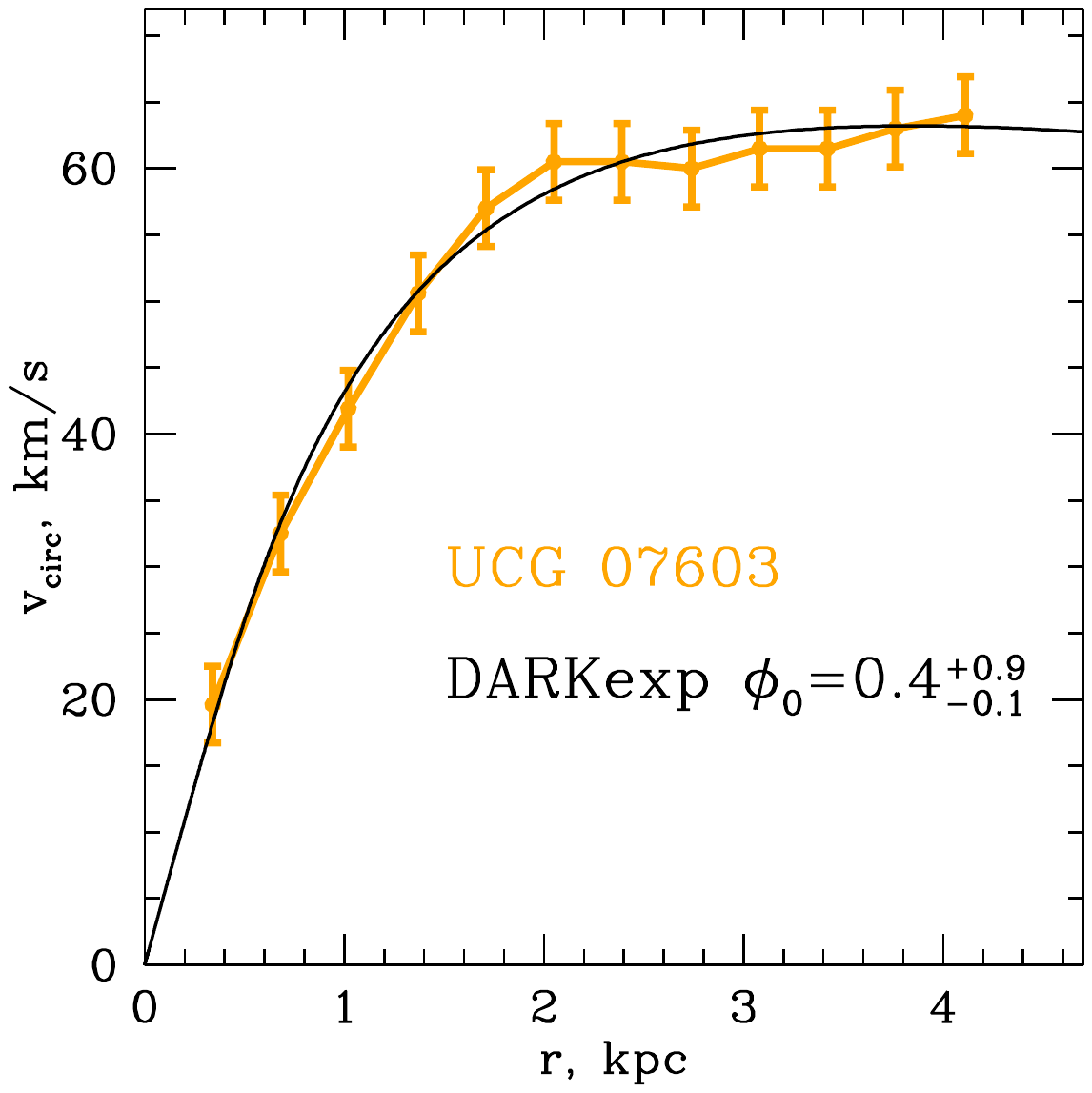}
         \vskip-1.65cm
    \includegraphics[width=0.237\linewidth]{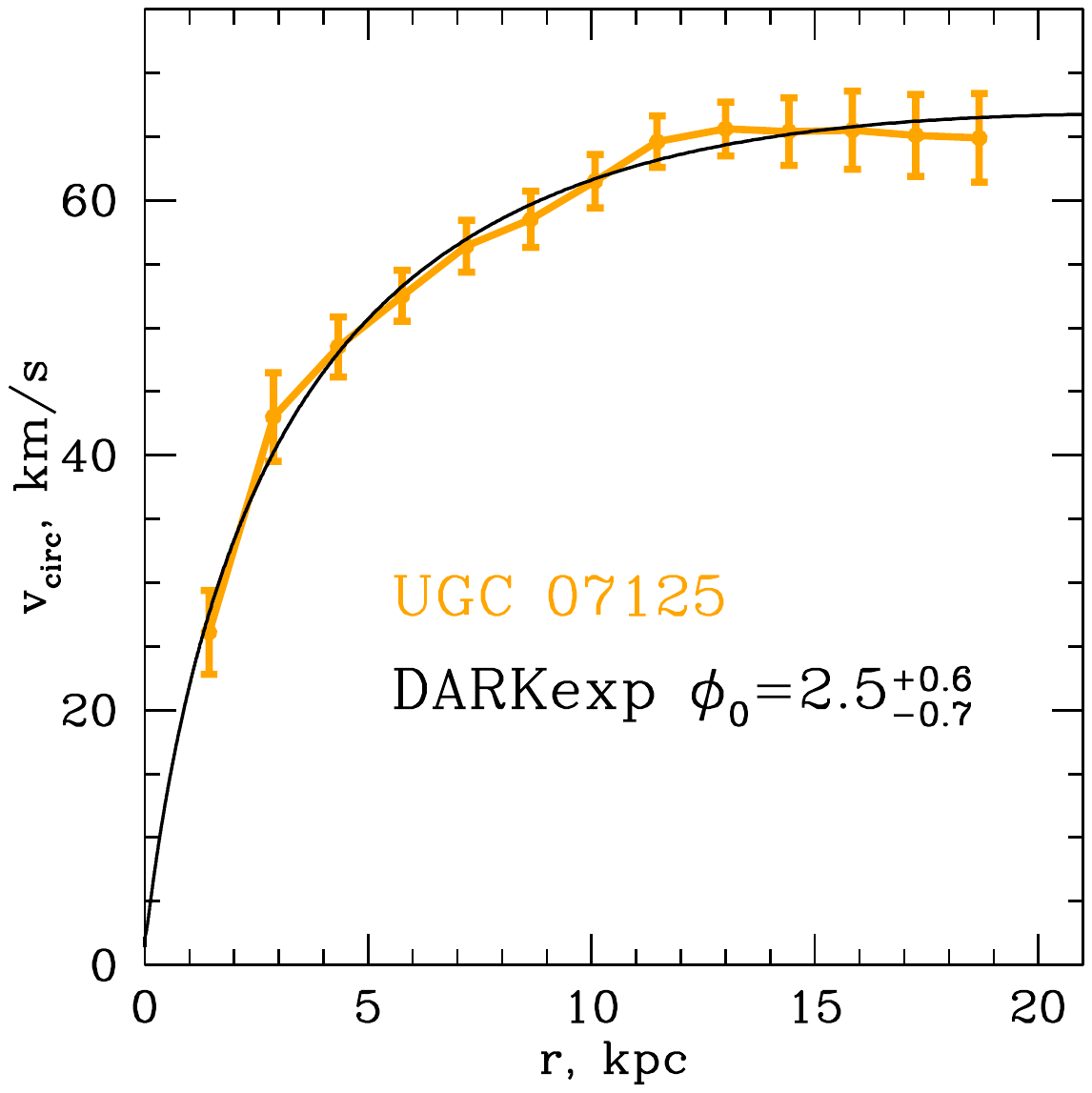}
    \includegraphics[width=0.237\linewidth]{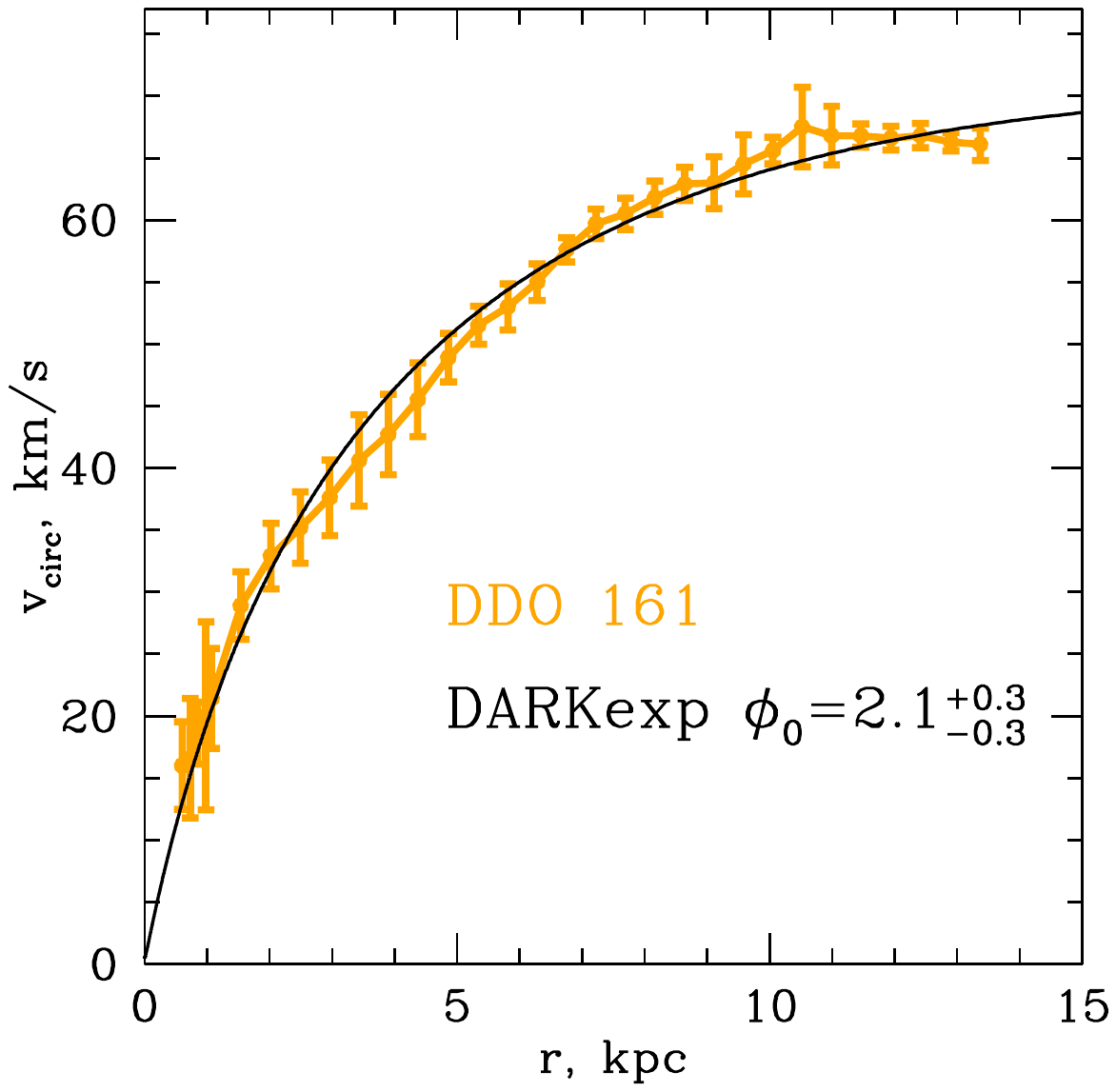}
    \includegraphics[width=0.237\linewidth]{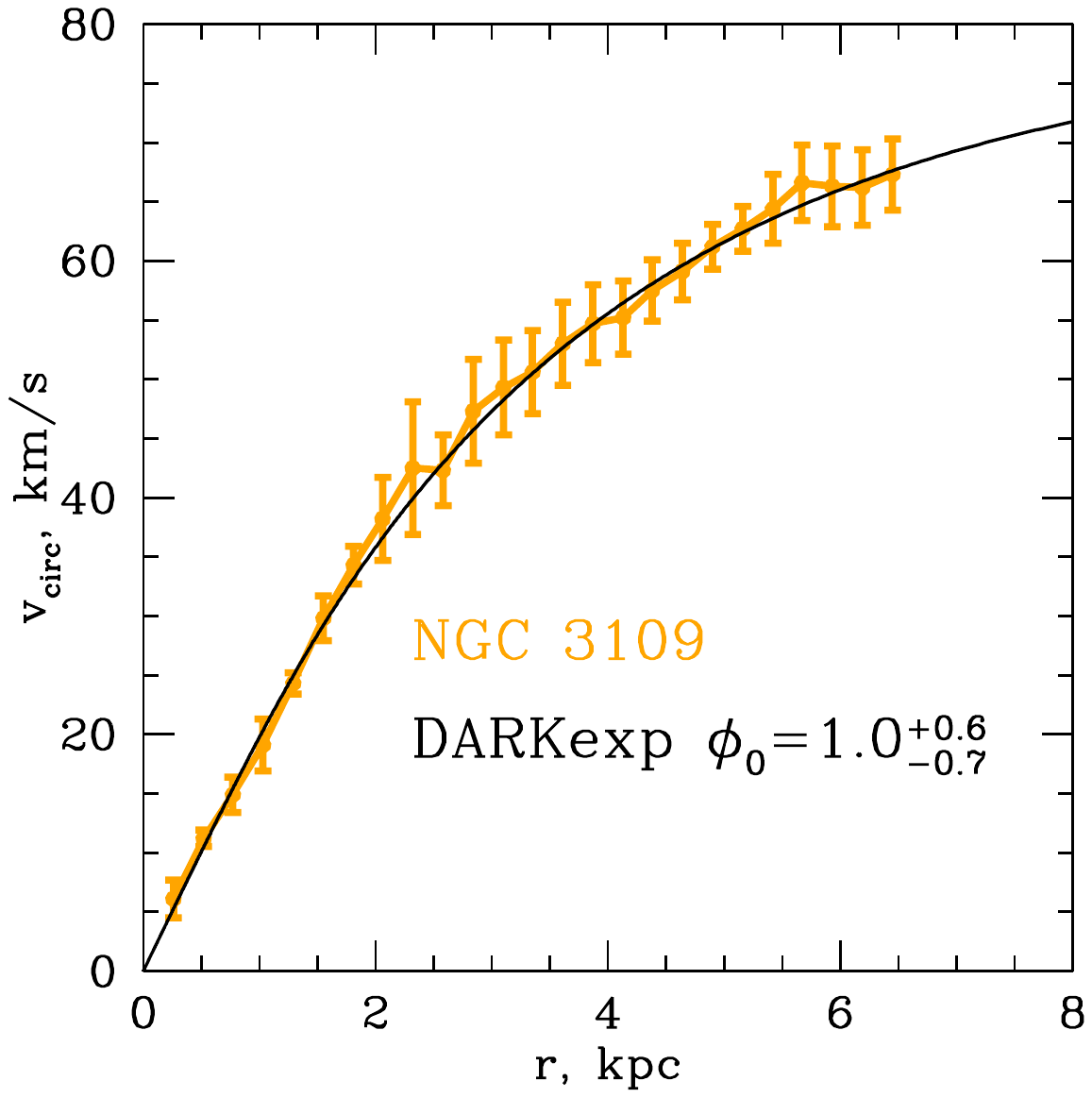}
    \includegraphics[width=0.237\linewidth]{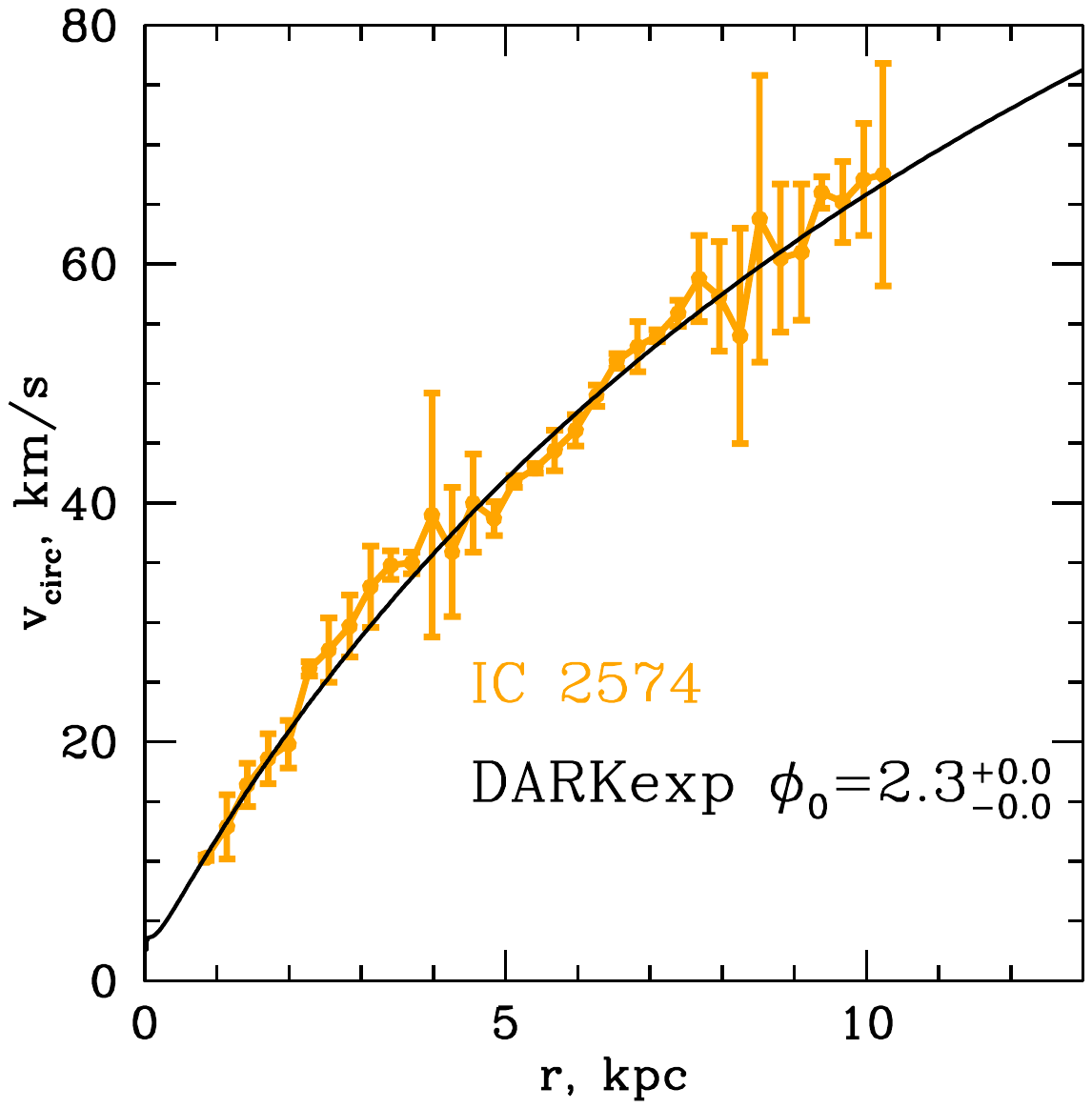}   
    \vskip-0.85cm
    \caption{Rotation curves of SPARC galaxies arranged in order of increasing $v_{\rm last}$. DARKexp best-fit models are black curves, and the corresponding $\phi_0$ values \newer{with uncertainties}, are shown in each panel. }
    \label{fig:rotcurve1}
\end{figure*}

\begin{figure*}
    \centering
    \vskip-1.75cm
    \includegraphics[width=0.237\linewidth]{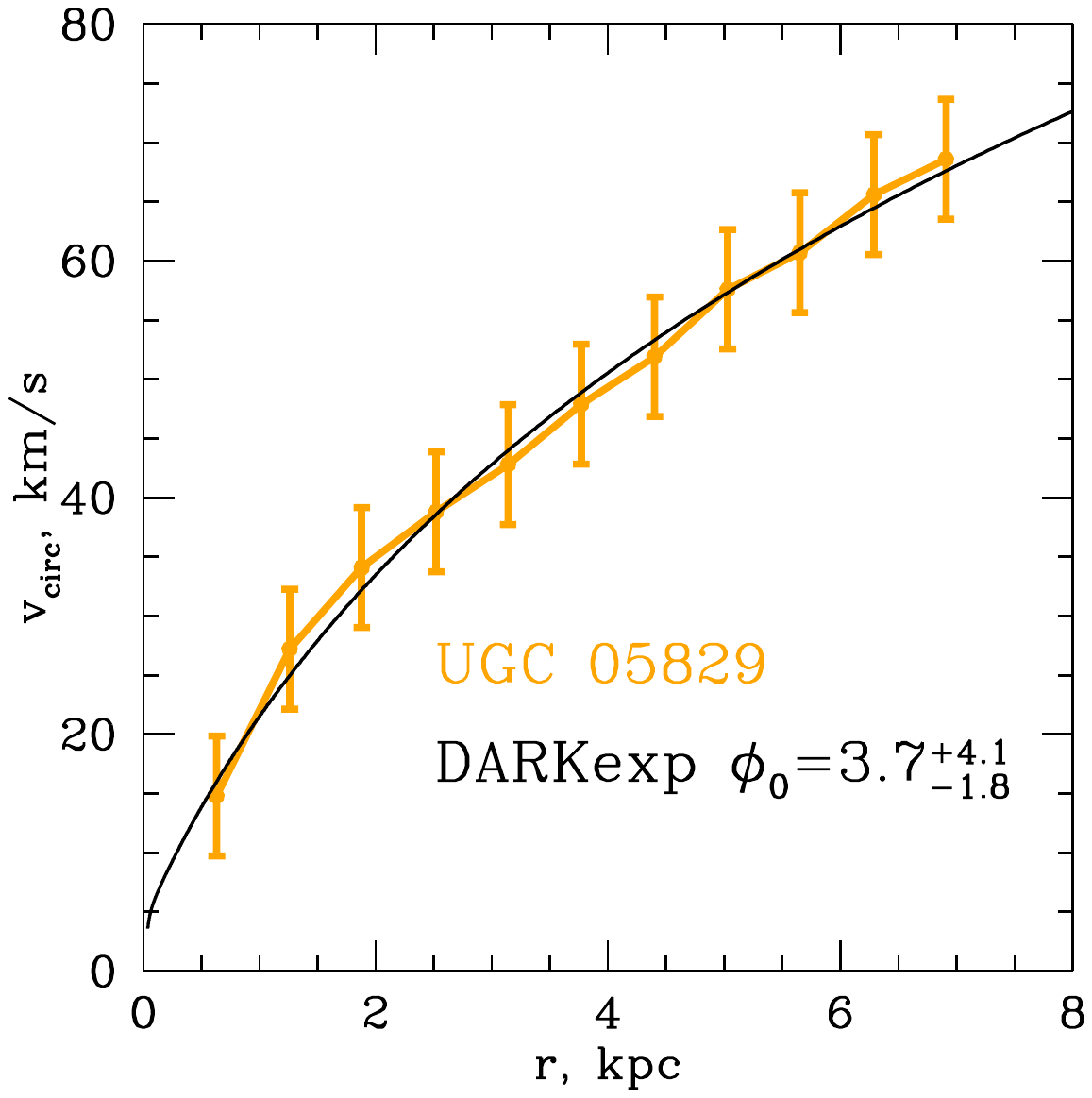}
    \includegraphics[width=0.237\linewidth]{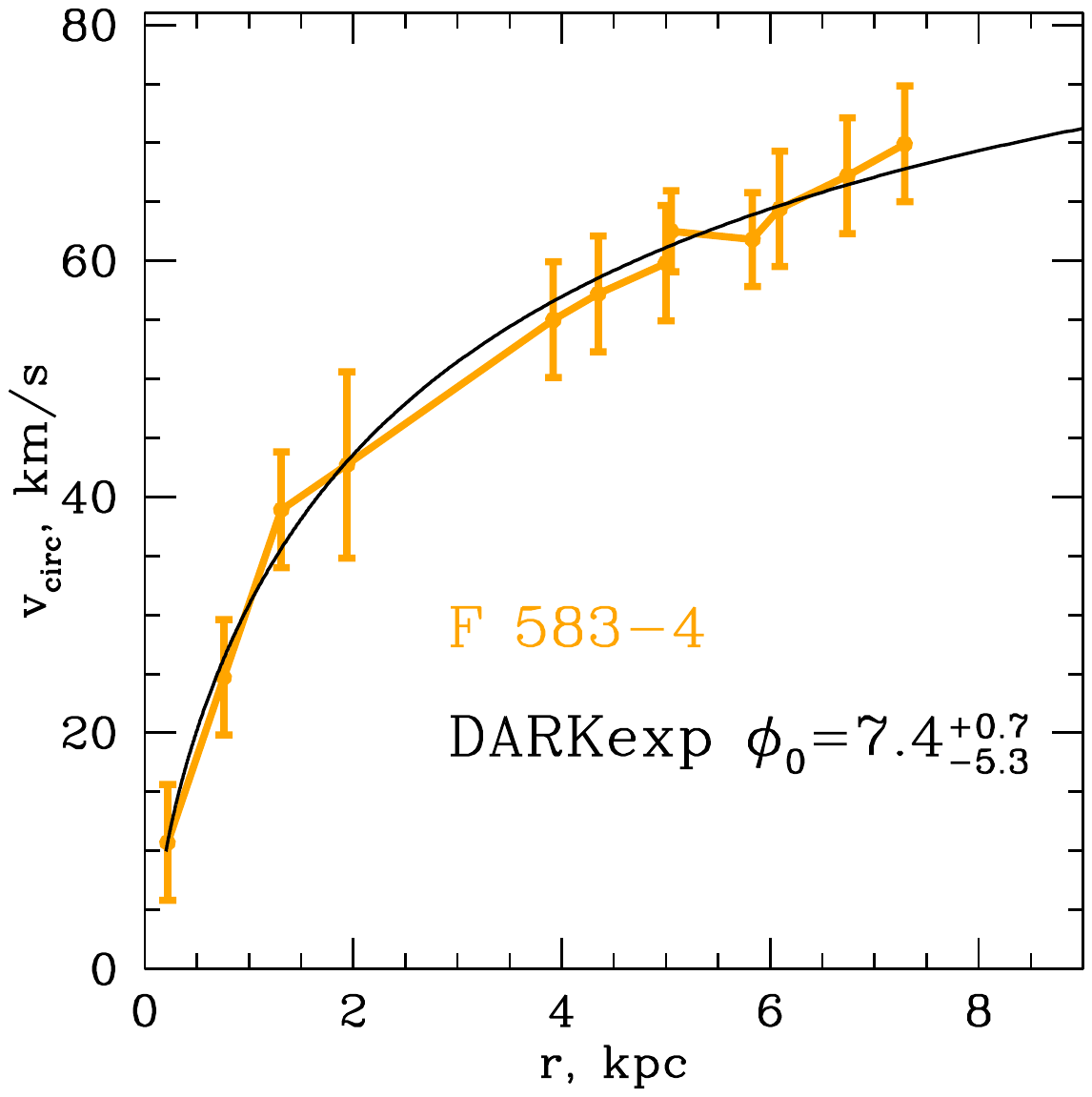}
    \includegraphics[width=0.237\linewidth]{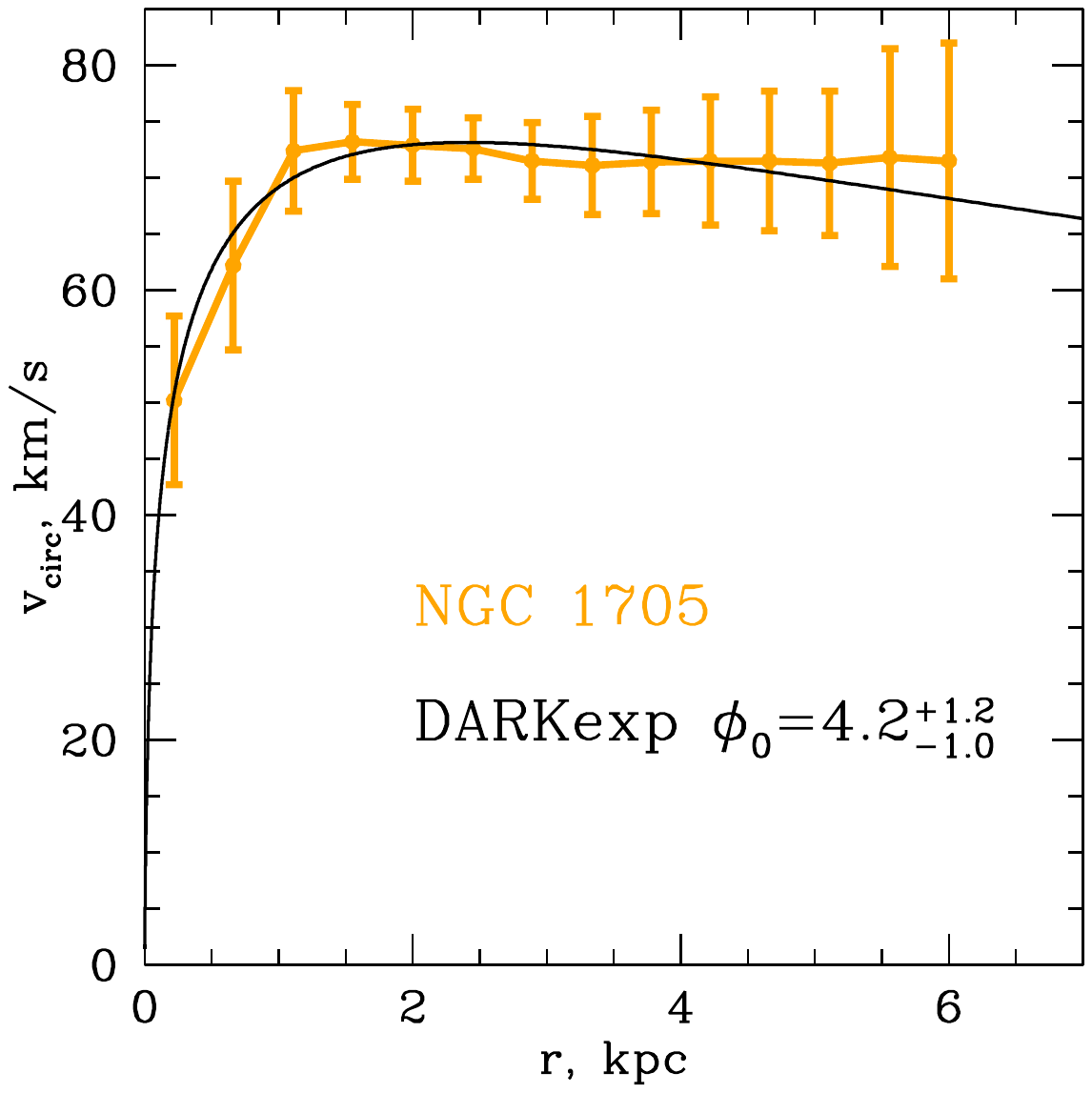}    \includegraphics[width=0.237\linewidth]{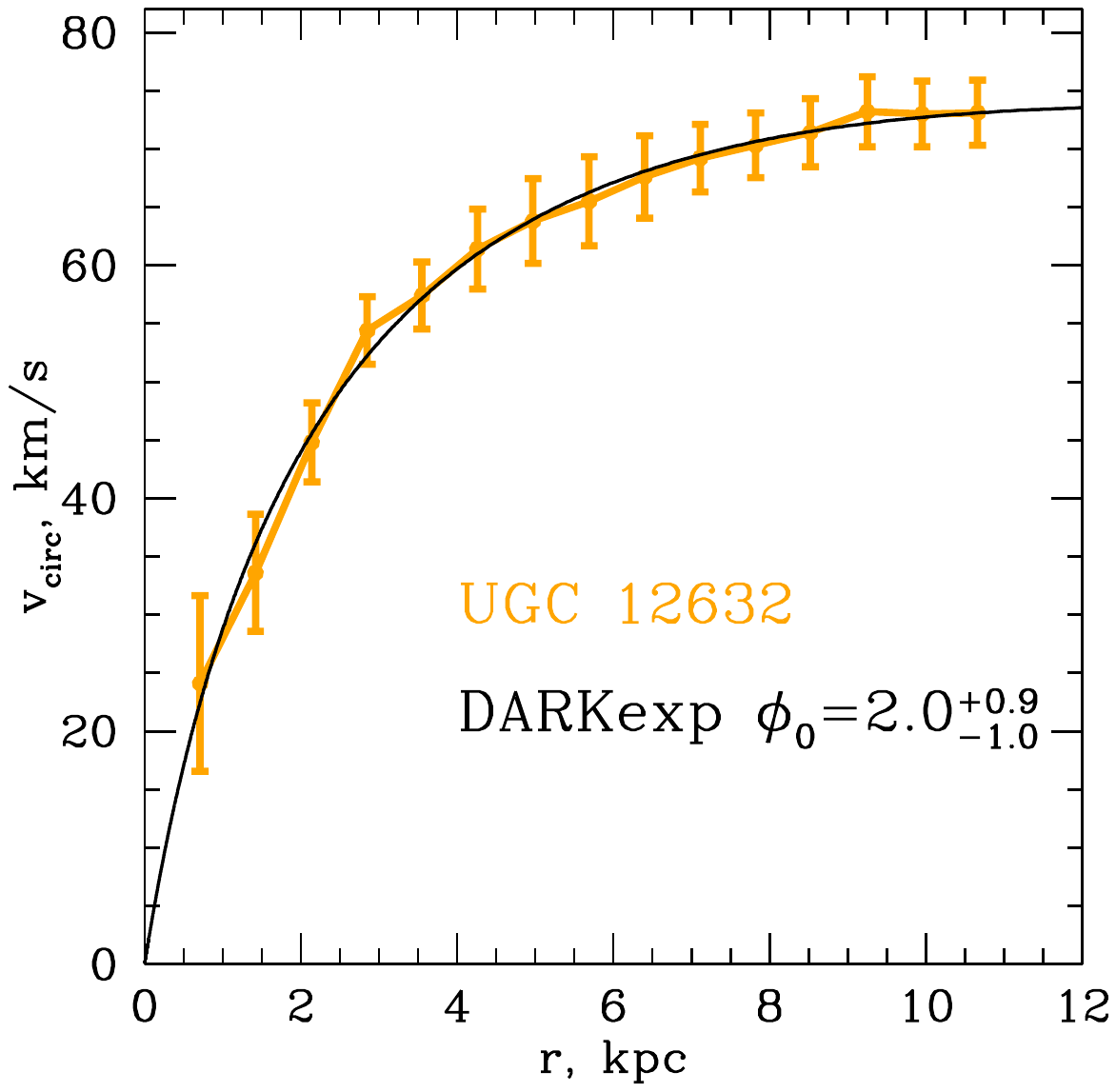}
            \vskip-1.65cm
    \includegraphics[width=0.237\linewidth]{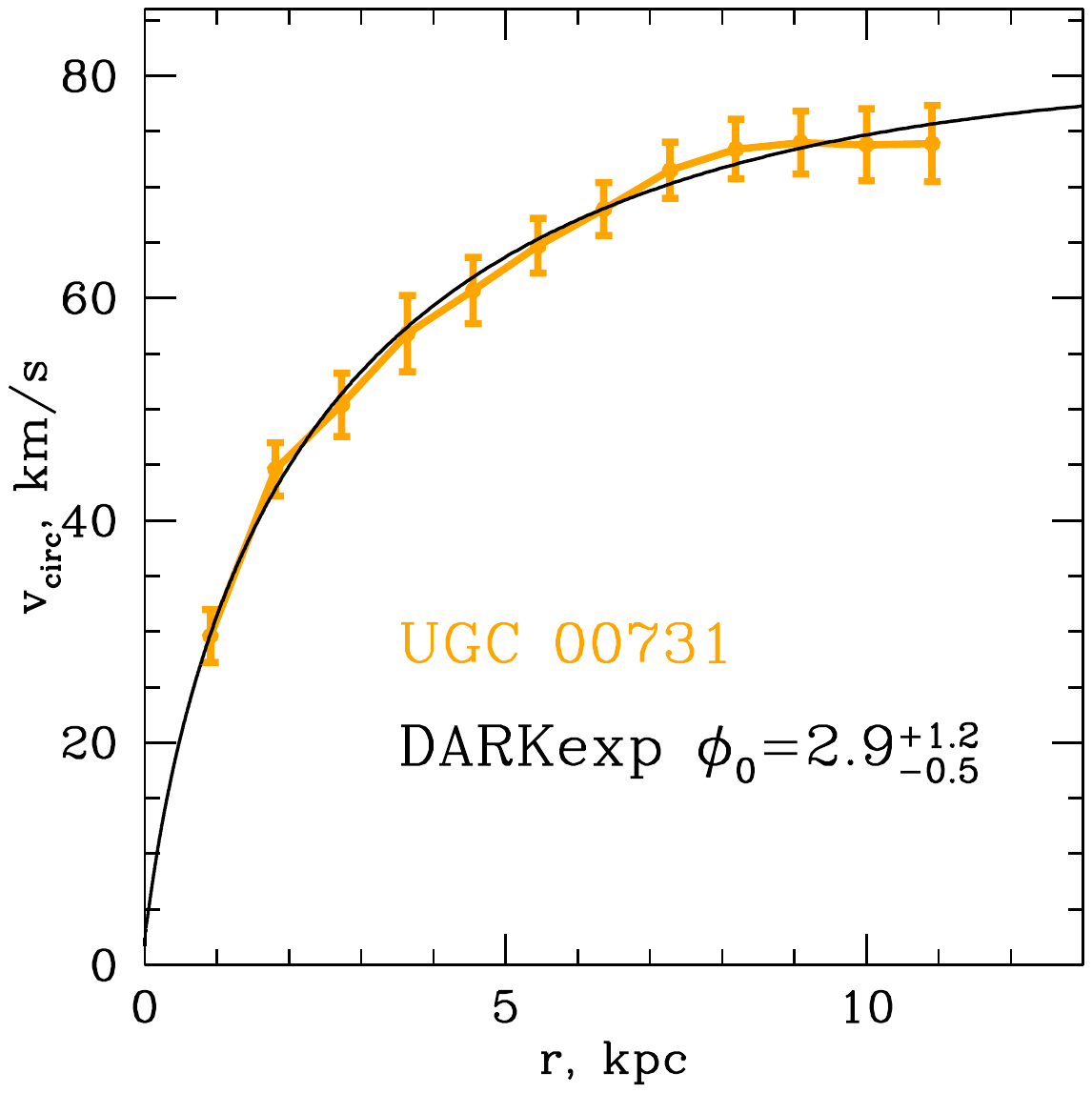}
    \includegraphics[width=0.237\linewidth]{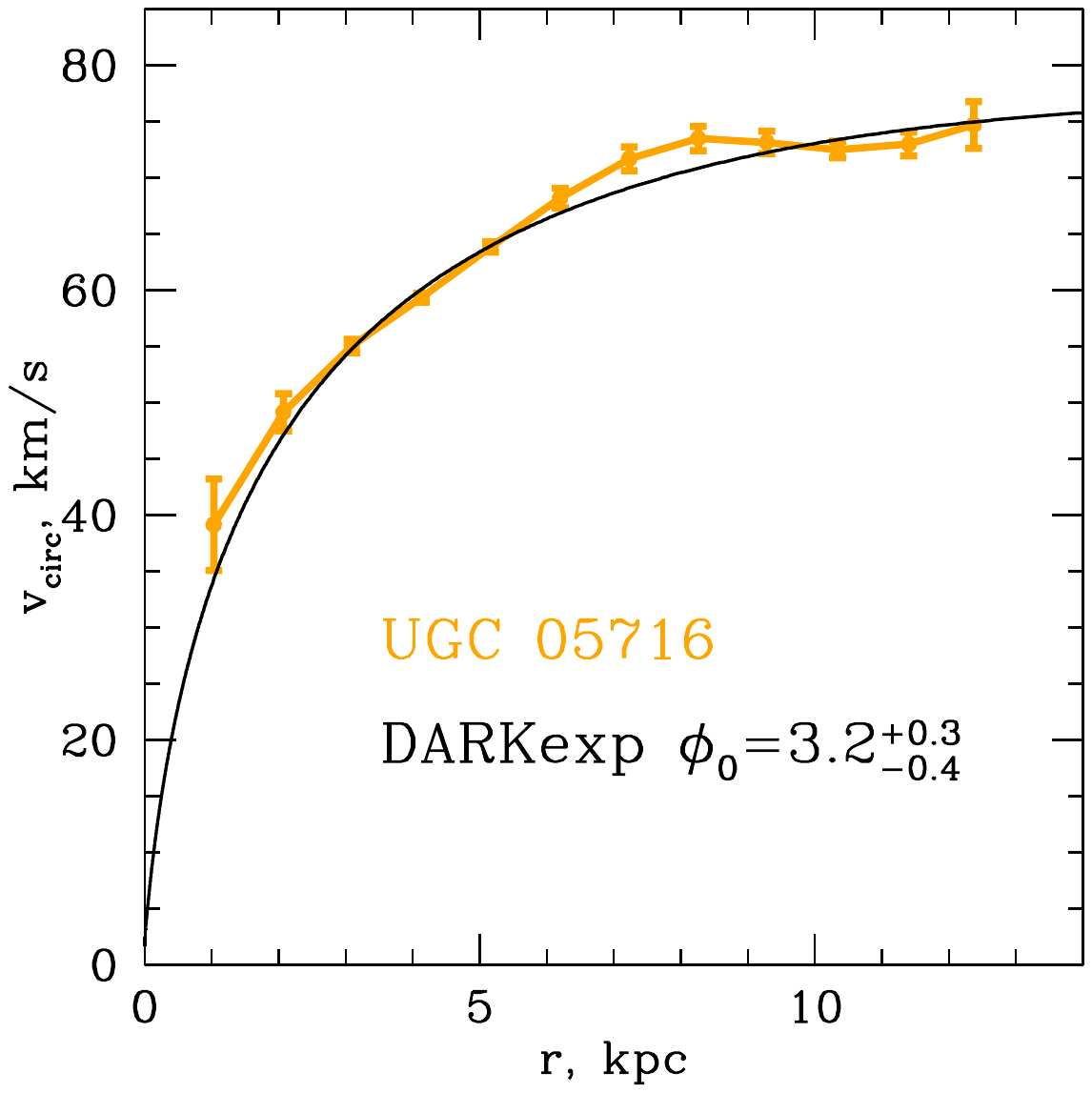}
    \includegraphics[width=0.237\linewidth]{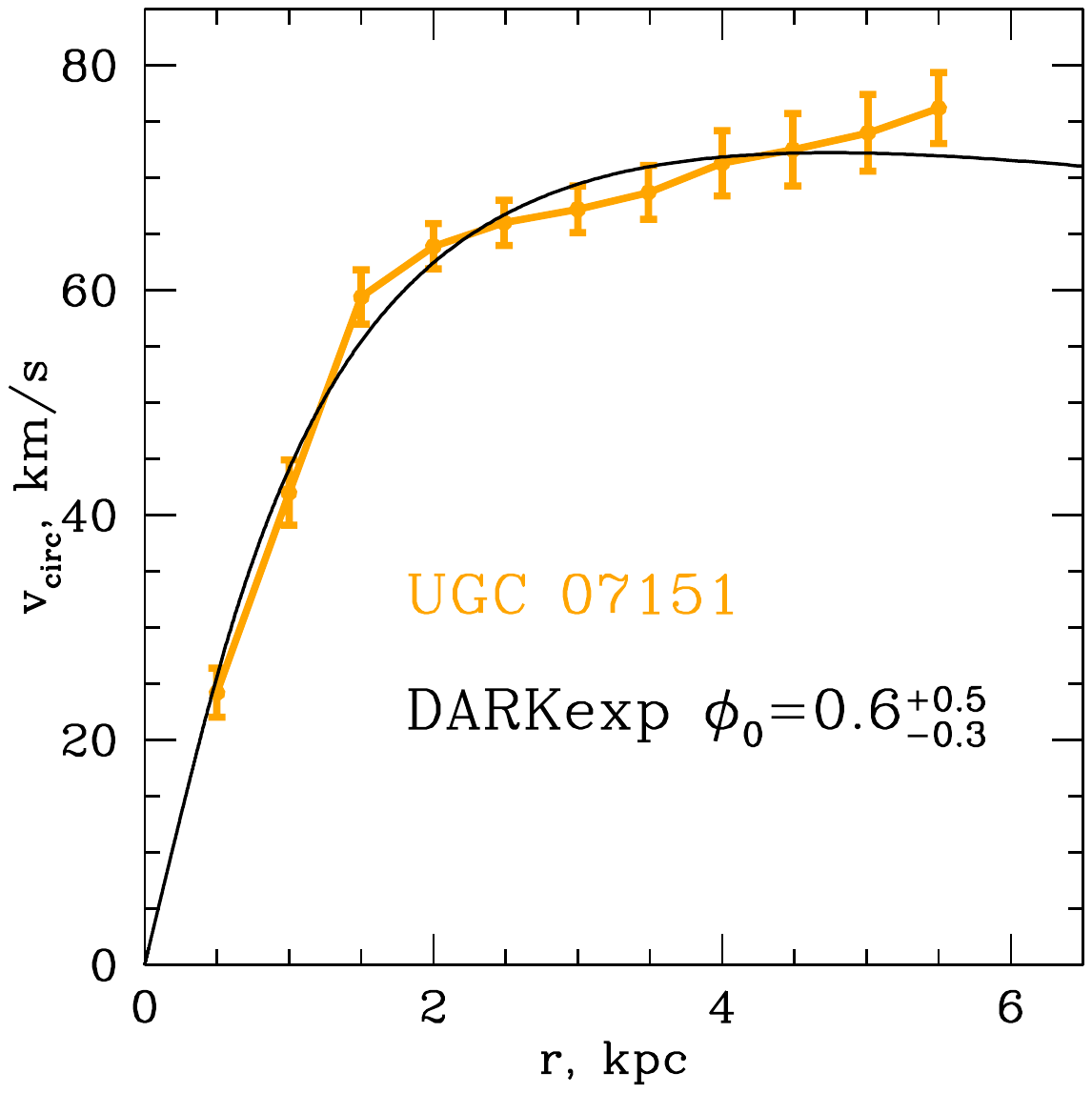}
    \includegraphics[width=0.237\linewidth]{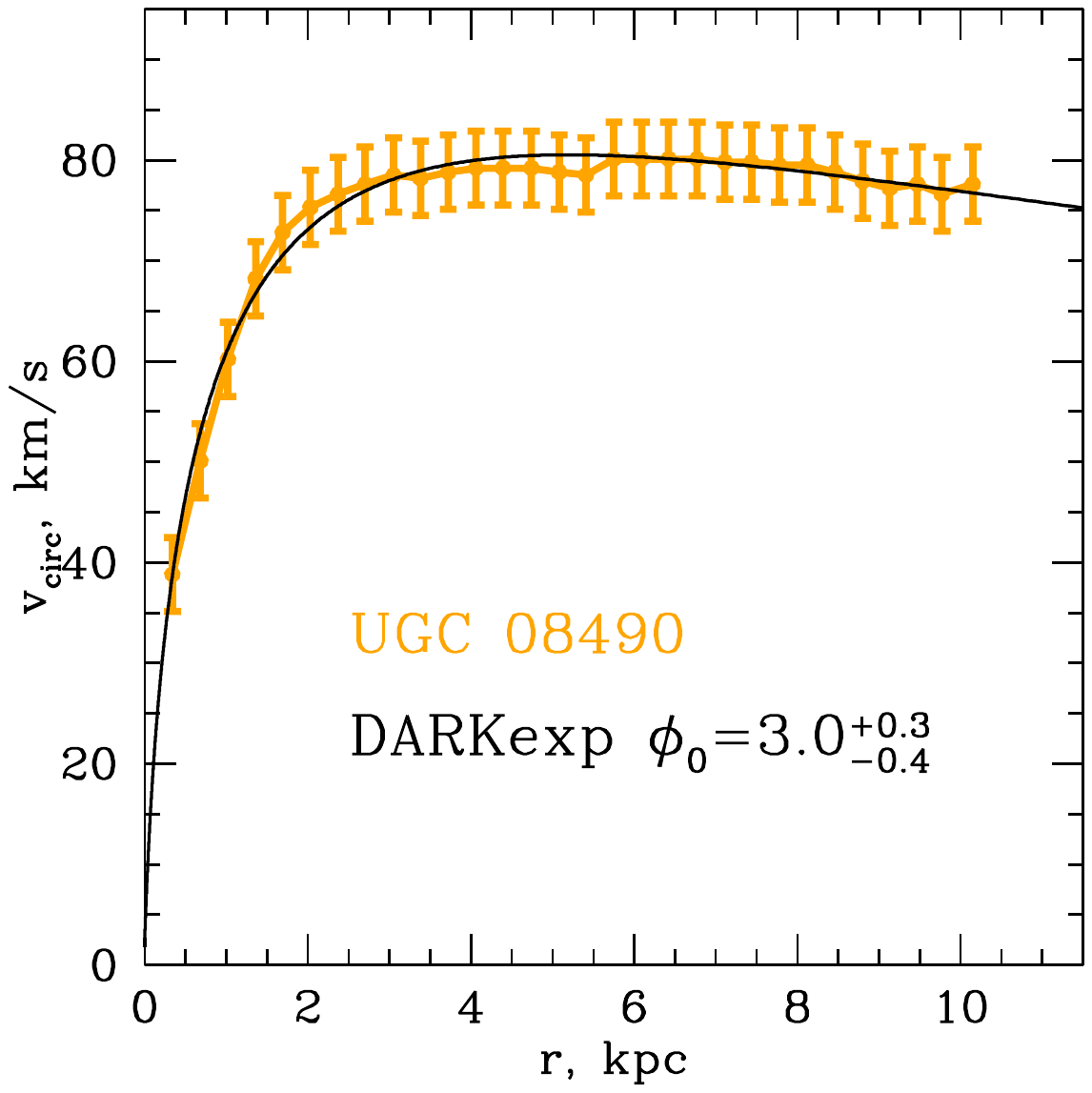}
         \vskip-1.65cm
    \includegraphics[width=0.237\linewidth]{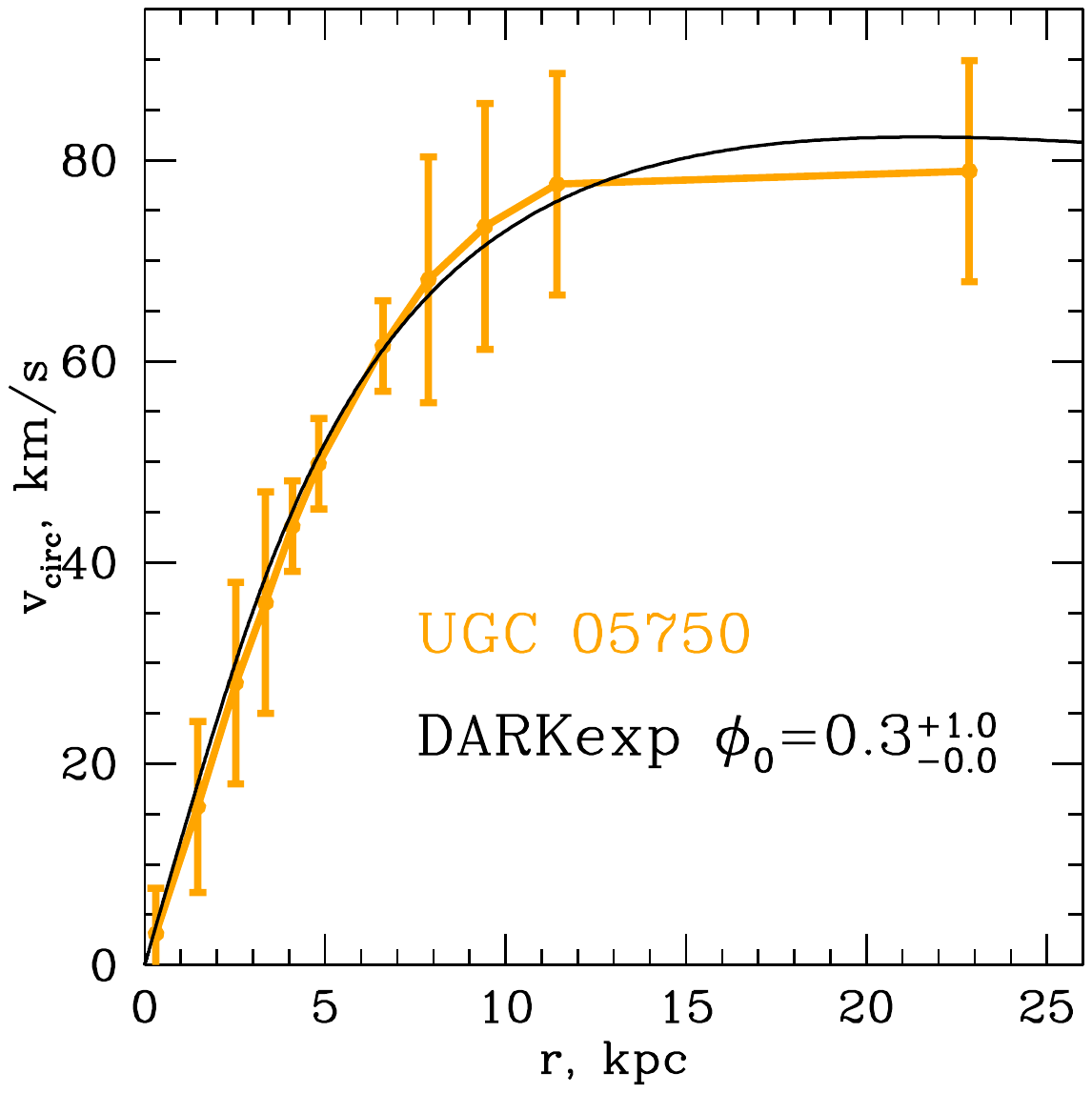}
    \includegraphics[width=0.237\linewidth]{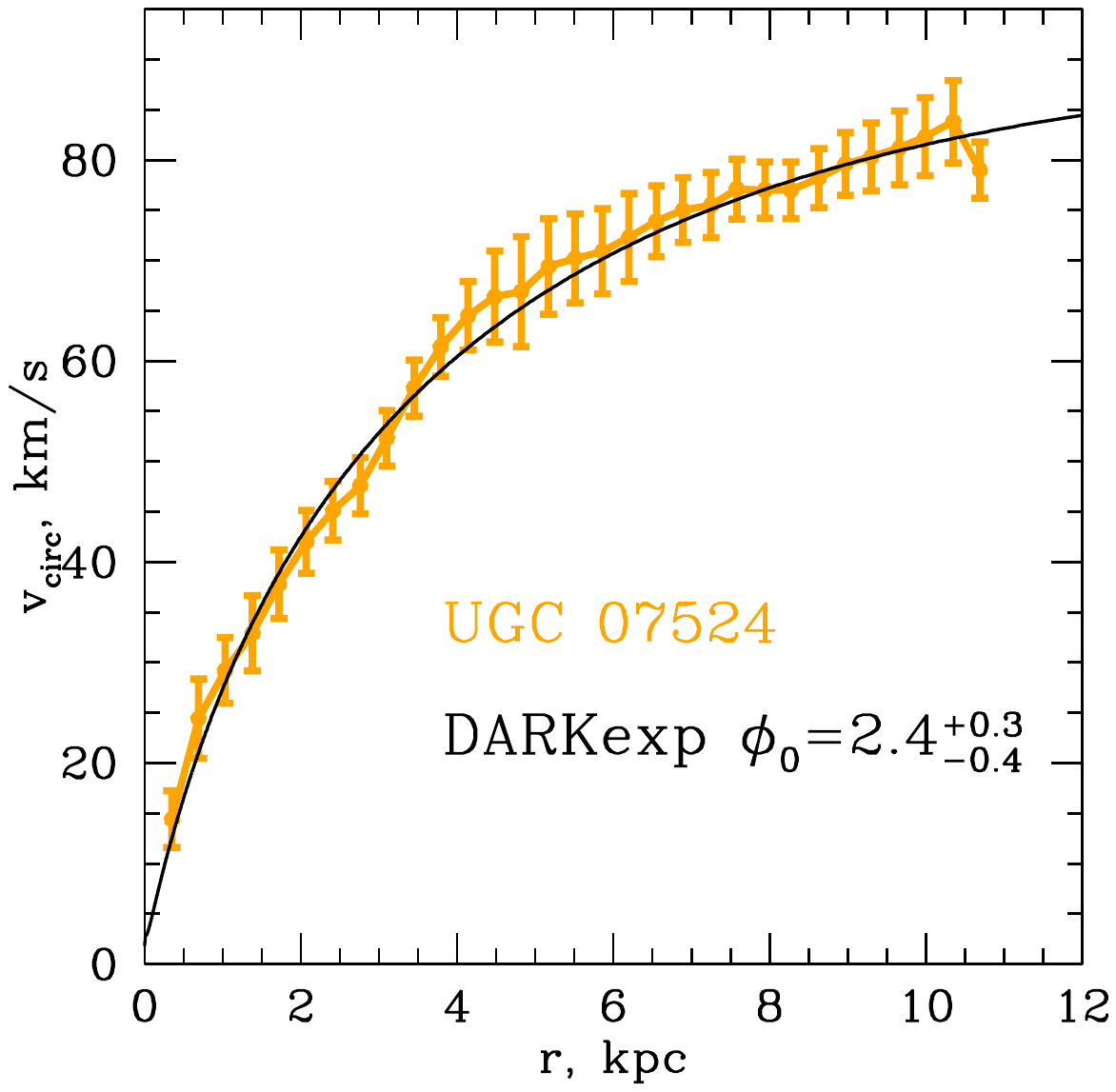}
    \includegraphics[width=0.237\linewidth]{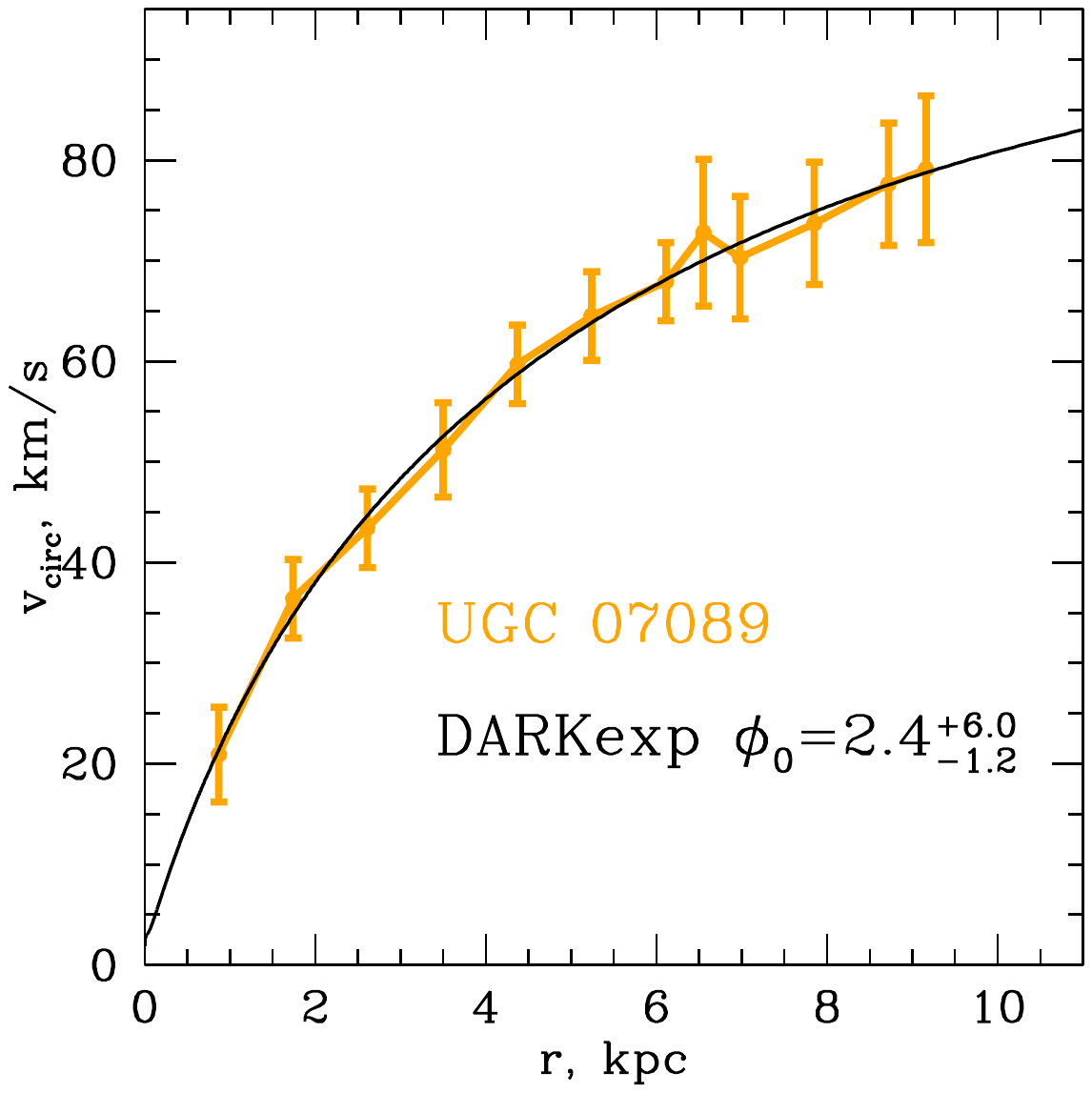}
    \includegraphics[width=0.237\linewidth]{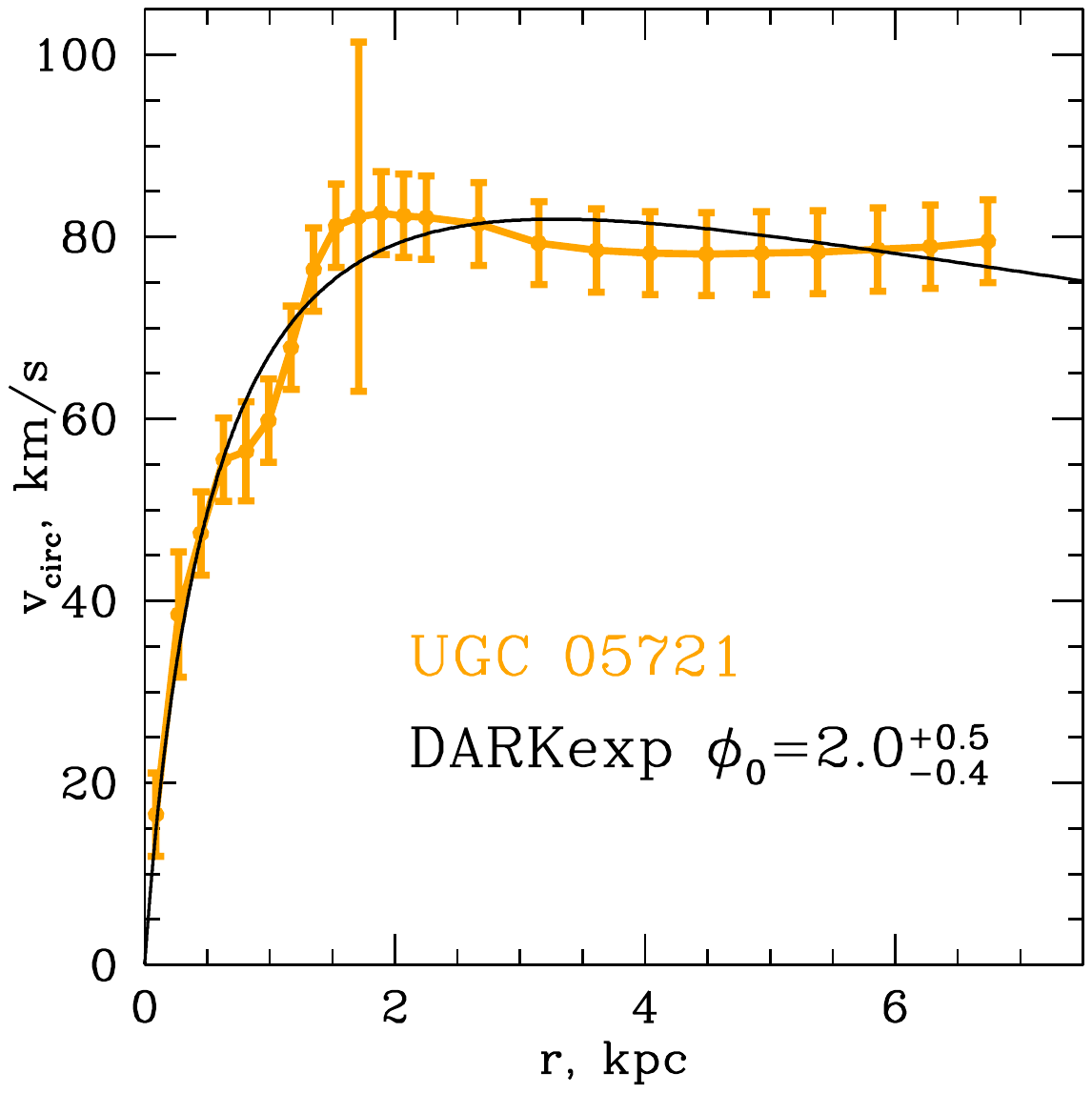}  
        \vskip-1.65cm
    \includegraphics[width=0.237\linewidth]{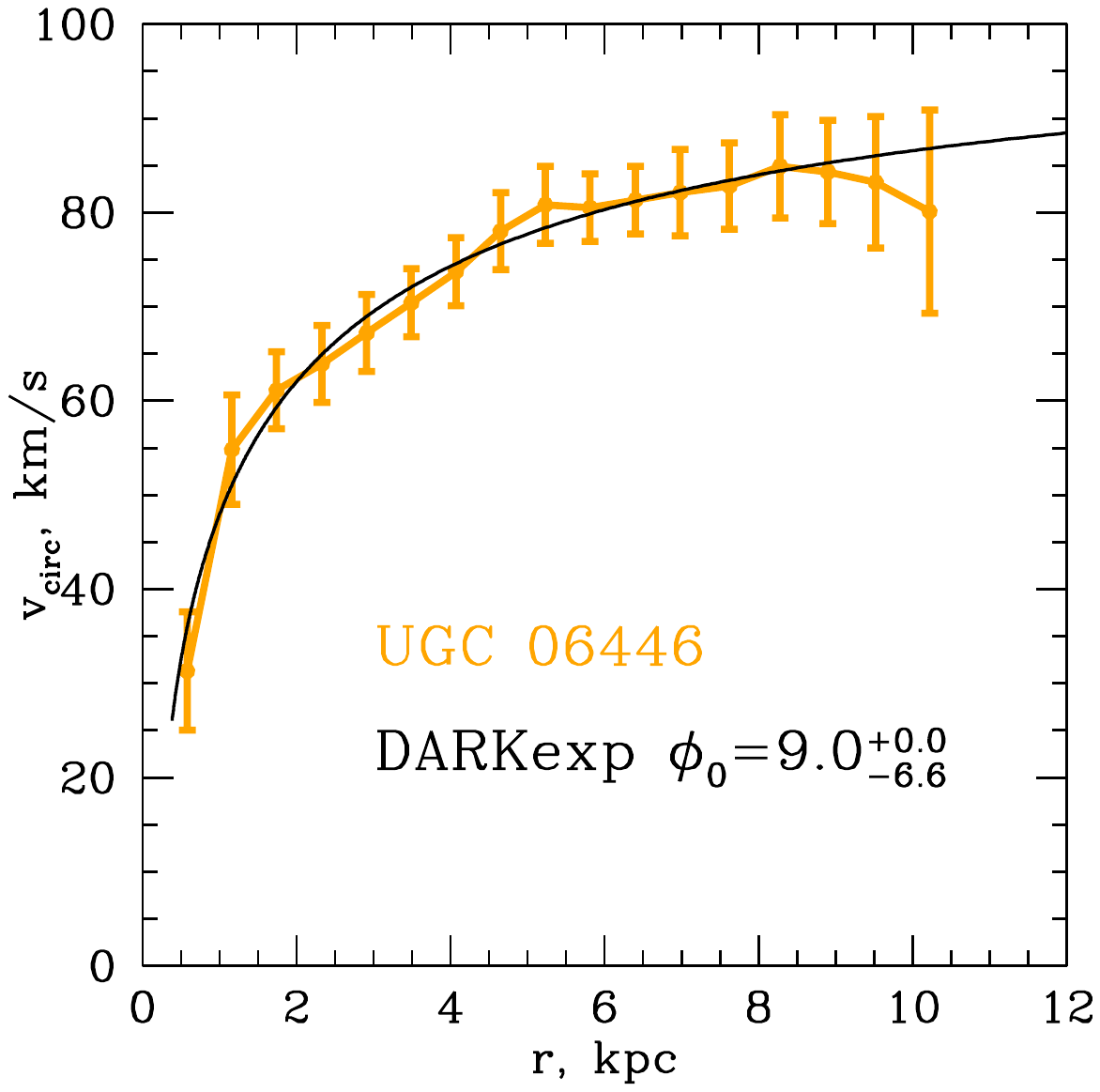}
    \includegraphics[width=0.237\linewidth]{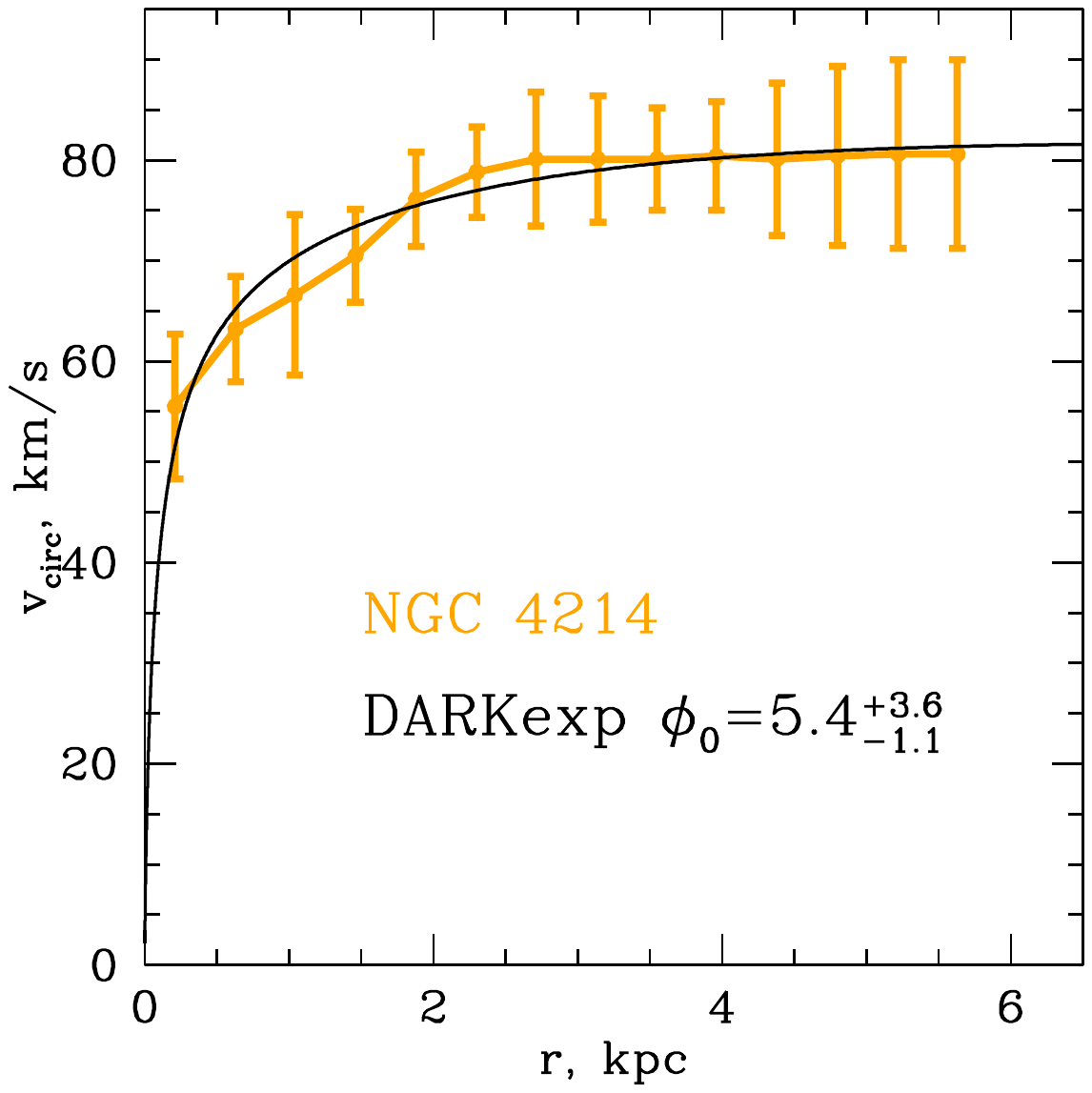}
    \includegraphics[width=0.237\linewidth]{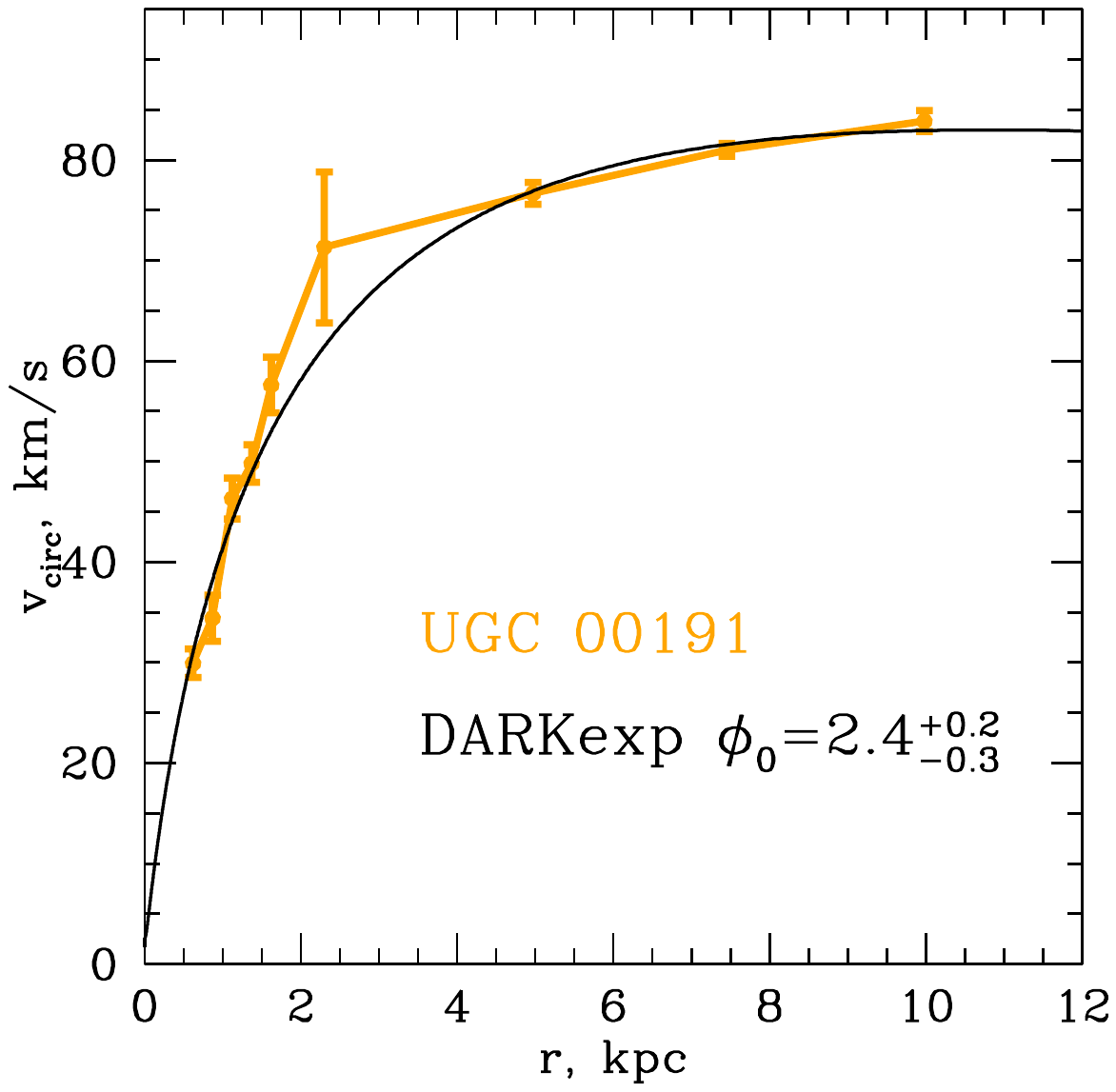}    \includegraphics[width=0.237\linewidth]{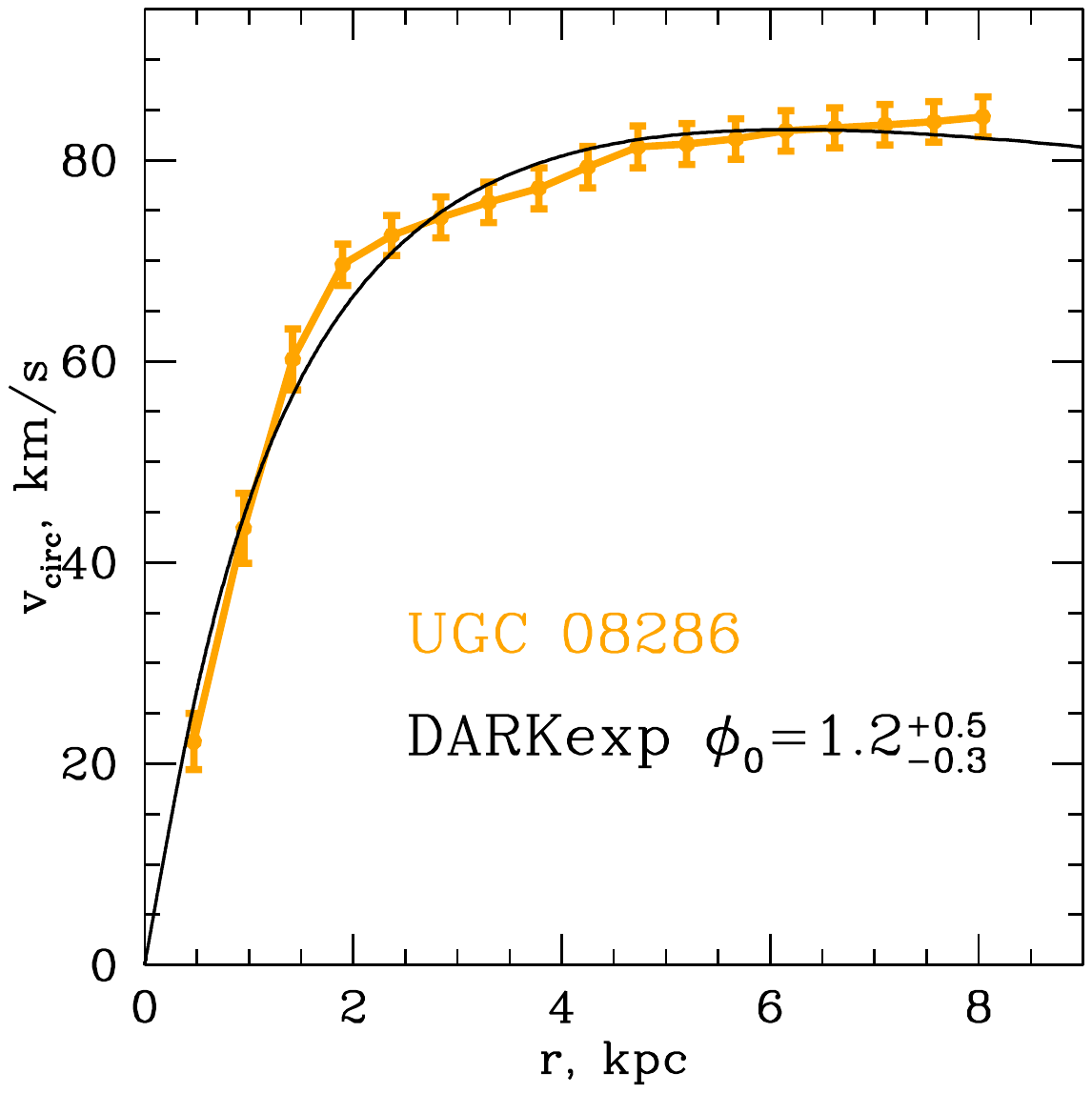}
            \vskip-1.65cm
    \includegraphics[width=0.237\linewidth]{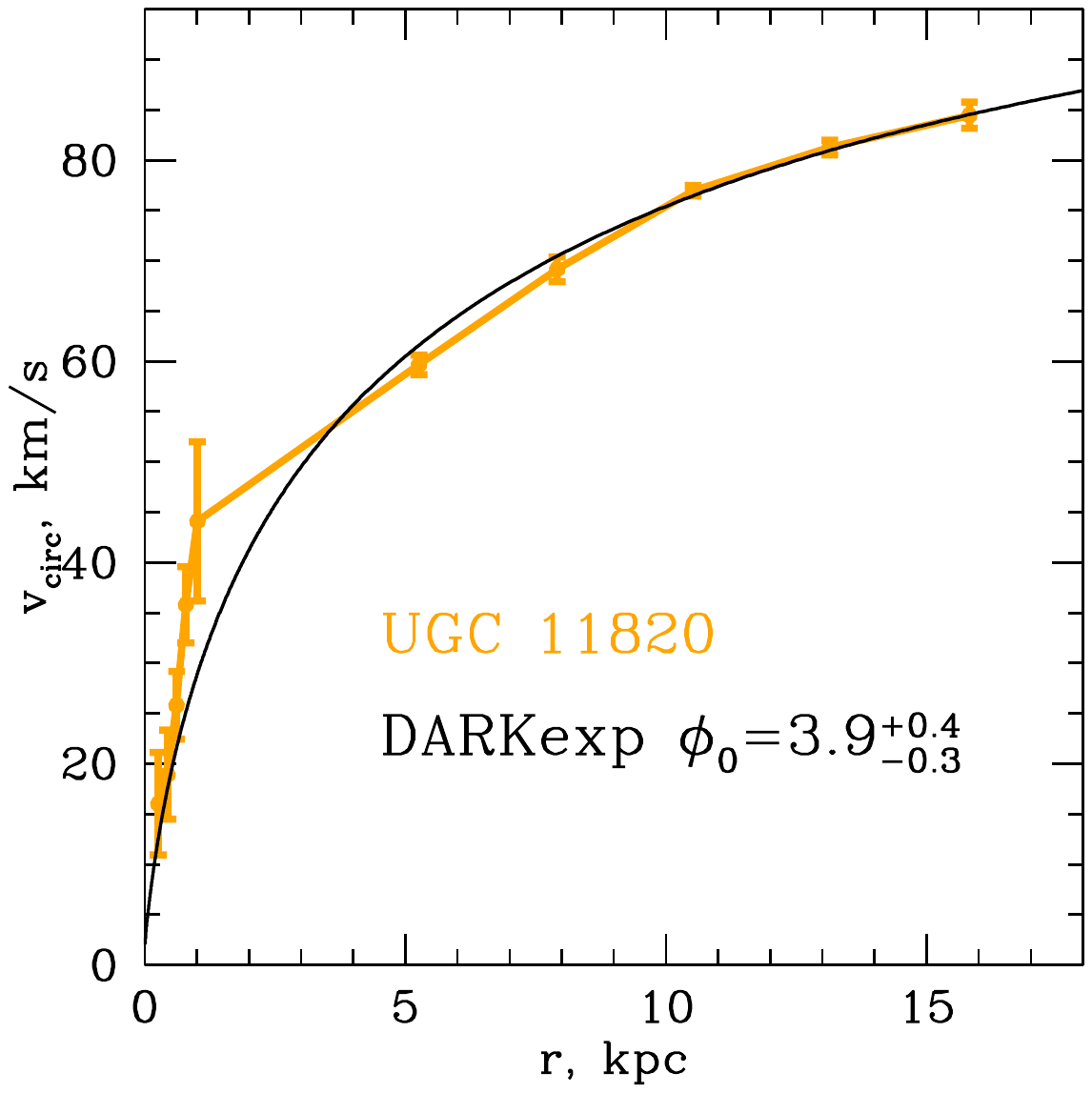}
    \includegraphics[width=0.237\linewidth]{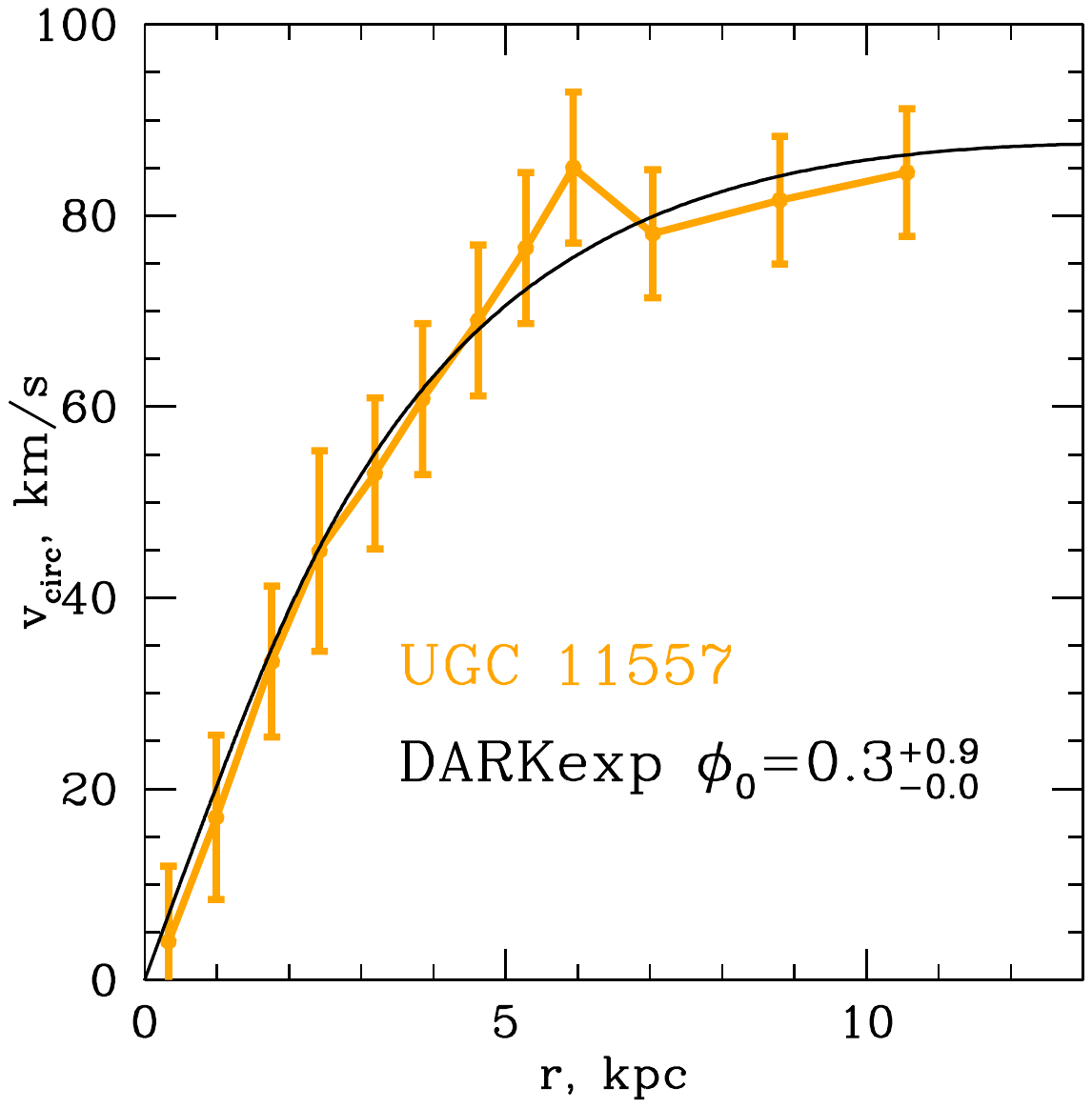}
    \includegraphics[width=0.237\linewidth]{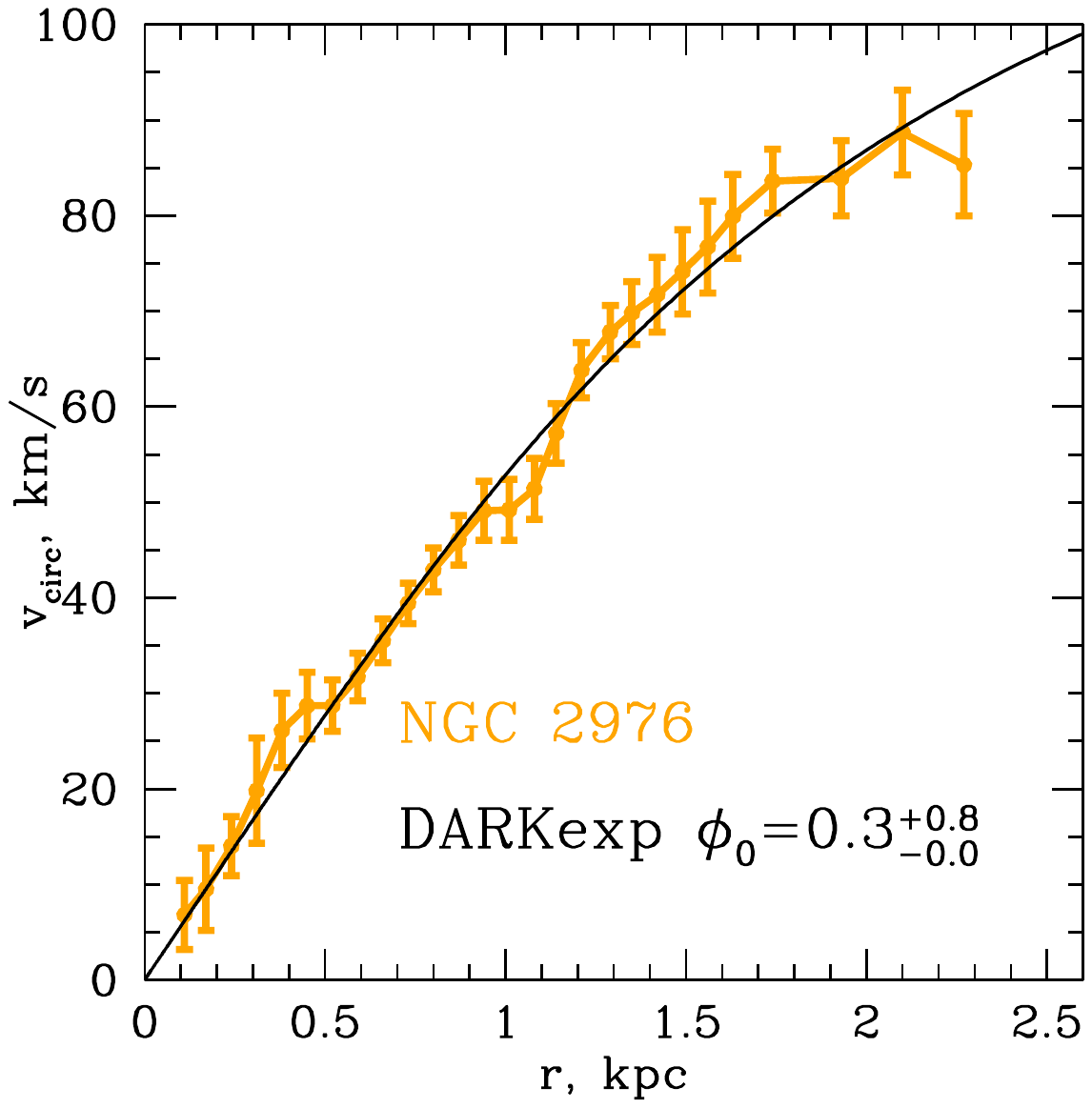}
    \includegraphics[width=0.237\linewidth]{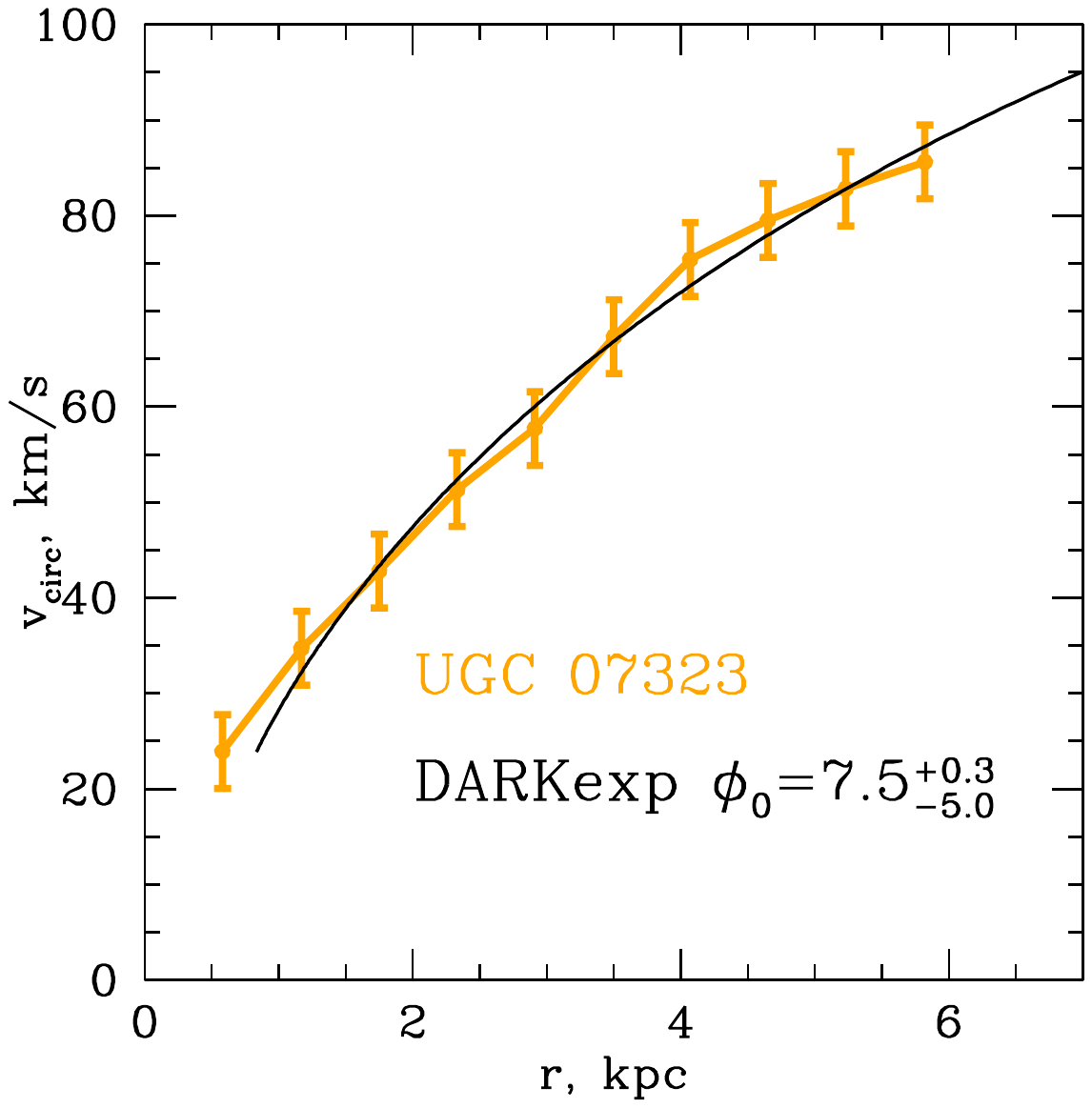}
         \vskip-1.65cm
    \includegraphics[width=0.237\linewidth]{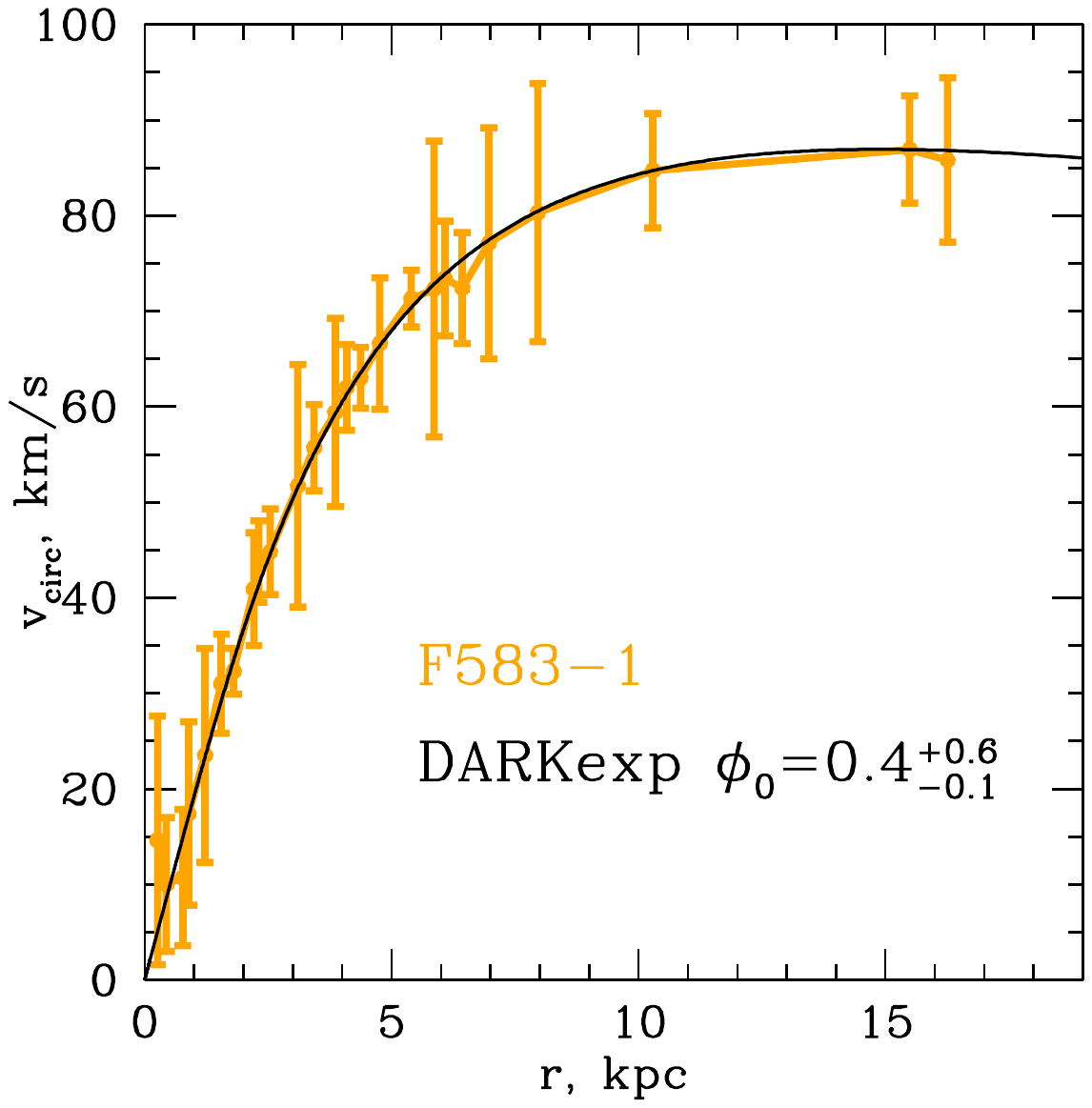}
    \includegraphics[width=0.237\linewidth]{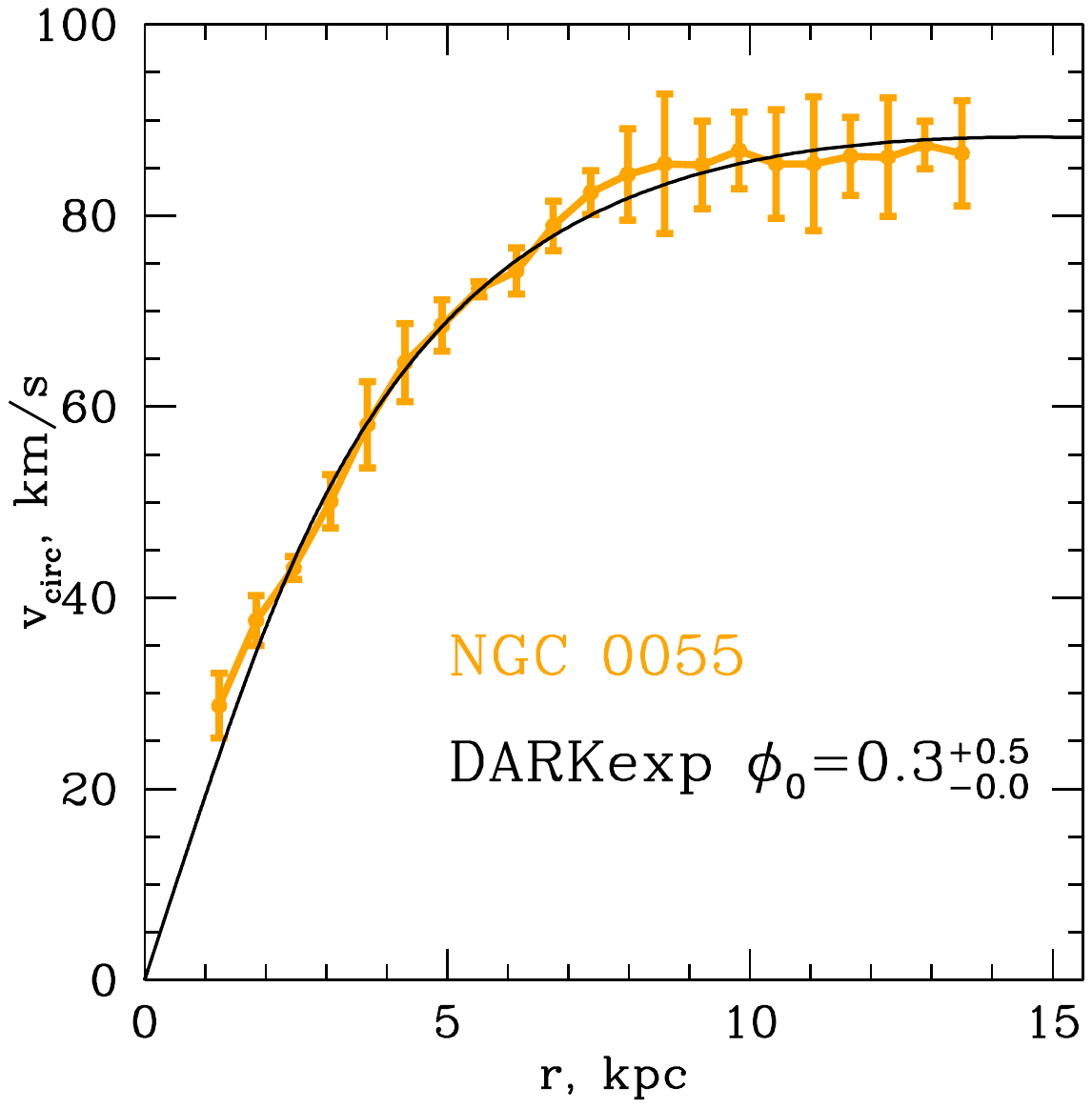}
    \includegraphics[width=0.237\linewidth]{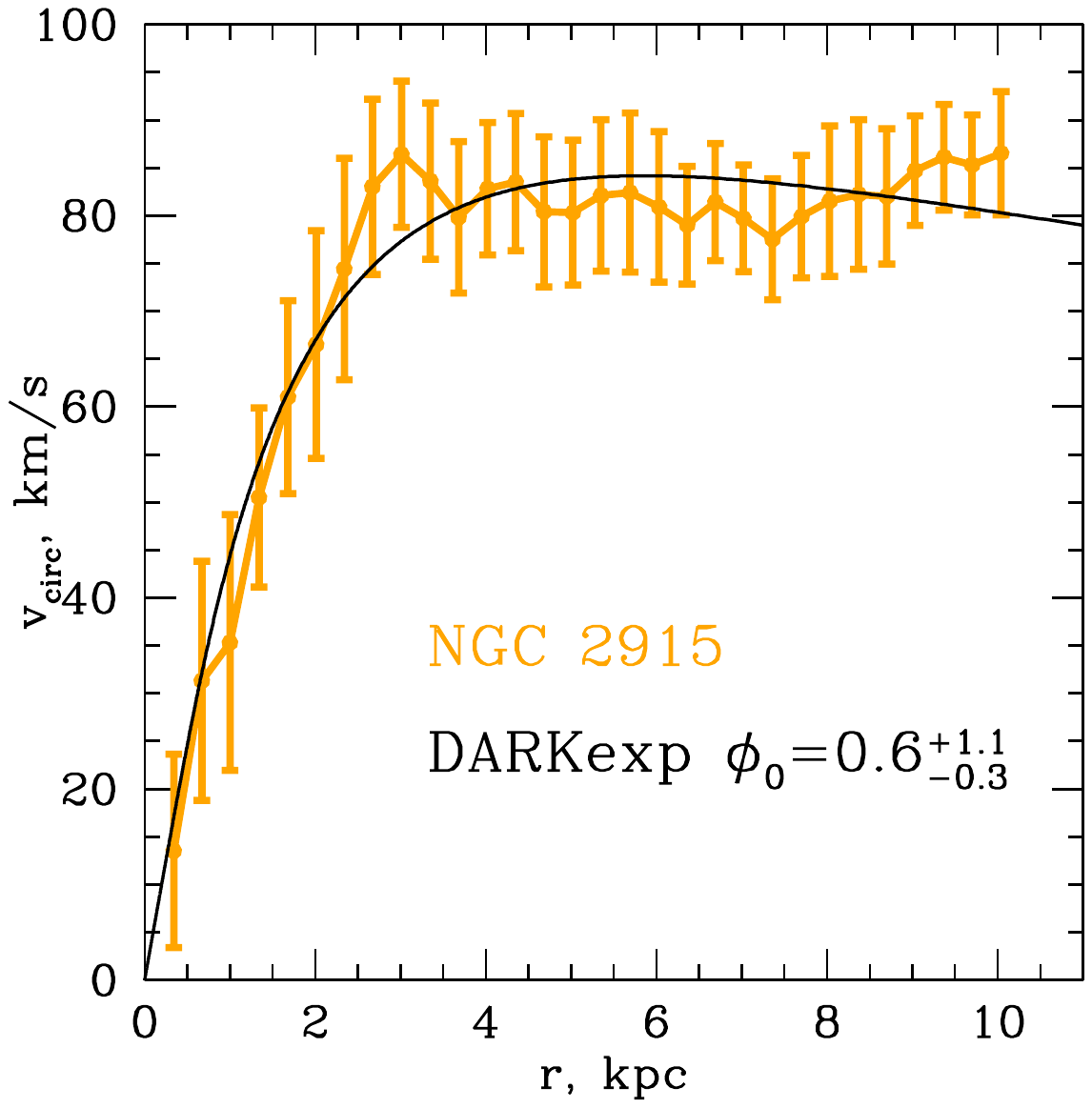}
    \includegraphics[width=0.237\linewidth]{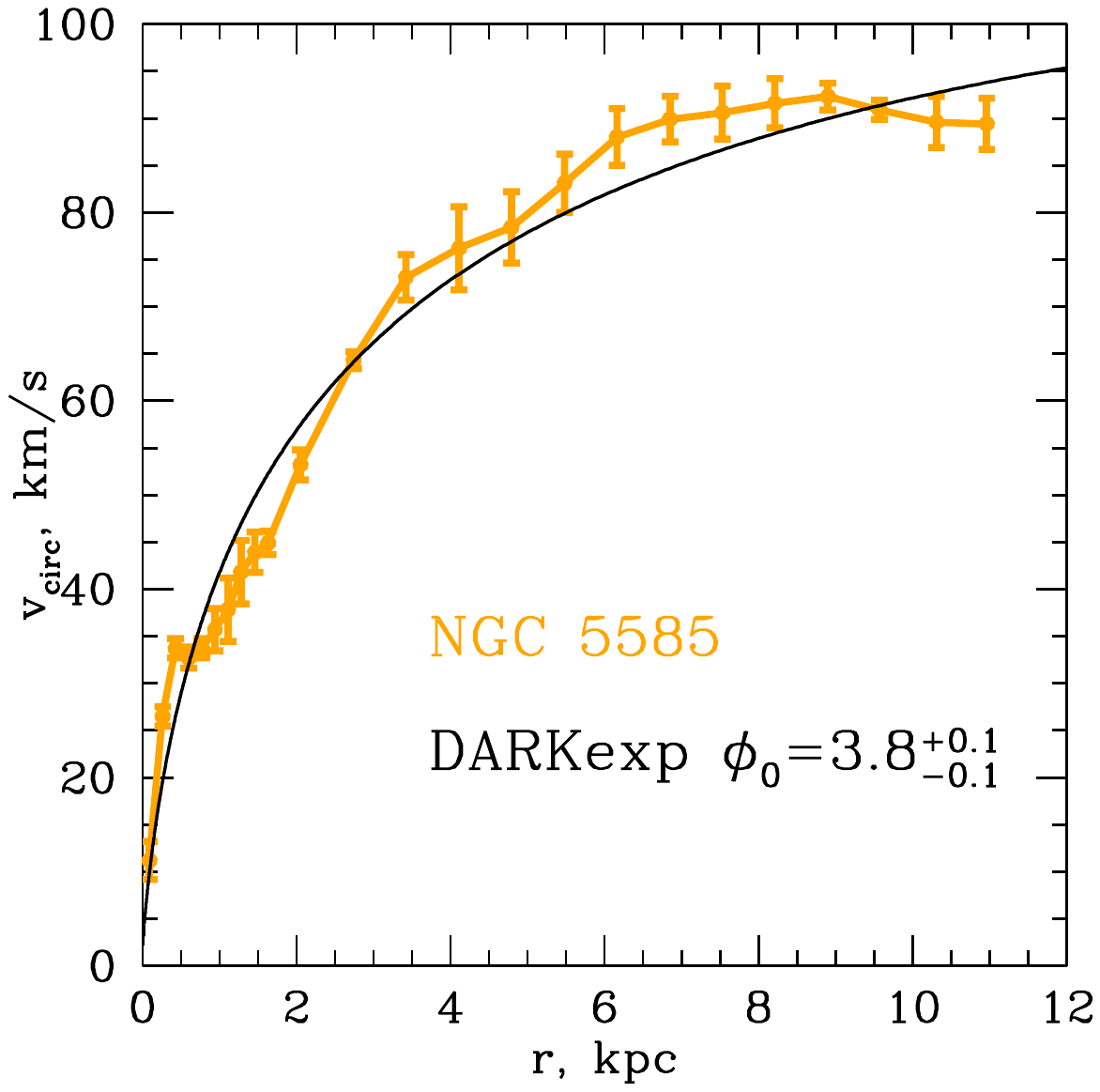}   
    \vskip-0.85cm
    \caption{Figure~\ref{fig:rotcurve1} continued. }
    \label{fig:rotcurve2}
\end{figure*}

\begin{figure*}
    \centering
    \vskip-1.75cm
    \includegraphics[width=0.237\linewidth]{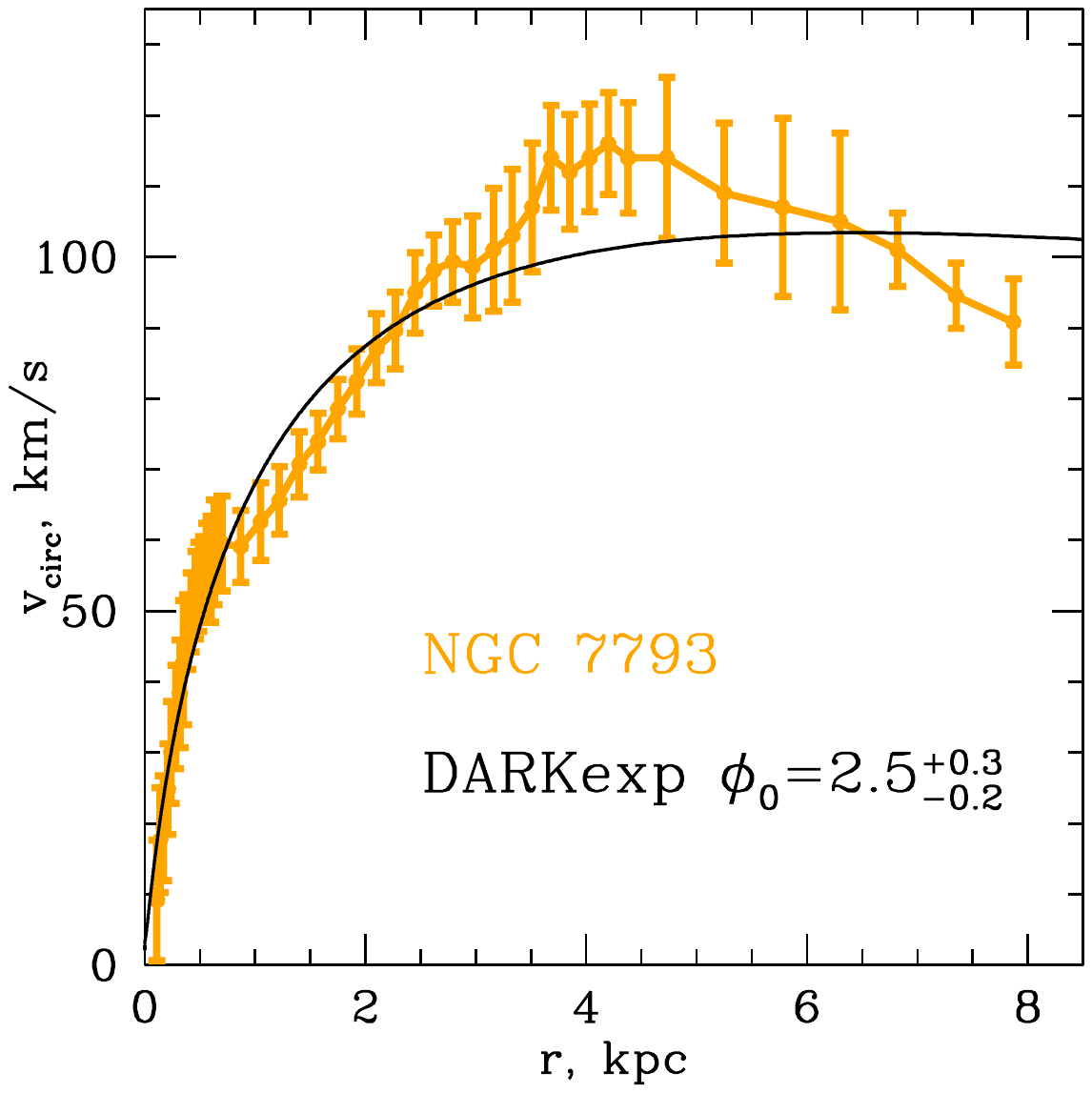}
    \includegraphics[width=0.237\linewidth]{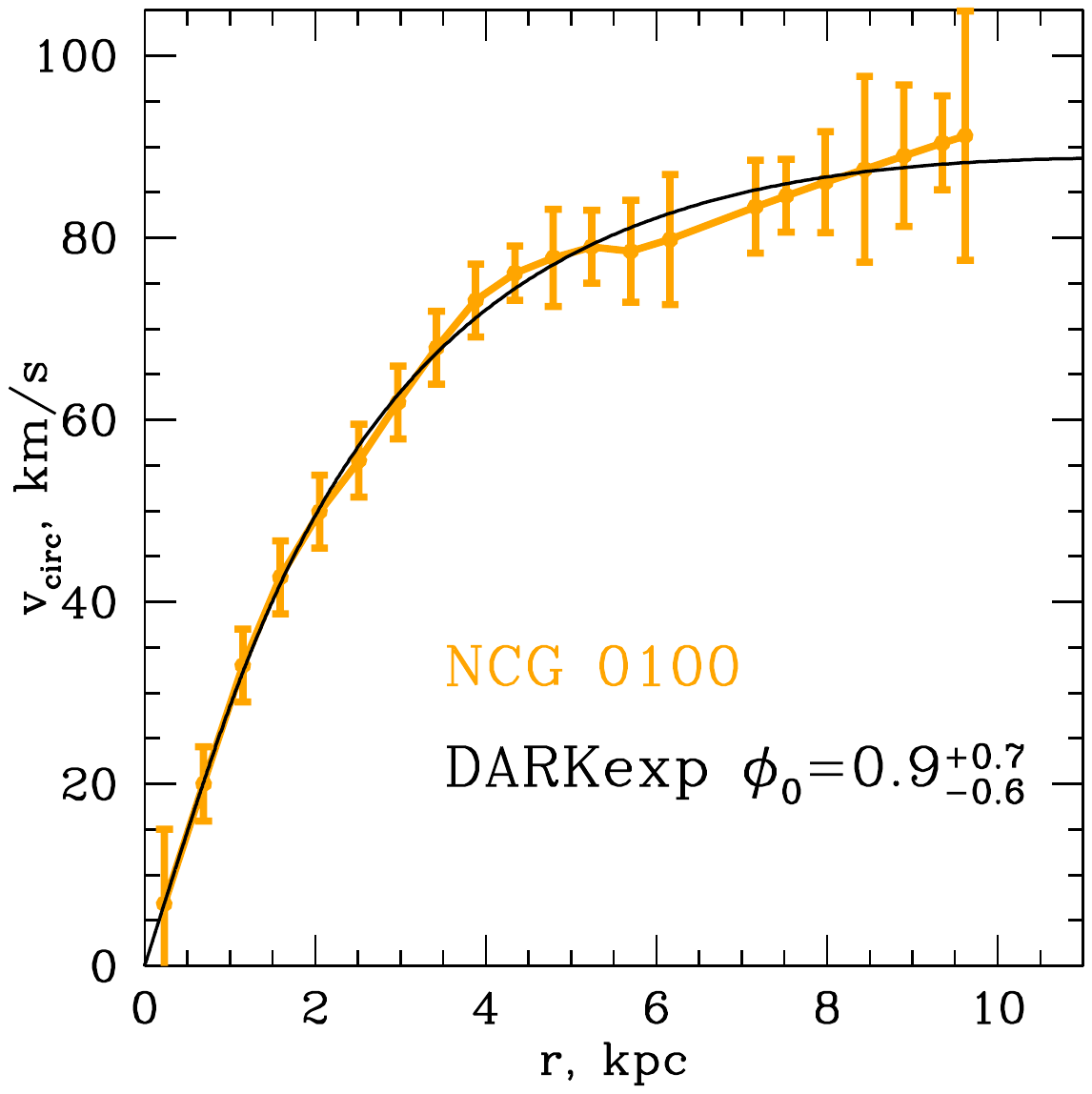}
    \includegraphics[width=0.237\linewidth]{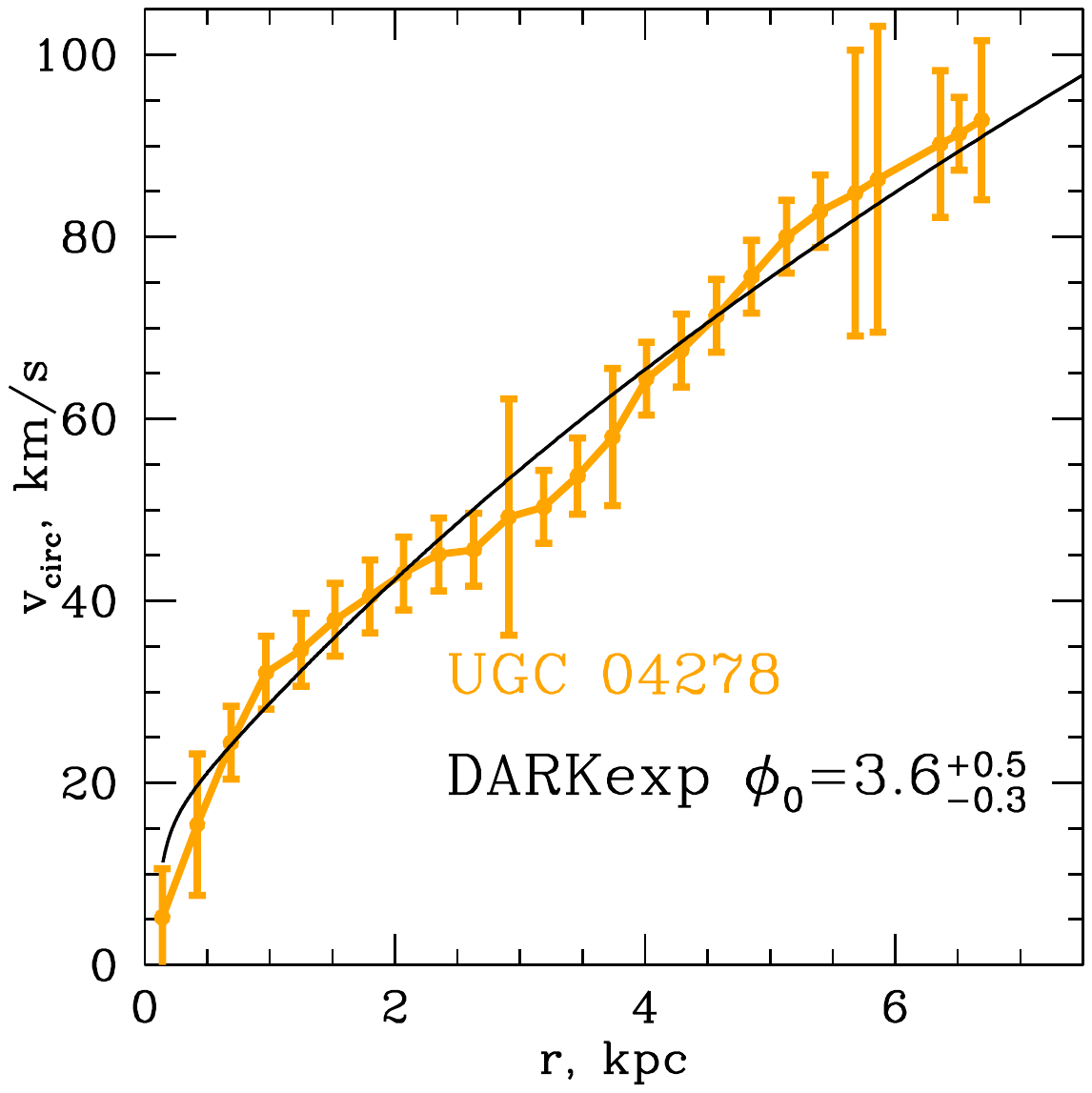}    \includegraphics[width=0.237\linewidth]{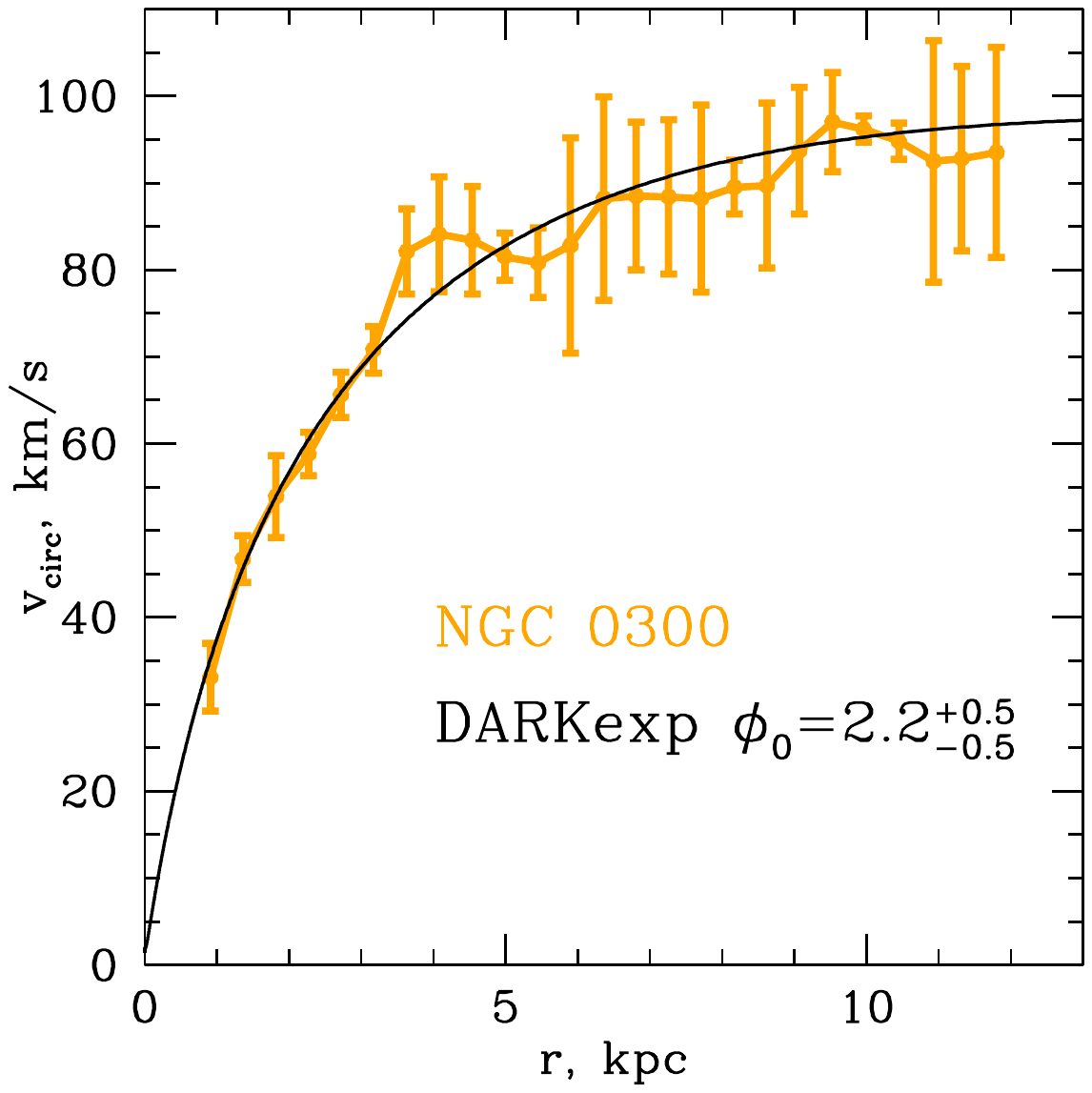}
            \vskip-1.65cm
    \includegraphics[width=0.237\linewidth]{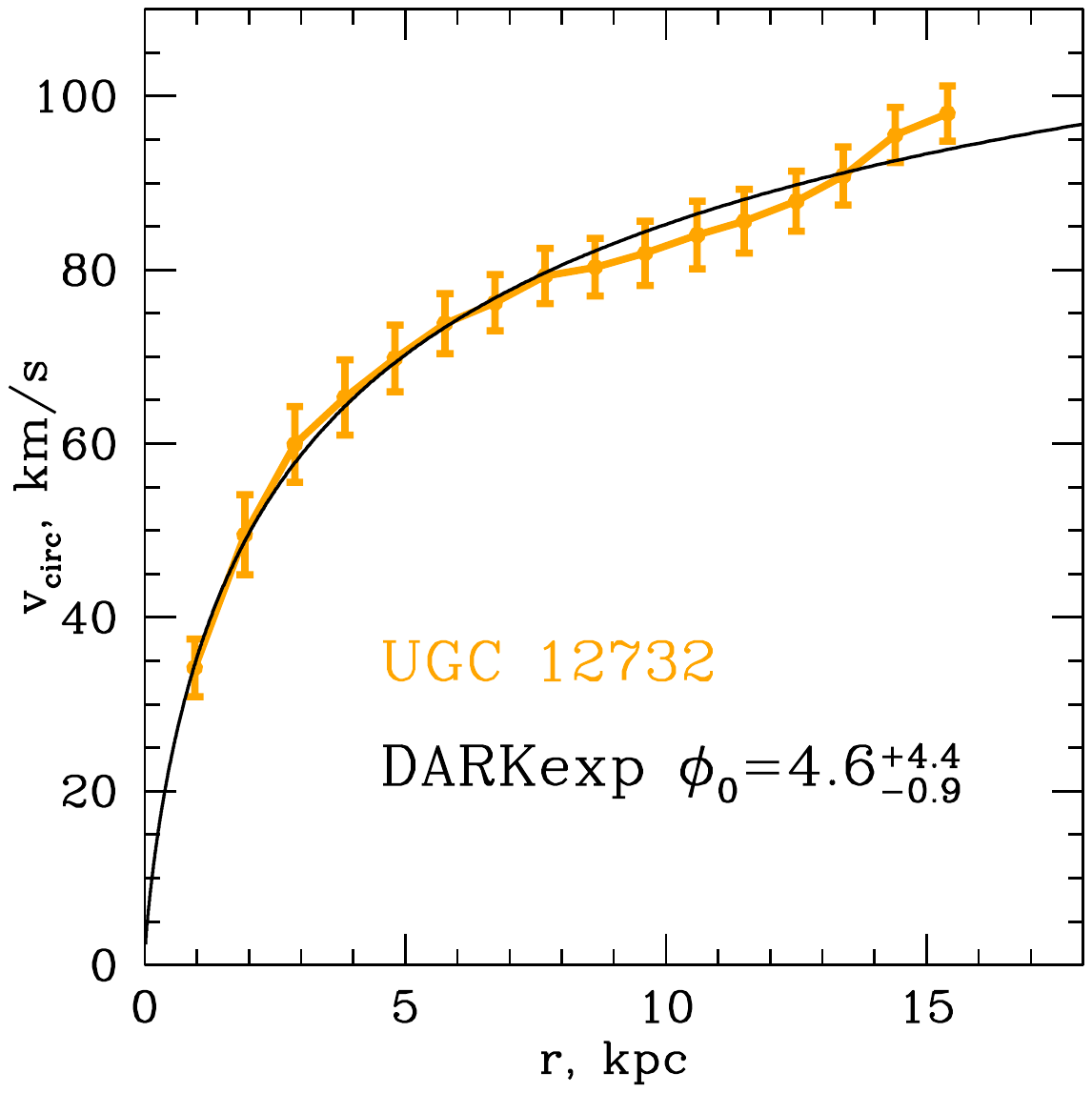}
    \includegraphics[width=0.237\linewidth]{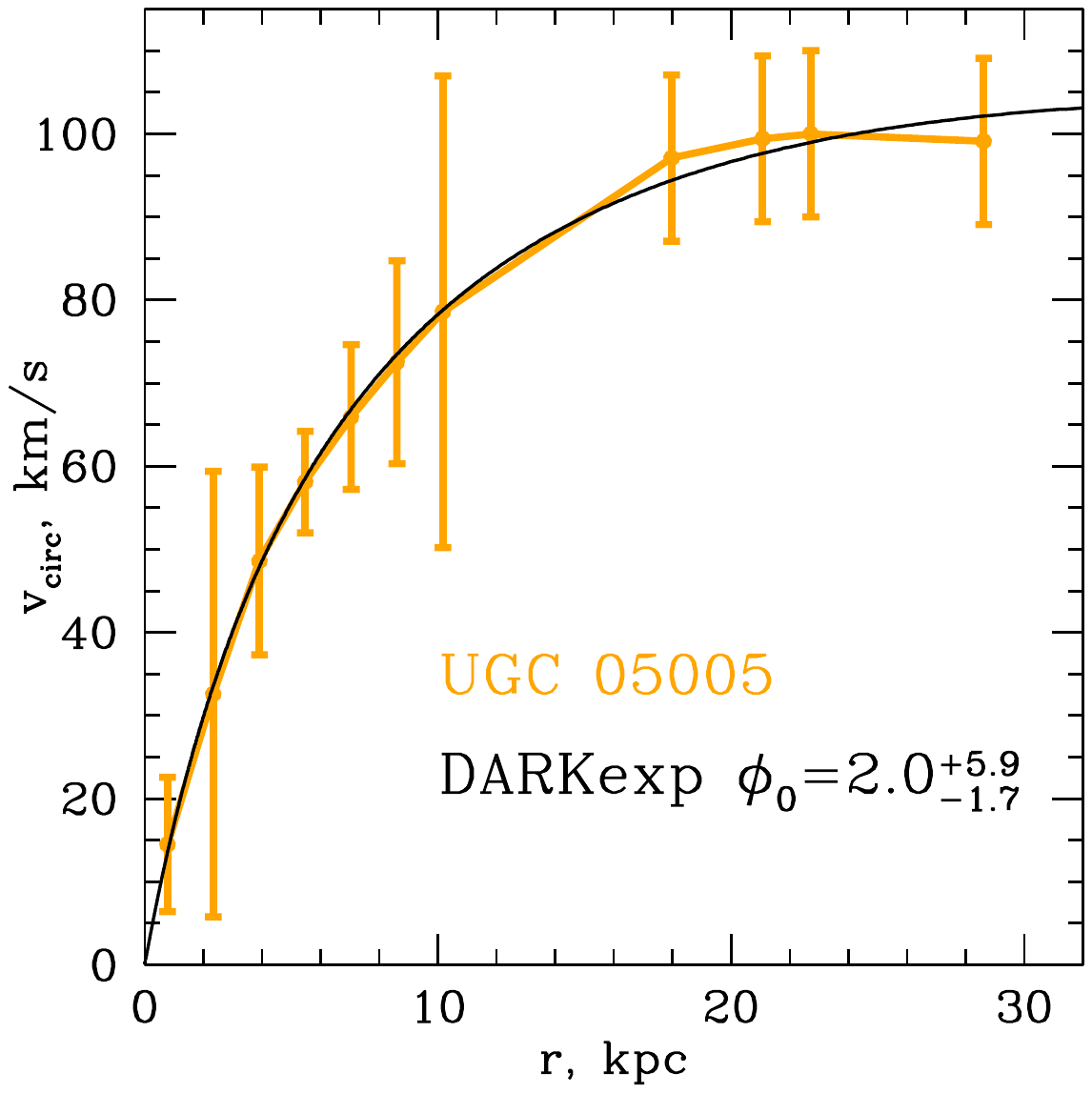}
    \includegraphics[width=0.237\linewidth]{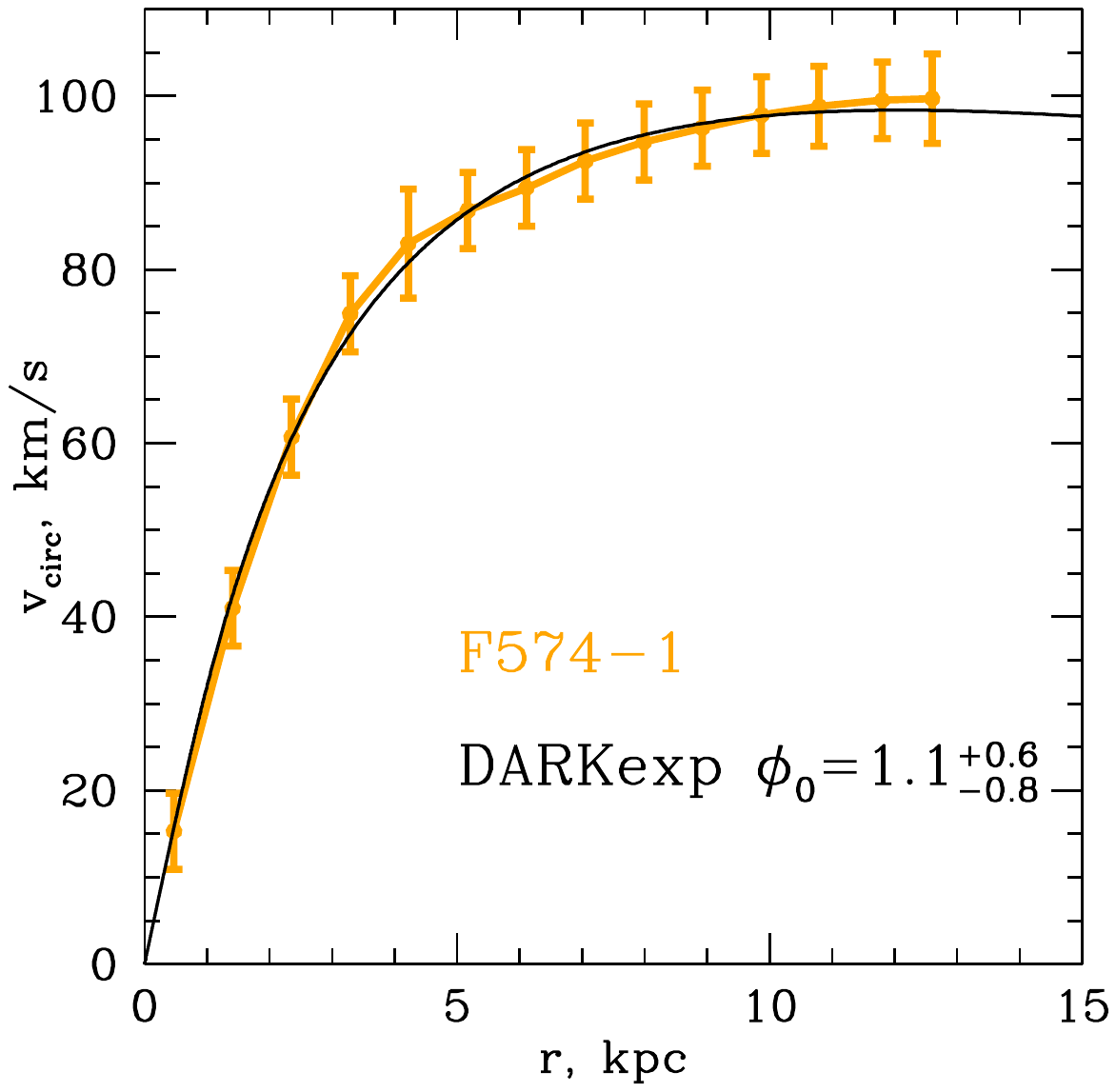}
    \includegraphics[width=0.237\linewidth]{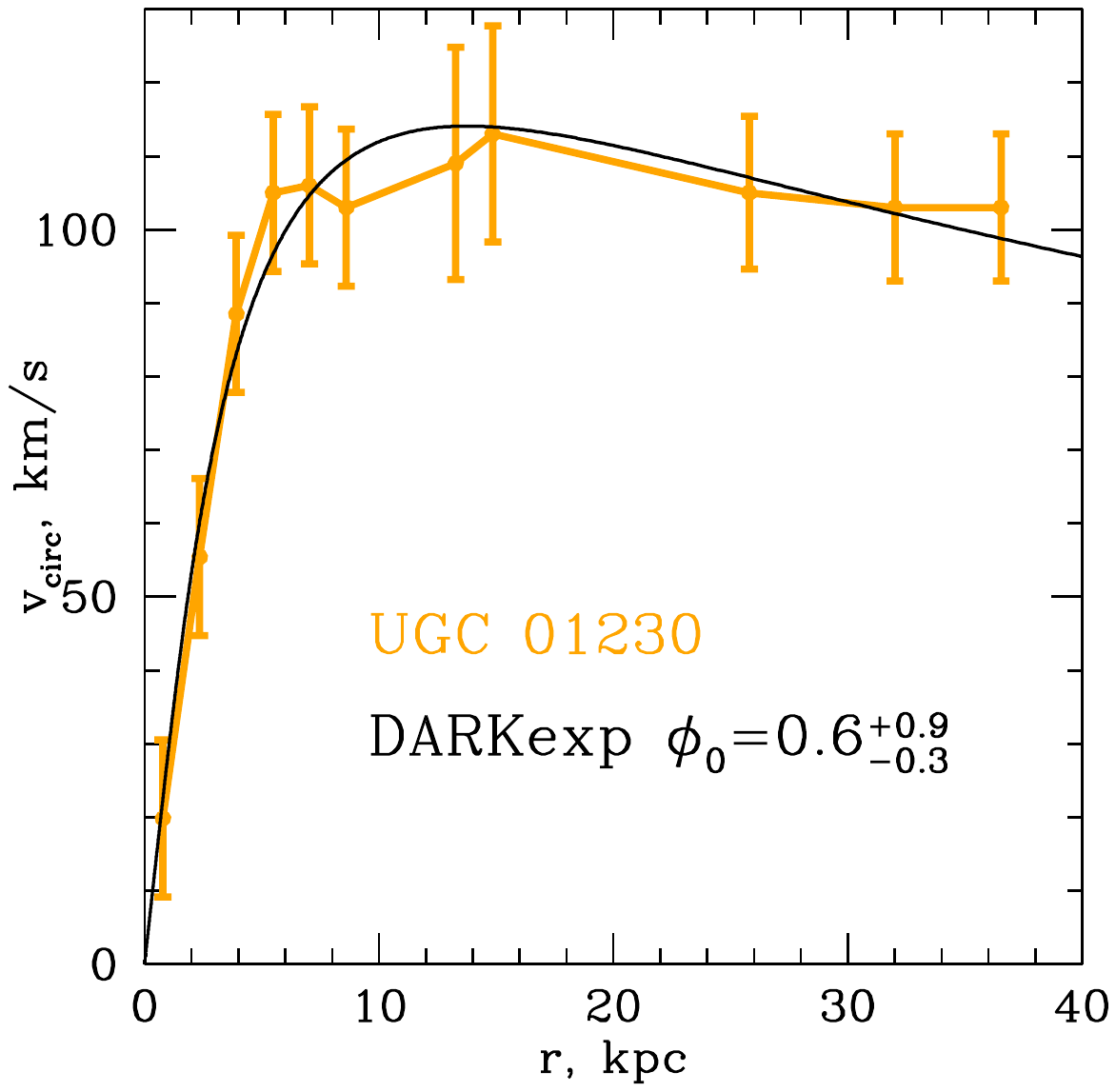}
         \vskip-1.65cm
    \includegraphics[width=0.237\linewidth]{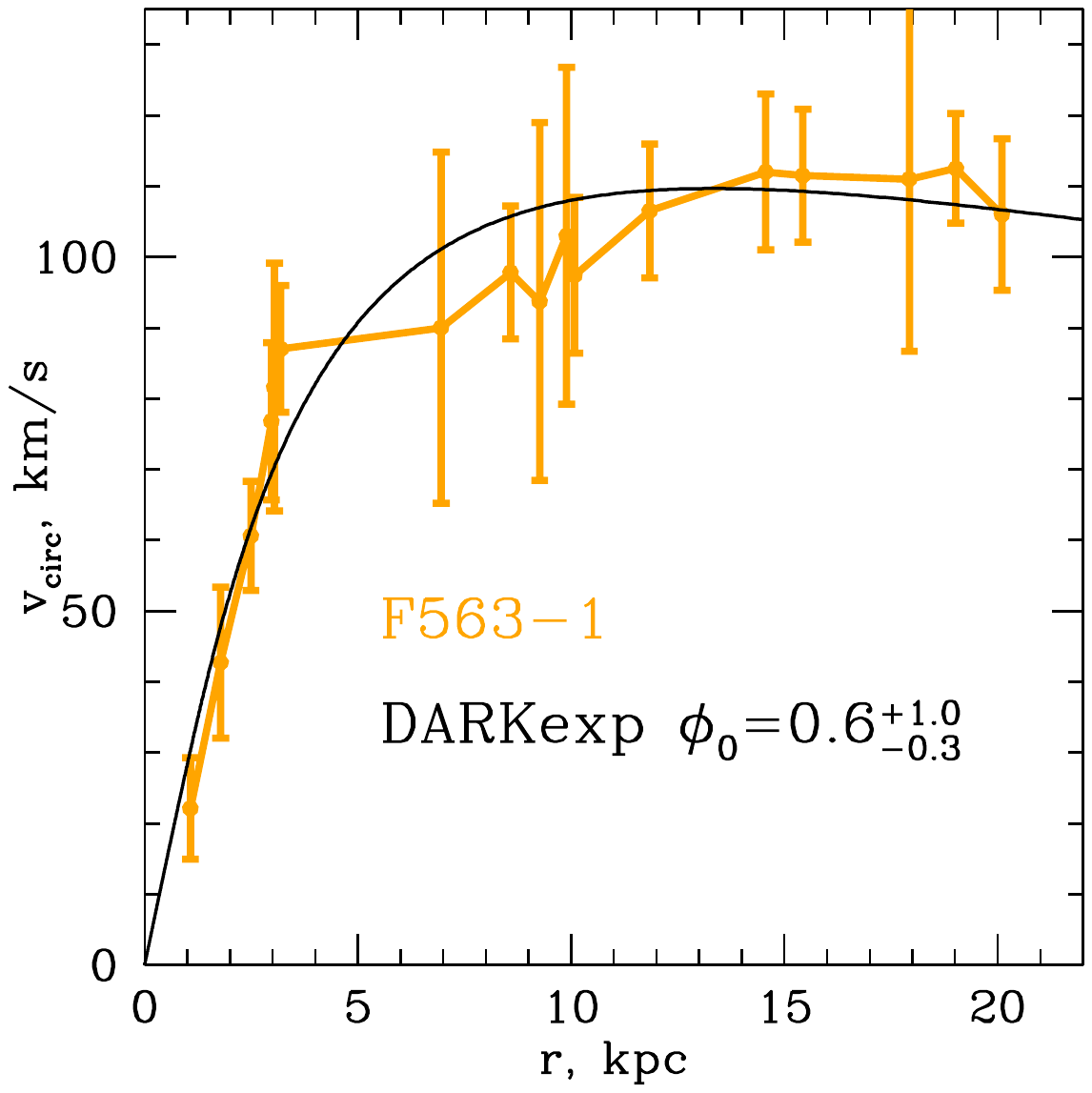}
    \includegraphics[width=0.237\linewidth]{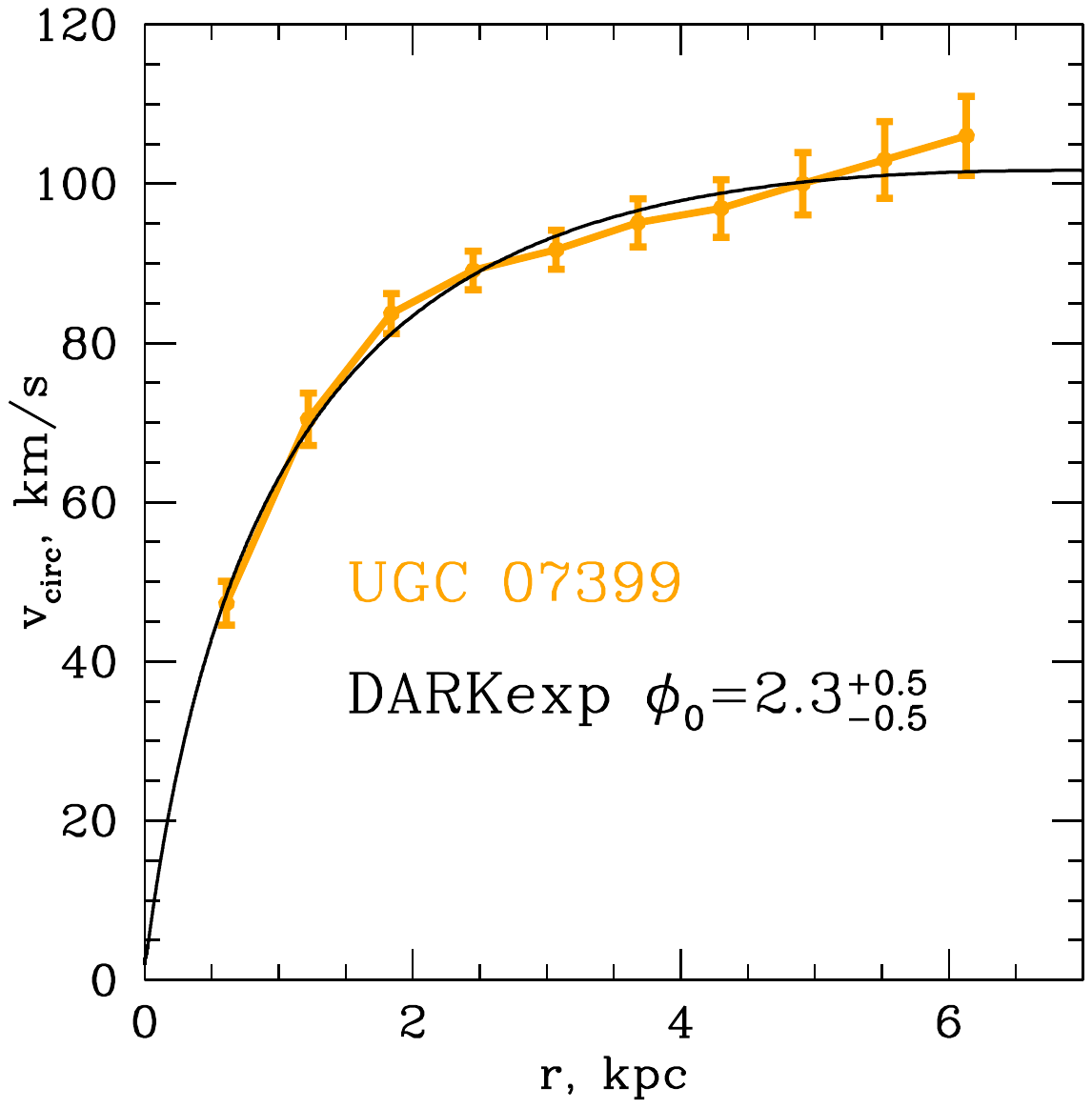}
    \includegraphics[width=0.237\linewidth]{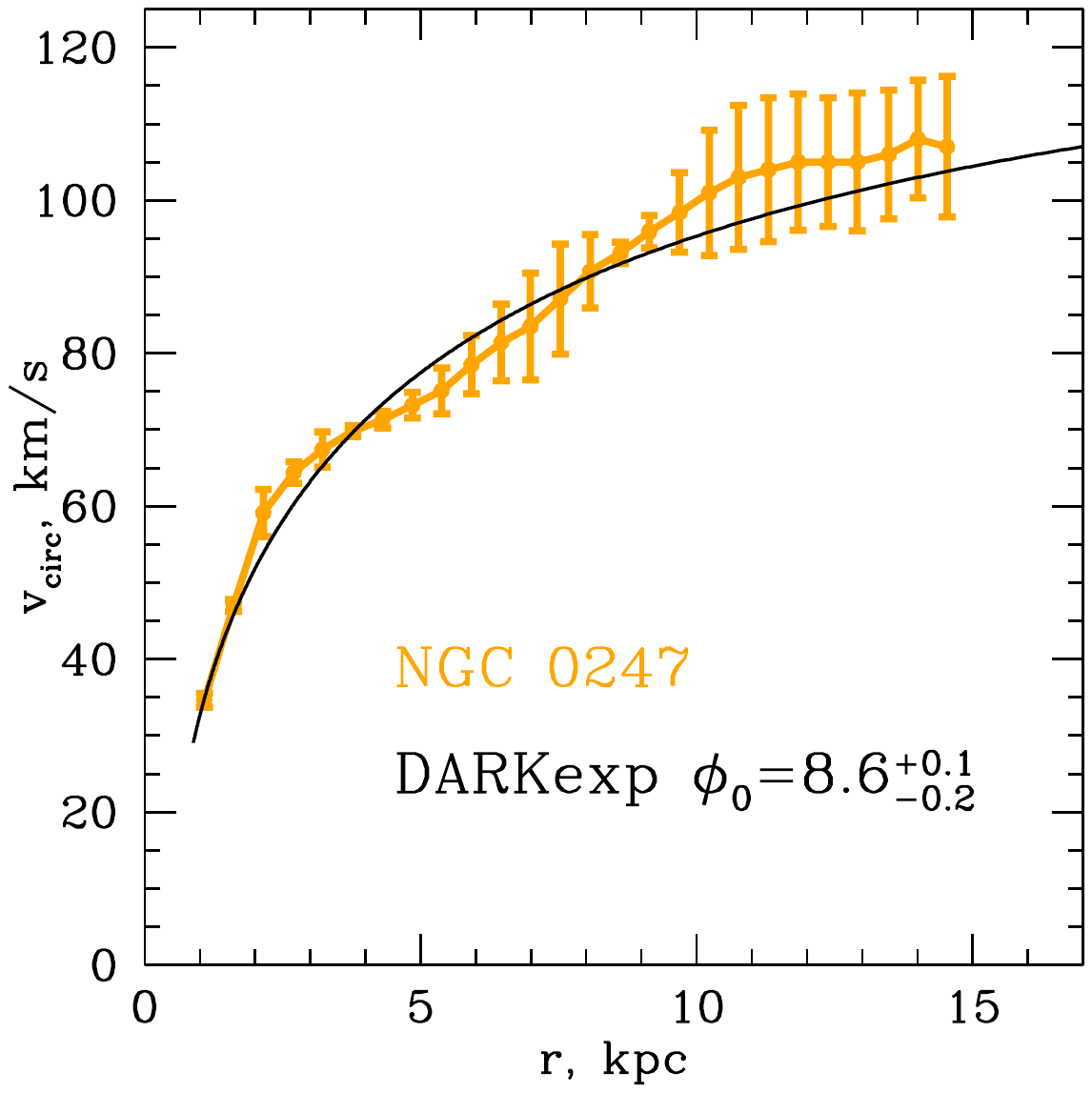}
    \includegraphics[width=0.237\linewidth]{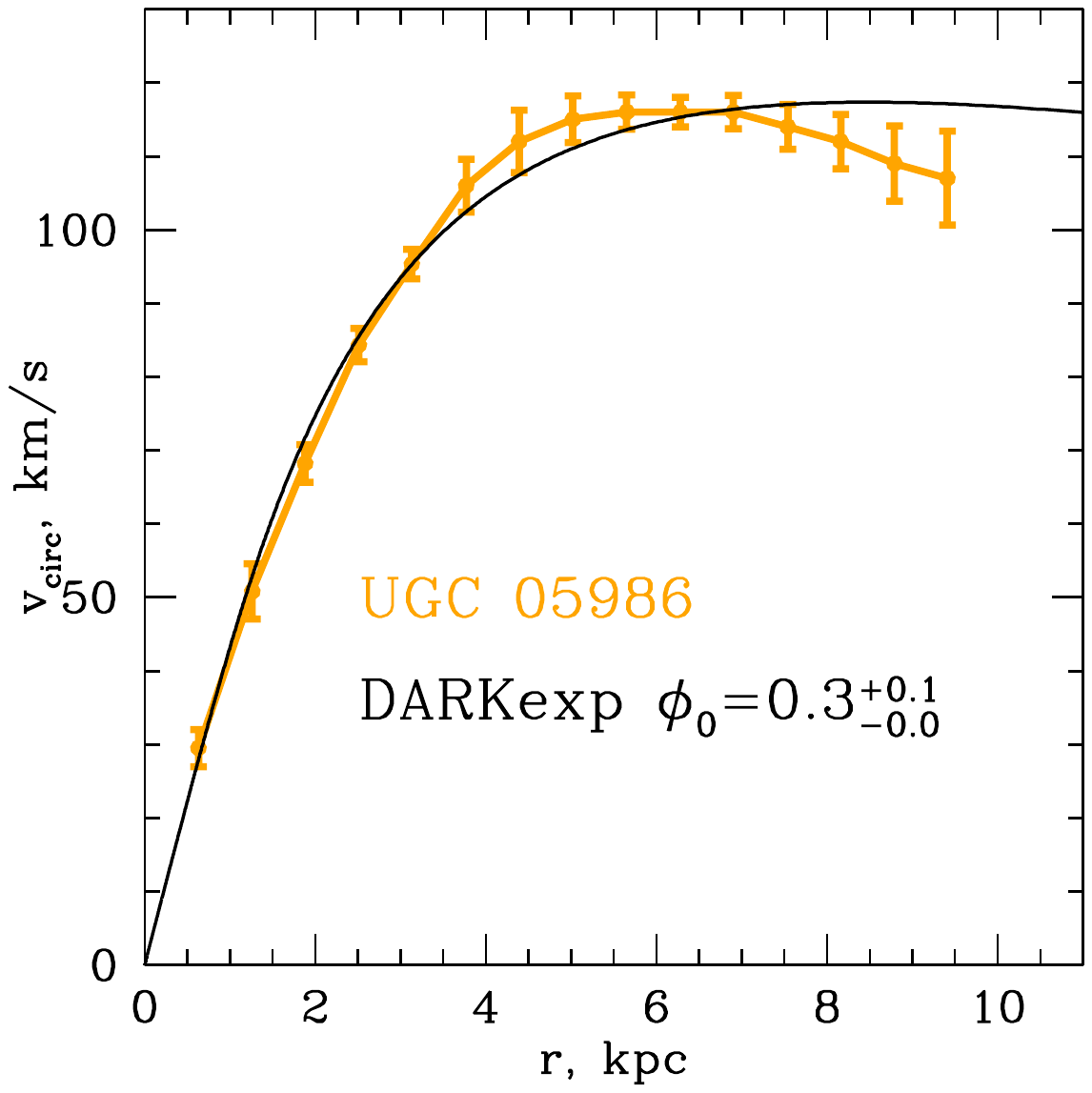}  
        \vskip-1.65cm
    \includegraphics[width=0.237\linewidth]{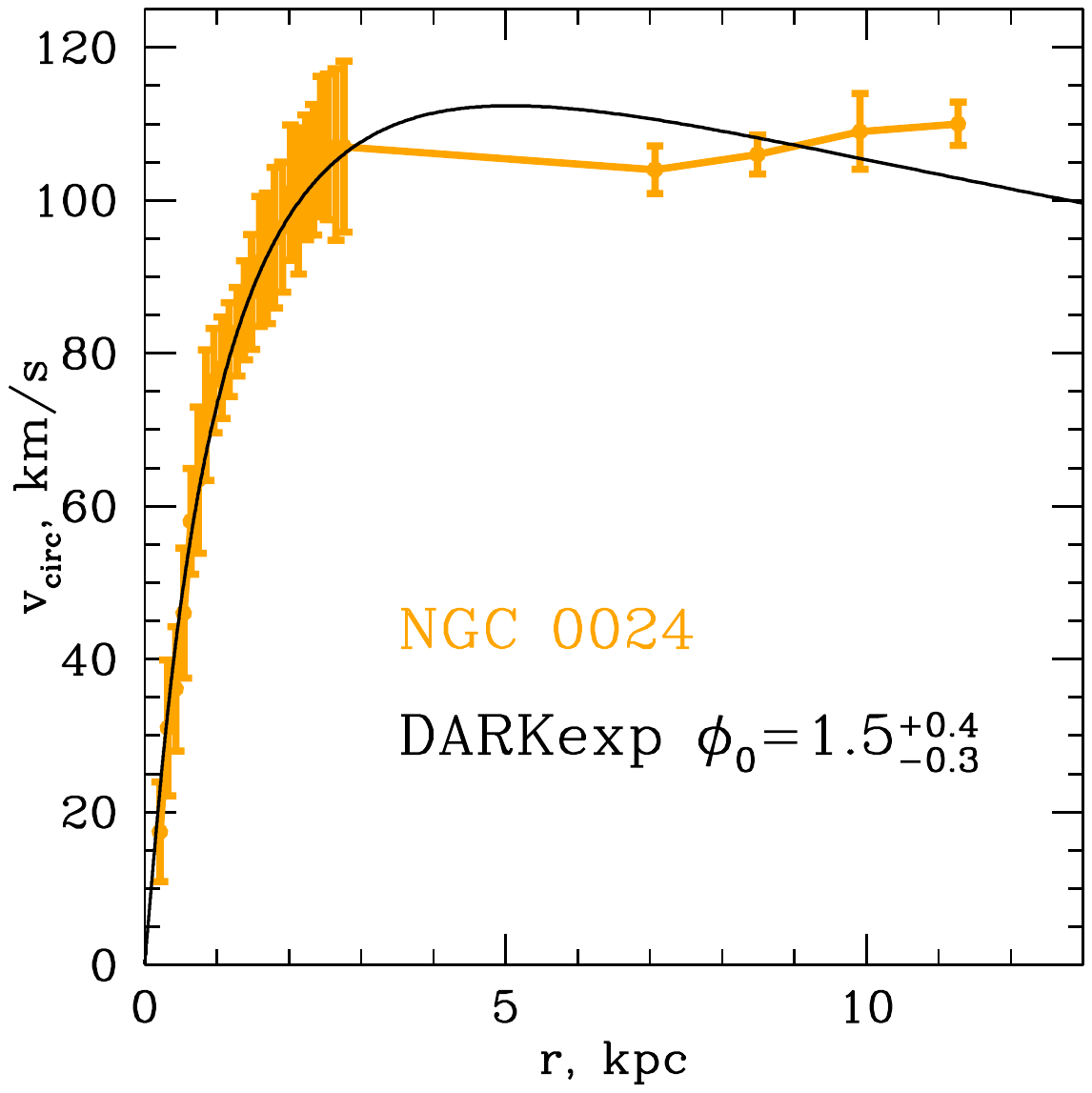}
    \includegraphics[width=0.237\linewidth]{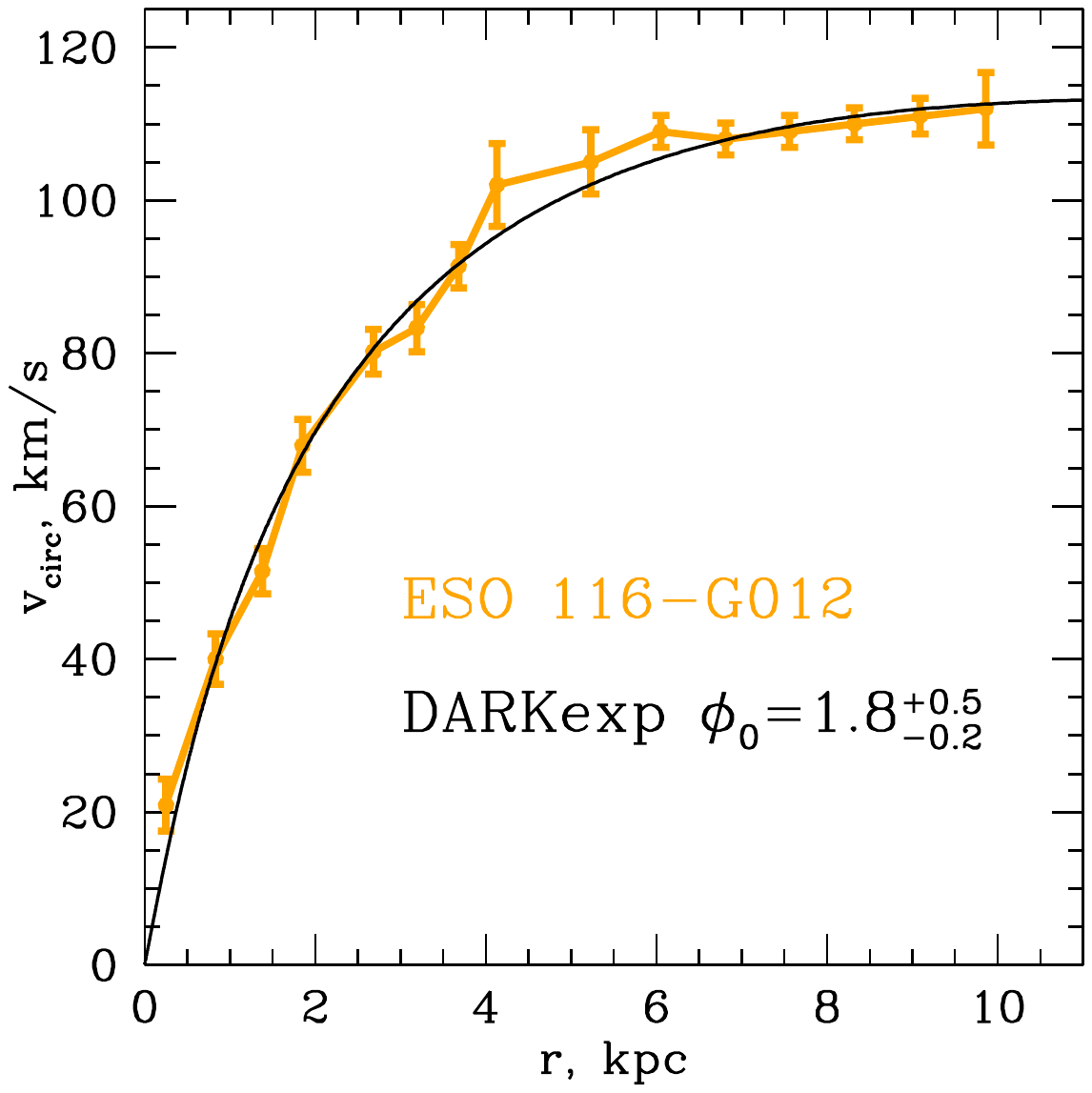}
    \includegraphics[width=0.237\linewidth]{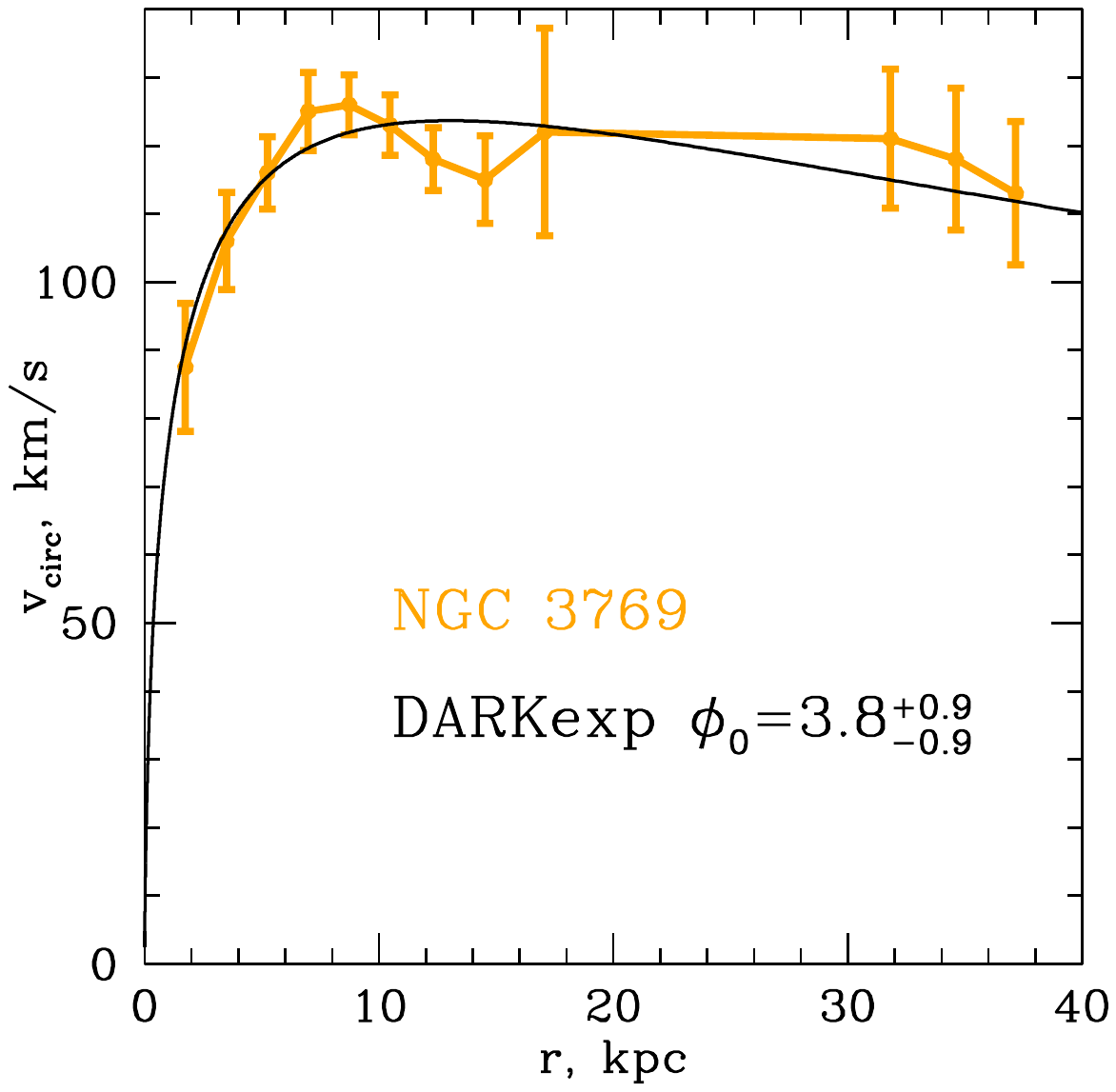}    \includegraphics[width=0.237\linewidth]{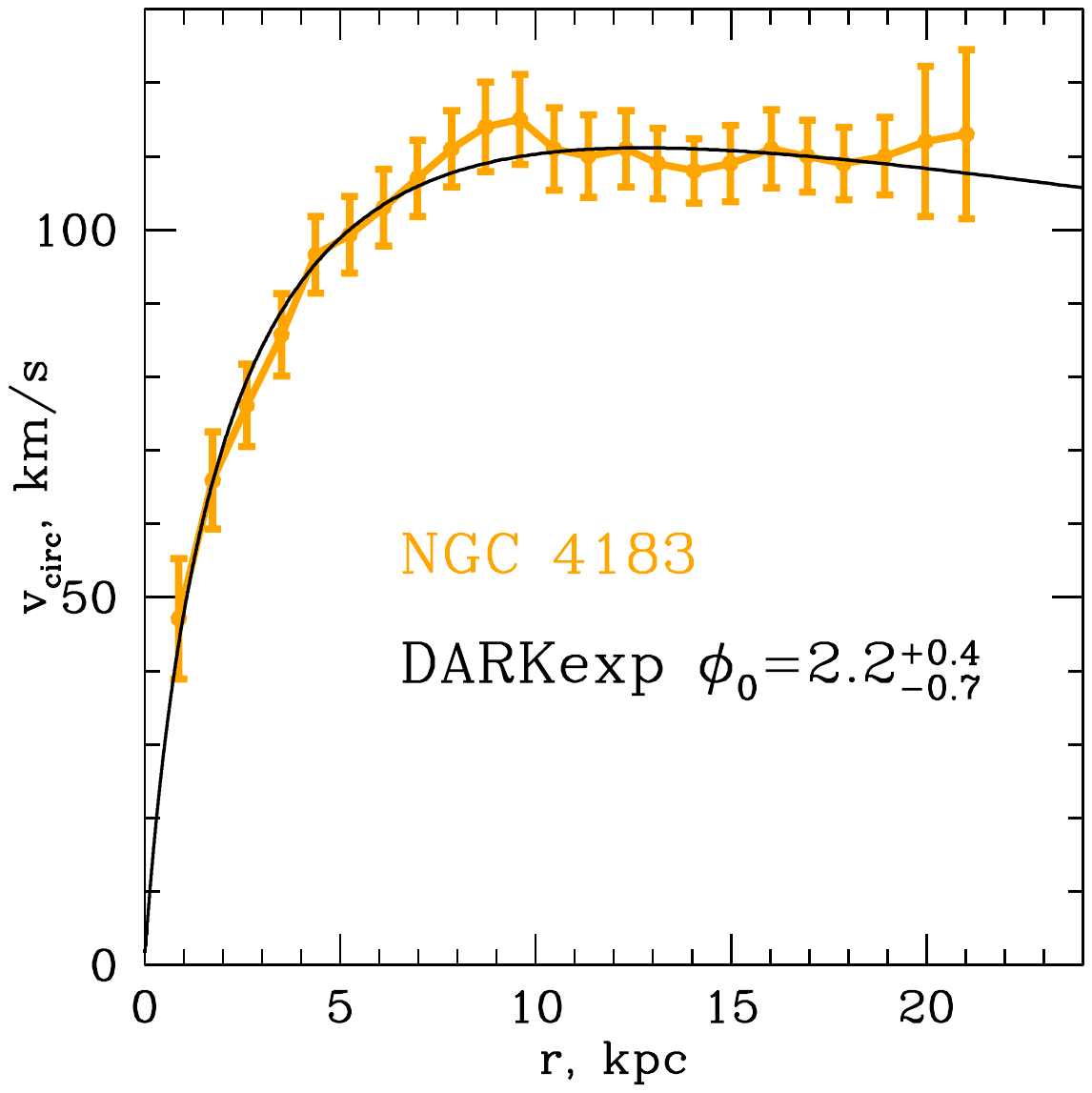}
            \vskip-1.65cm
    \includegraphics[width=0.237\linewidth]{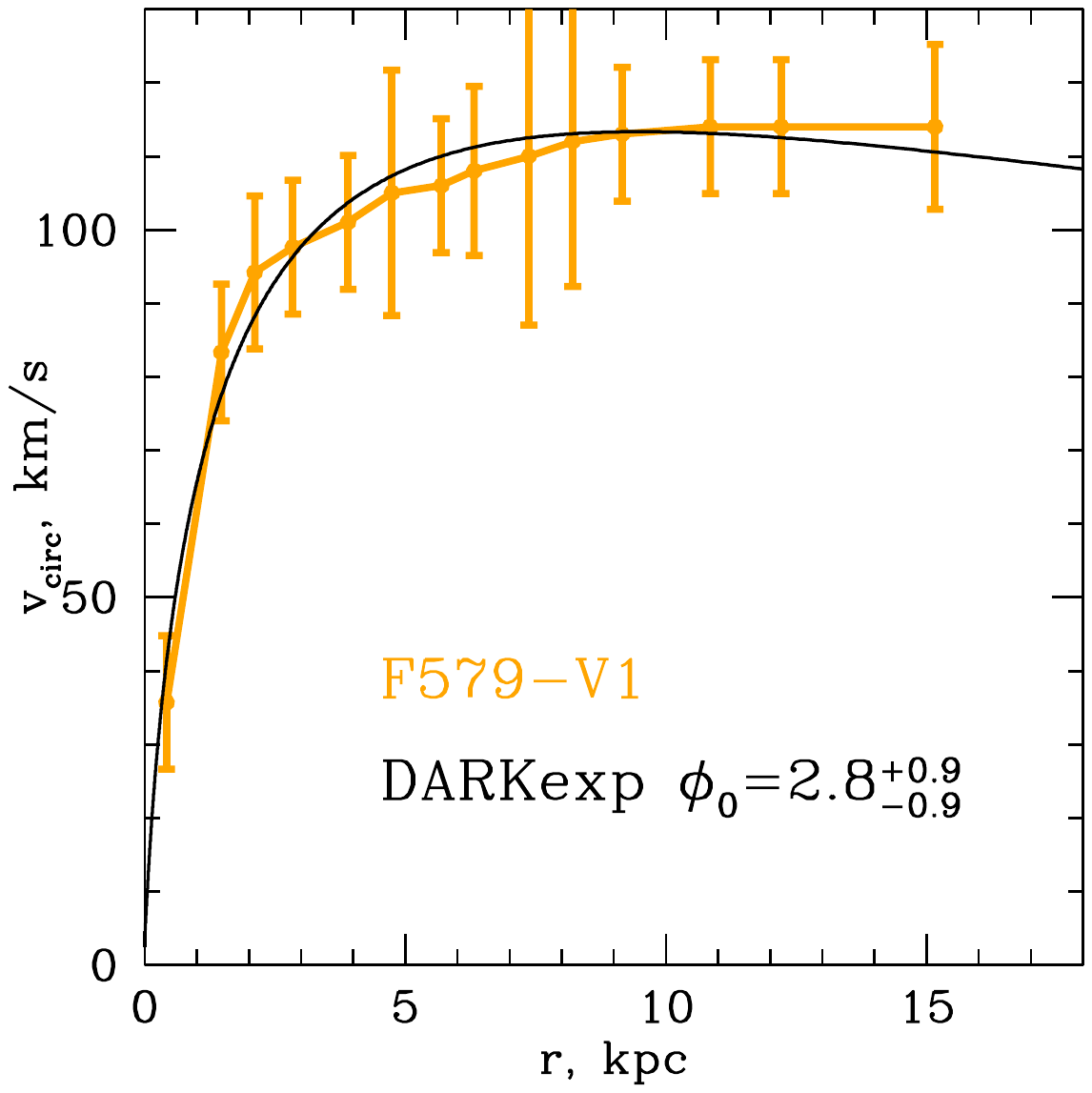}
    \includegraphics[width=0.237\linewidth]{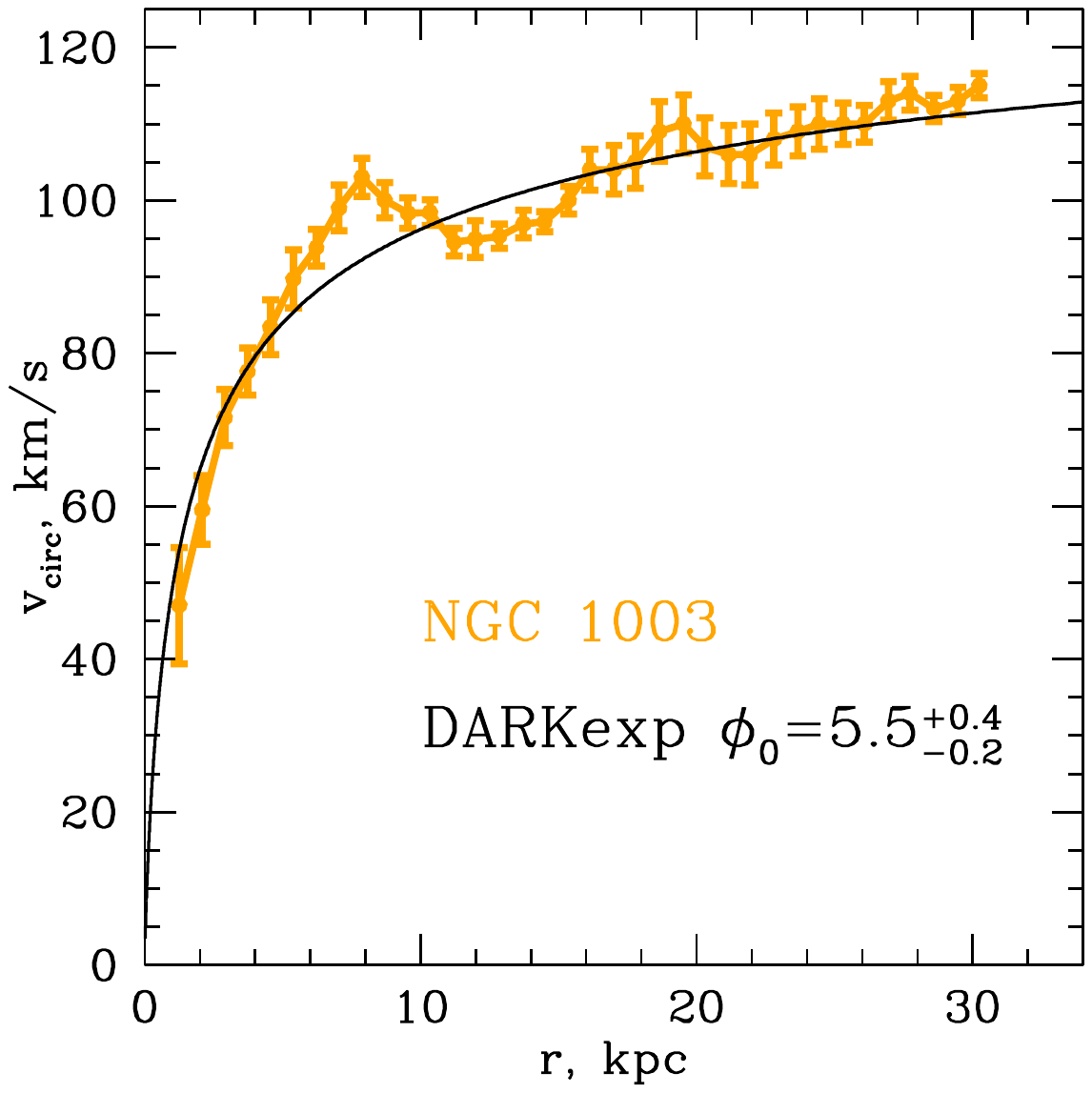}
    \includegraphics[width=0.237\linewidth]{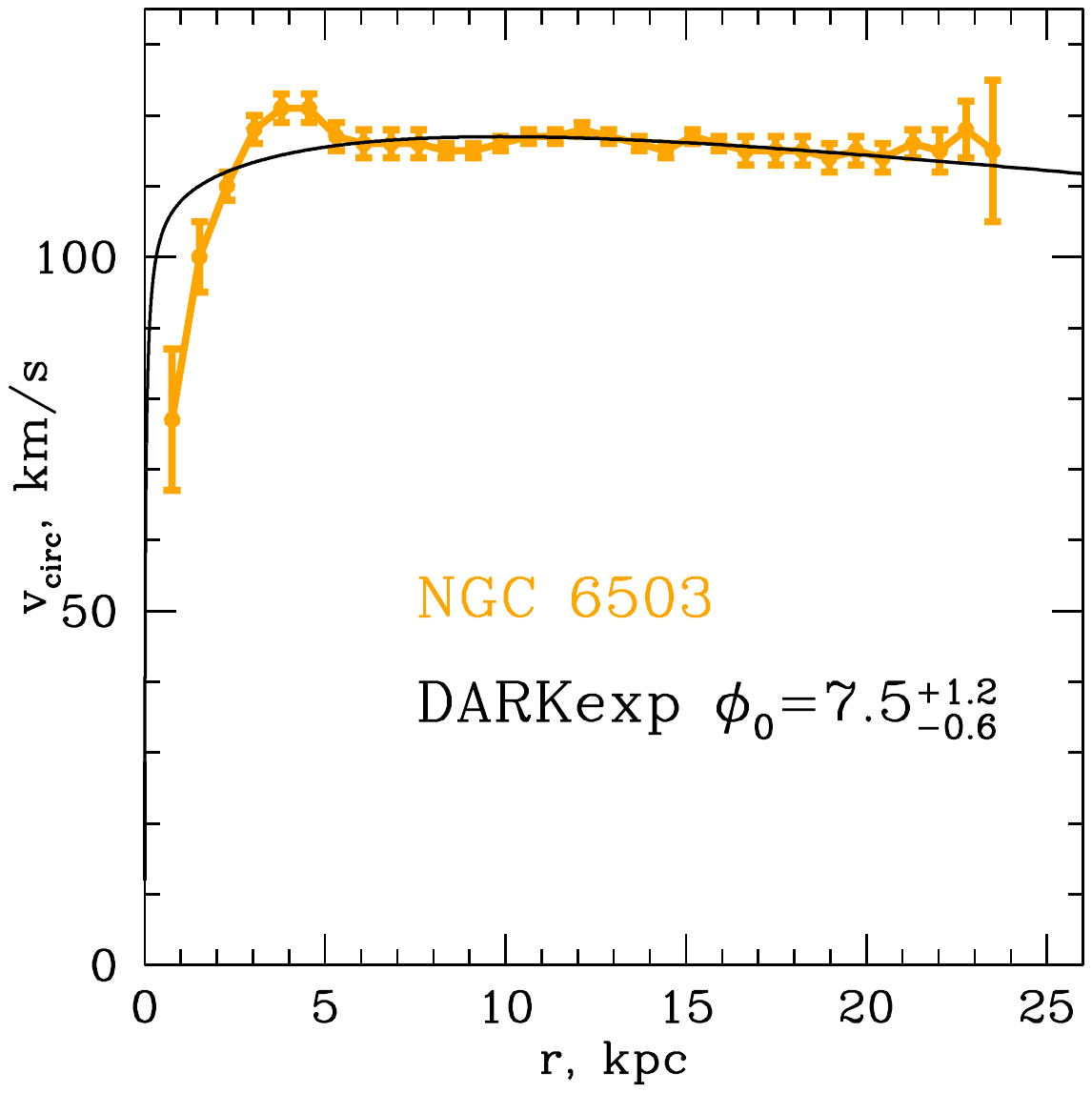}
    \includegraphics[width=0.237\linewidth]{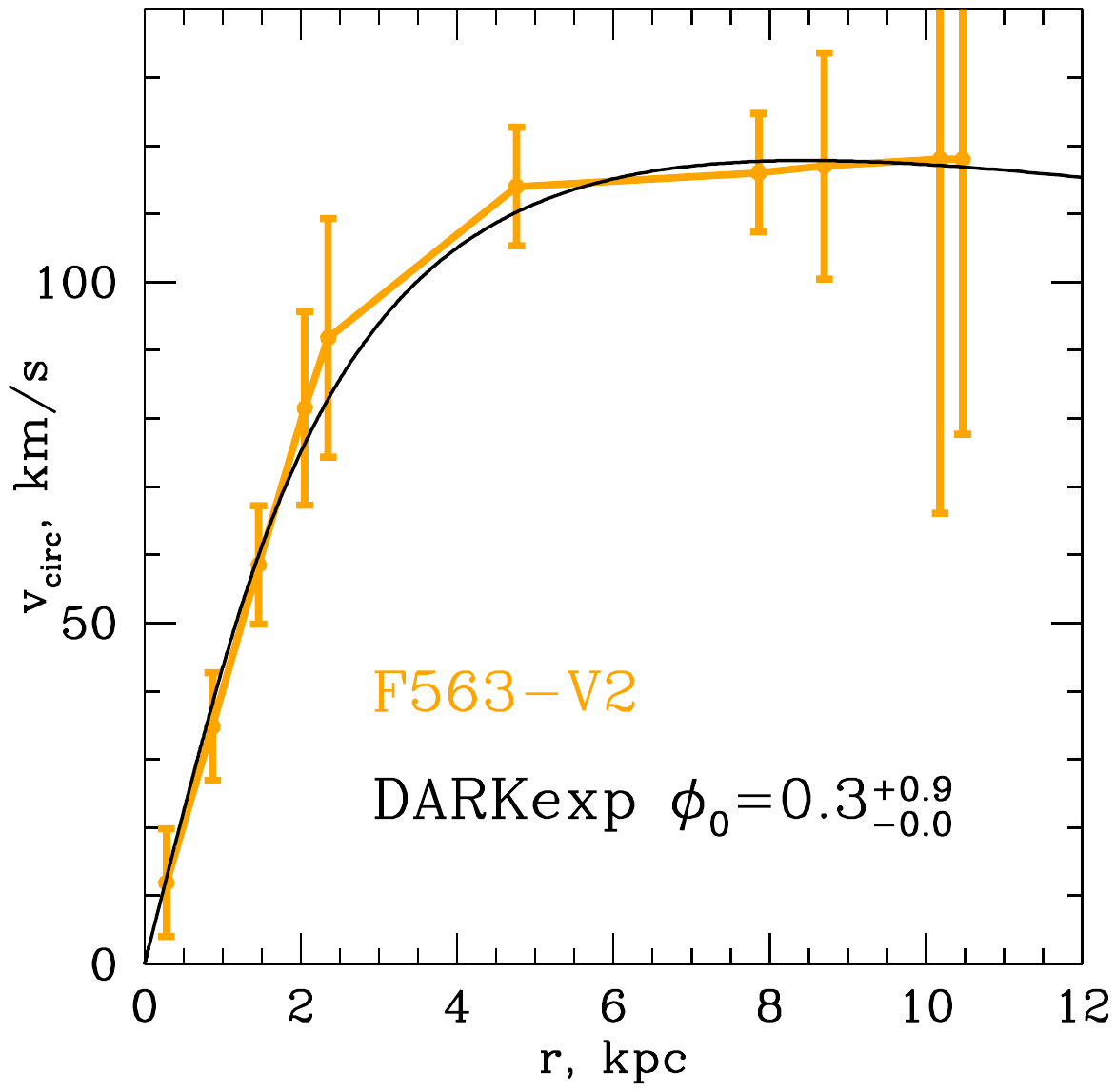}
         \vskip-1.65cm
    \includegraphics[width=0.237\linewidth]{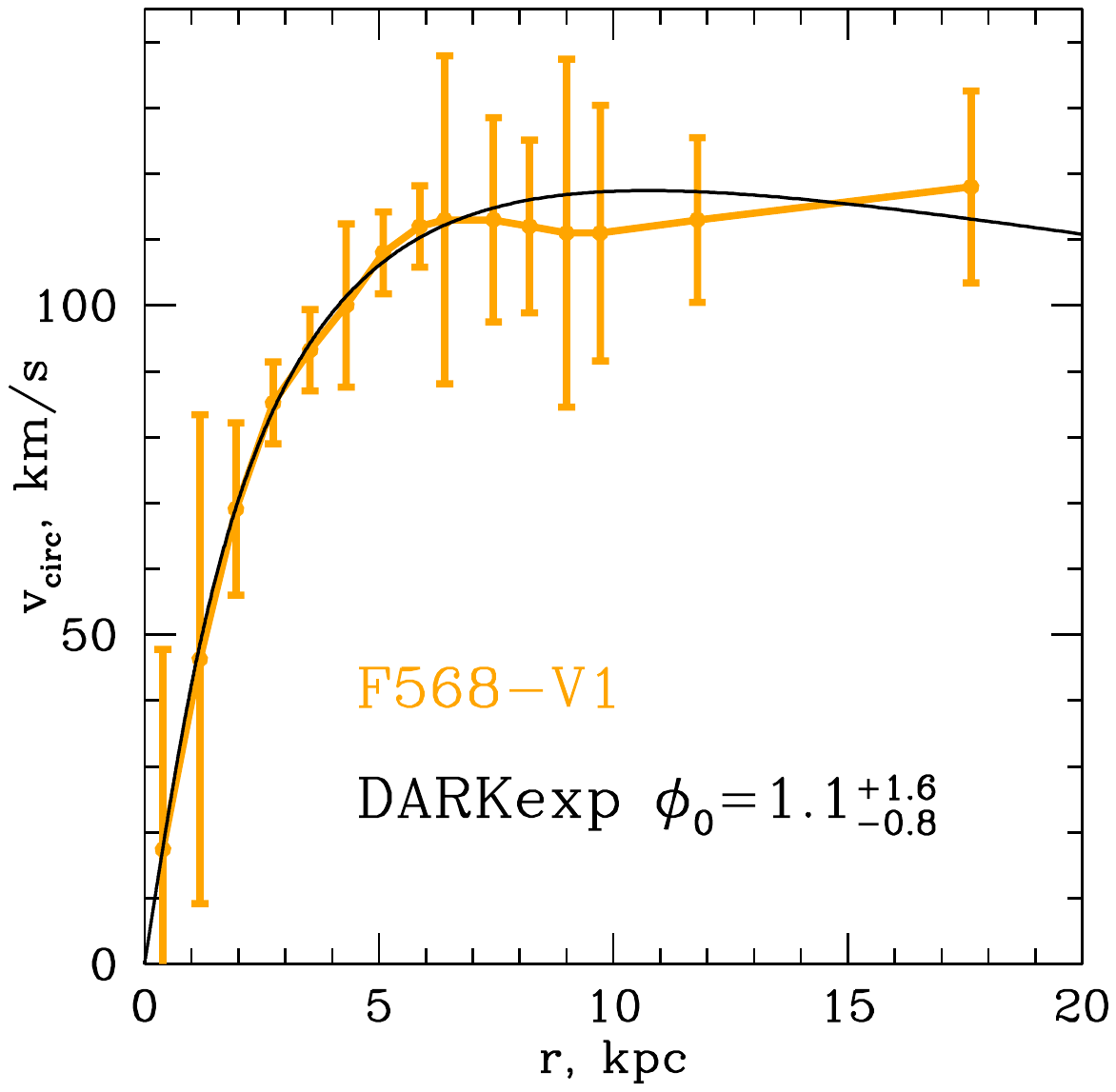}
    \includegraphics[width=0.237\linewidth]{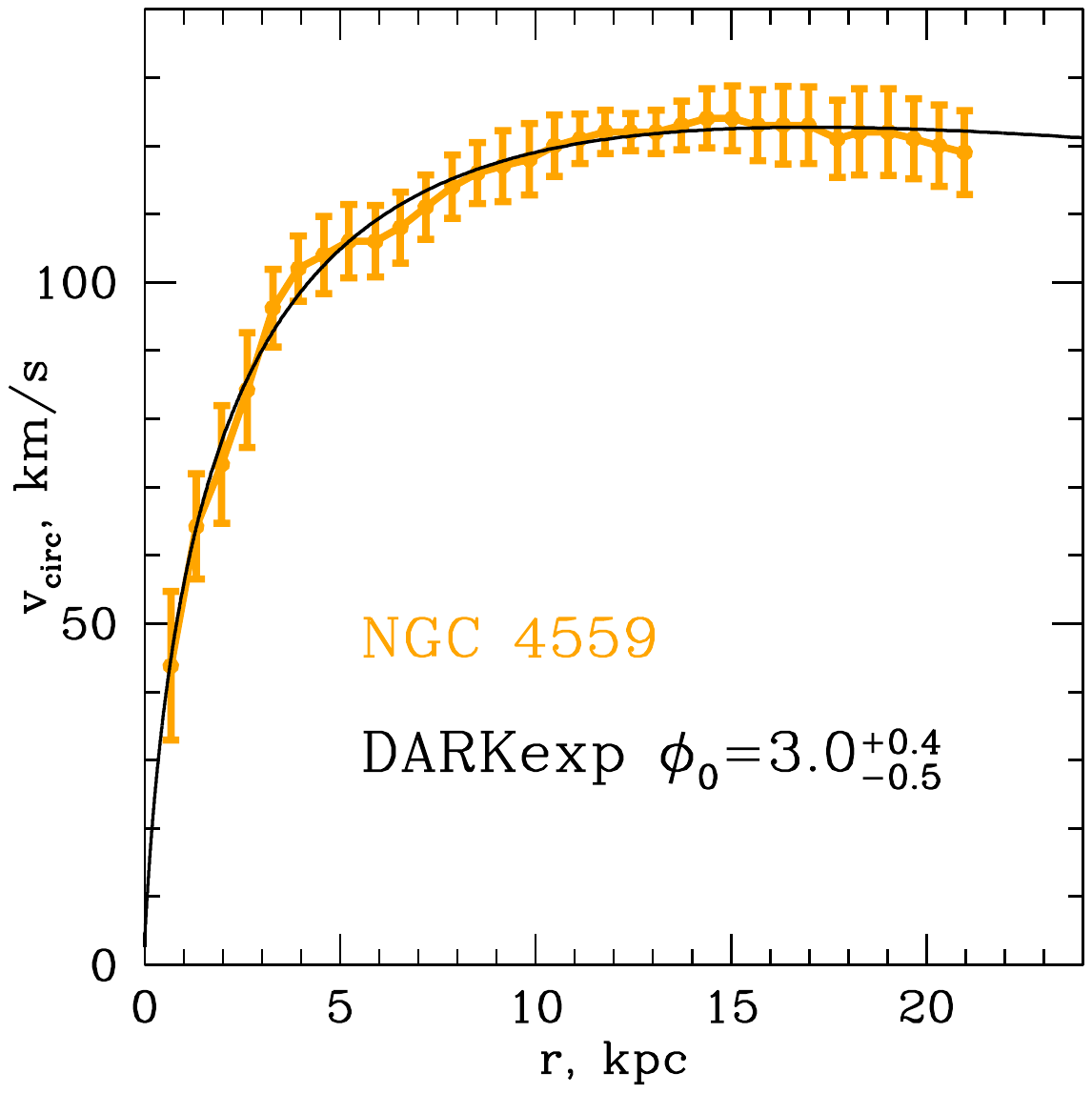}
    \includegraphics[width=0.237\linewidth]{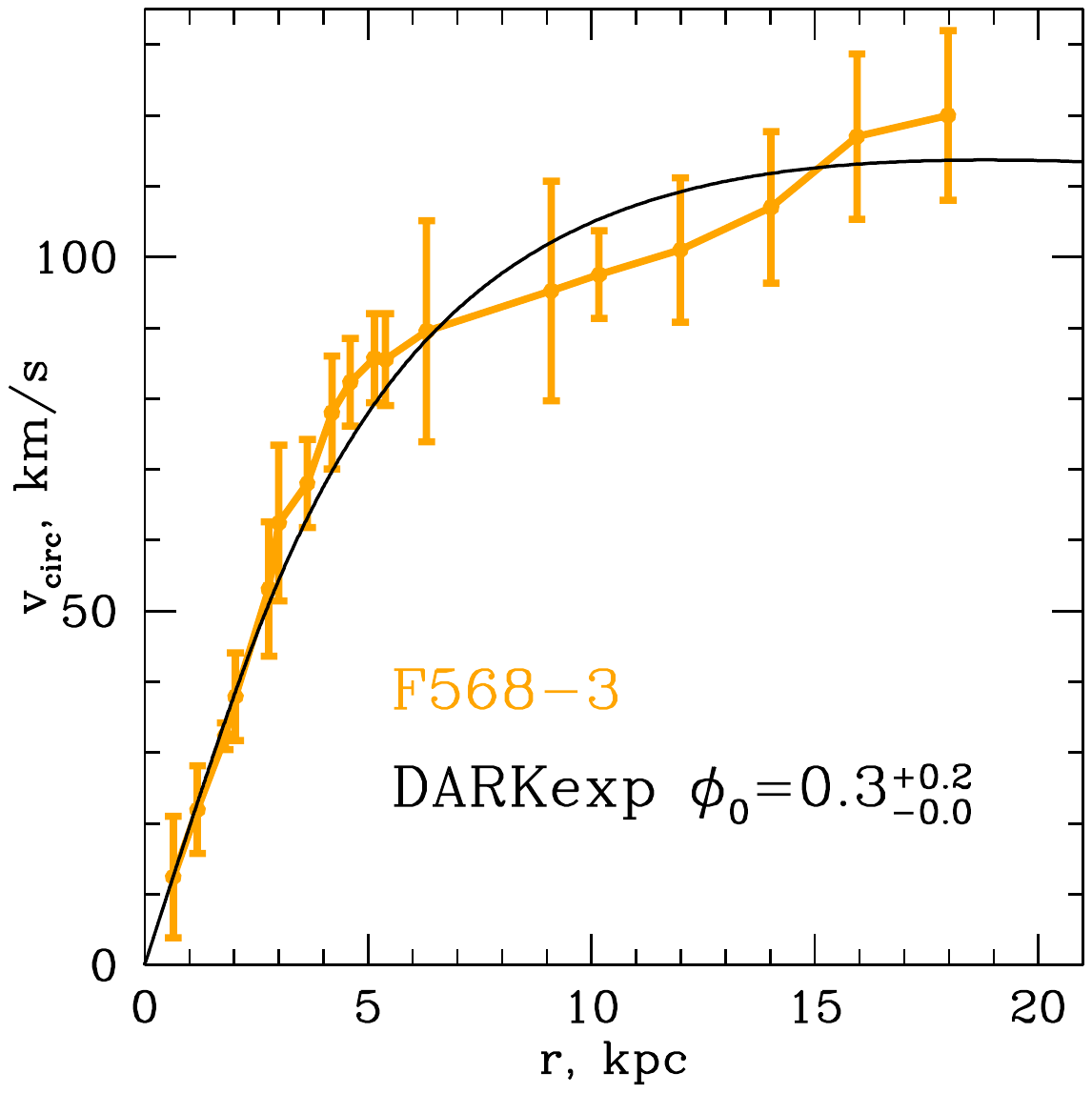}
    \includegraphics[width=0.237\linewidth]{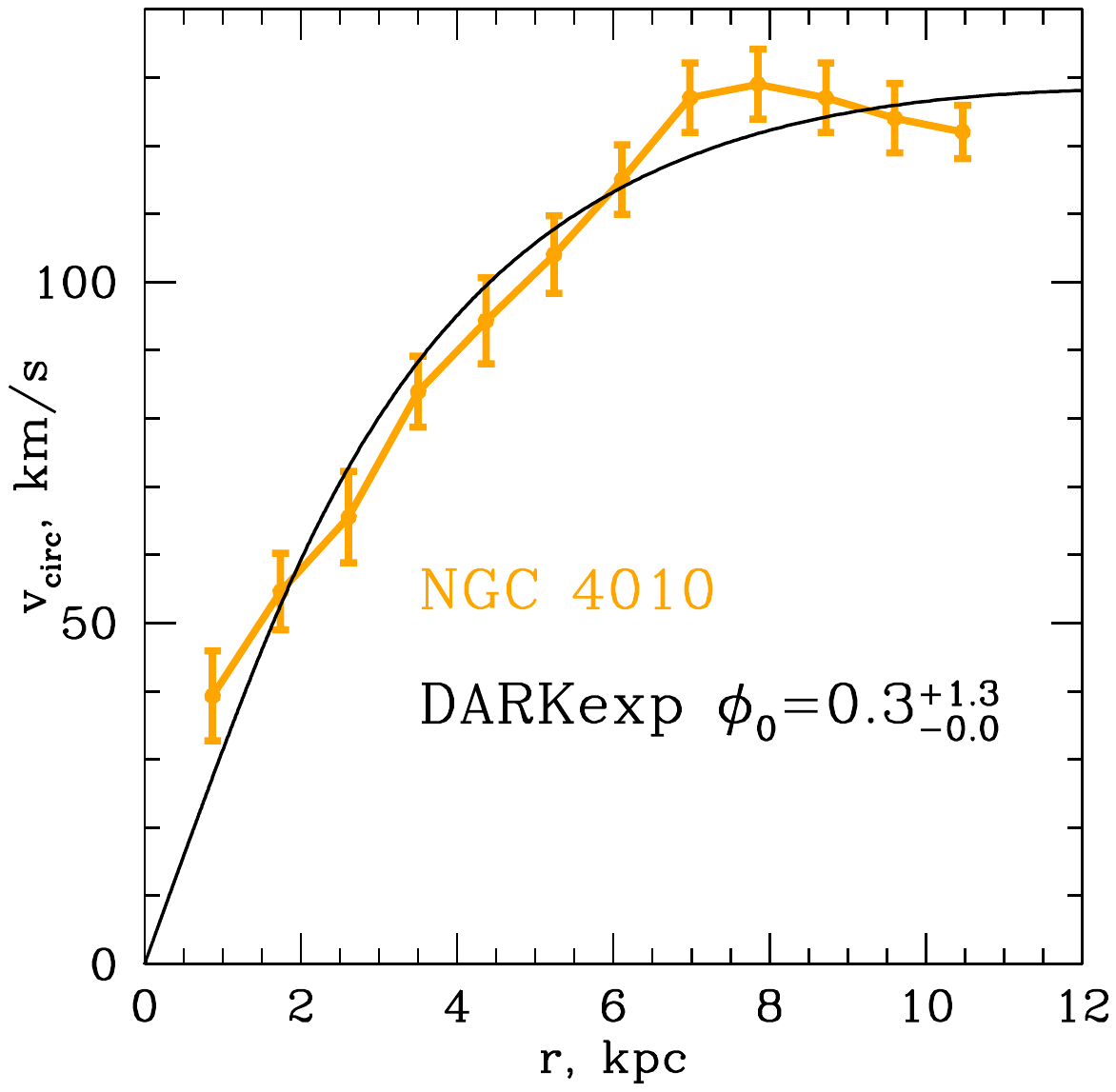}   
    \vskip-0.85cm
    \caption{Figure~\ref{fig:rotcurve1} continued. }
    \label{fig:rotcurve3}
\end{figure*}

\begin{figure*}
    \centering
    \vskip-1.75cm
    \includegraphics[width=0.237\linewidth]{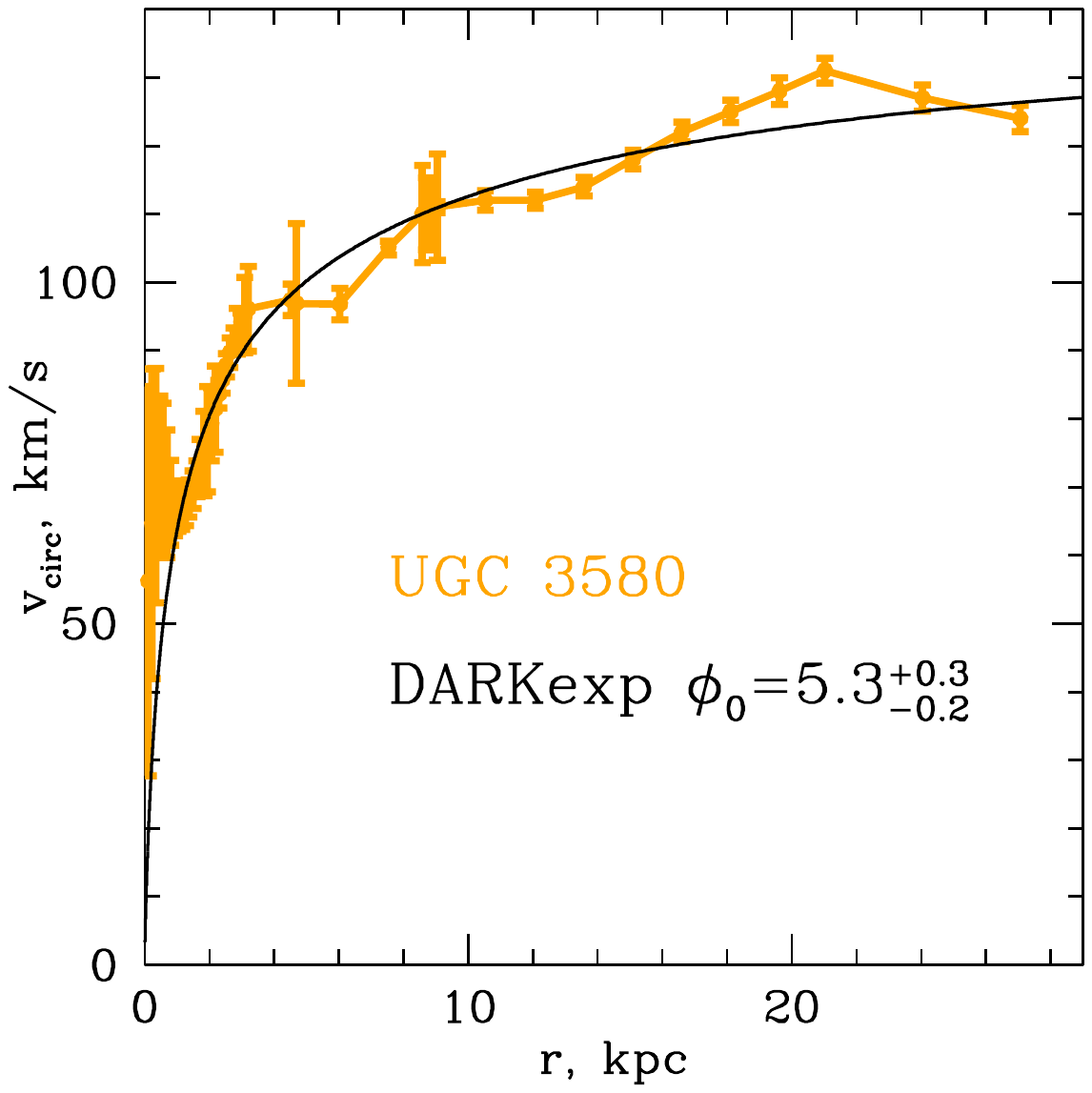}
    \includegraphics[width=0.237\linewidth]{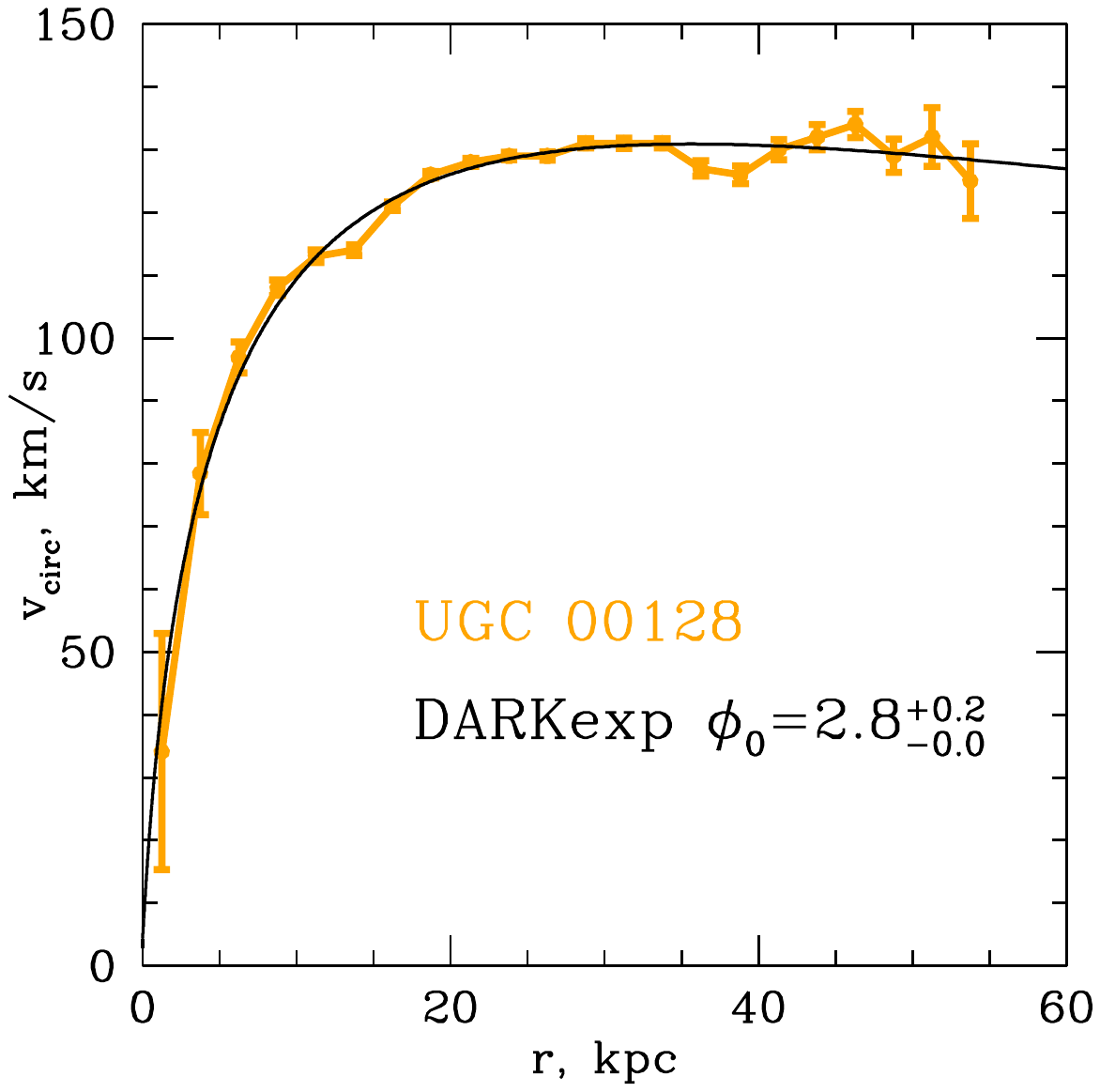}
    \includegraphics[width=0.237\linewidth]{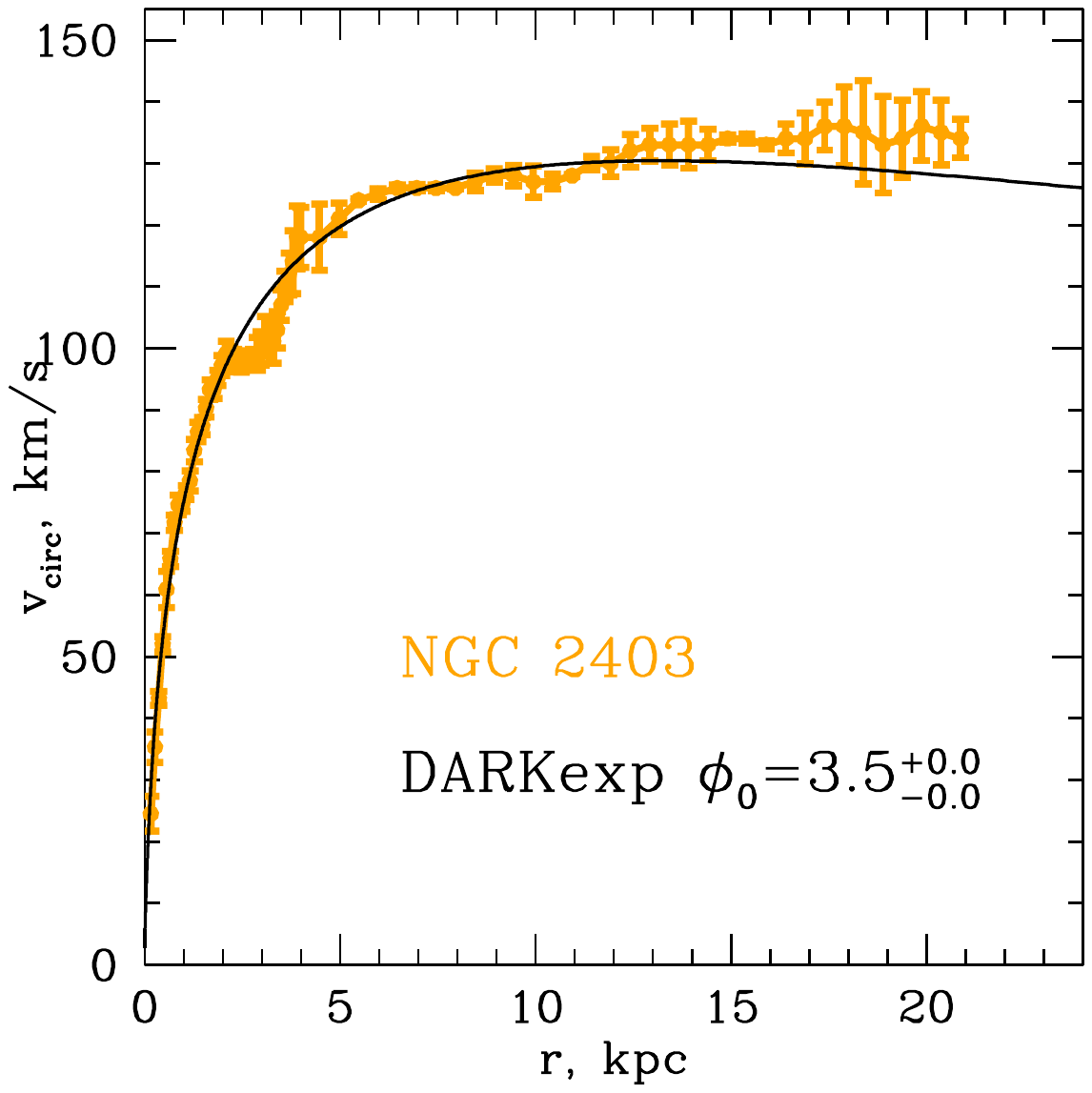}    \includegraphics[width=0.237\linewidth]{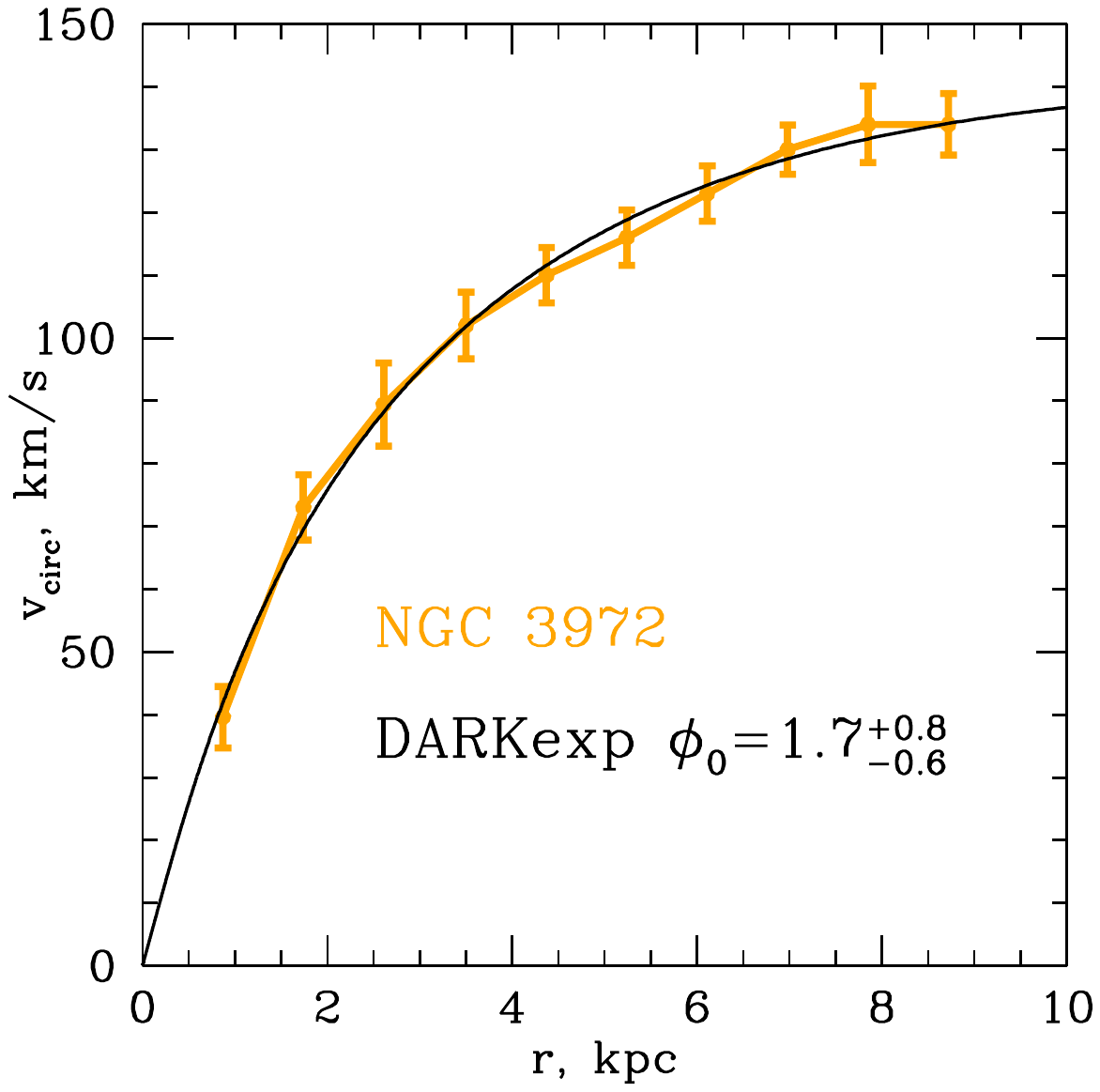}
            \vskip-1.65cm
    \includegraphics[width=0.237\linewidth]{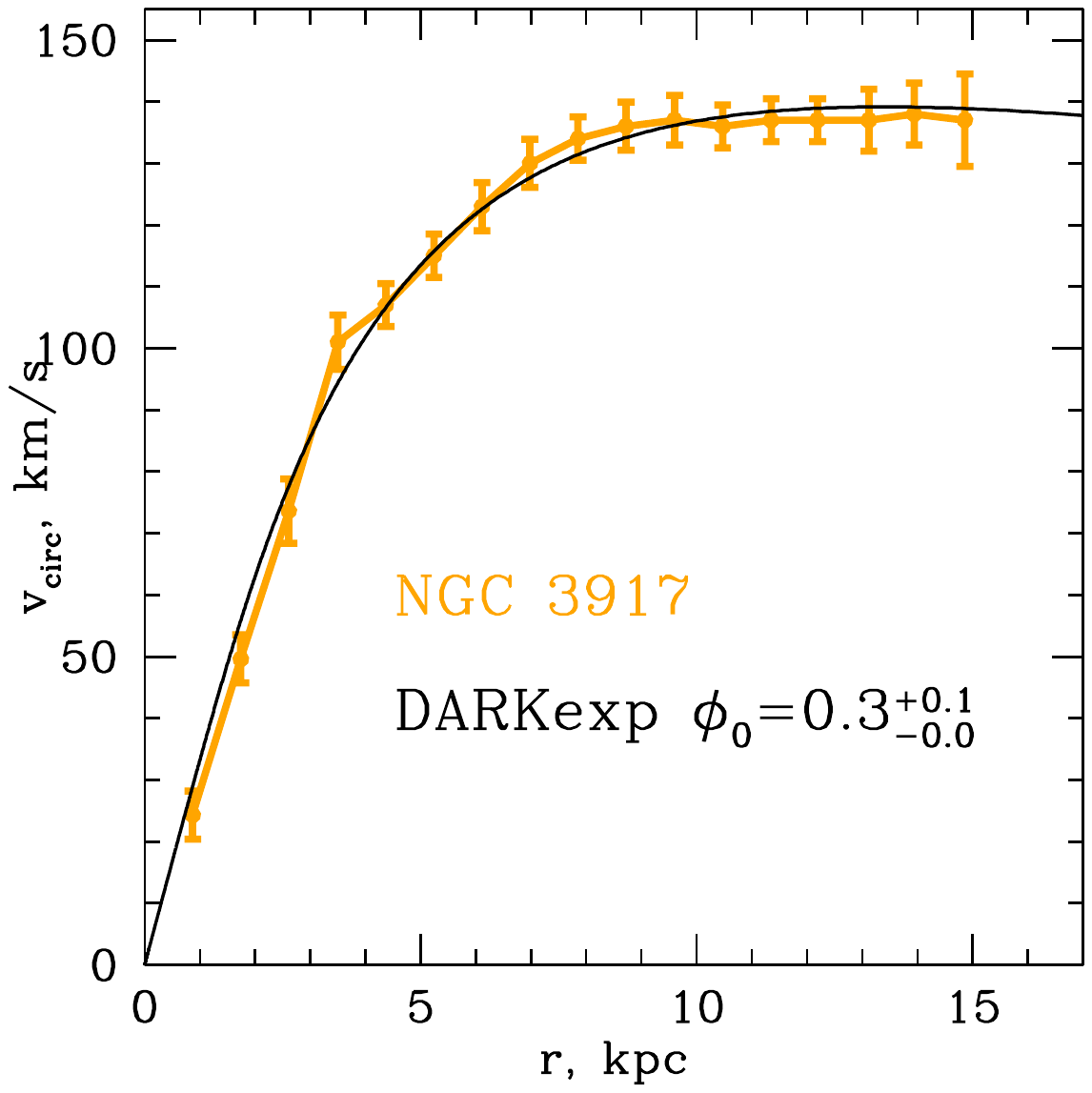}
    \includegraphics[width=0.237\linewidth]{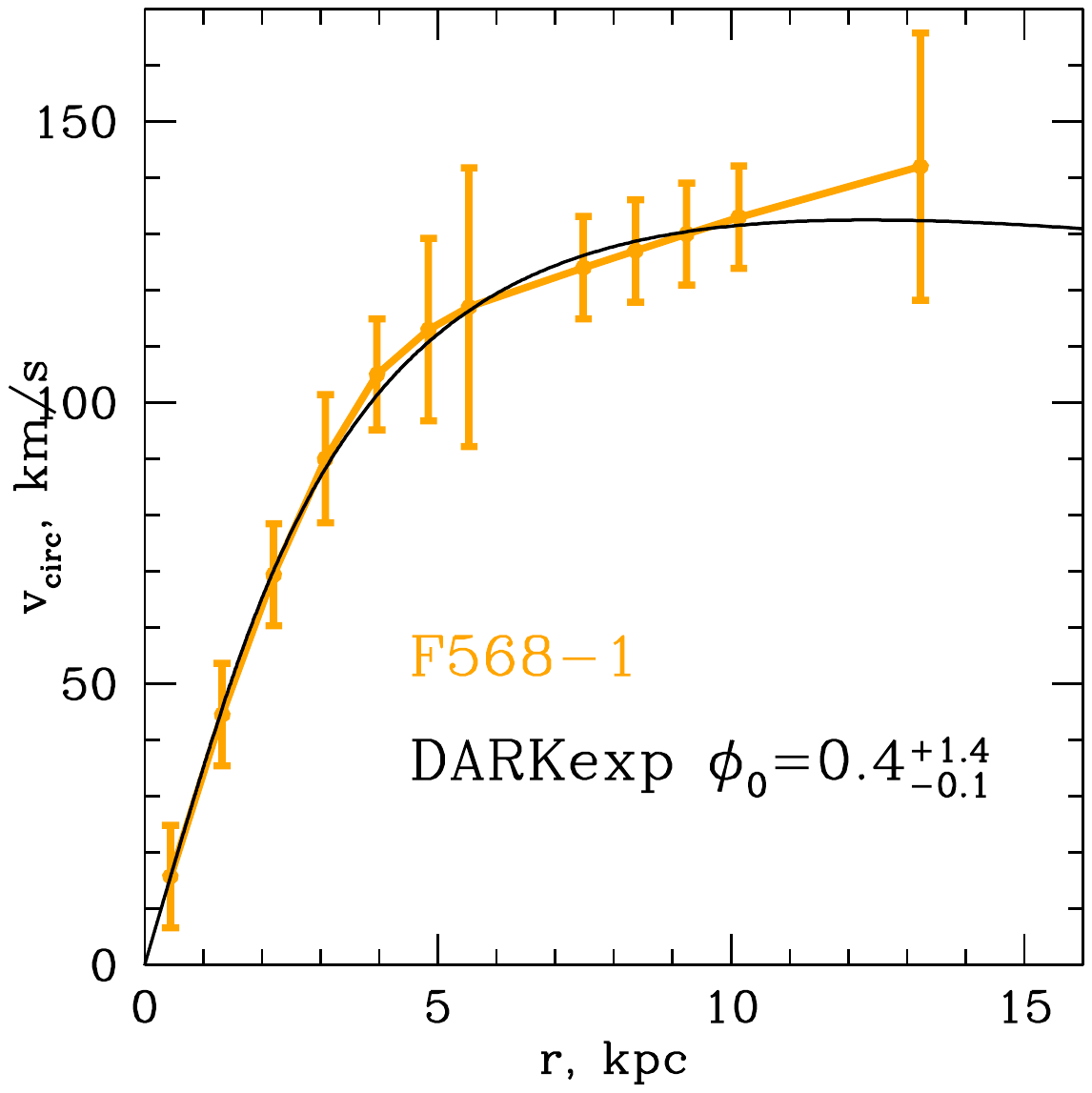}
    \includegraphics[width=0.237\linewidth]{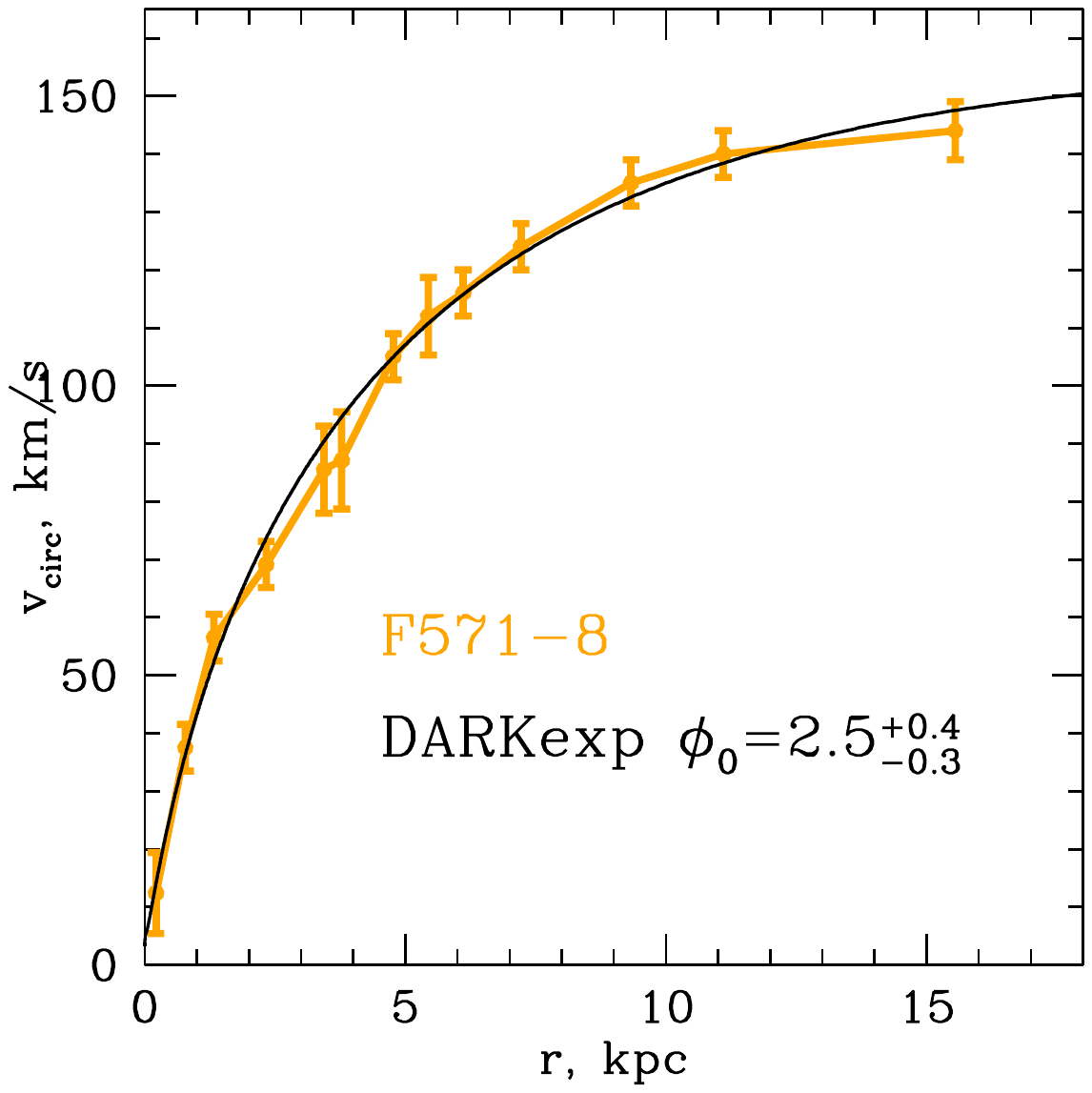}
    \includegraphics[width=0.237\linewidth]{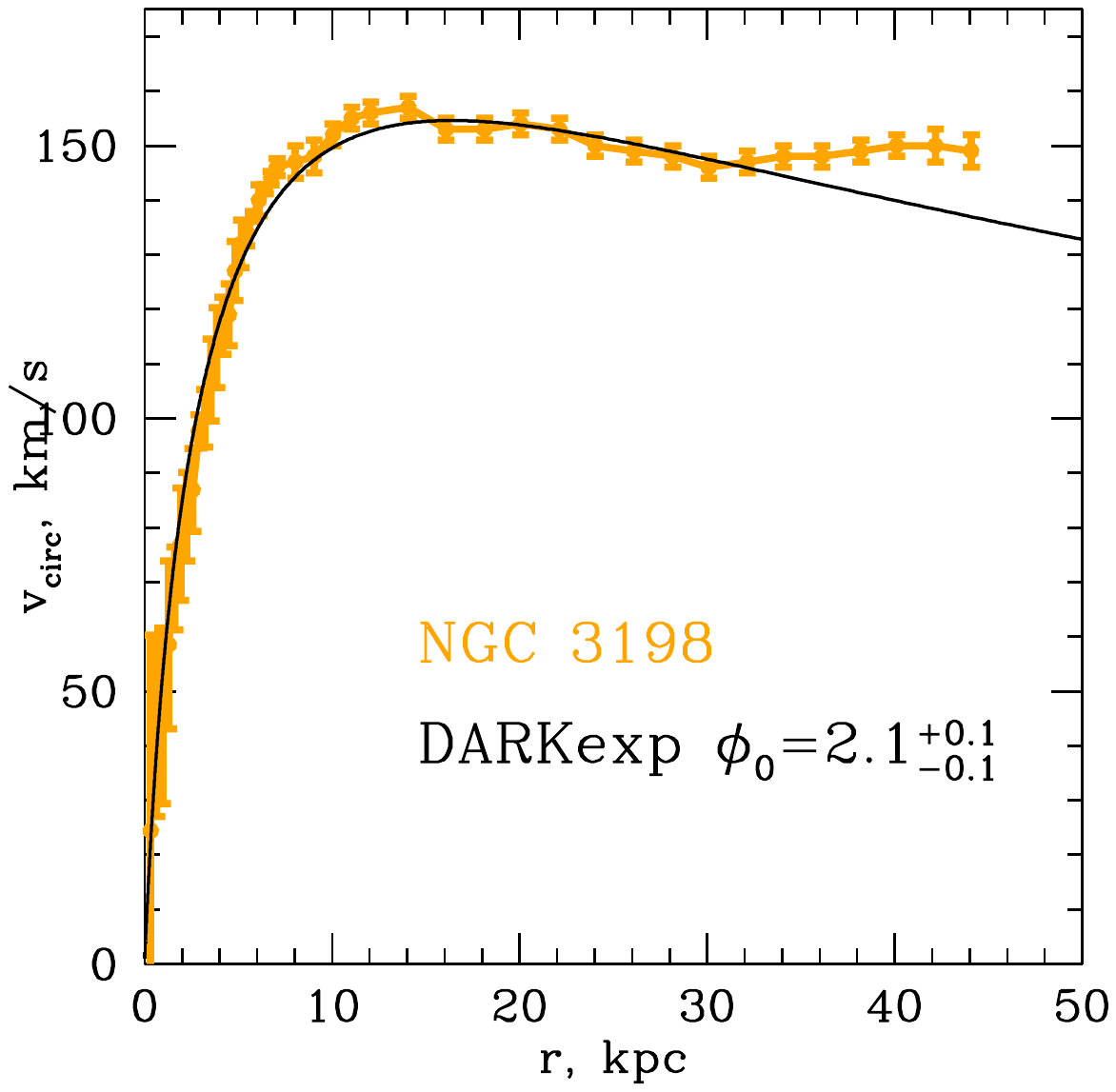}
         \vskip-1.65cm
    \includegraphics[width=0.237\linewidth]{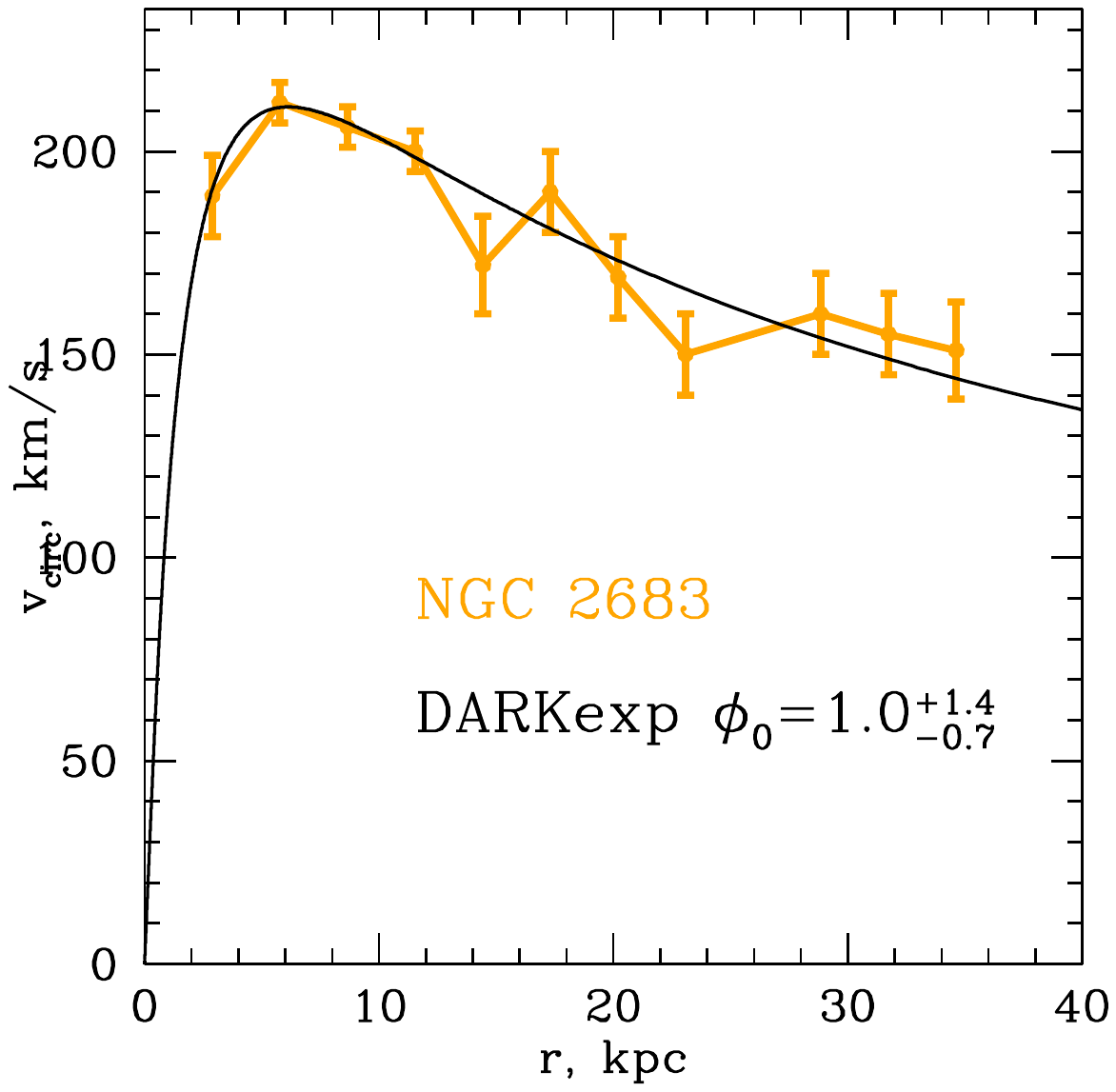}
    \includegraphics[width=0.237\linewidth]{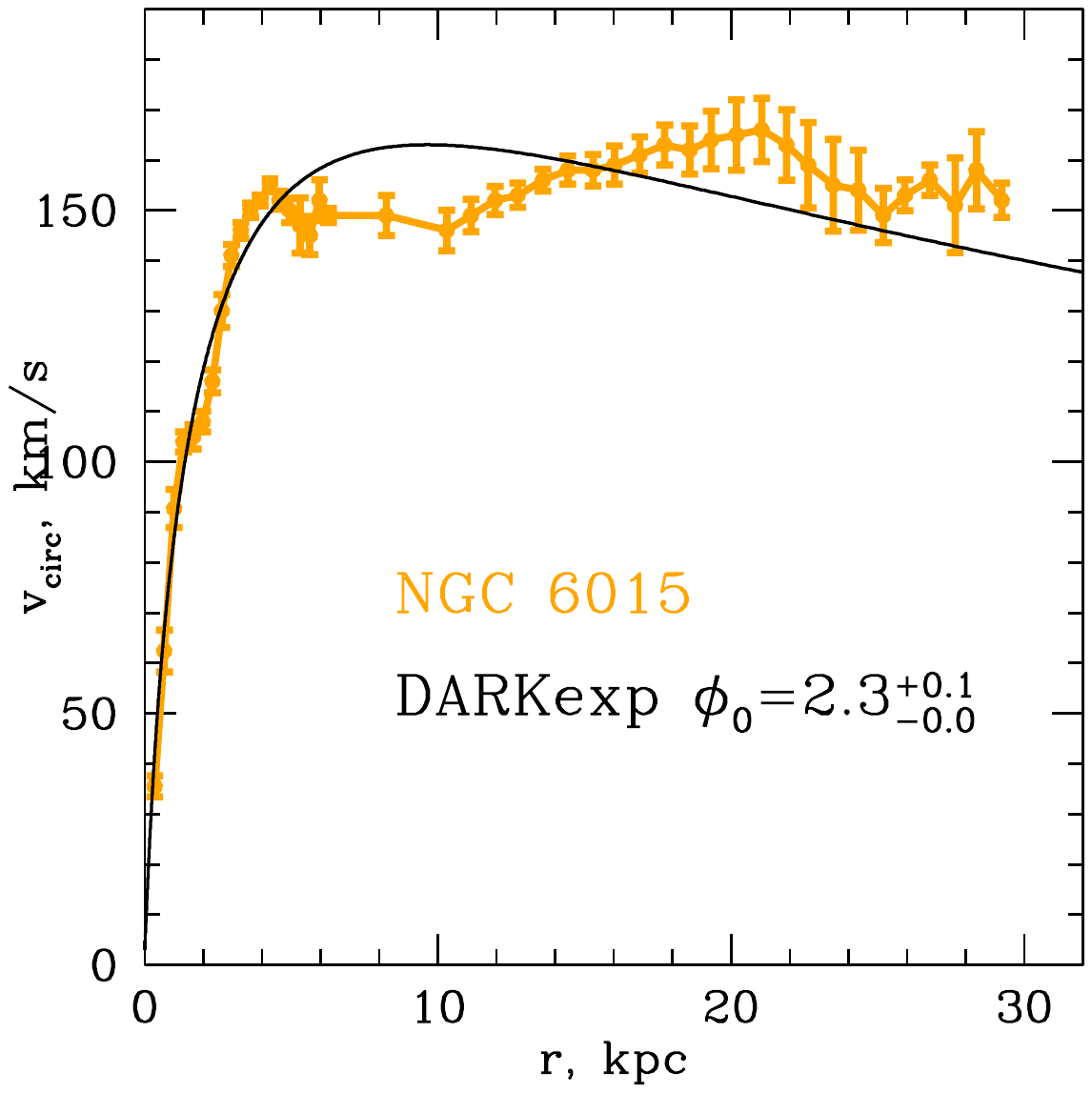}
    \includegraphics[width=0.237\linewidth]{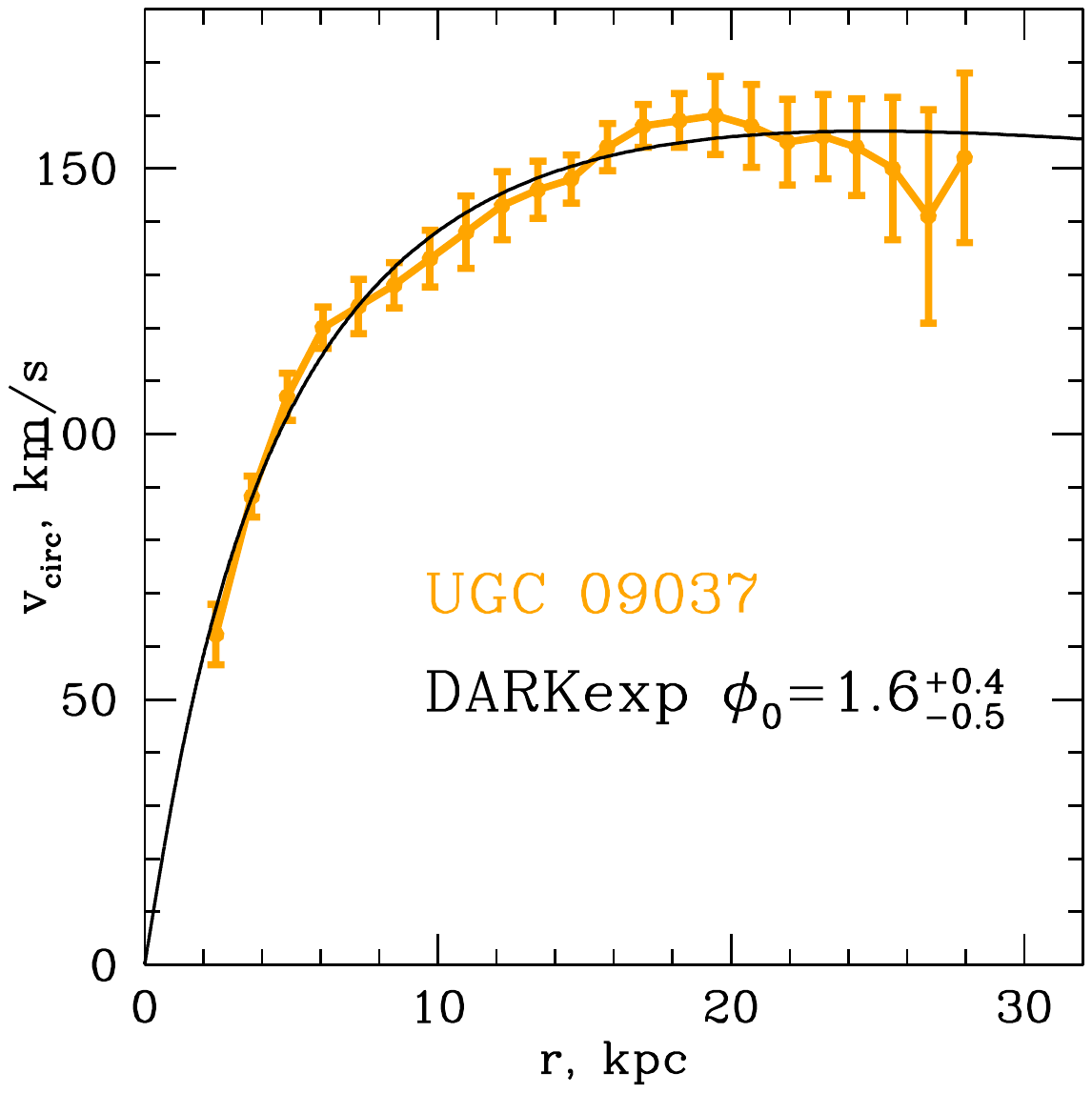}
    \includegraphics[width=0.237\linewidth]{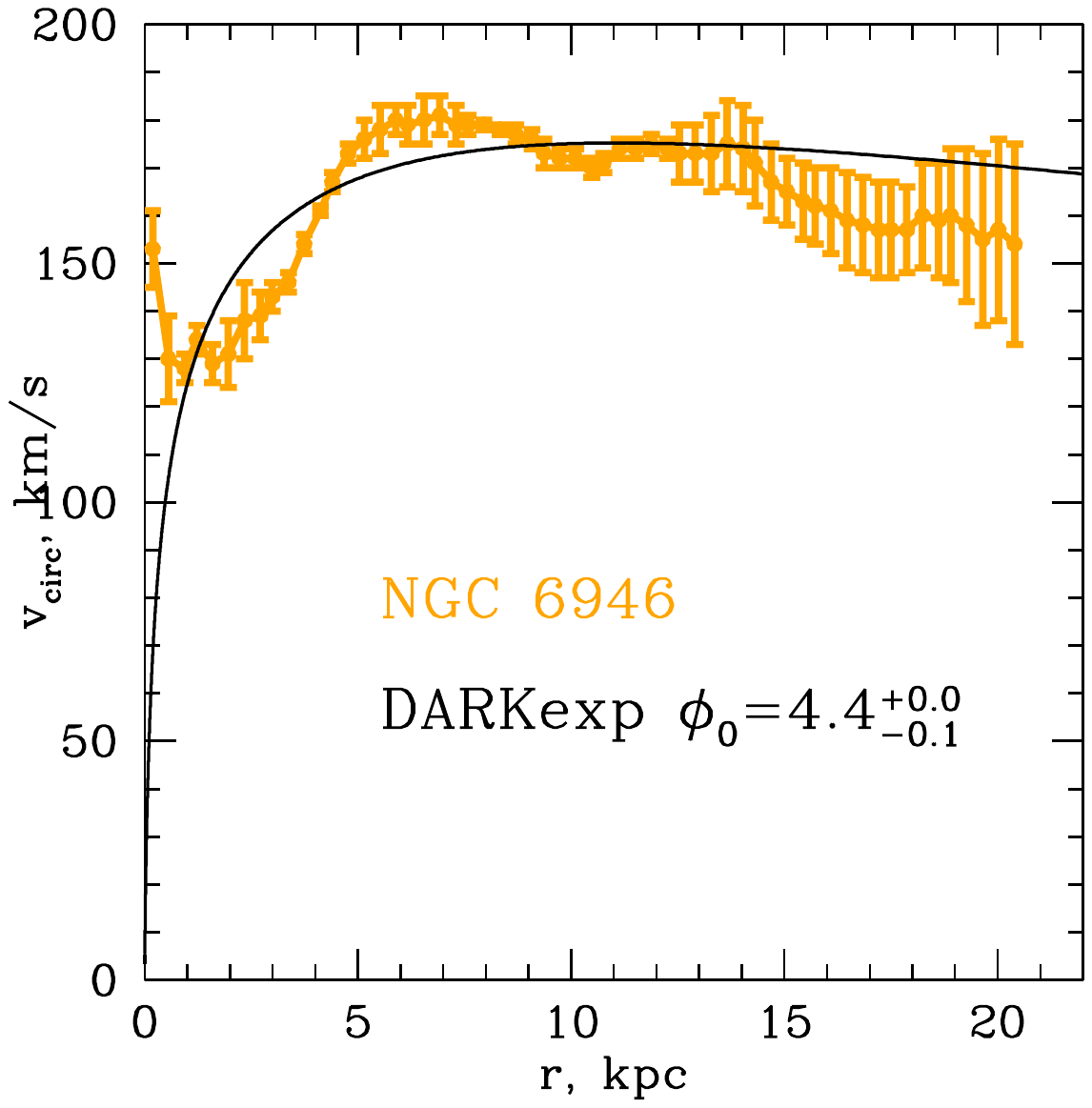}  
        \vskip-1.65cm
    \includegraphics[width=0.237\linewidth]{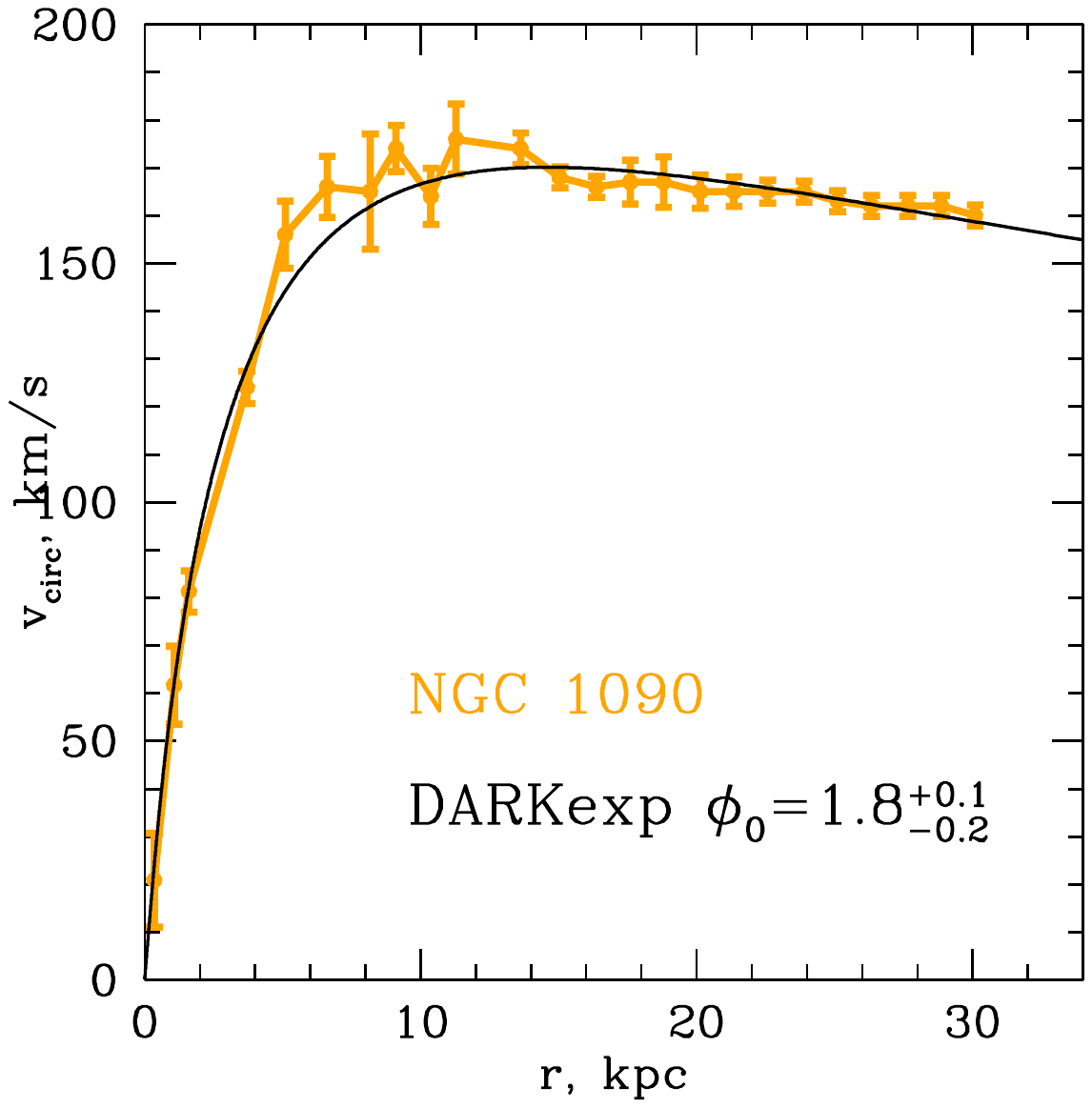}
    \includegraphics[width=0.237\linewidth]{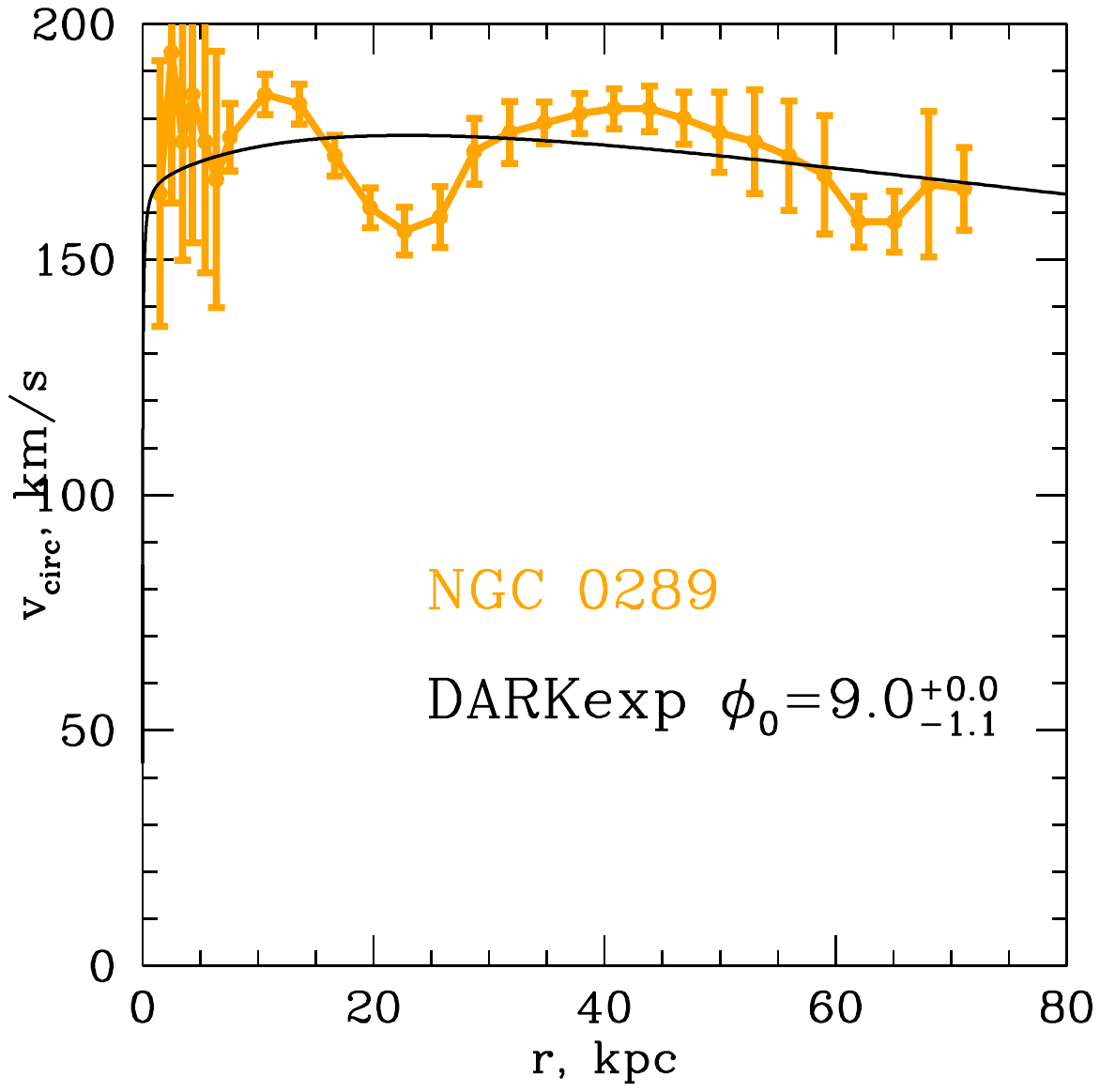}
    \includegraphics[width=0.237\linewidth]{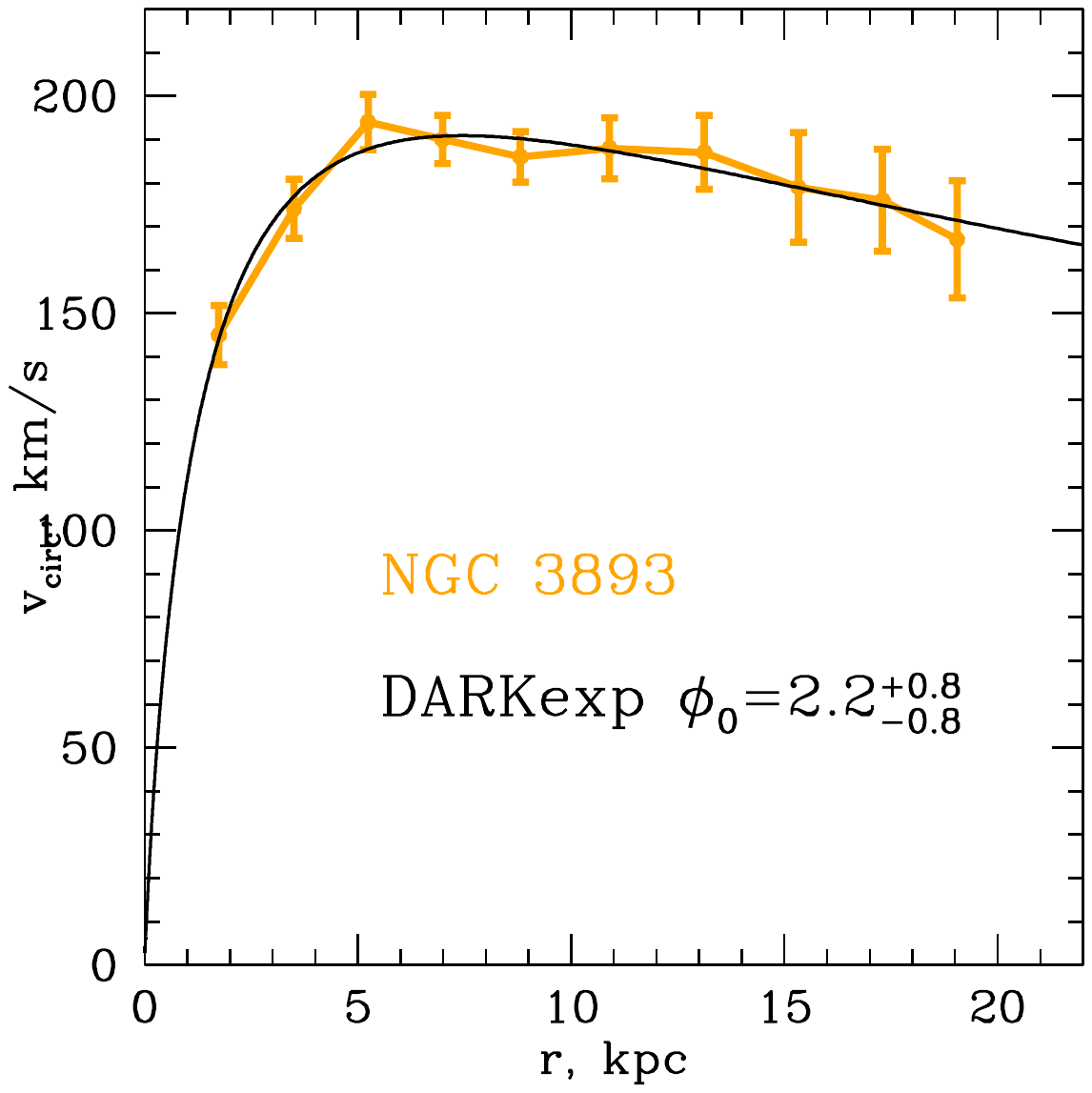}    \includegraphics[width=0.237\linewidth]{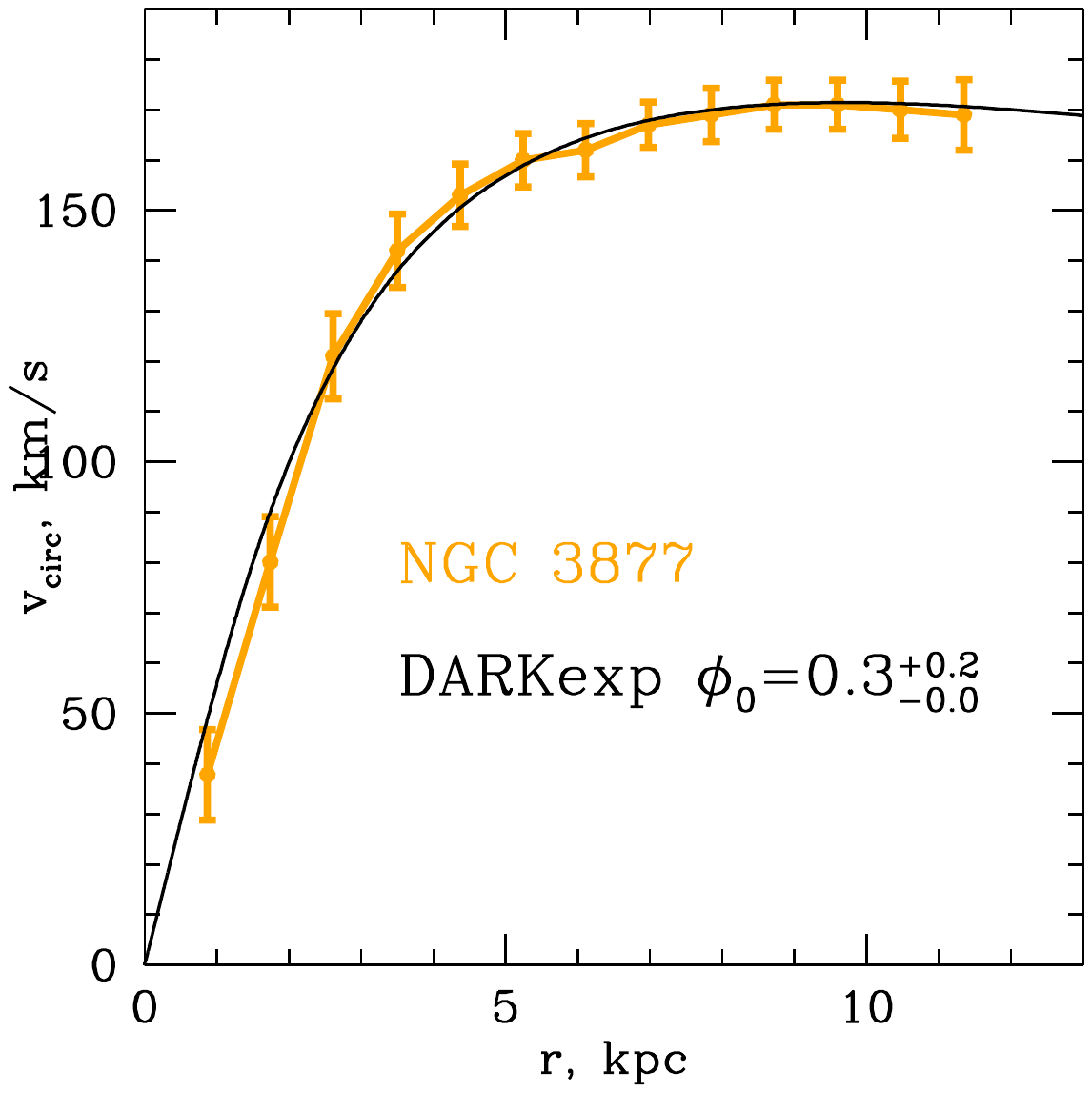}
            \vskip-1.65cm
    \includegraphics[width=0.237\linewidth]{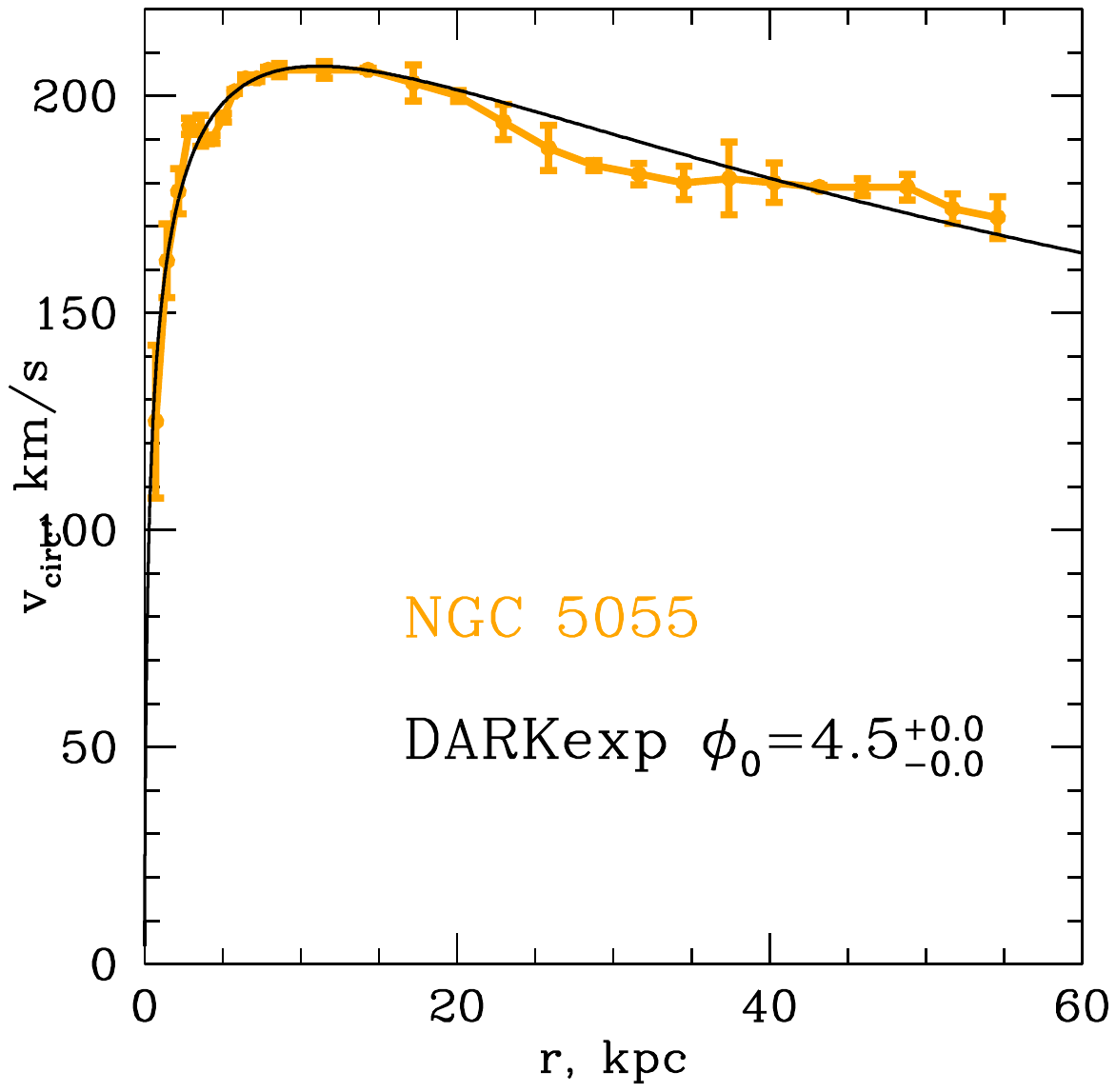}
    \includegraphics[width=0.237\linewidth]{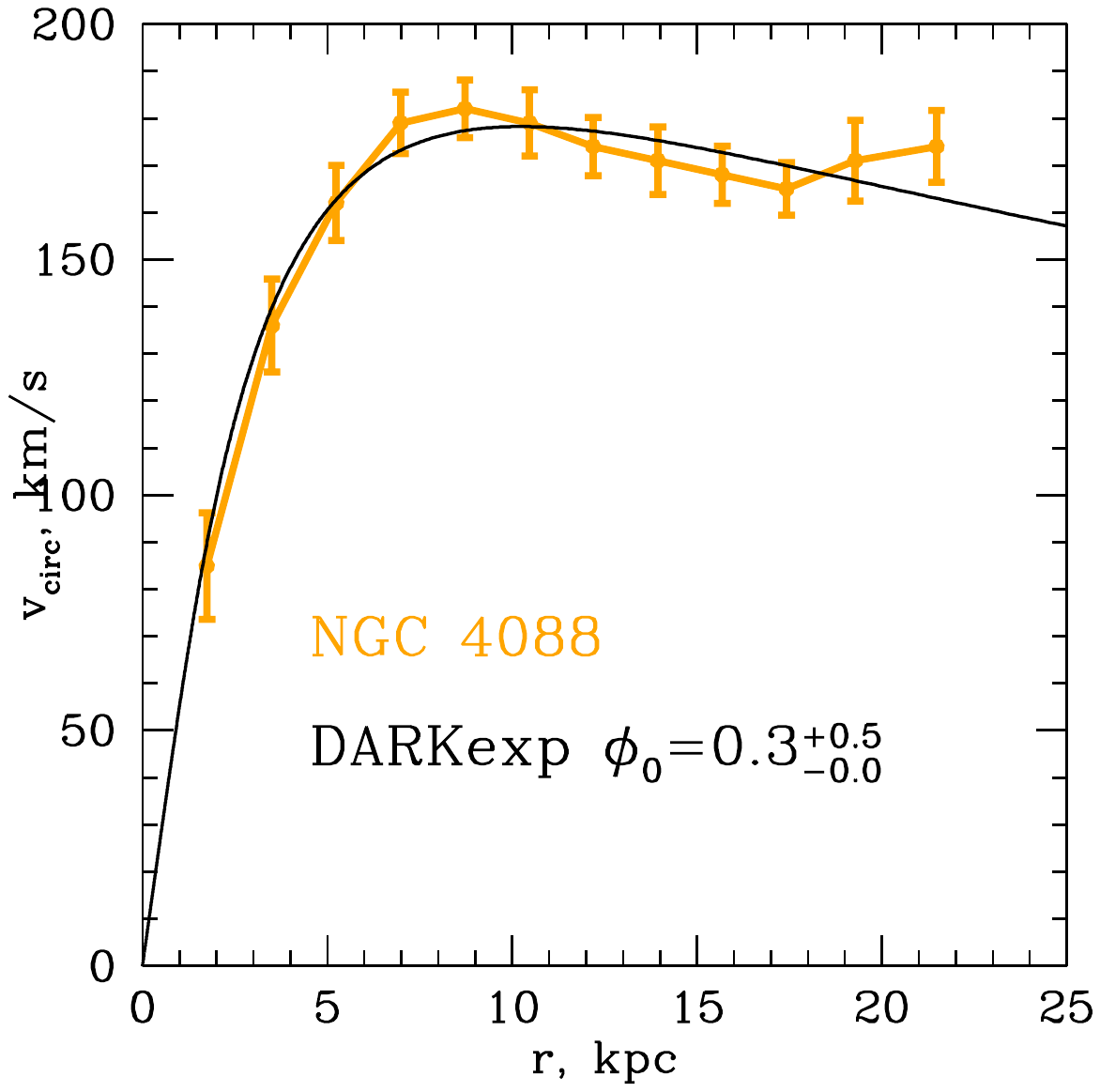}
    \includegraphics[width=0.237\linewidth]{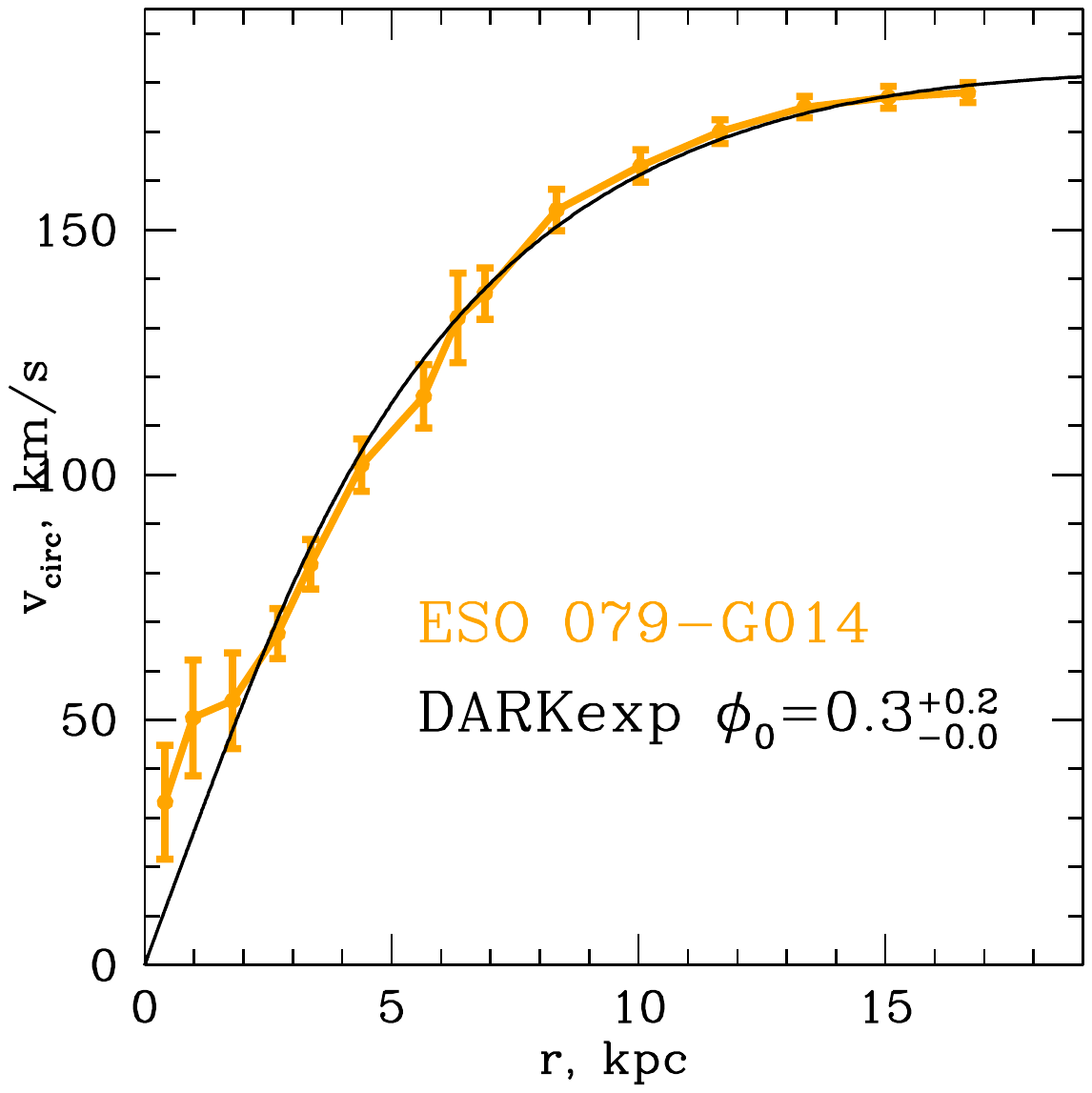}
    \includegraphics[width=0.237\linewidth]{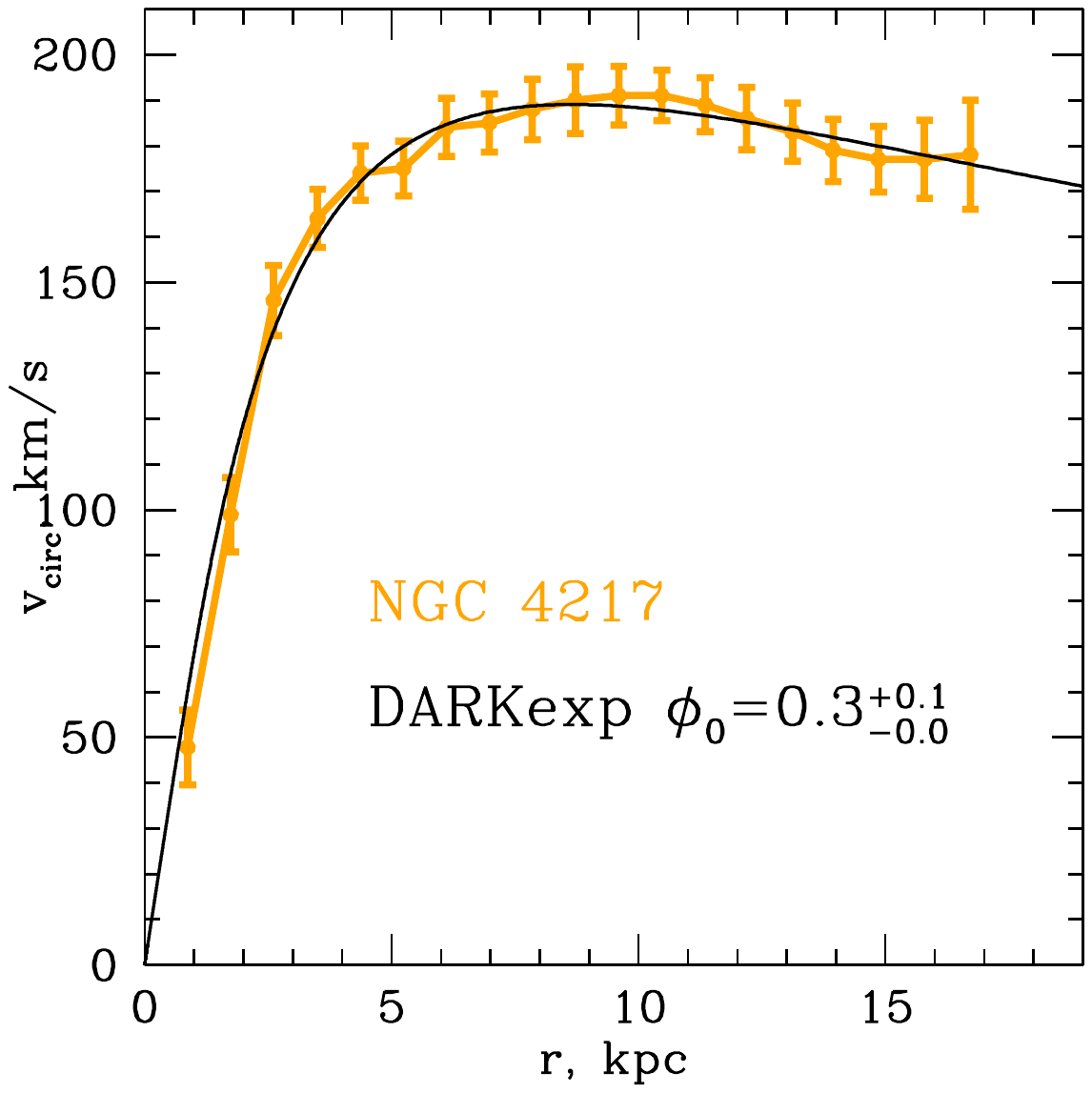}
         \vskip-1.65cm
    \includegraphics[width=0.237\linewidth]{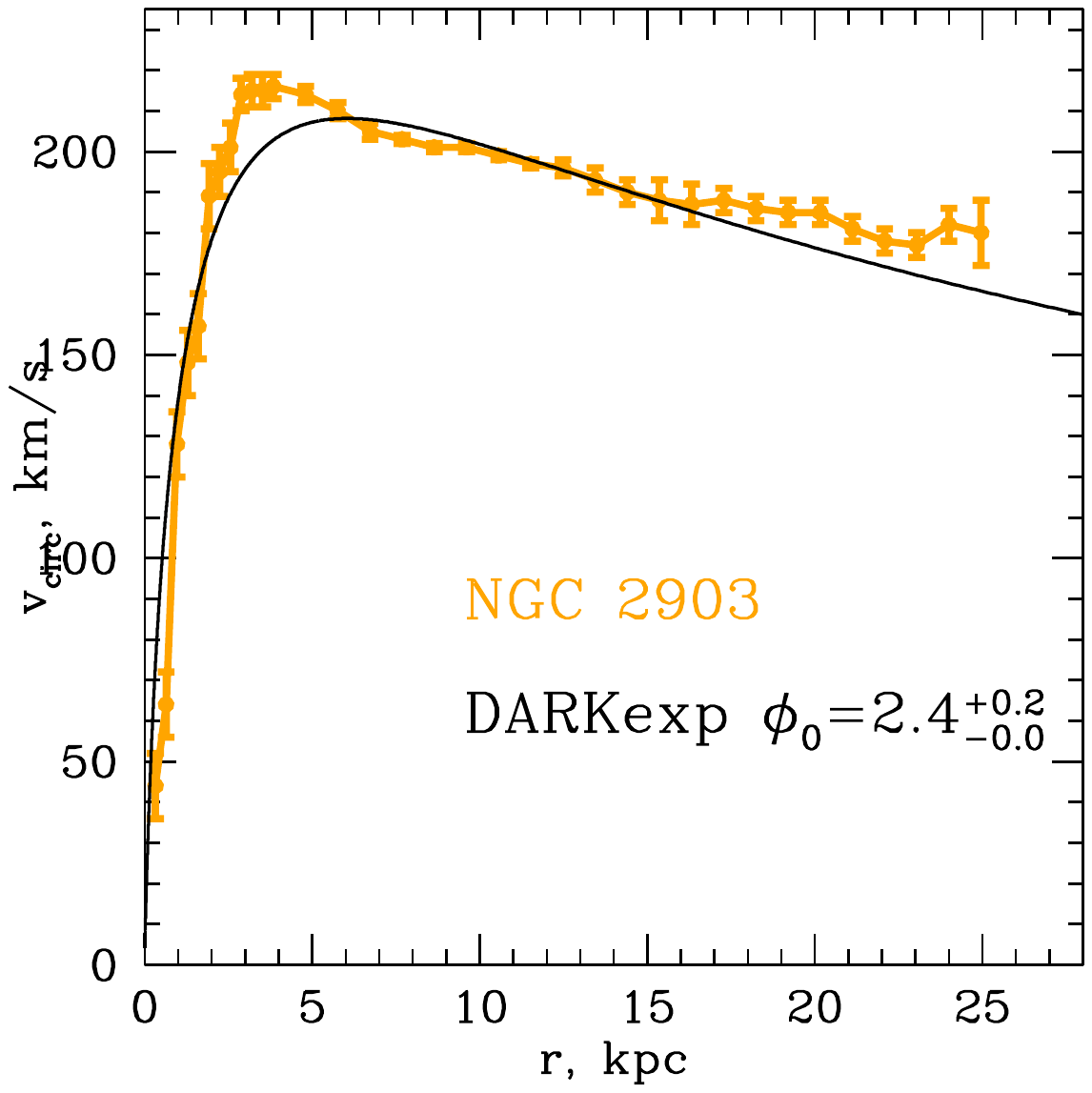}
    \includegraphics[width=0.237\linewidth]{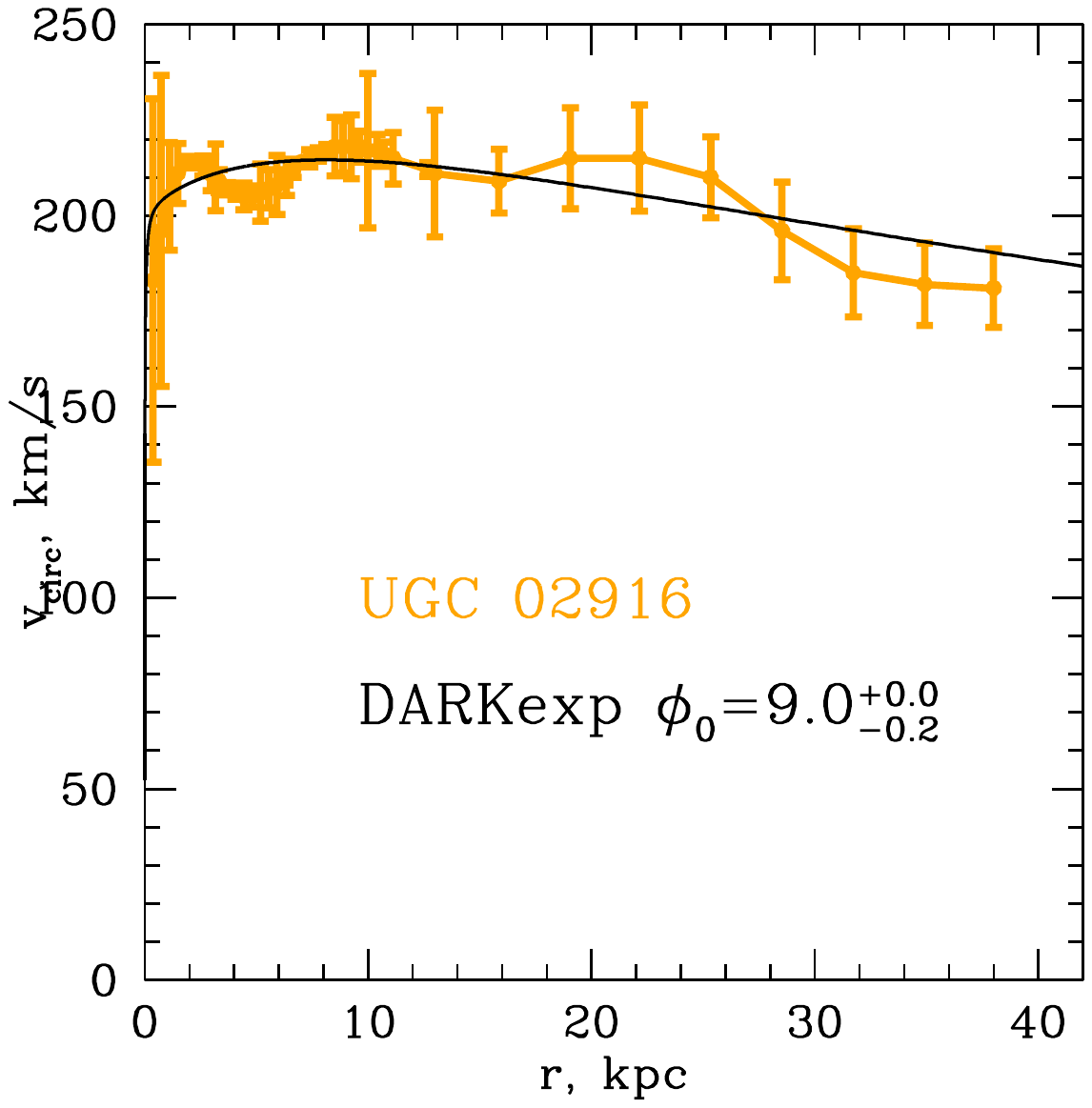}
    \includegraphics[width=0.237\linewidth]{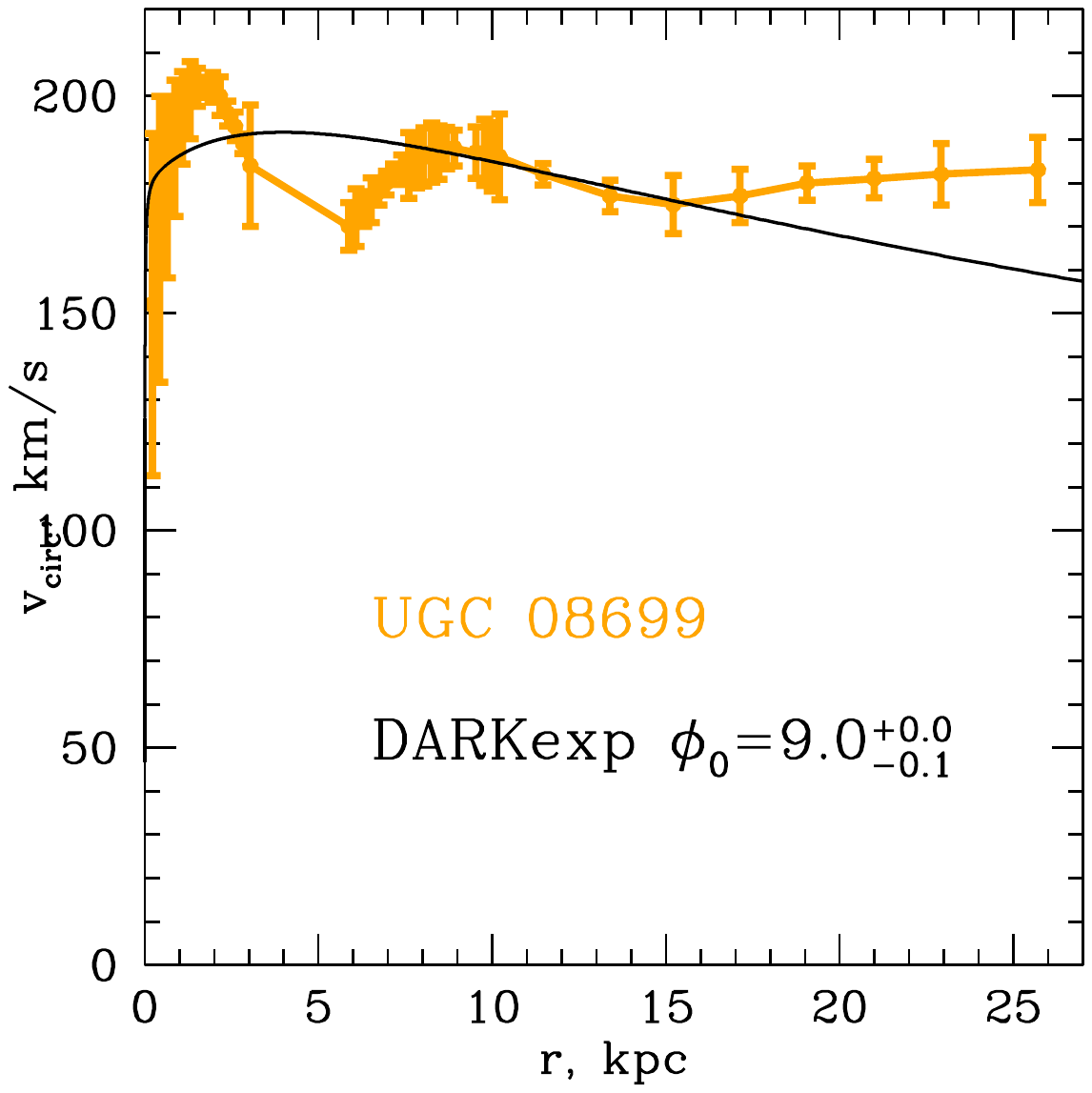}
    \includegraphics[width=0.237\linewidth]{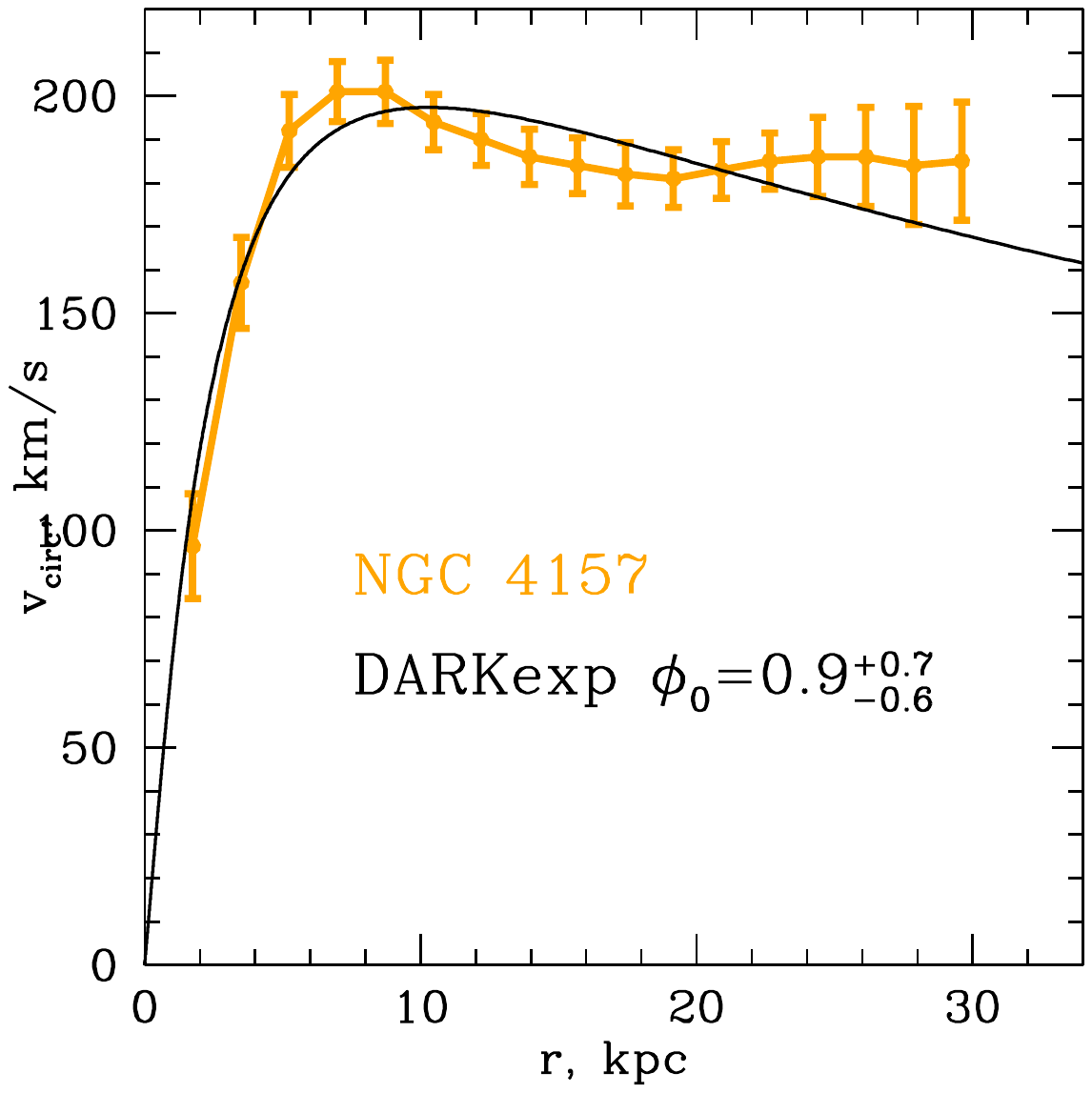}   
    \vskip-0.85cm
    \caption{Figure~\ref{fig:rotcurve1} continued. }
    \label{fig:rotcurve4}
\end{figure*}

\begin{figure*}
    \centering
    \vskip-1.5cm
    \includegraphics[width=0.48\linewidth]{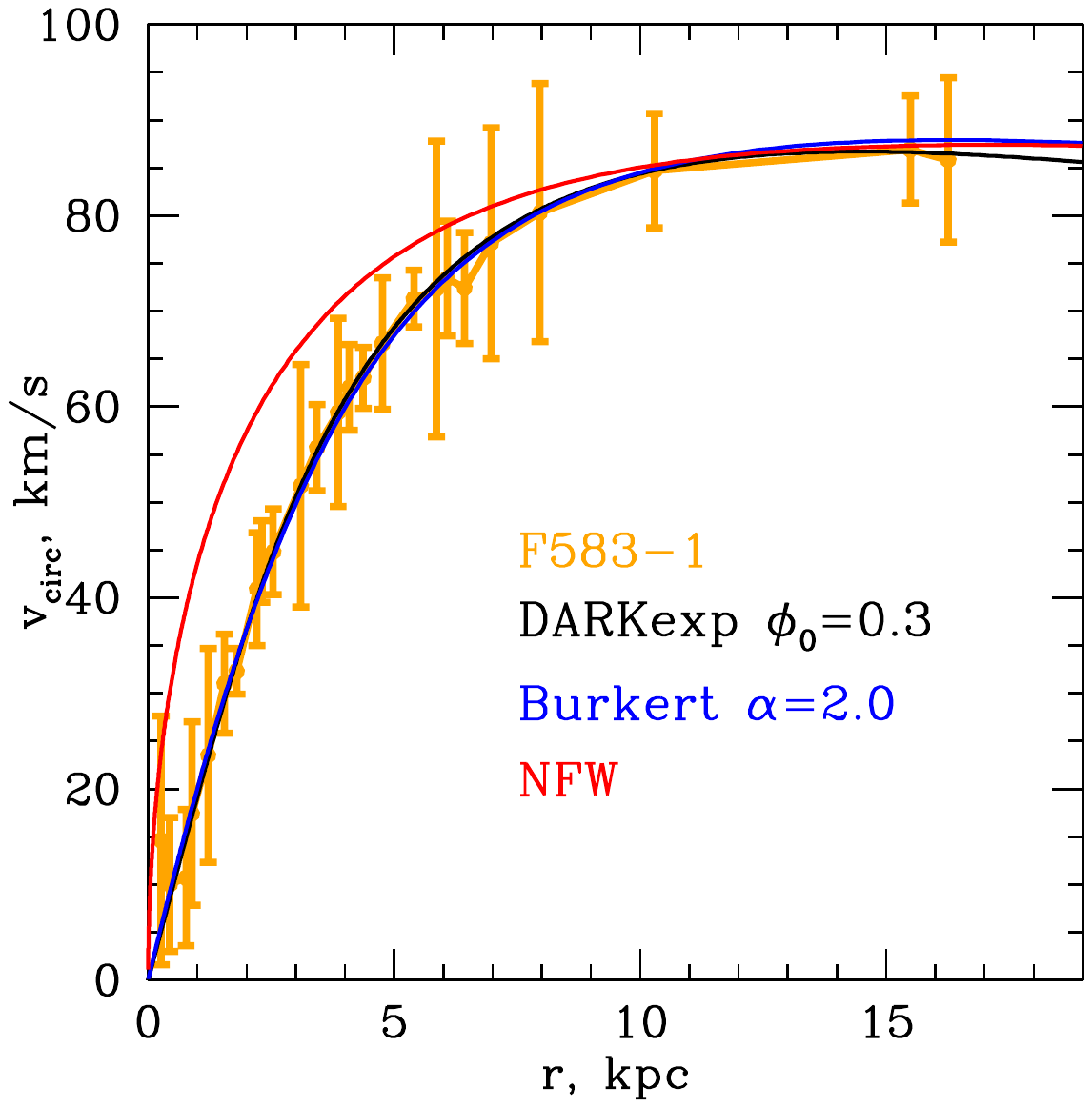}
    \includegraphics[width=0.48\linewidth]{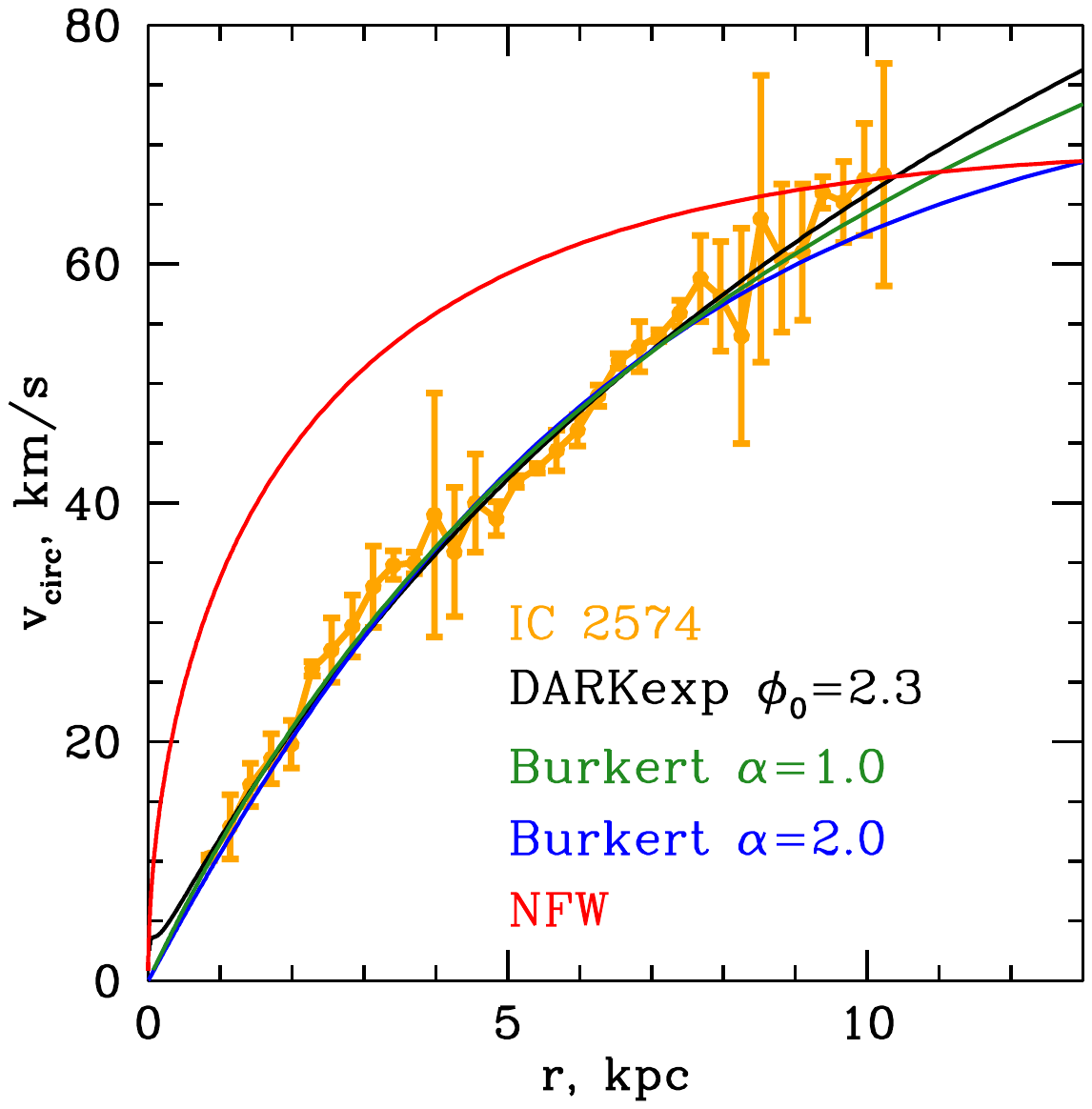}    
    \vskip-2cm
    \caption{Two examples of galaxies and fits with various models. We chose these particular galaxies because \cite{Oman2015} uses these as examples where neither NFW nor hydrodynamic simulations can fit the observed rotation curves. DARKexp is shows in black, and provides an excellent fit in both cases. Blue is the standard Burkert profile, while green (right panel) is generalized Burkert, $\rho\propto [(r+r_c)(r^\alpha+r_c^\alpha)]^{-1}$, with $\alpha=1$.}
    \label{fig:xyz}
\end{figure*}

\section{Theoretical prediction for dark matter density profiles}\label{sec:theory}

Results of N-body simulations are widely used to obtain empirical characteristics of dark-matter halos, but a derivation of their properties from fundamental physics would significantly improve our understanding. While merging, interacting and substructured systems are probably too complex for a simple theoretical analysis, the structure of isolated, relaxed systems is within reach of a theory. \cite{HW2010} applied statistical mechanics to collisionless self-gravitating systems, and found that their \llrw{maximum entropy} energy distributions are described by, $n(E)\propto\exp(-\beta[E-\Phi_0]-1)$, called DARKexp, which has one shape parameter, $\phi_0=\beta\Phi_0$, interpreted as a dimensionless depth of the potential well. Here, $E$ is the energy---kinetic and potential---of a particle, and $\epsilon=\beta E$ is the dimensionless energy. 

\llrw{There are two critical differences between other statistical mechanics models and DARKexp: (1) it accounts for the collisionless nature of the systems by interpreting the occupation number as the distribution in energy, $n(E)$, and (2) because the halo center, i.e., the bottom of the potential well can accommodate only a handful of particles, those with most bound energies, the Stirling approximation is replaced by a better one. These changes eliminate the common problems of other approaches, notably infinite total mass and energy \citep{lyn67}. DARKexp has been recently extended beyond the equilibrium state; \cite{WH2022} obtained an entropy functional that increases during the evolution of idealized collisionless systems, and also N-body simulated dark-matter halos from IllustrisTNG \citep{Francis2025}.}

When translated into density profiles, $\phi_0\approx 3.2$--3.3 systems resemble NFW, and $\phi_0\approx 4$ are well approximated by Einasto profiles of index $0.17$ \citep{WH2010}. Low $\phi_0$ systems, $\lesssim 2$, have flat density cores, while $\phi_0\gtrsim 8$ have an extended radial range where $\rho\propto r^{-2}$. This is illustrated in Figure~\ref{fig:phi0comp}, for the 3 dimensional density profiles (left panel), as well sky-projected profiles (right panel).

DARKexp was derived from fundamental theory and as such, it is not an {\it ad hoc} fitting function. Despite this, it has been shown to give excellent fits to $n(E)$ and density profiles of simulated CDM halos of massive galaxies and galaxy clusters \citep{WHW2010,Hjorth2015,Nolting2016}, and density profiles of observed galaxy clusters \citep{Beraldo2013, Umetsu2016}. DARKexp also provides better fits to globular cluster surface brightness profiles than single King models \citep{Williams2012}.

\newer{Note that, to date, DARKexp has not been tested against rotation curves of galaxies. Although it has been shown that DARKexp can produce cored profiles, this does not necessarily mean that it can provide acceptable fits to a range of rotation curves.  For example, in Appendix~\ref{sec:FDM}, we show that wave (fuzzy) dark matter, in its present form, can produce cored profiles but fails to fit the range of profiles observed in low-mass galaxies.}

We assume that the observed rotation curves are dominated by dark matter, and/or collisionless stellar distributions, i.e., DARKexp is applicable to these systems. For UFD galaxies we assume that the stellar distribution traces that of the dark matter.

\begin{figure*}
    \centering
    \vskip-0.85cm
    \includegraphics[trim={0.8cm 5cm 0.85cm 2cm},clip,width=0.32\linewidth]{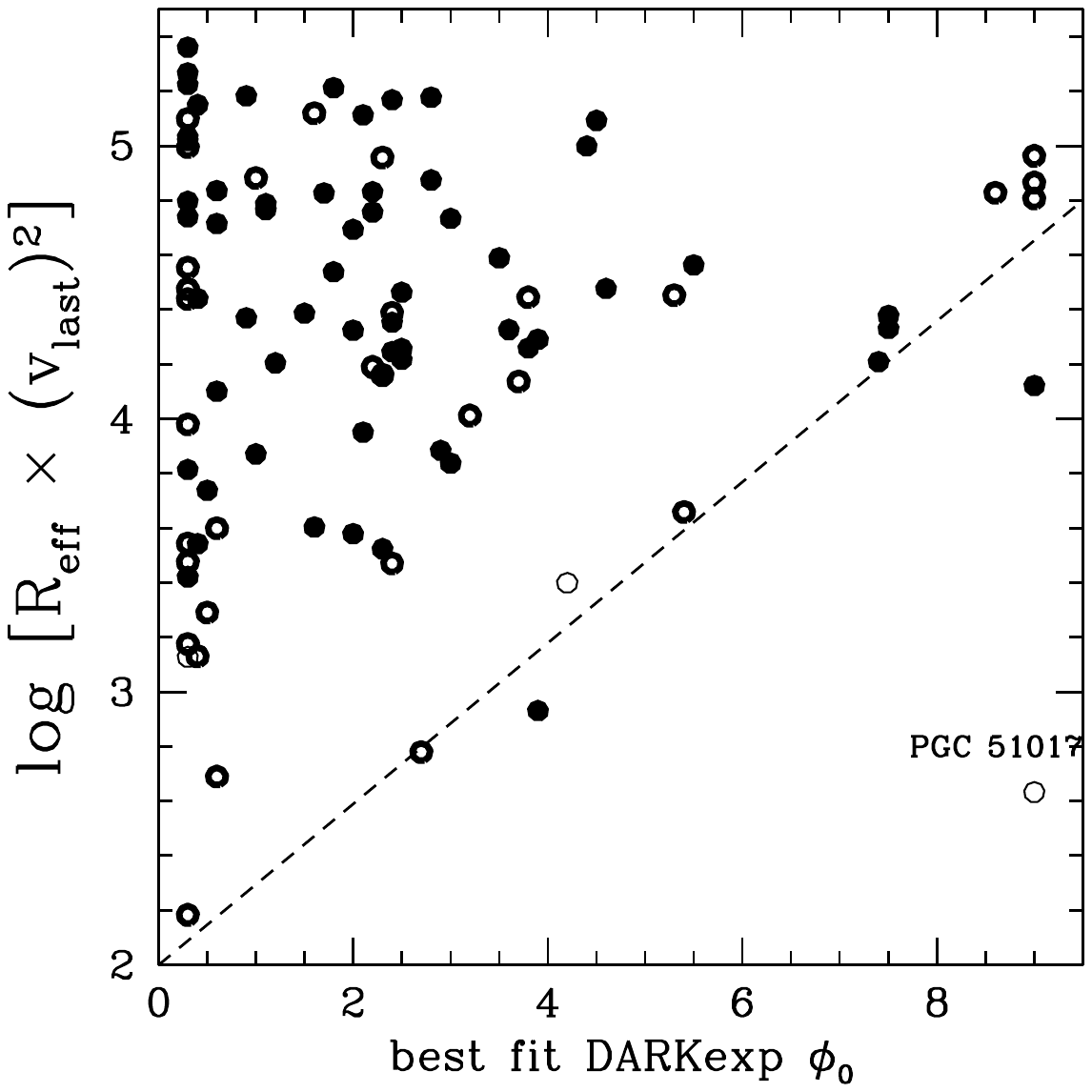}
    \includegraphics[trim={0.8cm 5cm 0.85cm 2cm},clip,width=0.32\linewidth]{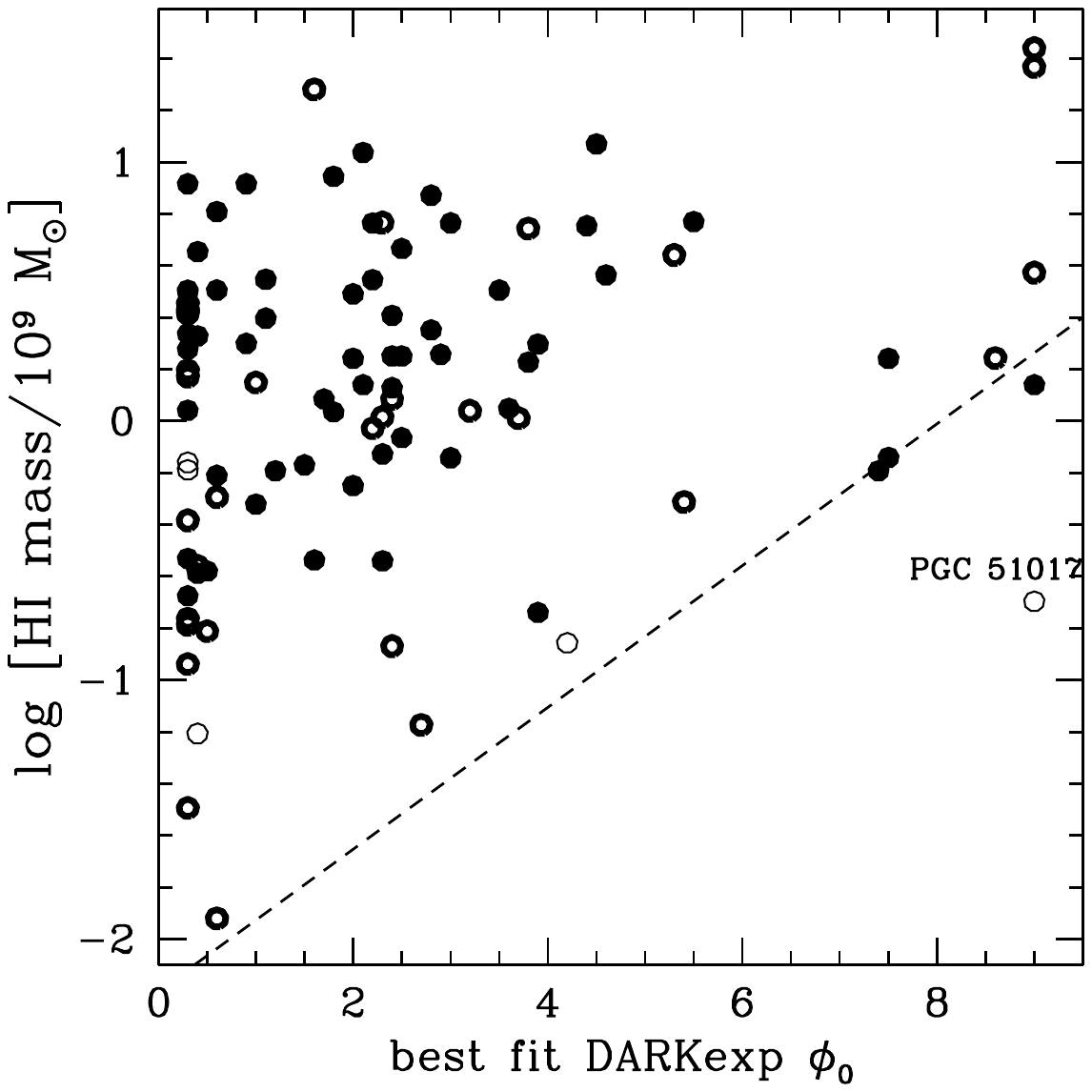}
    \includegraphics[trim={0.8cm 5cm 0.85cm 2cm},clip,width=0.32\linewidth]
    {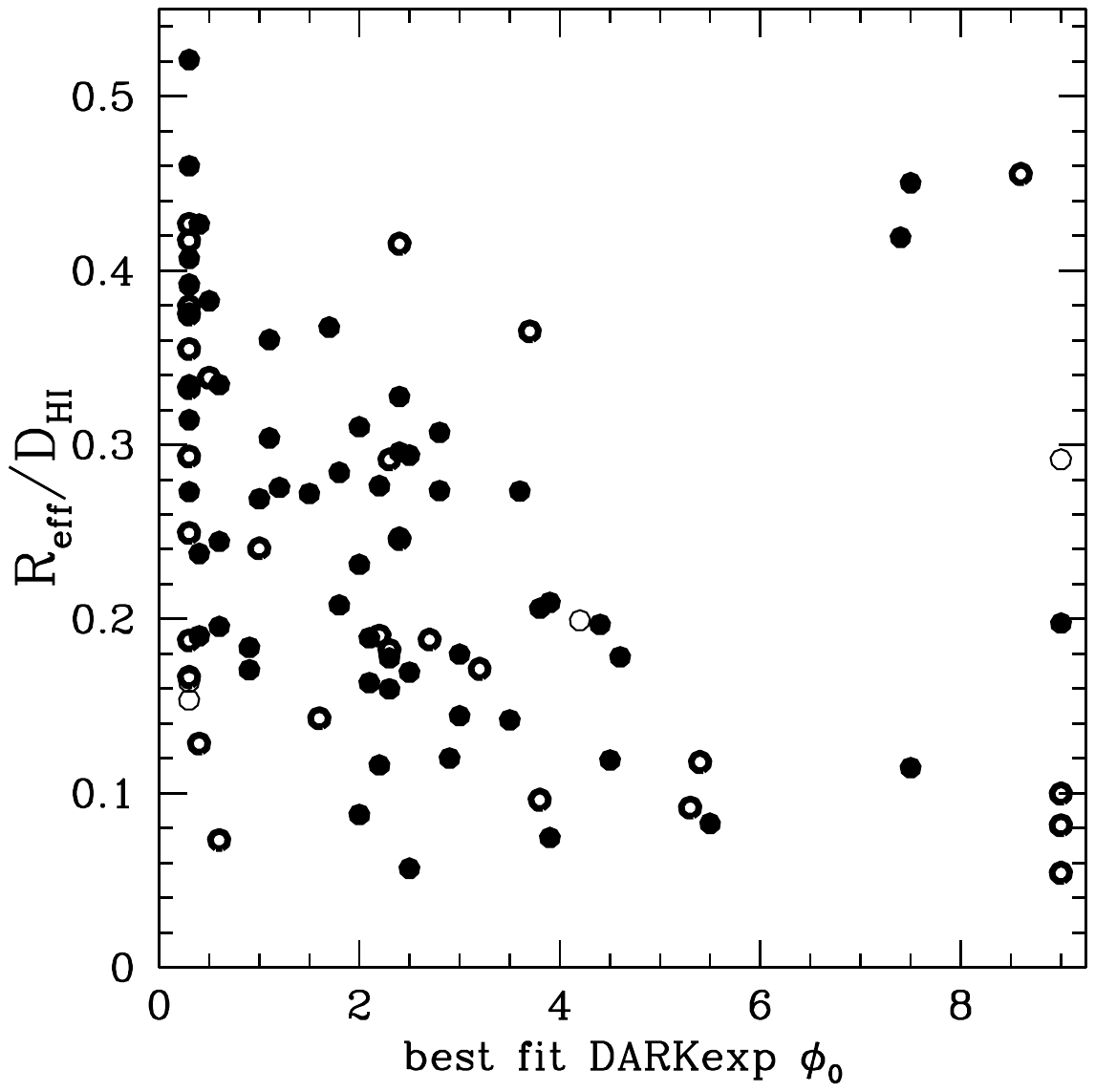}    \\
    \includegraphics[trim={0.8cm 5cm 0.85cm 2cm},clip,width=0.32\linewidth]{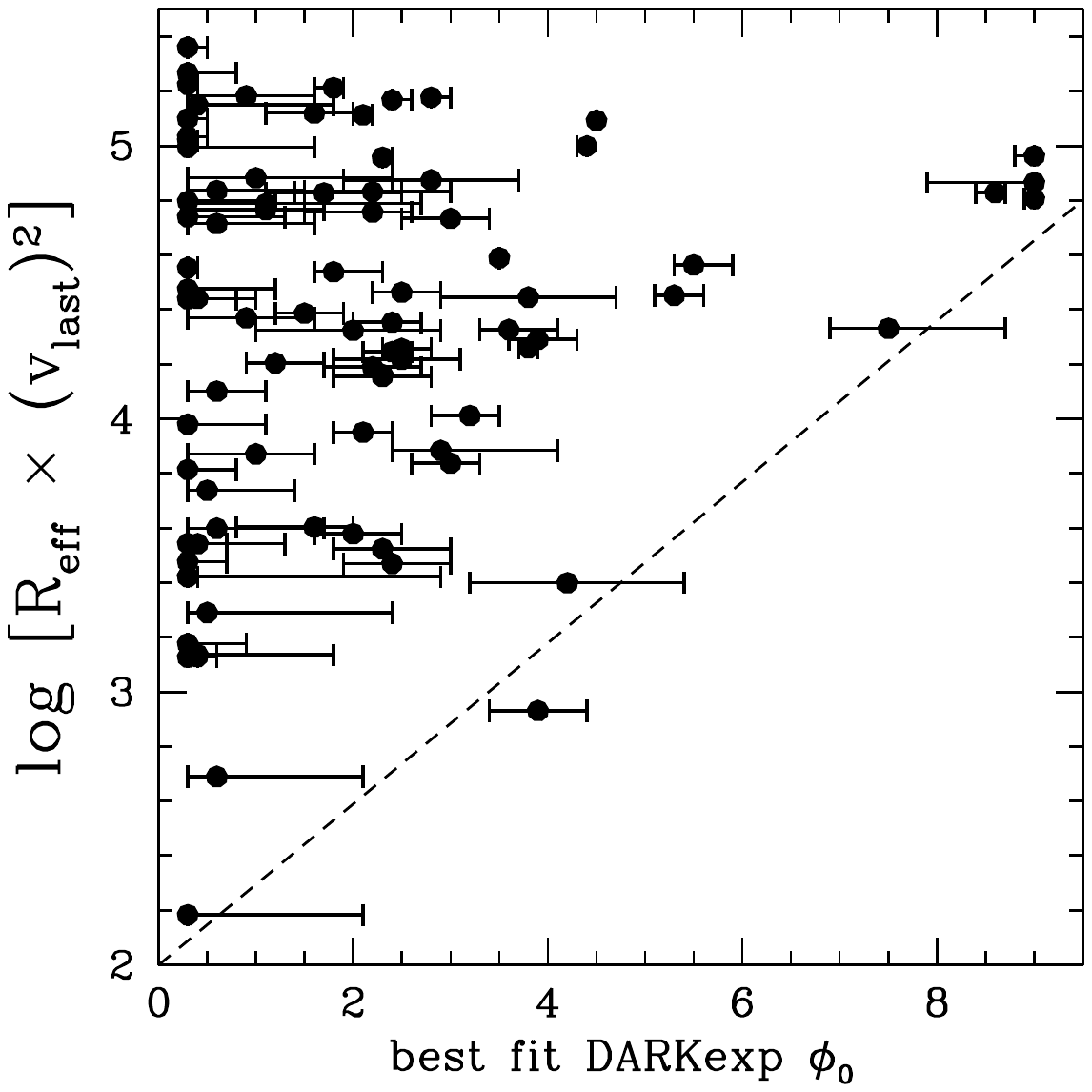}
    \includegraphics[trim={0.8cm 5cm 0.85cm 2cm},clip,width=0.32\linewidth]{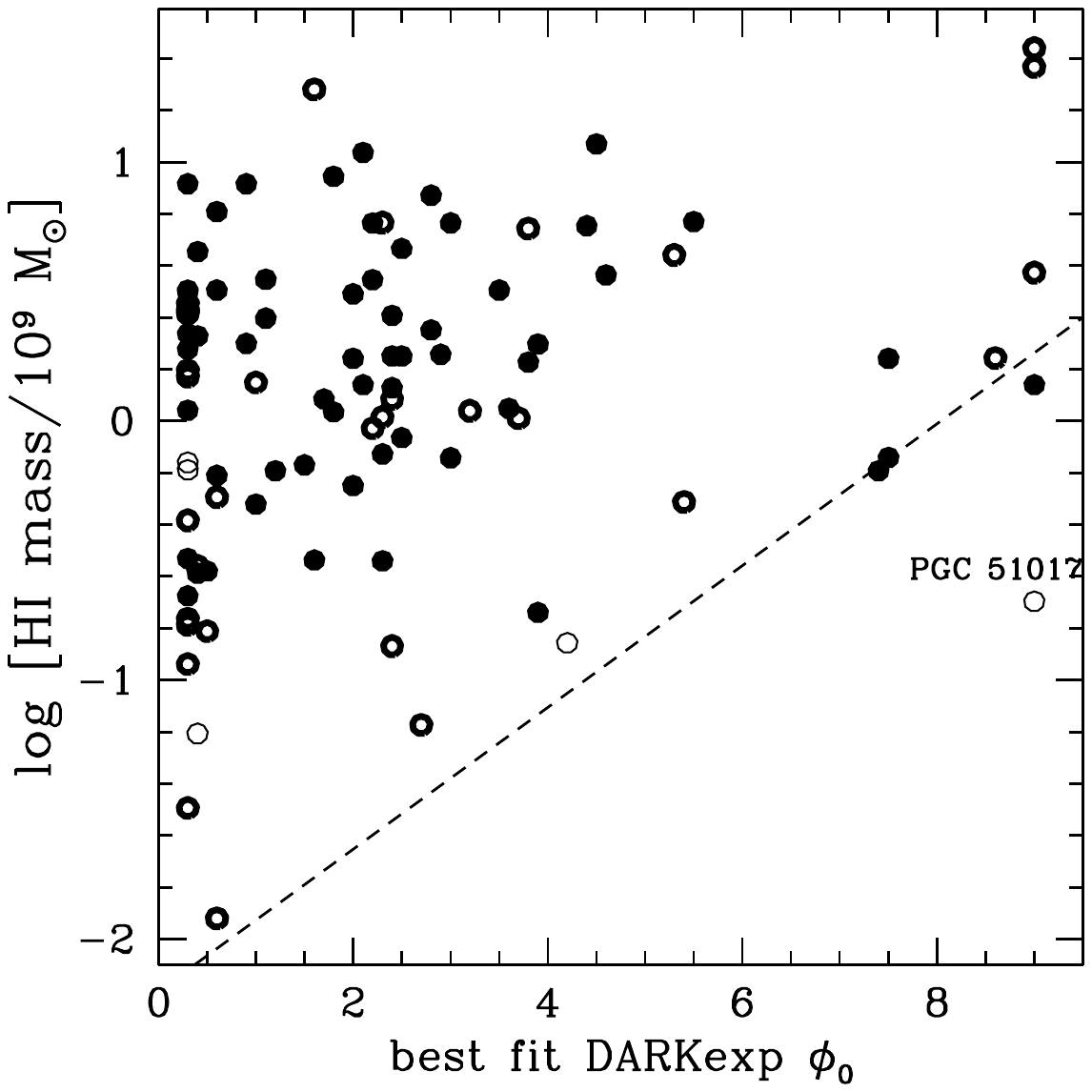}
    \includegraphics[trim={0.8cm 5cm 0.85cm 2cm},clip,width=0.32\linewidth]
    {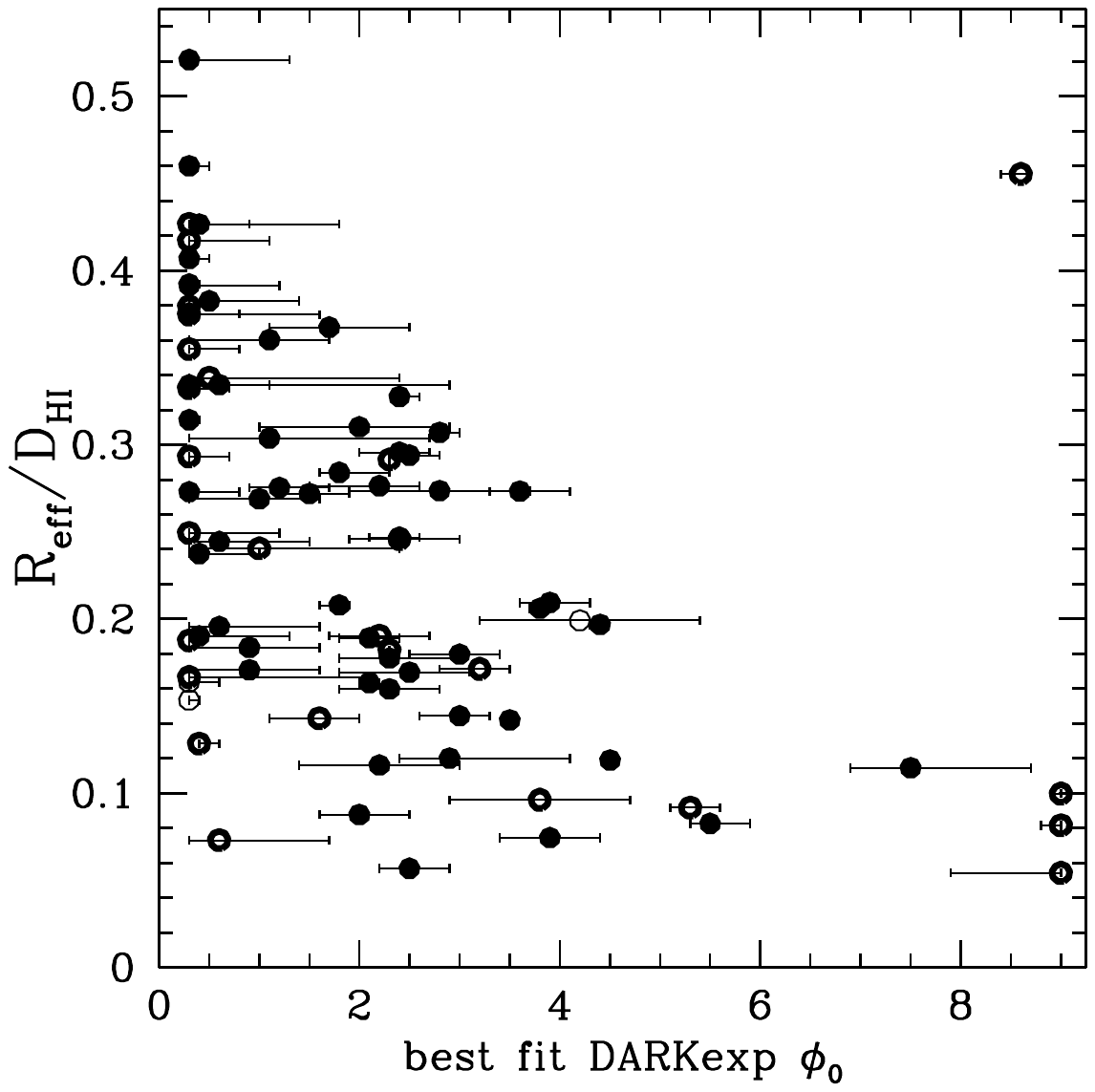} 
    \caption{\new{Three physical properties (three columns) of galaxies vs.\ best-fit DARKexp $\phi_0$. The top row shows all 96 galaxies, while the bottom row contains galaxies where the uncertainty on $\phi_0$ is less than $3$. There are 86 of these, and their uncertainties are shown as well.} 
    {\it Left:} Total dynamical mass ($GM\propto rv^2$), approximated as effective radius times the square of the velocity. 
    {\it Middle:} Total HI gas mass, as given in the SPARC catalog.
    {\it Right:} 
    \newer{The ratio of the effective radius in $3.6\mu$m, $R_{\rm eff}$, and the radius where HI gas attains 1 $M_\odot$/pc$^2$, $D_{\rm HI}$.}
    In all panels, the three types of points indicate quality flags from the SPARC catalog: filled circles, thick empty circles, and thin empty circles have high, medium, and low quality, respectively. The dashed lines delineate regions that galaxies appear to avoid. The outlier discussed in Section~\ref{sec:fits} (top row) is labeled.}
    \label{fig:correl}
\end{figure*}

In the following sections we show that both sets of data---galaxy rotation curves and UFD stellar density---are consistent with DARKexp, which admits a range of central slopes, including cores and cusps. Therefore, it is possible that the core-cusp problem, widely discussed in the literature, may not be a problem, and that cold collisionless dark matter is not necessarily ruled out by these observations.

\section{Fitting DARKexp}\label{sec:fits}

The DARKexp $n(E)$ has a simple analytic form, but the 3D density and hence the projected 2D density profiles do not, and have to be numerically calculated from $n(E)$. Therefore we first generated 3D, and projected 2D DARKexp density profiles, spaced by $\Delta\phi_0=0.1$. Each $\phi_0$ was fit to the data separately, using 2 parameters (radial and density normalization) and the minimum reduced $\chi_\nu^2$ for each galaxy was used as the final best-fit. Finding $\chi_\nu^2$ for SPARC and UFD galaxies was done using
$\chi_\nu^2=\frac{1}{I-3}\sum_{i=1}^I(v_{{\rm obs},i}-v_{{\rm mod},i})^2/\sigma_{v,i}^2$, and 
$\chi_\nu^2=\frac{1}{I-3}\sum_{i=1}^I(N_{{\rm obs},i}-N_{{\rm mod},i})^2/\sigma_{N,i}^2$, respectively.
Here, $I$ is the total number of radial bins for that galaxy, $v_{{\rm obs},i}$ and $v_{{\rm mod},i}$ are the observed and DARKexp predicted circular velocities in radial bin $i$, while $N_{{\rm obs},i}$ and $N_{{\rm mod},i}$ are the observed and predicted star counts in radial bins $i$. Uncertainties $\sigma_{v,i}$ and $\sigma_{N,i}$ were taken from the SPARC catalog and \cite{Richstein2024}, respectively.
We limit the $\phi_0$ range to $0.3-9.0$, because DARKexp of $\phi_0<0.3$ ($\phi_0>9$) look similar to those of $\phi_0=0.3$ ($\phi_0=9$), as illustrated in Figure~\ref{fig:phi0comp}.

Figures~\ref{fig:rotcurve1}--\ref{fig:rotcurve4} show the results for the 96 SPARC galaxies, arranged in order of increasing $v_{\rm last}$. \new{The upper and lower uncertainties on fitted $\phi_0$, \newer{displayed in each panel}, were calculated from $\phi_{0, \rm hi}$ and $\phi_{0, \rm lo}$, the values of $\phi_0$ where $\chi^2$ (not reduced $\chi_\nu^2$) just exceeds $\chi_{\rm min}^2+1$, and $\chi_{\rm min}^2$ is the minimum value, corresponding to the best fitting $\phi_0$. The top row of Figure~\ref{fig:correl} shows all 96 galaxies without uncertainties, while the bottom row excludes galaxies with large error bars ($>3$), and plots error bars for the other 86.
Note that for galaxies with sudden discontinuities in the rotation curves $\chi_{\rm min}^2$ was often greater than 1; no smooth model will be able to fit such galaxies with $\chi_\nu^2<1$. }

The large majority of the fits are very good. Some rotation curves have features that deviate from the fits; these can be due to various reasons. For example, uncertainties in the inclination angle of the outer disk of NGC 7793 (1st row in Fig.~\ref{fig:rotcurve3}) can affect its rotation curve determination \citep{Koribalski2018}, NGC 6015 (3rd row in Fig.~\ref{fig:rotcurve4}) has a warped disk beyond $\sim 14$ kpc \citep{Verdes1997}, and NGC 289 (4th row in Fig.~\ref{fig:rotcurve4}) has a satellite perturber $\sim 20$ kpc away \citep{Walsh1997}. 

The main conclusion from these figures is that a wide observed range of rotation curves shapes, and hence central density slopes, is well reproduced by DARKexp and its single shape parameter, $\phi_0$. This may suggest that galaxies from the least massive to about Milky Way mass  are dominated by collisionless dark matter. The exceptions are the central few kpc of massive galaxies in Figure~\ref{fig:rotcurve4}, for example, UGC 3580, UGC 2916, and UGC 8699, where baryons dominate and dynamics during evolution may not have been  collisionless.

A puzzling aspect regarding galaxy rotation curves that is often discussed in the literature is that galaxies having similar last recorded circular velocity (which is often the asymptotic flat rotation curve velocity), can display a wide range of rotation curve shapes at smaller radii. \cite{Oman2015} compares eight galaxies with rotational velocities $\sim 80$ km s$^{-1}$, to results of hydrodynamic simulations with baryonic feedback (their figures 4 and 5). They show that while some are well fit with simulated galaxies, others are not. Our Figure~\ref{fig:xyz} displays two examples of the latter. DARKexp fits are in black, and the NFW fits, similar to the ones in \cite{Oman2015}, are in red, and do not fit observations.

This fact that observed rotation curves show a wider range of shapes compared to simulated ones is called the diversity problem \citep{Oman2015,Sales2022}, with some papers proposing solutions that question the assumptions of equilibrium and circular motion in observed galaxies \citep[e.g.,][]{Sands2024}. DARKexp easily accounts for the diversity problem without the need to rely on a non-equilibrium state of observed rotation curves, or any baryonic physics. 

Figure~\ref{fig:xyz} also shows fits with a generalized Burkert profile \citep{Burkert1995}, $\rho\propto [(r+r_c)(r^\alpha+r_c^\alpha)]^{-1}$, in blue ($\alpha=2.0$), and green ($\alpha=1.0$). This commonly used profile has one shape parameter, just as DARKexp. Burkert profile fits are as good as DARKexp. However, the major difference between the two is that the Burkert profile is a fitting function, while DARKexp is derived from fundamental physics.

Related to the problem of diversity is the question of whether the shape of the rotation curve is correlated with any observed galaxy property. \llrw{Correlations of this sort have been found before, notably between the rotation curve velocities and galaxy luminosities \cite[e.g.,][]{Persic1991,Persic1996,DiPaolo2019}.} Figure~\ref{fig:correl} plots three quantities as a function of best fit DARKexp $\phi_0$: an approximate dynamical mass (left panels), total HI gas mass (middle), and \newer{the ratio of the effective radius in $3.6\mu$m, $R_{\rm eff}$, and the radius where HI gas reaches surface density of  $1\,M_\odot$/pc$^2$, $D_{\rm HI}$ (right). The last quantity was inspired by the work of \cite{Chamba2024}, who find that a closely related quantity is useful in exploring quenching and feedback in dwarf galaxies.} The three types of points indicate quality flags from the SPARC catalog: filled circles, thick empty circles, and thin empty circles have high, medium, and low quality, respectively. The dashed line in the right and middle panels delineates the region that galaxies appear to avoid. These two panels indicate that galaxies of low mass tend to avoid high $\phi_0$ values, whereas high mass galaxies can have a wide range of $\phi_0$'s. 

\newer{The third set of panels shows that $\phi_0$ is related to a parameter that quantifies the structure of galaxies, namely the ratio of two types of scale lengths. It may make sense that $\phi_0$, which is a structural parameter that describes the distribution of total mass, is related to the structural parameter of the baryonic distribution.}
\new{Excluding galaxies with large uncertainties on $\phi_0$ (lower panels have 86, instead of all 96 galaxies), preserves these trends.} Though the full significance of these trends is not yet clear, it is noteworthy that there appear to be connections between the observed galaxy properties and $\phi_0$.

Figure~\ref{fig:correl} has one outlier: PGC 51017 (top row), also known as SBS 1415+437, a blue compact dwarf (the galaxy in the very first panel of Figure~\ref{fig:rotcurve1}). Its SPARC quality flag is `low', and it has a rather short radial range in the SPARC catalog, but \cite{Thuan1999} obtained a more extended rotation curve that has maximum velocity $\sim 80$ km s$^{-1}$, instead of $\sim 20$ km s$^{-1}$. These data suggest that the galaxy could be placed at higher $v_{\rm last}$, and smaller $\phi_0$.  The issue of how the limited radial extent of the rotation curve affects the fitted $\phi_0$ value is briefly discussed in Appendix~\ref{sec:MW}, in relation to the Milky Way.

\begin{figure*}[t!]
    \centering
    \vskip-1.0cm
    \includegraphics[width=0.3\linewidth]{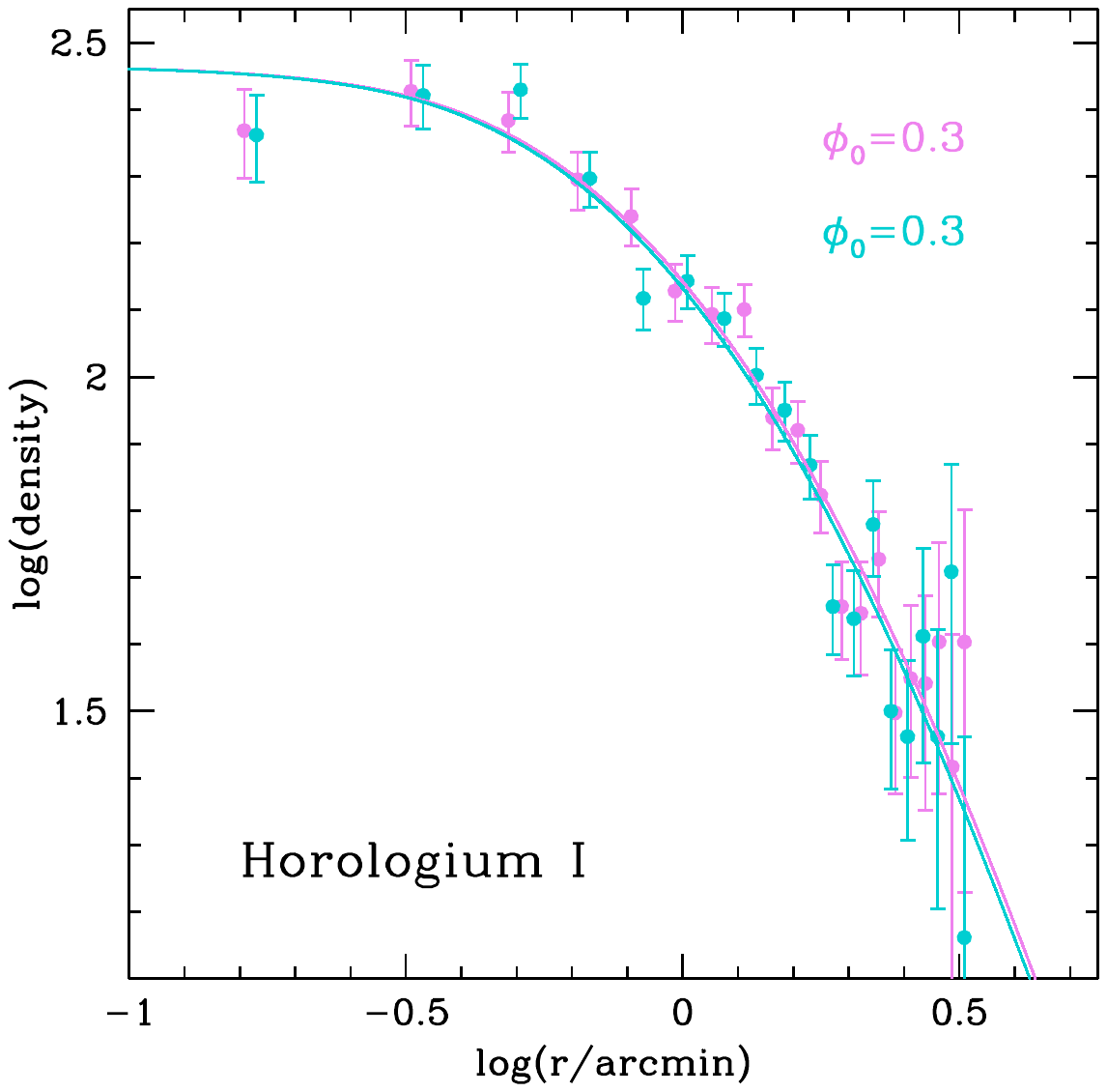}
    \includegraphics[width=0.3\linewidth]{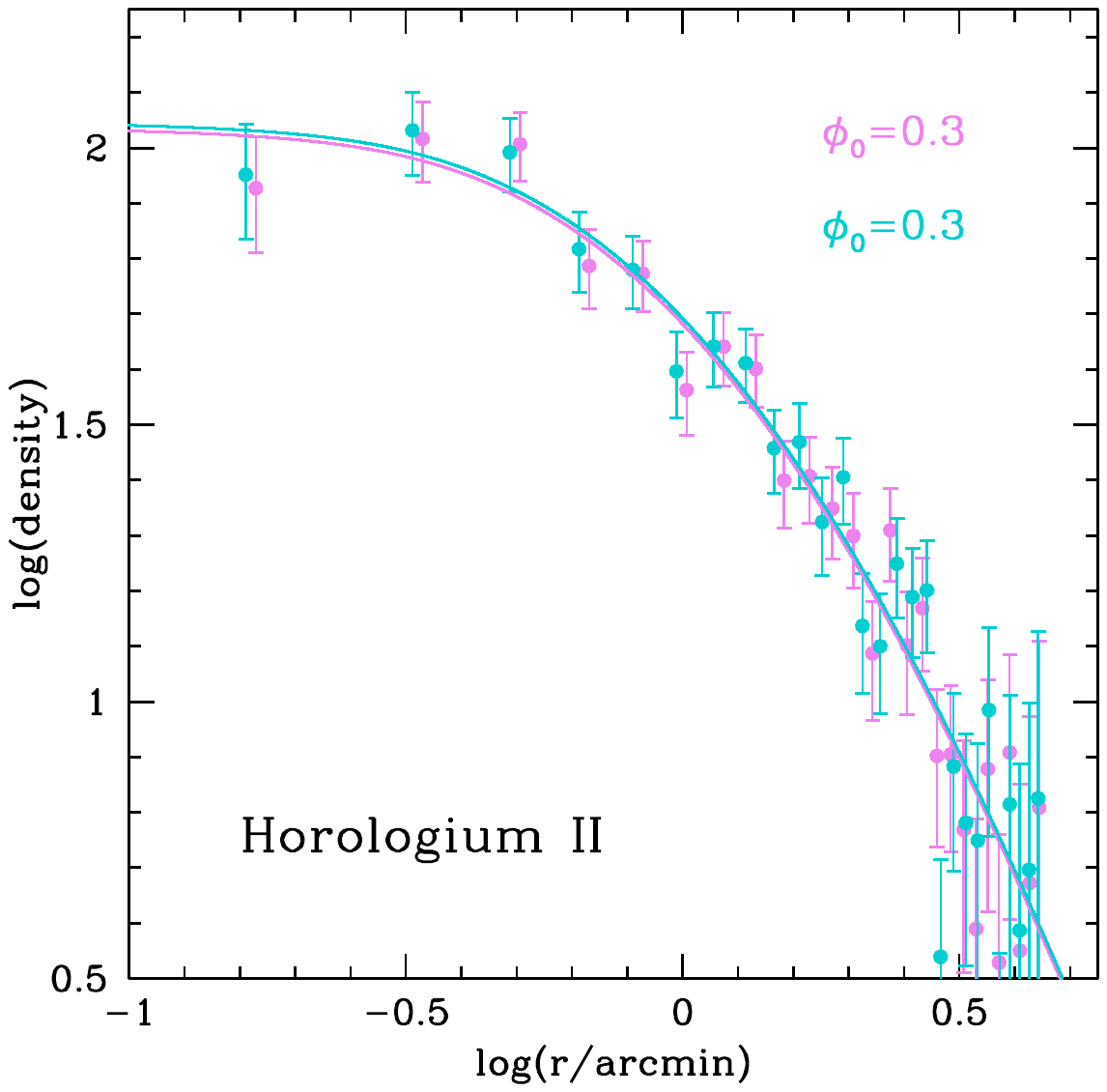}
    \includegraphics[width=0.3\linewidth]{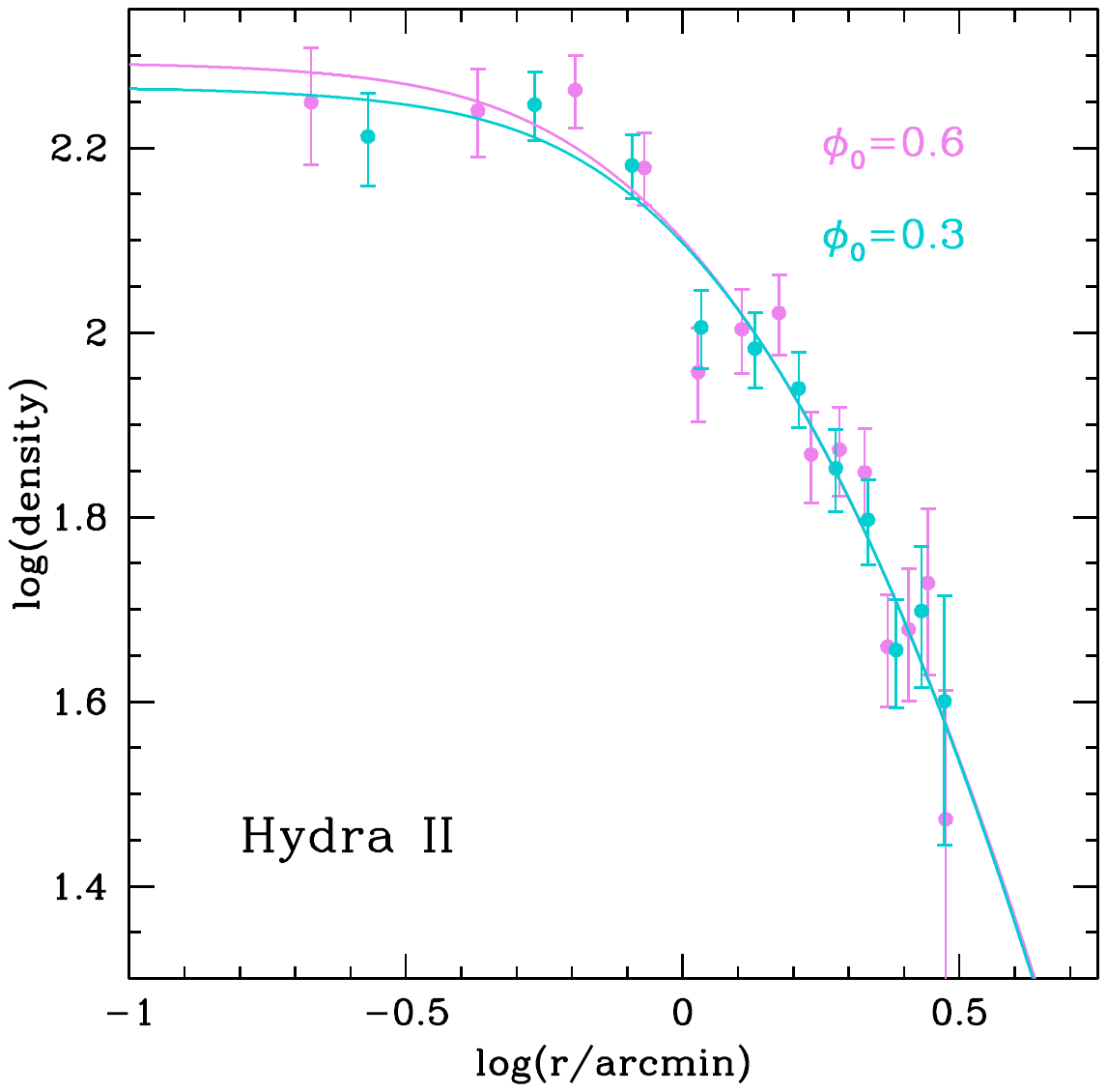}
    \vskip-2.2cm
    \includegraphics[width=0.3\linewidth]{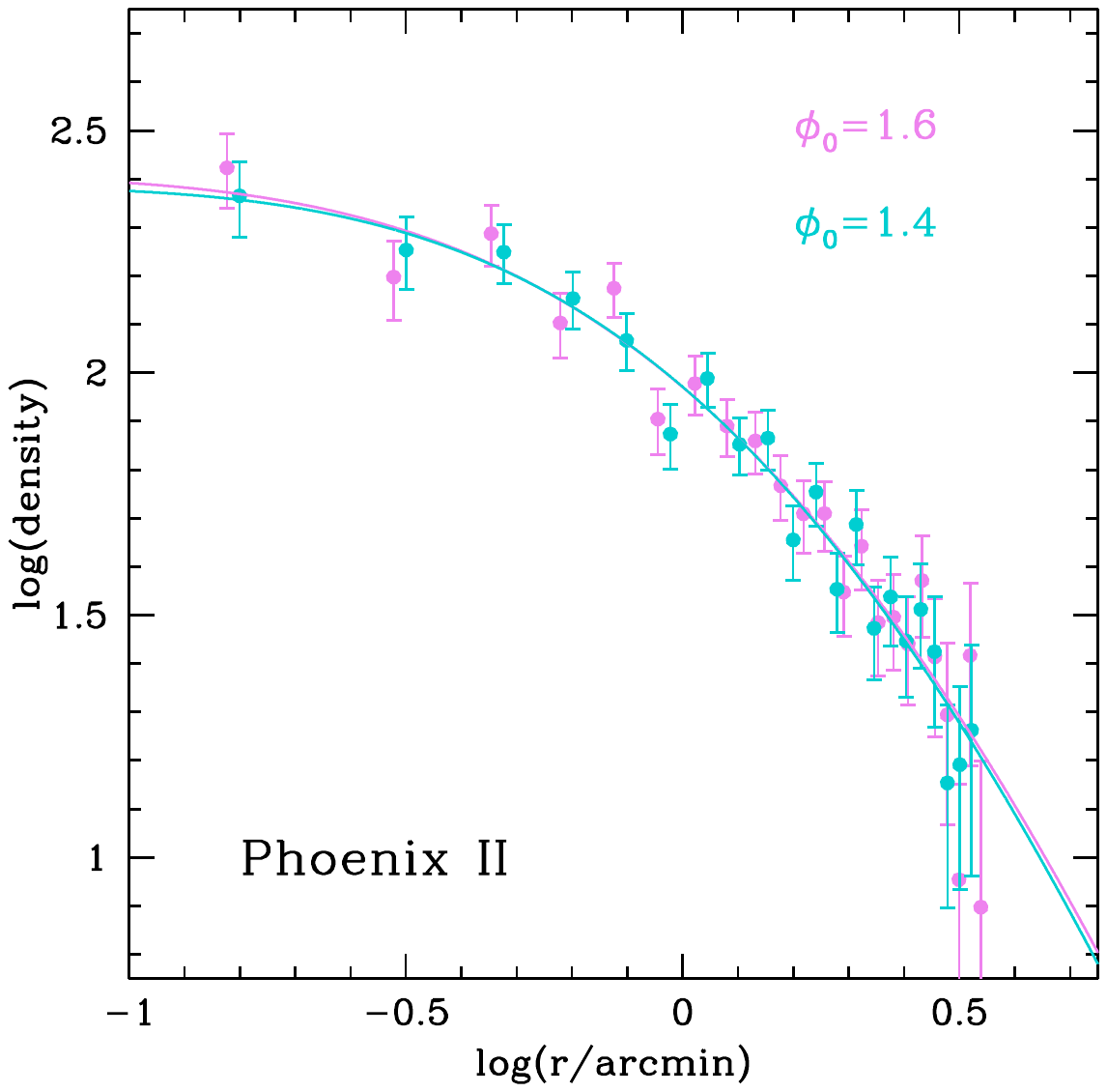}
    \includegraphics[width=0.3\linewidth]{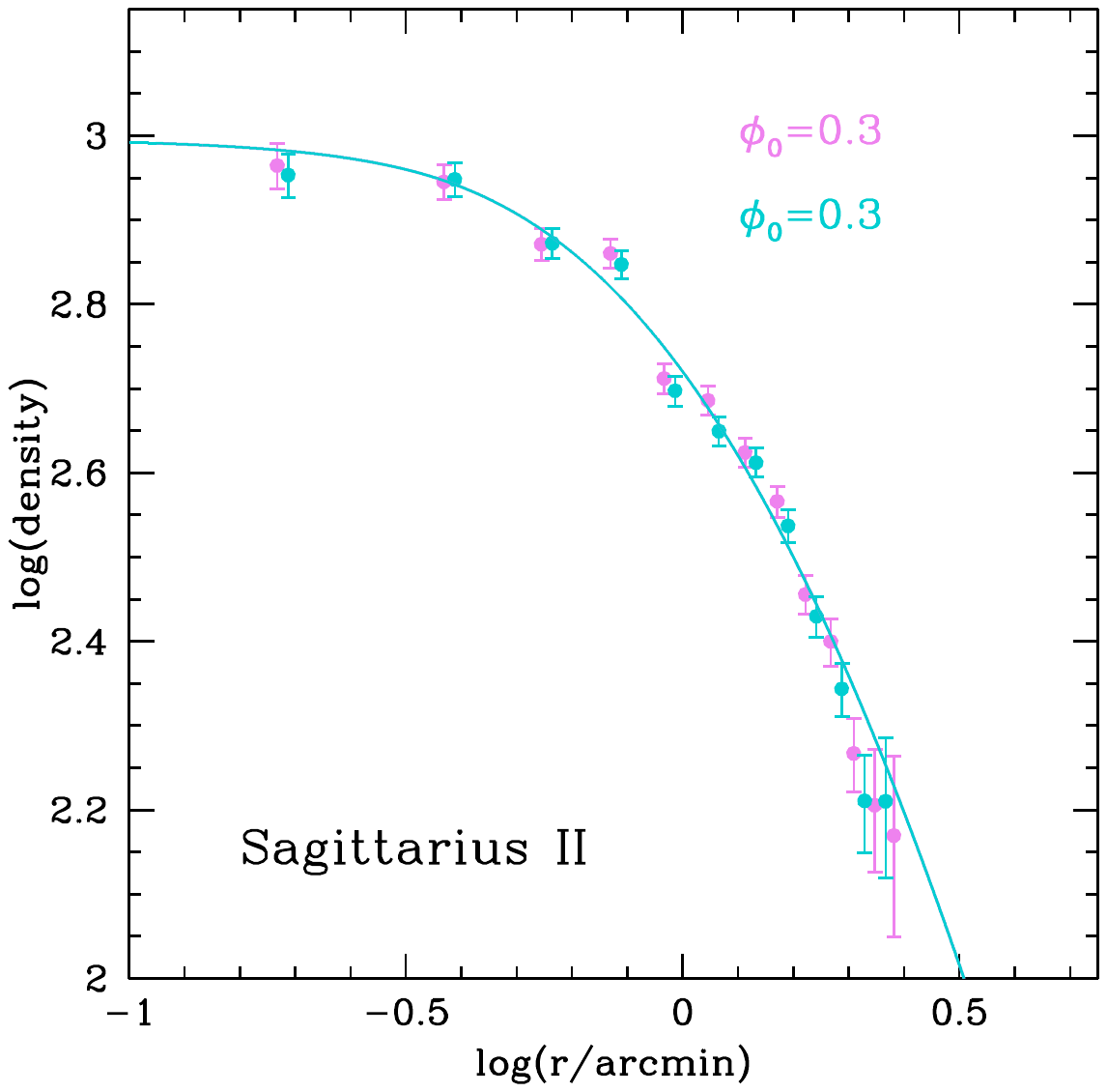}
    \includegraphics[width=0.3\linewidth]{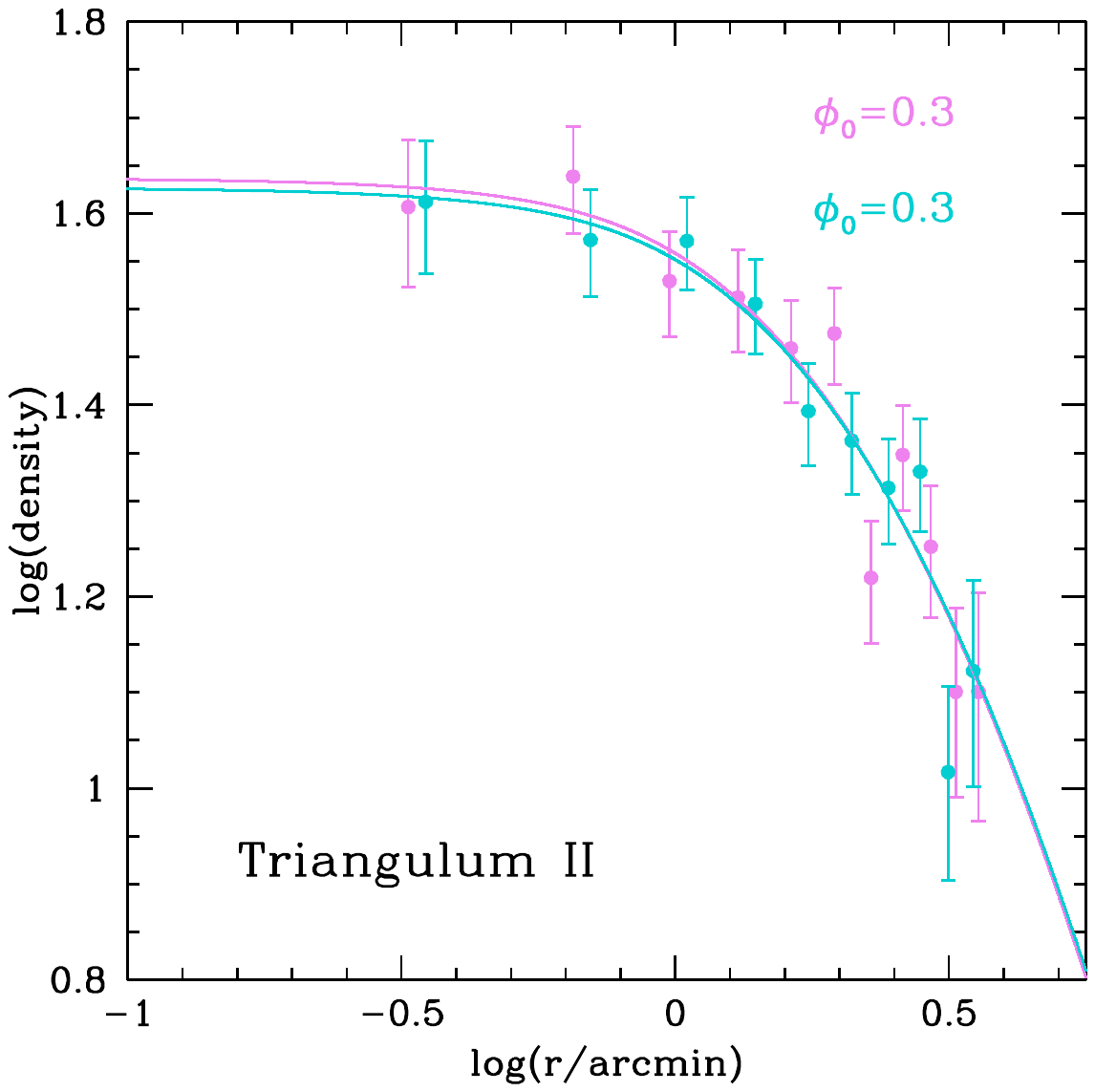}
    \vskip-1.25cm
    \caption{DARKexp fits to the elliptically averaged 2D projected stellar density distribution of 6 UFDs from \cite{Richstein2024}. The two sets (pink and teal) of data points and corresponding fits represent two methods of identifying UFD galaxies.}
    \label{fig:fits}
\end{figure*}

The DARKexp fits to projected stellar density profiles of UFDs are shown in Figure~\ref{fig:fits}: the pink and teal sets correspond to two methods of identifying UFDs: using Plummer and exponential profiles, respectively \citep[see][for details]{Richstein2024}. All fits are very good, with flat density cores well modeled with low $\phi_0$ values.

\section{Discussion and Conclusions}\label{sec:sum}

The diversity of rotation curve shapes of disk galaxies, including the flat density cores at the centers of dwarf and UFD galaxies, where stellar feedback is inefficient or even negligible, are often presented as evidence against the cold collisionless nature of dark matter \citep[e.g.,][]{Bullock2017,Boldrini2021,Sanchez2024}. These conclusions are based on the results of N-body simulations, which show that CDM halos follow cuspy NFW profiles, with only a modest dispersion in shape, and are unable to account for observations. Adding stellar feedback in hydrodynamic simulations widens the range of rotation curve shapes, but still does not fully reconcile simulations and observations \citep{Santos2018,Santos2020}.

Our analytical theory, DARKexp \citep{HW2010,WH2022}, predicts a wider range of central density slopes, including central cores, and suggests that there may not be a core-cusp problem. DARKexp is a maximum entropy statistical mechanics theory of cold collisionless dark matter. The inner slope depends on the value of its sole shape parameter, $\phi_0$. In this paper we demonstrate that DARKexp dark-matter halos provide very good fits to the observed rotation curves of disk galaxies over a wide range of circular velocities (Fig.~\ref{fig:rotcurve1}-\ref{fig:rotcurve4}), and hence masses, and to the stellar density distribution of UFD galaxies (Fig.~\ref{fig:fits}). Thus, DARKexp appears to alleviate the core-cusp problem, and the diversity problem, at least to the extent that it allows a wide range of central density slopes.

\newer{We cannot rule out that the success of DARKexp in fitting a range of galaxy rotation curves and light profiles (as well as globular clusters \citep{Williams2012}) may simply be related to the fact that its one adjustable parameter, $\phi_0$, essentially controls the strength of the inner cusp. In other words, while DARKexp has been derived from basic physical principles, its success and versatility in fitting the centers of dynamical systems may be just a coincidence. On the other hand, as we show in Appendix~\ref{sec:FDM}, it is not a given that a theory that allows cores will automatically fit all rotation curves.}

\newer{Alternatively, it is also possible that DARKexp does provide a physical description of most galaxies. If so, this seems to be at odds with the body of literature based on N-body simulations which shows that dark-matter halos invariably produce cusps, as parametrized by the NFW profile and its variants.  We note, however, that simulated sub-halos do appear to have a range of central cusps and that dark-matter halo profiles are not universal \citep{Hjorth2015}. Notably, a few studies have claimed that some simulated halos are well fitted by profiles with more shallow cusps or no cusps at all (\citep{Vera-Ciro2013}, \citep{DiCintio2013}), and \citep{Springel2008} find that sub-halo profiles do not converge to a central power law but exhibit curving shapes well described by the Einasto form.}

It is unclear what would determine the central slope or DARKexp $\phi_0$ in this case: whether it is the galaxy's environment, its accretion rate \cite{Diemer2022,Diemer2025}, or something else.  While stellar feedback is likely important in disk galaxies, the fact that nearly all of the 96 SPARC galaxies are very well fit with DARKexp may indicate that these galaxies are largely in equilibrium and their evolution was dominated by collisionless dynamics.

\newer{The central potential, $\phi_0$, could be set early in the history of the Universe, and would not correlate with present-day properties of galaxies. \cite{Fitts2017} showed that the large dispersion in the halo-mass vs.\ stellar-mass relationship could be explained by the range in formation times, particularly whether the dominant star formation happened before or after reionization.
Alternatively, \cite{Londrillo1991}  showed that the dimensionless central potential of simulated halos resulting from dissipationless collapse depends on initial conditions, in particular the coldness of the collapse. Such initial conditions may indeed be different in massive halos vs.\ halos within halos, such as those leading to satellite UFD galaxies of the kind studied here. Finally, the formation and evolution of sub-halos may be different than those of massive halos which are well-fitted by cuspy density profiles. We found that there appears to be a relation between the fitted $\phi_0$ and approximate total mass, HI mass, as well as $R_{\rm eff}/D_{\rm HI}$, but the significance of these is yet to be explored.}

The apparent disagreement between the results of dark matter only N-body simulations, which do not show density cores, and the predictions of statistical mechanics that do yield cores for low $\phi_0$ values, does not have an explanation yet. Future work should examine the differences, in order to arrive at the fuller understanding of galaxies and dark matter.

\vspace{0.5cm}\noindent
\begin{acknowledgments}
We are grateful to Hannah Richstein and Nitya Kallivayalil for sharing their data on the six LMC ultrafaint dwarfs. We thank Claudia Scarlata, Sarah Pearson, and Cecilia Bacchini for useful suggestions.
JH was supported by a research grant (VIL54489) from VILLUM FONDEN.
\end{acknowledgments} 

\appendix

\section{\newer{Wave/Fuzzy Dark Matter}}\label{sec:FDM}

\newer{Wave, or fuzzy dark matter (FDM) was originally proposed as a solution to some astrophysical problems, mainly the core-cusp problem in low mass galaxies. FDM creates flat density cores, which could resolve the discrepancy between the density profiles of the observed and simulated cold dark-matter halos. For example, \cite{Broadhurst2020} modeled five dSph Local Group galaxies (Antlia II, Fornax, Sextants, Sculptor and Draco) with FDM density profiles, and showed that these are consistent with FDM in how their core radii relate to core mass (see also \citep{Pozo2024}). Here we compare the predictions of FDM to those of DARKexp. }

\newer{The density profile of an FDM halo depends on the mass of the axion-like bosonic particle, $m_\psi$, and the virial mass of the halo, $M_{200}$ The latter is uniquely related to the particle velocity dispersion ($GM_{200}/r_{200}=\sigma^2$, and $M_{200}=200[4\pi/3]{r_{200}}^3\rho_{\rm crit}$), and $\sigma$ sets the typical particle momentum, $m_\psi\,\sigma$. The uncertainty principle then determines the de Broglie wavelength, i.e., the typical scale of the density perturbations that affect the whole halo, and of the central soliton. It is the soliton that produces the density core; the random density perturbations are not relevant here.}

\newer{FDM density profiles have two free parameters, but one of them, the particle mass, is pretty much set at $m_\psi\approx 10^{-22}$eV by many recent papers \citep{Eberhardt2025,Broadhurst2025,Elgamal2024,Broadhurst2020}. The remaining one can be the core radius, $r_c,$ or the core mass, $m_c$, or the total halo mass, $M_{200}$. Once any of these is set, the rest can be calculated from theory, or simulations of FDM \citep[e.g.,][]{Schive2014,Liao2024}.}

\begin{figure}
    \centering
    \includegraphics[trim={0cm 5cm 0cm 2cm},clip,width=0.75\linewidth]{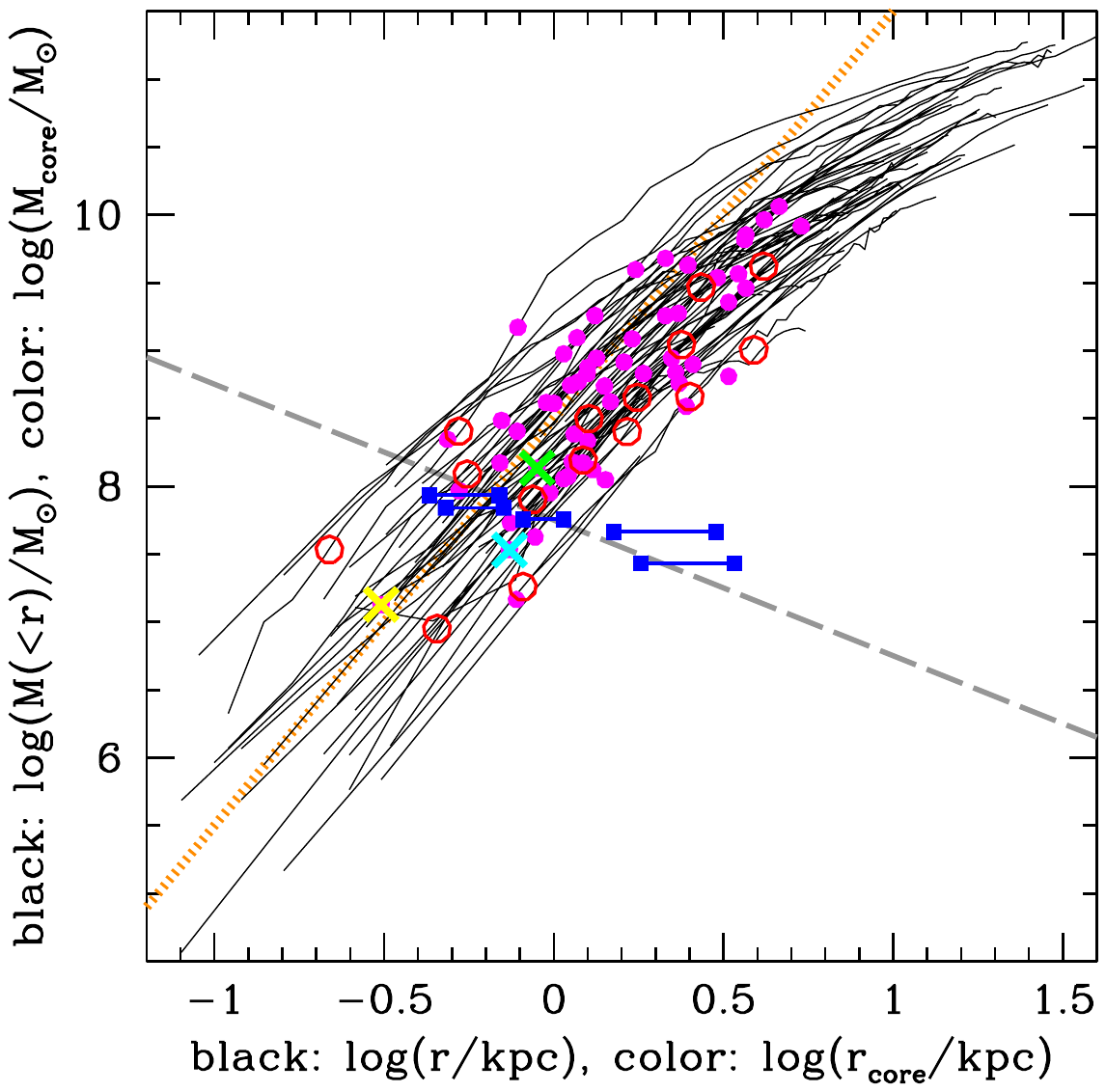}
    \caption{\newer{Black lines show enclosed mass profile of a subset of 56 SPARC galaxies used in Appendix~\ref{sec:FDM}. Magenta and red point indicate core radii and core masses of SPARC galaxies, estimated from rotation curves using prescriptions (i) and (ii) as described in the text. The slope of the orange dotted line corresponds to a flat density profile, or $\Gamma=0$. Blue horizontal lines span core radius uncertainties of 5 dSPh galaxies analyzed in \cite{Broadhurst2020}. Gray dashed line is the prediction of FDM. The  green, cyan, and yellow crosses, respectively, mark the 3 SPARC galaxies shown in Figure~\ref{fig:solitonRC}.}}
    \label{fig:CmassCrad}
\end{figure}

\newer{We estimate $r_c$ and $m_c$ for SPARC galaxies, as follows. We translate observed rotation curves to the enclosed mass as a function of radius, using $GM(<r)=r{v^2(r)}$. These profiles are shown as black thin lines in Figure~\ref{fig:CmassCrad}. (Similar results were obtained by \cite{Benito2025}.) Because these profiles are not monotonic, it is calculated as an average of the smallest and largest radii where the density slope becomes larger than $\Gamma$, where $\rho\propto r^{\Gamma}$, or the enclosed mass becomes steeper than $M(<r)\propto r^{\Gamma+3}$. We do that measurement in two different ways to encompass uncertainties associated with discrete rotation curve data: (i) using the $(i-1)$th and $(i+1)$th rotation curve points to get the slope, and assuming that slope values larger than $\Gamma=-0.5$ correspond to cores, (ii) using the $(i-2)$th and $(i+2)$th points, and $\Gamma=0$.  The second way is more conservative because it assumes a flat density core, and more widely separated rotation curve points for measuring the slope.}

\newer{Many SPARC galaxies have no core at all---and hence present a possible problem for FDM by themselves---these were excluded from Figure~\ref{fig:CmassCrad}. The remaining core radii are plotted as magenta solid points for choice (i), and as red empty points for (ii). There are 56 and 15 of these, respectively, out of an initial set of 96 we use in this paper. The cumulative mass profiles are shown for 56 galaxies as black thin lines. The orange dotted line corresponds to $\Gamma=0$, i.e., a flat density profile. One can see that even if our choices (i) and (ii) are not optimal, the observed core radii are still confined to the black lines.}

\newer{The five dSph from \cite{Broadhurst2020} are shown on the same figure in blue (the values were extracted from their Table 1, columns 5 and 6). The horizontal lines span the uncertainty in $r_c$.  The gray dashed line is the predicted relation for FDM, and the authors point out that the five dSph lie on this line. }

\newer{The first point to note is that the two sets of galaxies (blue vs. magenta or red) appear to belong to the same relation. The second point is that the five dSph may be a small subset of a broader distribution that does not follow FDM predictions, but goes quite orthogonally to it (orange dotted vs. gray dashed lines).}

\begin{figure}
    \centering
    \includegraphics[trim={0cm 5cm 0cm 2cm},clip,width=0.315\linewidth]{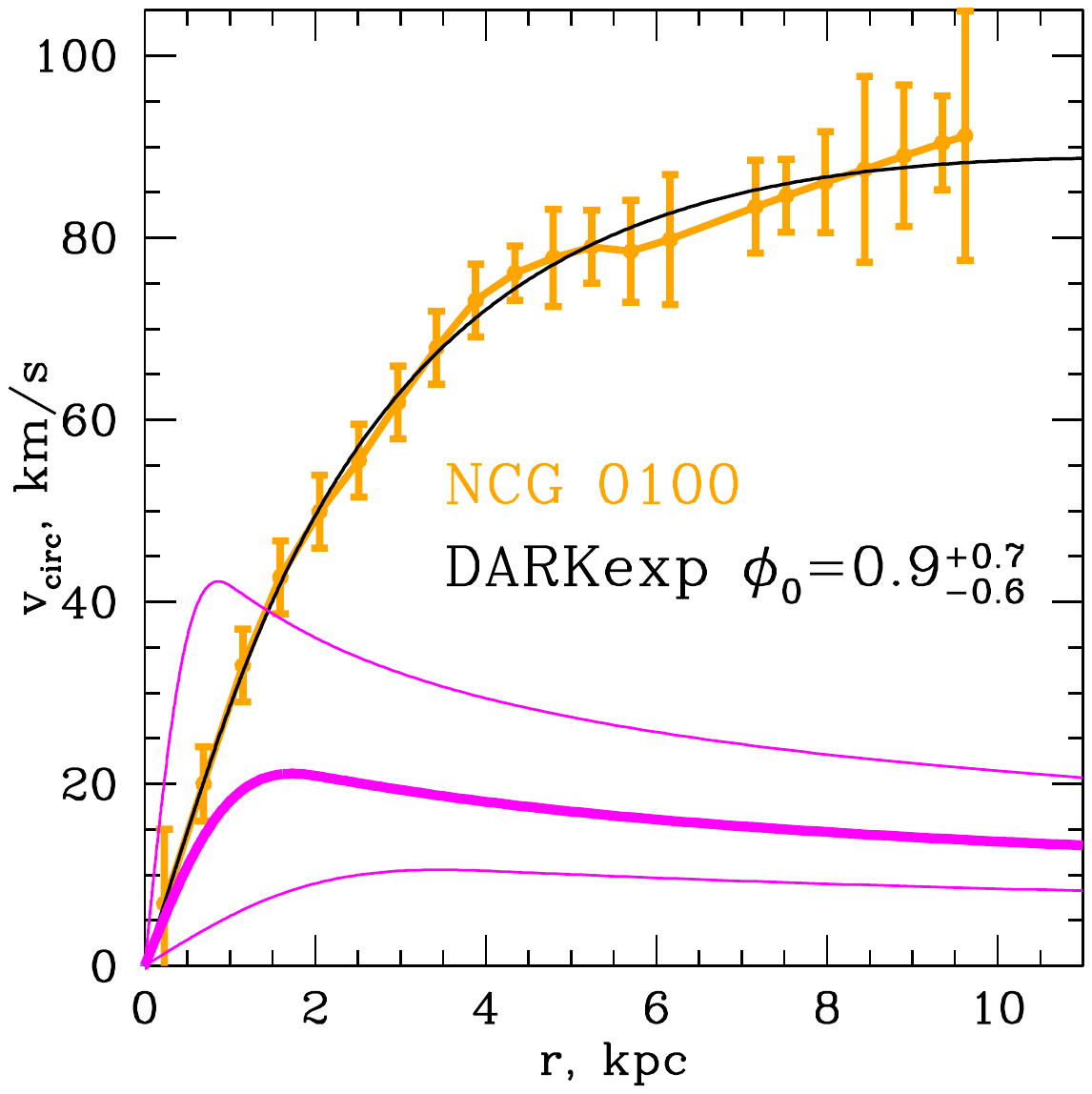}
    \includegraphics[trim={0cm 5cm 0cm 2cm},clip,width=0.315\linewidth]
    {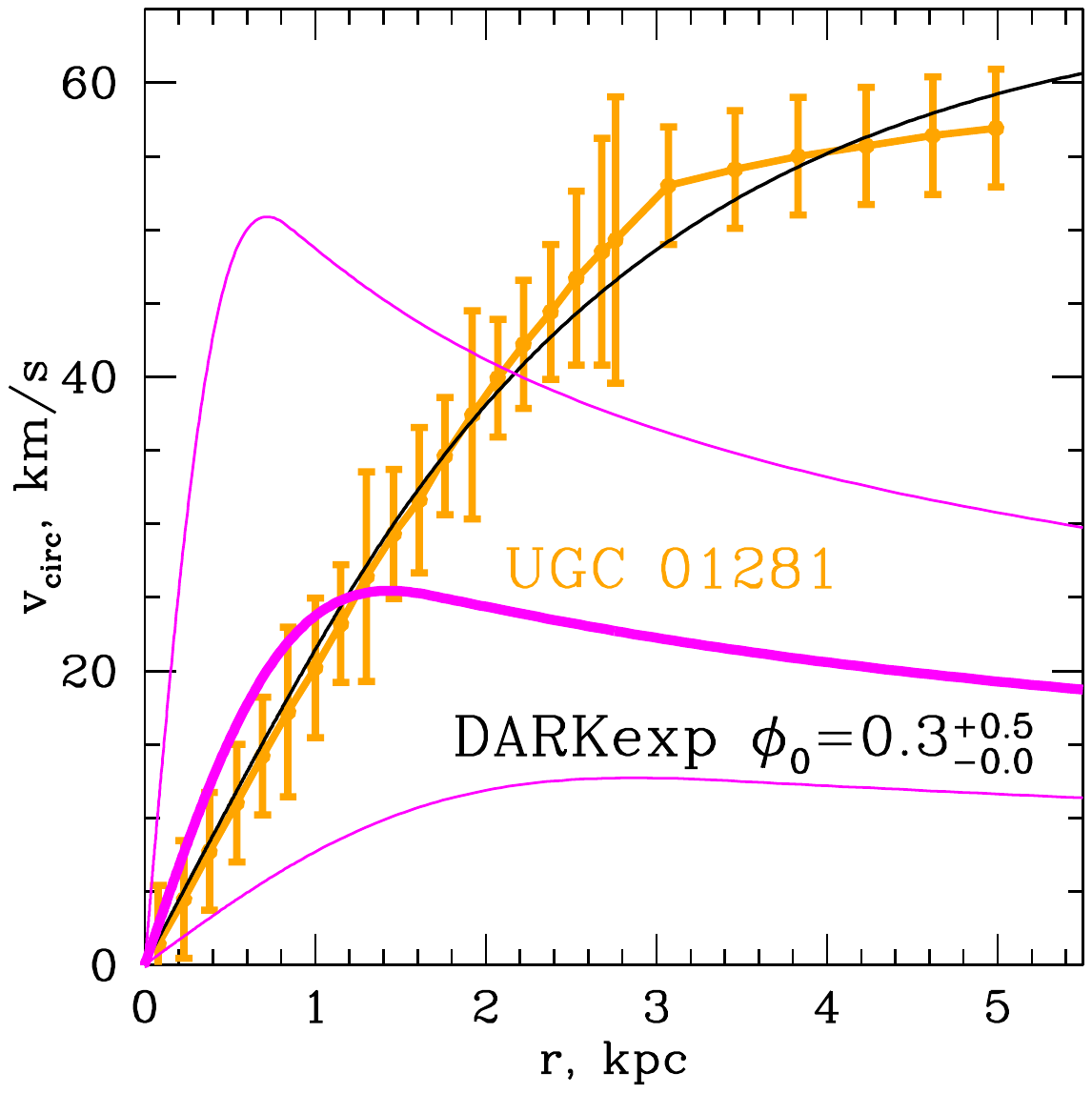}
    \includegraphics[trim={0cm 5cm 0cm 2cm},clip,width=0.315\linewidth]{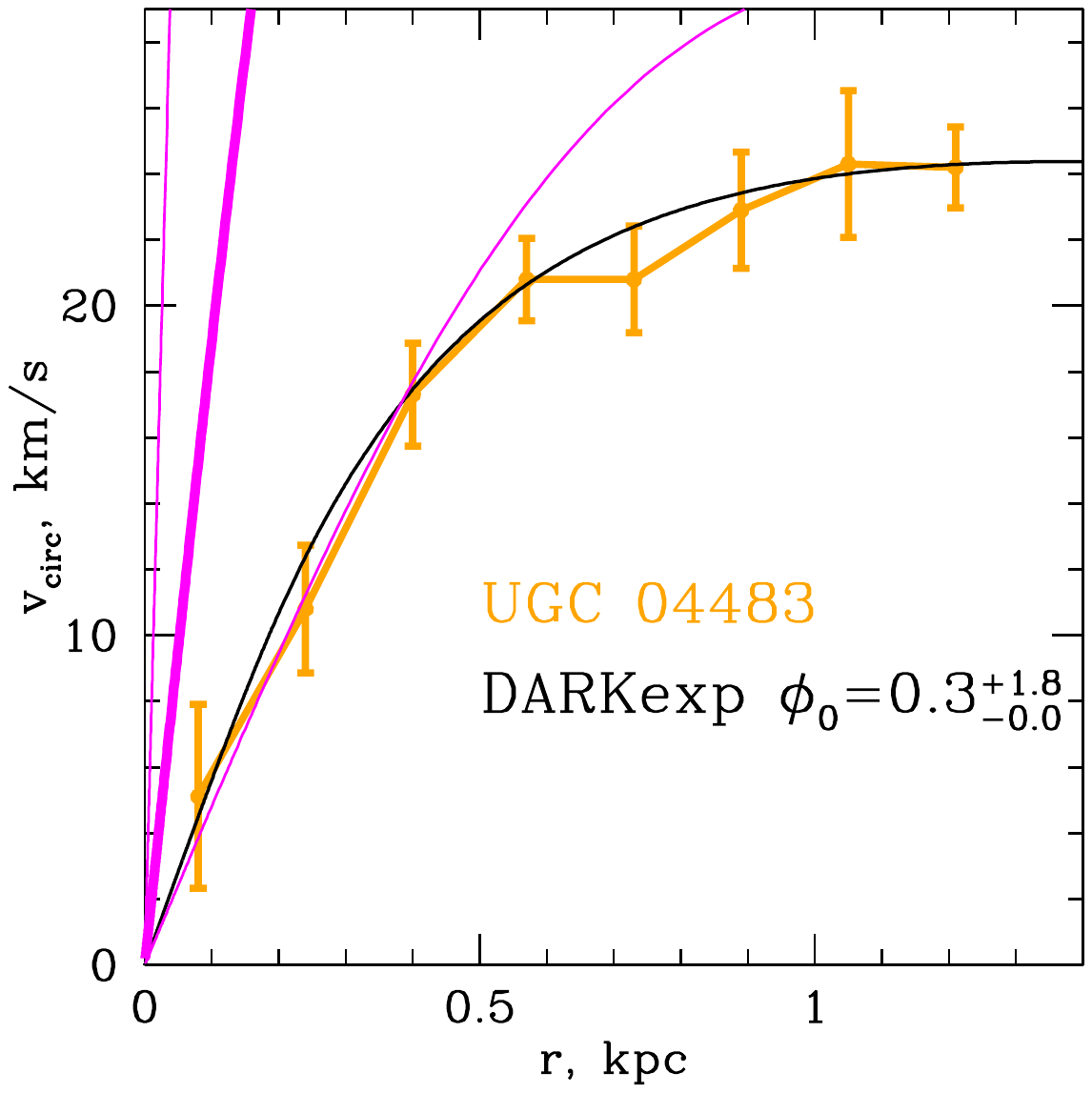}
    \caption{\newer{FDM rotation curve fits to 3 SPARC galaxies whose $r_c$ and $m_c$ are roughly comparable to those of the 5 dSph from \cite{Broadhurst2020}. The core radii we measured for these 3 galaxies are, $r_c=0.89, 0.74, 0.31$ kpc, respectively, and they are marked with green, cyan, and yellow crosses in Figure~\ref{fig:CmassCrad}. Dashed and dotted orange lines show the rotational velocity of the disk (i.e., stars) and gas components, as reported in the SPARC catalogue.  The thick magenta lines show FDM rotation curves, while the thin lines show what these would look like if $r_c$ were increased or decreased by a factor of 2. Black curves are the same DARKexp fits as shown in Figure~\ref{fig:rotcurve1} and \ref{fig:rotcurve2}.}}
    \label{fig:solitonRC}
\end{figure}

\newer{Next, we calculate the rotation curves predicted by FDM, for 3 SPARC galaxies whose $r_c$ and $m_c$ are similar to the 5 dSph galaxies. The 3 are marked with green, cyan and yellow crosses in Figure~\ref{fig:CmassCrad}. Given $r_c$ one can directly calculate the soliton density profile, using eq.(14--15) of \cite{Palencia2025}, assuming a boson mass of $m_\psi=10^{-22}$eV, a common choice in FDM analysis. At larger radii, the NFW profile describes the dark-matter halo, as shown in Fig.~1 of \cite{Broadhurst2020}.  Guided by that figure we assumed that the NFW profile starts at $2.25\,r_c$, and the NFW density slope for these low mass galaxies has a log-log slope of $-3$, though slopes of $-2$ have been reported in some simulations \citep{Chan2022}. (The inner portion of the NFW profile has been replaced by the soliton.) Figure~\ref{fig:solitonRC} presents the results of these calculations for the 3 SPARC galaxies. As already shown above, they are well fit by DARKexp (black lines).  But FDM fits are not satisfactory (thick magenta lines), even if the core radius is changed by a factor of 2 in both directions (thin magenta lines).  }

\newer{FDM halos have 3 free parameters: core radius $r_c$, radius where the soliton gives way to NFW, and the slope of the NFW beyond that radius. However, according to FDM, these parameters are related in the way described above. Considering an unphysical scenario where all 3 are allowed to vary, then many SPARC rotation curves can be fit with FDM, but not all by far.}

\newer{Figure~\ref{fig:solitonRC} gives us an interesting insight. It shows that within $\sim 1$ kpc from the center, FDM does fit rotation curves reasonably well. Disagreements with observations appear at larger radii. This hints at the reason why FDM fits some dwarf galaxies and not others. Galaxies that are well fit are dSph \citep{Broadhurst2020,Pozo2024}, whose dynamical mass is measured using velocity dispersion. Velocity dispersion measurements do not extend much beyond $1-2$ kpc. SPARC galaxies are disk systems, with rotation curves measured at a wide range of radii, allowing for a more extended determination of their mass distribution, and this is where it becomes apparent that FDM does not fit. (A recent paper did fit rotation curves with FDM, but they require two components: a scalar as well as a vector boson to model dark matter \citep{Mourelle2025}.) }

\newer{To sum up, we contrast the predictions by FDM and DARKexp: both density profiles predict central cores, therefore one might conclude that both FDM and DARKexp would fit low mass galaxies equally well.  However, quantitative fits of galaxies that have a wide radial range of dynamical data make it clear that DARKexp, and not FDM, can fit rotation curves of galaxies over a wide range of masses and rotation curve shapes.}

\section{Milky Way Galaxy}\label{sec:MW}

\begin{figure}
    \centering
    \includegraphics[trim={0cm 5cm 0cm 2cm},clip,width=0.67\linewidth]{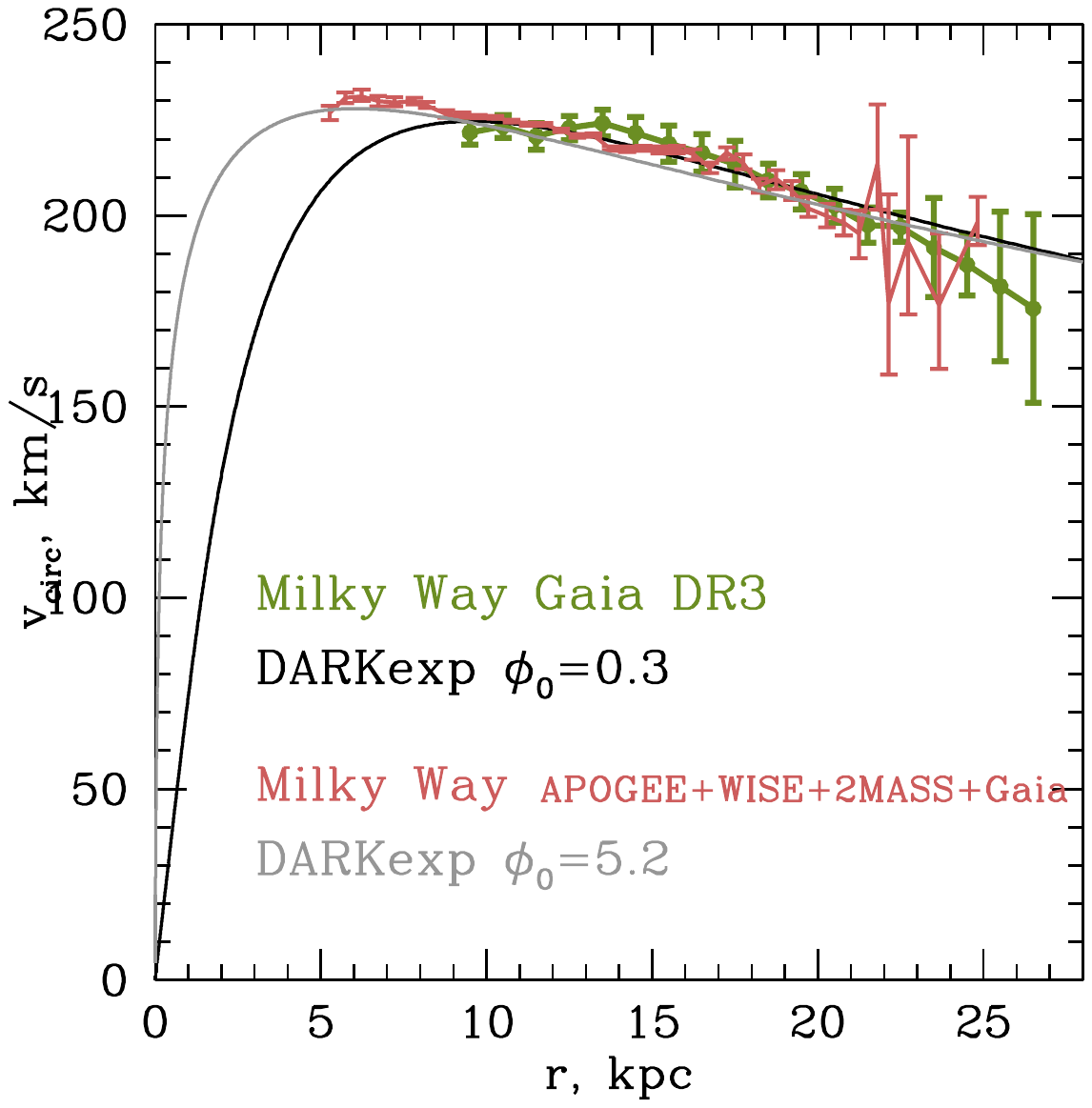}
    \caption{Milky Way rotation curve fit with the same cold collisionless DARKexp one-shape parameter model as galaxies in Figures~\ref{fig:rotcurve1}--\ref{fig:rotcurve4}, and \ref{fig:fits}. The radial range of rotation curves for the Milky Way are narrow, so there is no unique value of $\phi_0$ that fits best. The data from \cite{Eilers2019} (red) is best fit with $\phi_0=5.2$, while that from \cite{Jiao2023} (green) prefers $\phi_0=0.3$. The region interior to $\approx 5$ kpc is dominated by baryons and perturbations from the bar, so is not expected to follow equilibrium collisionless dynamics.}
    \label{fig:MW}
\end{figure}

In this Appendix we fit the rotation curve of the Milky Way Galaxy. The two rotation curves we use, from \cite{Eilers2019} and \cite{Jiao2023}, are based on the data from {\it Gaia}, APOGEE, WISE, and 2MASS, and are plotted in Figure~\ref{fig:MW}. The two estimates are very similar. The rotation curve peaks around $220$ km s$^{-1}$, but its $v_{\rm last}$ is $<200$ km s$^{-1}$, so in that regard it is comparable to several galaxies in Figure~\ref{fig:rotcurve4}. At radii $\gtrsim 10-14$ kpc the Milky Way is dark matter dominated.  

The data exclude the central region where mass is dominated by baryons, and velocities are perturbed by the action of the central bar. As a result, the radial range of these datasets, $r_{\rm outer}/r_{\rm inner}$, are 4.7 and 2.8, respectively, considerably smaller than a factor of 10, which was the criterion applied to SPARC galaxies. Because of the short radial span of the rotation curves, a range of $\phi_0$ values can be fit: $5.2$ and $0.3$. The fact that two somewhat different observational determinations of the same rotation curve are fitted by very different values of $\phi_0$ implies that it is imperative to have data with a long radial range. 

\bibliographystyle{unsrt}
\bibliography{paper_dwarfs3_rev2}{}


\end{document}